\RequirePackage[]{lineno}
\documentclass[floatfix,aps,prd,twocolumn,letterpaper,lengthcheck,superscriptaddress,showpacs,amssymb,amsmath,amsfonts,nofootinbib,altaffilletter,nopreprintnumbers,showpacs,longbibliography]{revtex4-1}
\pdfoutput=1
\usepackage[utf8]{inputenc}
\usepackage{xspace}
\usepackage{acronym}
\usepackage{booktabs}
\usepackage{amsmath,amssymb,graphicx}
\usepackage{color}
\usepackage[colorlinks, linkcolor=blue, pdfborder={0 0 0}, breaklinks=true]{hyperref}
\usepackage[table]{xcolor}
\definecolor{lightgray}{gray}{0.9} 
\usepackage{tabularx}
\usepackage{multirow}

\usepackage{makecell}

\newcolumntype{Y}{>{\centering\arraybackslash}X}
\newcolumntype{Z}{>{\raggedright\arraybackslash}X}
\newcolumntype{K}{>{\raggedleft\arraybackslash}X}
\newcolumntype{U}{>{\hsize=1.01\hsize}Y}
\newcolumntype{V}{>{\hsize=1.2\hsize}Y}
\newcolumntype{W}{>{\hsize=0.71\hsize}Y}

\def\mystrut{\vrule height 9.3pt depth 3.1pt width 0pt}
\newenvironment{PE_table}{\setlength{\tabcolsep}{0pt}}{}
\newenvironment{event_table}{\setlength{\tabcolsep}{0pt}\setlength{\aboverulesep}{1pt}\setlength{\belowrulesep}{1pt}}{}

\usepackage{scrextend}
\usepackage{textgreek}
\usepackage{graphicx}
\usepackage[caption=false]{subfig}

\usepackage{makerobust}

\renewcommand{\today}{\number\day\space\ifcase\month\or
  January\or February\or March\or April\or May\or June\or
  July\or August\or September\or October\or November\or December\fi
  \space\number\year}

\definecolor{NOTECOLOR}{rgb}{0.4, 0.2, 0.1}
\definecolor{notecolor}{rgb}{0.4, 0.2, 0.1}

\newcommand{\fixme}[1]{\textcolor{black}{#1}}

\begin{document}



\DeclareRobustCommand{\NUMEVENTSMAR}{2}

\DeclareRobustCommand{\TIMEMAR}[1]{\IfEqCase{#1}{{GW190426H_o3asub}{15:21:55}{GW190531B_o3asub}{02:36:48}}}

\DeclareRobustCommand{\DATEMAR}[1]{\IfEqCase{#1}{{GW190426H_o3asub}{2019-04-26}{GW190531B_o3asub}{2019-05-31}}}

\DeclareRobustCommand{\NAMEMAR}[1]{\IfEqCase{#1}{{GW190426H_o3asub}{GW190426\_152155}{GW190531B_o3asub}{GW190531\_023648}}}

\DeclareRobustCommand{\SIDMAR}[1]{\IfEqCase{#1}{{GW190426H_o3asub}{S190426a}}}

\DeclareRobustCommand{\PUBLICMAR}[1]{\IfEqCase{#1}{{GW190426H_o3asub}{True}{GW190531B_o3asub}{False}}}

\DeclareRobustCommand{\INSTRUMENTSMAR}[1]{\IfEqCase{#1}{{GW190426H_o3asub}{HLV}{GW190531B_o3asub}{HLV}}}

\DeclareRobustCommand{\PARTINSTRUMENTSMAR}[1]{\IfEqCase{#1}{{GW190426H_o3asub}{HL}{GW190531B_o3asub}{HL}}}

\DeclareRobustCommand{\FARTHRESHYRMAR}{2.0}

\DeclareRobustCommand{\NUMSNGLSMAR}{0}

\DeclareRobustCommand{\PREVIOUSLYREPORTEDMAR}{1}

\DeclareRobustCommand{\NEWEVENTSMAR}{1}

\DeclareRobustCommand{\EVENTNAMEBOLDMAR}[1]{\IfEqCase{#1}{{GW190426H_o3asub}{}{GW190531B_o3asub}{\bf}}}


\DeclareRobustCommand{\MBTAALLSKYPBBHMAR}[1]{\IfEqCase{#1}{{GW190426H_o3asub}{$0.00$}{GW190531B_o3asub}{$0.00$}}}

\DeclareRobustCommand{\MBTAALLSKYPBNSMAR}[1]{\IfEqCase{#1}{{GW190426H_o3asub}{$0.00$}{GW190531B_o3asub}{$0.05$}}}

\DeclareRobustCommand{\MBTAALLSKYPNSBHMAR}[1]{\IfEqCase{#1}{{GW190426H_o3asub}{$0.01$}{GW190531B_o3asub}{$0.03$}}}

\DeclareRobustCommand{\MBTAALLSKYPTERRESMAR}[1]{\IfEqCase{#1}{{GW190426H_o3asub}{$0.99$}{GW190531B_o3asub}{$0.92$}}}

\DeclareRobustCommand{\MBTAALLSKYPASTROMAR}[1]{\IfEqCase{#1}{{GW190426H_o3asub}{$0.01$}{GW190531B_o3asub}{$0.08$}}}

\DeclareRobustCommand{\MBTAALLSKYPCATMAXMAR}[1]{\IfEqCase{#1}{{GW190426H_o3asub}{$\pnsbh{}=0.01$ }{GW190531B_o3asub}{$\pbns{}=0.05$ }}}

\DeclareRobustCommand{\PYCBCHIGHMASSPBBHMAR}[1]{\IfEqCase{#1}{{GW190426H_o3asub}{--}{GW190531B_o3asub}{--}}}

\DeclareRobustCommand{\PYCBCHIGHMASSPBNSMAR}[1]{\IfEqCase{#1}{{GW190426H_o3asub}{--}{GW190531B_o3asub}{--}}}

\DeclareRobustCommand{\PYCBCHIGHMASSPNSBHMAR}[1]{\IfEqCase{#1}{{GW190426H_o3asub}{--}{GW190531B_o3asub}{--}}}

\DeclareRobustCommand{\PYCBCHIGHMASSPTERRESMAR}[1]{\IfEqCase{#1}{{GW190426H_o3asub}{--}{GW190531B_o3asub}{--}}}

\DeclareRobustCommand{\PYCBCHIGHMASSPASTROMAR}[1]{\IfEqCase{#1}{{GW190426H_o3asub}{--}{GW190531B_o3asub}{--}}}

\DeclareRobustCommand{\PYCBCHIGHMASSPCATMAXMAR}[1]{\IfEqCase{#1}{{GW190426H_o3asub}{--}{GW190531B_o3asub}{--}}}

\DeclareRobustCommand{\PYCBCALLSKYPBBHMAR}[1]{\IfEqCase{#1}{{GW190426H_o3asub}{$0.00$}{GW190531B_o3asub}{$0.00$}}}

\DeclareRobustCommand{\PYCBCALLSKYPBNSMAR}[1]{\IfEqCase{#1}{{GW190426H_o3asub}{$0.00$}{GW190531B_o3asub}{$0.01$}}}

\DeclareRobustCommand{\PYCBCALLSKYPNSBHMAR}[1]{\IfEqCase{#1}{{GW190426H_o3asub}{$0.01$}{GW190531B_o3asub}{$0.01$}}}

\DeclareRobustCommand{\PYCBCALLSKYPTERRESMAR}[1]{\IfEqCase{#1}{{GW190426H_o3asub}{$0.99$}{GW190531B_o3asub}{$0.98$}}}

\DeclareRobustCommand{\PYCBCALLSKYPASTROMAR}[1]{\IfEqCase{#1}{{GW190426H_o3asub}{$0.01$}{GW190531B_o3asub}{$0.02$}}}

\DeclareRobustCommand{\PYCBCALLSKYPCATMAXMAR}[1]{\IfEqCase{#1}{{GW190426H_o3asub}{$\pnsbh{}=0.01$ }{GW190531B_o3asub}{$\pnsbh{}=0.01$ }}}

\DeclareRobustCommand{\GSTLALALLSKYPBBHMAR}[1]{\IfEqCase{#1}{{GW190426H_o3asub}{$0.00$}{GW190531B_o3asub}{$0.00$}}}

\DeclareRobustCommand{\GSTLALALLSKYPBNSMAR}[1]{\IfEqCase{#1}{{GW190426H_o3asub}{$0.00$}{GW190531B_o3asub}{$0.00$}}}

\DeclareRobustCommand{\GSTLALALLSKYPNSBHMAR}[1]{\IfEqCase{#1}{{GW190426H_o3asub}{$0.14$}{GW190531B_o3asub}{$0.28$}}}

\DeclareRobustCommand{\GSTLALALLSKYPTERRESMAR}[1]{\IfEqCase{#1}{{GW190426H_o3asub}{$0.86$}{GW190531B_o3asub}{$0.72$}}}

\DeclareRobustCommand{\GSTLALALLSKYPASTROMAR}[1]{\IfEqCase{#1}{{GW190426H_o3asub}{$0.14$}{GW190531B_o3asub}{$0.28$}}}

\DeclareRobustCommand{\GSTLALALLSKYPCATMAXMAR}[1]{\IfEqCase{#1}{{GW190426H_o3asub}{$\pnsbh{}=0.14$ }{GW190531B_o3asub}{$\pnsbh{}=0.28$ }}}


\DeclareRobustCommand{\NUMGSTLALMAR}{2}

\DeclareRobustCommand{\NUMMBTAMAR}{0}

\DeclareRobustCommand{\NUMPYCBCMAR}{0}

\DeclareRobustCommand{\NUMPYCBCBBHMAR}{0}

\DeclareRobustCommand{\NUMPYCBCHYPERBANKMAR}{0}

\DeclareRobustCommand{\NUMUNIQUEGSTLALEVENTSMAR}{2}

\DeclareRobustCommand{\NUMUNIQUEMBTAEVENTSMAR}{0}

\DeclareRobustCommand{\NUMUNIQUEPYCBCEVENTSMAR}{0}

\DeclareRobustCommand{\TWOPIPESMAR}{0}

\DeclareRobustCommand{\THREEPIPESMAR}{0}

\DeclareRobustCommand{\ALLCBCPIPESMAR}{0}

\DeclareRobustCommand{\ALLPIPESMAR}{0}


\DeclareRobustCommand{\MINFARMAR}[1]{\IfEqCase{#1}{{GW190426H_o3asub}{\ensuremath{0.91} yr$^{-1}$}{GW190531B_o3asub}{\ensuremath{0.41} yr$^{-1}$}}}

\DeclareRobustCommand{\MBTAALLSKYFARMAR}[1]{\IfEqCase{#1}{{GW190426H_o3asub}{\ensuremath{32}}{GW190531B_o3asub}{\ensuremath{8.1}}}}

\DeclareRobustCommand{\MBTAALLSKYIFARMAR}[1]{\IfEqCase{#1}{{GW190426H_o3asub}{-1.5}{GW190531B_o3asub}{-0.9}}}

\DeclareRobustCommand{\MBTAALLSKYSNRMAR}[1]{\IfEqCase{#1}{{GW190426H_o3asub}{9.8}{GW190531B_o3asub}{9.8}}}

\DeclareRobustCommand{\PYCBCHIGHMASSFARMAR}[1]{\IfEqCase{#1}{{GW190426H_o3asub}{--}{GW190531B_o3asub}{--}}}

\DeclareRobustCommand{\PYCBCHIGHMASSIFARMAR}[1]{\IfEqCase{#1}{{GW190426H_o3asub}{--}{GW190531B_o3asub}{--}}}

\DeclareRobustCommand{\PYCBCHIGHMASSSNRMAR}[1]{\IfEqCase{#1}{{GW190426H_o3asub}{--}{GW190531B_o3asub}{--}}}

\DeclareRobustCommand{\PYCBCALLSKYFARMAR}[1]{\IfEqCase{#1}{{GW190426H_o3asub}{\ensuremath{43}}{GW190531B_o3asub}{\ensuremath{29}}}}

\DeclareRobustCommand{\PYCBCALLSKYIFARMAR}[1]{\IfEqCase{#1}{{GW190426H_o3asub}{-1.6}{GW190531B_o3asub}{-1.5}}}

\DeclareRobustCommand{\PYCBCALLSKYSNRMAR}[1]{\IfEqCase{#1}{{GW190426H_o3asub}{8.8}{GW190531B_o3asub}{9.2}}}

\DeclareRobustCommand{\GSTLALALLSKYFARMAR}[1]{\IfEqCase{#1}{{GW190426H_o3asub}{\ensuremath{0.91}}{GW190531B_o3asub}{\ensuremath{0.41}}}}

\DeclareRobustCommand{\GSTLALALLSKYIFARMAR}[1]{\IfEqCase{#1}{{GW190426H_o3asub}{0.0}{GW190531B_o3asub}{0.4}}}

\DeclareRobustCommand{\GSTLALALLSKYSNRMAR}[1]{\IfEqCase{#1}{{GW190426H_o3asub}{10.1}{GW190531B_o3asub}{10.0}}}



\DeclareRobustCommand{\NUMEVENTS}{44}

\DeclareRobustCommand{\TOTALEVENTS}{1201}

\DeclareRobustCommand{\NUMADDCANDIDATES}{8}

\DeclareRobustCommand{\NUMREMCANDIDATES}{3}

\DeclareRobustCommand{\TIME}[1]{\IfEqCase{#1}{{GW190403B_o3afin}{05:15:19}{GW190408H_o3afin}{18:18:02}{GW190412B_o3afin}{05:30:44}{GW190413A_o3afin}{05:29:54}{GW190413E_o3afin}{13:43:08}{GW190421I_o3afin}{21:38:56}{GW190425B_o3afin}{08:18:05}{GW190426N_o3afin}{19:06:42}{GW190503E_o3afin}{18:54:04}{GW190512G_o3afin}{18:07:14}{GW190513E_o3afin}{20:54:28}{GW190514E_o3afin}{06:54:16}{GW190517B_o3afin}{05:51:01}{GW190519J_o3afin}{15:35:44}{GW190521B_o3afin}{03:02:29}{GW190521E_o3afin}{07:43:59}{GW190527H_o3afin}{09:20:55}{GW190602E_o3afin}{17:59:27}{GW190620B_o3afin}{03:04:21}{GW190630E_o3afin}{18:52:05}{GW190701E_o3afin}{20:33:06}{GW190706F_o3afin}{22:26:41}{GW190707E_o3afin}{09:33:26}{GW190708M_o3afin}{23:24:57}{GW190719H_o3afin}{21:55:14}{GW190720A_o3afin}{00:08:36}{GW190725F_o3afin}{17:47:28}{GW190727B_o3afin}{06:03:33}{GW190728D_o3afin}{06:45:10}{GW190731E_o3afin}{14:09:36}{GW190803B_o3afin}{02:27:01}{GW190805J_o3afin}{21:11:37}{GW190814H_o3afin}{21:10:39}{GW190828A_o3afin}{06:34:05}{GW190828B_o3afin}{06:55:09}{GW190910B_o3afin}{11:28:07}{GW190915K_o3afin}{23:57:02}{GW190916K_o3afin}{20:06:58}{GW190917B_o3afin}{11:46:30}{GW190924A_o3afin}{02:18:46}{GW190925J_o3afin}{23:28:45}{GW190926C_o3afin}{05:03:36}{GW190929B_o3afin}{01:21:49}{GW190930C_o3afin}{13:35:41}}}

\DeclareRobustCommand{\DATE}[1]{\IfEqCase{#1}{{GW190403B_o3afin}{2019-04-03}{GW190408H_o3afin}{2019-04-08}{GW190412B_o3afin}{2019-04-12}{GW190413A_o3afin}{2019-04-13}{GW190413E_o3afin}{2019-04-13}{GW190421I_o3afin}{2019-04-21}{GW190425B_o3afin}{2019-04-25}{GW190426N_o3afin}{2019-04-26}{GW190503E_o3afin}{2019-05-03}{GW190512G_o3afin}{2019-05-12}{GW190513E_o3afin}{2019-05-13}{GW190514E_o3afin}{2019-05-14}{GW190517B_o3afin}{2019-05-17}{GW190519J_o3afin}{2019-05-19}{GW190521B_o3afin}{2019-05-21}{GW190521E_o3afin}{2019-05-21}{GW190527H_o3afin}{2019-05-27}{GW190602E_o3afin}{2019-06-02}{GW190620B_o3afin}{2019-06-20}{GW190630E_o3afin}{2019-06-30}{GW190701E_o3afin}{2019-07-01}{GW190706F_o3afin}{2019-07-06}{GW190707E_o3afin}{2019-07-07}{GW190708M_o3afin}{2019-07-08}{GW190719H_o3afin}{2019-07-19}{GW190720A_o3afin}{2019-07-20}{GW190725F_o3afin}{2019-07-25}{GW190727B_o3afin}{2019-07-27}{GW190728D_o3afin}{2019-07-28}{GW190731E_o3afin}{2019-07-31}{GW190803B_o3afin}{2019-08-03}{GW190805J_o3afin}{2019-08-05}{GW190814H_o3afin}{2019-08-14}{GW190828A_o3afin}{2019-08-28}{GW190828B_o3afin}{2019-08-28}{GW190910B_o3afin}{2019-09-10}{GW190915K_o3afin}{2019-09-15}{GW190916K_o3afin}{2019-09-16}{GW190917B_o3afin}{2019-09-17}{GW190924A_o3afin}{2019-09-24}{GW190925J_o3afin}{2019-09-25}{GW190926C_o3afin}{2019-09-26}{GW190929B_o3afin}{2019-09-29}{GW190930C_o3afin}{2019-09-30}}}

\DeclareRobustCommand{\NAME}[1]{\IfEqCase{#1}{{GW190403B_o3afin}{GW190403\_051519}{GW190408H_o3afin}{GW190408\_181802}{GW190412B_o3afin}{GW190412}{GW190413A_o3afin}{GW190413\_052954}{GW190413E_o3afin}{GW190413\_134308}{GW190421I_o3afin}{GW190421\_213856}{GW190425B_o3afin}{GW190425}{GW190426N_o3afin}{GW190426\_190642}{GW190503E_o3afin}{GW190503\_185404}{GW190512G_o3afin}{GW190512\_180714}{GW190513E_o3afin}{GW190513\_205428}{GW190514E_o3afin}{GW190514\_065416}{GW190517B_o3afin}{GW190517\_055101}{GW190519J_o3afin}{GW190519\_153544}{GW190521B_o3afin}{GW190521}{GW190521E_o3afin}{GW190521\_074359}{GW190527H_o3afin}{GW190527\_092055}{GW190602E_o3afin}{GW190602\_175927}{GW190620B_o3afin}{GW190620\_030421}{GW190630E_o3afin}{GW190630\_185205}{GW190701E_o3afin}{GW190701\_203306}{GW190706F_o3afin}{GW190706\_222641}{GW190707E_o3afin}{GW190707\_093326}{GW190708M_o3afin}{GW190708\_232457}{GW190719H_o3afin}{GW190719\_215514}{GW190720A_o3afin}{GW190720\_000836}{GW190725F_o3afin}{GW190725\_174728}{GW190727B_o3afin}{GW190727\_060333}{GW190728D_o3afin}{GW190728\_064510}{GW190731E_o3afin}{GW190731\_140936}{GW190803B_o3afin}{GW190803\_022701}{GW190805J_o3afin}{GW190805\_211137}{GW190814H_o3afin}{GW190814}{GW190828A_o3afin}{GW190828\_063405}{GW190828B_o3afin}{GW190828\_065509}{GW190910B_o3afin}{GW190910\_112807}{GW190915K_o3afin}{GW190915\_235702}{GW190916K_o3afin}{GW190916\_200658}{GW190917B_o3afin}{GW190917\_114630}{GW190924A_o3afin}{GW190924\_021846}{GW190925J_o3afin}{GW190925\_232845}{GW190926C_o3afin}{GW190926\_050336}{GW190929B_o3afin}{GW190929\_012149}{GW190930C_o3afin}{GW190930\_133541}}}

\DeclareRobustCommand{\FULLNAME}[1]{\IfEqCase{#1}{{GW190403B_o3afin}{GW190403\_051519}{GW190408H_o3afin}{GW190408\_181802}{GW190412B_o3afin}{GW190412\_053044}{GW190413A_o3afin}{GW190413\_052954}{GW190413E_o3afin}{GW190413\_134308}{GW190421I_o3afin}{GW190421\_213856}{GW190425B_o3afin}{GW190425\_081805}{GW190426N_o3afin}{GW190426\_190642}{GW190503E_o3afin}{GW190503\_185404}{GW190512G_o3afin}{GW190512\_180714}{GW190513E_o3afin}{GW190513\_205428}{GW190514E_o3afin}{GW190514\_065416}{GW190517B_o3afin}{GW190517\_055101}{GW190519J_o3afin}{GW190519\_153544}{GW190521B_o3afin}{GW190521\_030229}{GW190521E_o3afin}{GW190521\_074359}{GW190527H_o3afin}{GW190527\_092055}{GW190602E_o3afin}{GW190602\_175927}{GW190620B_o3afin}{GW190620\_030421}{GW190630E_o3afin}{GW190630\_185205}{GW190701E_o3afin}{GW190701\_203306}{GW190706F_o3afin}{GW190706\_222641}{GW190707E_o3afin}{GW190707\_093326}{GW190708M_o3afin}{GW190708\_232457}{GW190719H_o3afin}{GW190719\_215514}{GW190720A_o3afin}{GW190720\_000836}{GW190725F_o3afin}{GW190725\_174728}{GW190727B_o3afin}{GW190727\_060333}{GW190728D_o3afin}{GW190728\_064510}{GW190731E_o3afin}{GW190731\_140936}{GW190803B_o3afin}{GW190803\_022701}{GW190805J_o3afin}{GW190805\_211137}{GW190814H_o3afin}{GW190814\_211039}{GW190828A_o3afin}{GW190828\_063405}{GW190828B_o3afin}{GW190828\_065509}{GW190910B_o3afin}{GW190910\_112807}{GW190915K_o3afin}{GW190915\_235702}{GW190916K_o3afin}{GW190916\_200658}{GW190917B_o3afin}{GW190917\_114630}{GW190924A_o3afin}{GW190924\_021846}{GW190925J_o3afin}{GW190925\_232845}{GW190926C_o3afin}{GW190926\_050336}{GW190929B_o3afin}{GW190929\_012149}{GW190930C_o3afin}{GW190930\_133541}}}

\DeclareRobustCommand{\SID}[1]{\IfEqCase{#1}{{GW190408H_o3afin}{S190408a}{GW190412B_o3afin}{S190412a}{GW190413A_o3afin}{S190413a}{GW190413E_o3afin}{S190413b}{GW190421I_o3afin}{S190421a}{GW190425B_o3afin}{S190425a}{GW190503E_o3afin}{S190503a}{GW190512G_o3afin}{S190512a}{GW190513E_o3afin}{S190513a}{GW190514E_o3afin}{S190514a}{GW190517B_o3afin}{S190517a}{GW190519J_o3afin}{S190519a}{GW190521B_o3afin}{S190521a}{GW190521E_o3afin}{S190521b}{GW190527H_o3afin}{S190527a}{GW190602E_o3afin}{S190602a}{GW190620B_o3afin}{S190620a}{GW190630E_o3afin}{S190630a}{GW190701E_o3afin}{S190701a}{GW190706F_o3afin}{S190706a}{GW190707E_o3afin}{S190707a}{GW190708M_o3afin}{S190708a}{GW190719H_o3afin}{S190719a}{GW190720A_o3afin}{S190720a}{GW190727B_o3afin}{S190727a}{GW190728D_o3afin}{S190728a}{GW190731E_o3afin}{S190731a}{GW190803B_o3afin}{S190803a}{GW190814H_o3afin}{S190814a}{GW190828A_o3afin}{S190828a}{GW190828B_o3afin}{S190828b}{GW190910B_o3afin}{S190910a}{GW190915K_o3afin}{S190915a}{GW190924A_o3afin}{S190924a}{GW190929B_o3afin}{S190929a}{GW190930C_o3afin}{S190930a}}}

\DeclareRobustCommand{\PUBLIC}[1]{\IfEqCase{#1}{{GW190403B_o3afin}{False}{GW190408H_o3afin}{True}{GW190412B_o3afin}{True}{GW190413A_o3afin}{True}{GW190413E_o3afin}{True}{GW190421I_o3afin}{True}{GW190425B_o3afin}{True}{GW190426N_o3afin}{False}{GW190503E_o3afin}{True}{GW190512G_o3afin}{True}{GW190513E_o3afin}{True}{GW190514E_o3afin}{True}{GW190517B_o3afin}{True}{GW190519J_o3afin}{True}{GW190521B_o3afin}{True}{GW190521E_o3afin}{True}{GW190527H_o3afin}{True}{GW190602E_o3afin}{True}{GW190620B_o3afin}{True}{GW190630E_o3afin}{True}{GW190701E_o3afin}{True}{GW190706F_o3afin}{True}{GW190707E_o3afin}{True}{GW190708M_o3afin}{True}{GW190719H_o3afin}{True}{GW190720A_o3afin}{True}{GW190725F_o3afin}{False}{GW190727B_o3afin}{True}{GW190728D_o3afin}{True}{GW190731E_o3afin}{True}{GW190803B_o3afin}{True}{GW190805J_o3afin}{False}{GW190814H_o3afin}{True}{GW190828A_o3afin}{True}{GW190828B_o3afin}{True}{GW190910B_o3afin}{True}{GW190915K_o3afin}{True}{GW190916K_o3afin}{False}{GW190917B_o3afin}{False}{GW190924A_o3afin}{True}{GW190925J_o3afin}{False}{GW190926C_o3afin}{False}{GW190929B_o3afin}{True}{GW190930C_o3afin}{True}}}

\DeclareRobustCommand{\INSTRUMENTS}[1]{\IfEqCase{#1}{{GW190403B_o3afin}{HLV}{GW190408H_o3afin}{HLV}{GW190412B_o3afin}{HLV}{GW190413A_o3afin}{HLV}{GW190413E_o3afin}{HLV}{GW190421I_o3afin}{HL}{GW190425B_o3afin}{LV}{GW190426N_o3afin}{HLV}{GW190503E_o3afin}{HLV}{GW190512G_o3afin}{HLV}{GW190513E_o3afin}{HLV}{GW190514E_o3afin}{HL}{GW190517B_o3afin}{HLV}{GW190519J_o3afin}{HLV}{GW190521B_o3afin}{HLV}{GW190521E_o3afin}{HL}{GW190527H_o3afin}{HL}{GW190602E_o3afin}{HLV}{GW190620B_o3afin}{LV}{GW190630E_o3afin}{LV}{GW190701E_o3afin}{HLV}{GW190706F_o3afin}{HLV}{GW190707E_o3afin}{HL}{GW190708M_o3afin}{LV}{GW190719H_o3afin}{HL}{GW190720A_o3afin}{HLV}{GW190725F_o3afin}{HLV}{GW190727B_o3afin}{HLV}{GW190728D_o3afin}{HLV}{GW190731E_o3afin}{HL}{GW190803B_o3afin}{HLV}{GW190805J_o3afin}{HLV}{GW190814H_o3afin}{LV}{GW190828A_o3afin}{HLV}{GW190828B_o3afin}{HLV}{GW190910B_o3afin}{LV}{GW190915K_o3afin}{HLV}{GW190916K_o3afin}{HLV}{GW190917B_o3afin}{HLV}{GW190924A_o3afin}{HLV}{GW190925J_o3afin}{HV}{GW190926C_o3afin}{HLV}{GW190929B_o3afin}{HLV}{GW190930C_o3afin}{HL}}}

\DeclareRobustCommand{\PARTINSTRUMENTS}[1]{\IfEqCase{#1}{{GW190403B_o3afin}{HL}{GW190408H_o3afin}{HL}{GW190412B_o3afin}{HL}{GW190413A_o3afin}{HL}{GW190413E_o3afin}{HL}{GW190421I_o3afin}{HL}{GW190425B_o3afin}{L}{GW190426N_o3afin}{HL}{GW190503E_o3afin}{HL}{GW190512G_o3afin}{HL}{GW190513E_o3afin}{HLV}{GW190514E_o3afin}{HL}{GW190517B_o3afin}{HL}{GW190519J_o3afin}{HL}{GW190521B_o3afin}{HL}{GW190521E_o3afin}{HL}{GW190527H_o3afin}{HL}{GW190602E_o3afin}{HL}{GW190620B_o3afin}{L}{GW190630E_o3afin}{LV}{GW190701E_o3afin}{HLV}{GW190706F_o3afin}{HL}{GW190707E_o3afin}{HL}{GW190708M_o3afin}{L}{GW190719H_o3afin}{HL}{GW190720A_o3afin}{HLV}{GW190725F_o3afin}{HL}{GW190727B_o3afin}{HL}{GW190728D_o3afin}{HL}{GW190731E_o3afin}{HL}{GW190803B_o3afin}{HL}{GW190805J_o3afin}{HL}{GW190814H_o3afin}{LV}{GW190828A_o3afin}{HL}{GW190828B_o3afin}{HL}{GW190910B_o3afin}{L}{GW190915K_o3afin}{HL}{GW190916K_o3afin}{HL}{GW190917B_o3afin}{HL}{GW190924A_o3afin}{HL}{GW190925J_o3afin}{HV}{GW190926C_o3afin}{HL}{GW190929B_o3afin}{HL}{GW190930C_o3afin}{HL}}}

\DeclareRobustCommand{\FARTHRESHYR}{2.0}

\DeclareRobustCommand{\NUMSNGLS}{4}

\DeclareRobustCommand{\PREVIOUSLYREPORTED}{36}

\DeclareRobustCommand{\NEWEVENTS}{8}

\DeclareRobustCommand{\EVENTNAMEBOLD}[1]{\IfEqCase{#1}{{GW190403B_o3afin}{\bf}{GW190408H_o3afin}{}{GW190412B_o3afin}{}{GW190413A_o3afin}{}{GW190413E_o3afin}{}{GW190421I_o3afin}{}{GW190425B_o3afin}{}{GW190426N_o3afin}{\bf}{GW190503E_o3afin}{}{GW190512G_o3afin}{}{GW190513E_o3afin}{}{GW190514E_o3afin}{}{GW190517B_o3afin}{}{GW190519J_o3afin}{}{GW190521B_o3afin}{}{GW190521E_o3afin}{}{GW190527H_o3afin}{}{GW190602E_o3afin}{}{GW190620B_o3afin}{}{GW190630E_o3afin}{}{GW190701E_o3afin}{}{GW190706F_o3afin}{}{GW190707E_o3afin}{}{GW190708M_o3afin}{}{GW190719H_o3afin}{}{GW190720A_o3afin}{}{GW190725F_o3afin}{\bf}{GW190727B_o3afin}{}{GW190728D_o3afin}{}{GW190731E_o3afin}{}{GW190803B_o3afin}{}{GW190805J_o3afin}{\bf}{GW190814H_o3afin}{}{GW190828A_o3afin}{}{GW190828B_o3afin}{}{GW190910B_o3afin}{}{GW190915K_o3afin}{}{GW190916K_o3afin}{\bf}{GW190917B_o3afin}{\bf}{GW190924A_o3afin}{}{GW190925J_o3afin}{\bf}{GW190926C_o3afin}{\bf}{GW190929B_o3afin}{}{GW190930C_o3afin}{}}}

\DeclareRobustCommand{\EVENTNAMEASTERIX}[1]{\IfEqCase{#1}{{GW190403B_o3afin}{}{GW190408H_o3afin}{}{GW190412B_o3afin}{}{GW190413A_o3afin}{}{GW190413E_o3afin}{}{GW190421I_o3afin}{}{GW190425B_o3afin}{}{GW190426N_o3afin}{}{GW190503E_o3afin}{}{GW190512G_o3afin}{}{GW190513E_o3afin}{}{GW190514E_o3afin}{}{GW190517B_o3afin}{}{GW190519J_o3afin}{}{GW190521B_o3afin}{}{GW190521E_o3afin}{}{GW190527H_o3afin}{}{GW190602E_o3afin}{}{GW190620B_o3afin}{}{GW190630E_o3afin}{}{GW190701E_o3afin}{}{GW190706F_o3afin}{}{GW190707E_o3afin}{}{GW190708M_o3afin}{}{GW190719H_o3afin}{}{GW190720A_o3afin}{}{GW190725F_o3afin}{*}{GW190727B_o3afin}{}{GW190728D_o3afin}{}{GW190731E_o3afin}{}{GW190803B_o3afin}{}{GW190805J_o3afin}{}{GW190814H_o3afin}{}{GW190828A_o3afin}{}{GW190828B_o3afin}{}{GW190910B_o3afin}{}{GW190915K_o3afin}{}{GW190916K_o3afin}{*}{GW190917B_o3afin}{}{GW190924A_o3afin}{}{GW190925J_o3afin}{*}{GW190926C_o3afin}{*}{GW190929B_o3afin}{}{GW190930C_o3afin}{}}}


\DeclareRobustCommand{\MBTAALLSKYPBBH}[1]{\IfEqCase{#1}{{GW190403B_o3afin}{--}{GW190408H_o3afin}{1.00}{GW190412B_o3afin}{1.00}{GW190413A_o3afin}{--}{GW190413E_o3afin}{0.99}{GW190421I_o3afin}{0.99}{GW190425B_o3afin}{--}{GW190426N_o3afin}{--}{GW190503E_o3afin}{1.00}{GW190512G_o3afin}{0.99}{GW190513E_o3afin}{0.99}{GW190514E_o3afin}{--}{GW190517B_o3afin}{1.00}{GW190519J_o3afin}{1.00}{GW190521B_o3afin}{0.96}{GW190521E_o3afin}{1.00}{GW190527H_o3afin}{--}{GW190602E_o3afin}{1.00}{GW190620B_o3afin}{--}{GW190630E_o3afin}{--}{GW190701E_o3afin}{0.86}{GW190706F_o3afin}{0.99}{GW190707E_o3afin}{1.00}{GW190708M_o3afin}{--}{GW190719H_o3afin}{--}{GW190720A_o3afin}{1.00}{GW190725F_o3afin}{0.59}{GW190727B_o3afin}{1.00}{GW190728D_o3afin}{1.00}{GW190731E_o3afin}{0.80}{GW190803B_o3afin}{0.96}{GW190805J_o3afin}{--}{GW190814H_o3afin}{0.93}{GW190828A_o3afin}{1.00}{GW190828B_o3afin}{0.96}{GW190910B_o3afin}{--}{GW190915K_o3afin}{1.00}{GW190916K_o3afin}{0.66}{GW190917B_o3afin}{--}{GW190924A_o3afin}{0.92}{GW190925J_o3afin}{0.35}{GW190926C_o3afin}{--}{GW190929B_o3afin}{0.64}{GW190930C_o3afin}{0.87}}}

\DeclareRobustCommand{\MBTAALLSKYPBNS}[1]{\IfEqCase{#1}{{GW190403B_o3afin}{--}{GW190408H_o3afin}{0.00}{GW190412B_o3afin}{0.00}{GW190413A_o3afin}{--}{GW190413E_o3afin}{0.00}{GW190421I_o3afin}{0.00}{GW190425B_o3afin}{--}{GW190426N_o3afin}{--}{GW190503E_o3afin}{0.00}{GW190512G_o3afin}{0.00}{GW190513E_o3afin}{0.00}{GW190514E_o3afin}{--}{GW190517B_o3afin}{0.00}{GW190519J_o3afin}{0.00}{GW190521B_o3afin}{0.00}{GW190521E_o3afin}{0.00}{GW190527H_o3afin}{--}{GW190602E_o3afin}{0.00}{GW190620B_o3afin}{--}{GW190630E_o3afin}{--}{GW190701E_o3afin}{0.00}{GW190706F_o3afin}{0.00}{GW190707E_o3afin}{0.00}{GW190708M_o3afin}{--}{GW190719H_o3afin}{--}{GW190720A_o3afin}{0.00}{GW190725F_o3afin}{0.00}{GW190727B_o3afin}{0.00}{GW190728D_o3afin}{0.00}{GW190731E_o3afin}{0.00}{GW190803B_o3afin}{0.00}{GW190805J_o3afin}{--}{GW190814H_o3afin}{0.00}{GW190828A_o3afin}{0.00}{GW190828B_o3afin}{0.00}{GW190910B_o3afin}{--}{GW190915K_o3afin}{0.00}{GW190916K_o3afin}{0.00}{GW190917B_o3afin}{--}{GW190924A_o3afin}{0.00}{GW190925J_o3afin}{0.00}{GW190926C_o3afin}{--}{GW190929B_o3afin}{0.00}{GW190930C_o3afin}{0.00}}}

\DeclareRobustCommand{\MBTAALLSKYPNSBH}[1]{\IfEqCase{#1}{{GW190403B_o3afin}{--}{GW190408H_o3afin}{0.00}{GW190412B_o3afin}{0.00}{GW190413A_o3afin}{--}{GW190413E_o3afin}{0.00}{GW190421I_o3afin}{0.00}{GW190425B_o3afin}{--}{GW190426N_o3afin}{--}{GW190503E_o3afin}{0.00}{GW190512G_o3afin}{0.00}{GW190513E_o3afin}{0.00}{GW190514E_o3afin}{--}{GW190517B_o3afin}{0.00}{GW190519J_o3afin}{0.00}{GW190521B_o3afin}{0.00}{GW190521E_o3afin}{0.00}{GW190527H_o3afin}{--}{GW190602E_o3afin}{0.00}{GW190620B_o3afin}{--}{GW190630E_o3afin}{--}{GW190701E_o3afin}{0.00}{GW190706F_o3afin}{0.00}{GW190707E_o3afin}{0.00}{GW190708M_o3afin}{--}{GW190719H_o3afin}{--}{GW190720A_o3afin}{0.00}{GW190725F_o3afin}{0.00}{GW190727B_o3afin}{0.00}{GW190728D_o3afin}{0.00}{GW190731E_o3afin}{0.00}{GW190803B_o3afin}{0.00}{GW190805J_o3afin}{--}{GW190814H_o3afin}{0.07}{GW190828A_o3afin}{0.00}{GW190828B_o3afin}{0.00}{GW190910B_o3afin}{--}{GW190915K_o3afin}{0.00}{GW190916K_o3afin}{0.00}{GW190917B_o3afin}{--}{GW190924A_o3afin}{0.07}{GW190925J_o3afin}{0.00}{GW190926C_o3afin}{--}{GW190929B_o3afin}{0.00}{GW190930C_o3afin}{0.00}}}

\DeclareRobustCommand{\MBTAALLSKYPMASSGAP}[1]{\IfEqCase{#1}{{GW190403B_o3afin}{--}{GW190408H_o3afin}{--}{GW190412B_o3afin}{--}{GW190413A_o3afin}{--}{GW190413E_o3afin}{--}{GW190421I_o3afin}{--}{GW190425B_o3afin}{--}{GW190426N_o3afin}{--}{GW190503E_o3afin}{--}{GW190512G_o3afin}{--}{GW190513E_o3afin}{--}{GW190514E_o3afin}{--}{GW190517B_o3afin}{--}{GW190519J_o3afin}{--}{GW190521B_o3afin}{--}{GW190521E_o3afin}{--}{GW190527H_o3afin}{--}{GW190602E_o3afin}{--}{GW190620B_o3afin}{--}{GW190630E_o3afin}{--}{GW190701E_o3afin}{--}{GW190706F_o3afin}{--}{GW190707E_o3afin}{--}{GW190708M_o3afin}{--}{GW190719H_o3afin}{--}{GW190720A_o3afin}{--}{GW190725F_o3afin}{--}{GW190727B_o3afin}{--}{GW190728D_o3afin}{--}{GW190731E_o3afin}{--}{GW190803B_o3afin}{--}{GW190805J_o3afin}{--}{GW190814H_o3afin}{--}{GW190828A_o3afin}{--}{GW190828B_o3afin}{--}{GW190910B_o3afin}{--}{GW190915K_o3afin}{--}{GW190916K_o3afin}{--}{GW190917B_o3afin}{--}{GW190924A_o3afin}{--}{GW190925J_o3afin}{--}{GW190926C_o3afin}{--}{GW190929B_o3afin}{--}{GW190930C_o3afin}{--}}}

\DeclareRobustCommand{\MBTAALLSKYPTERRES}[1]{\IfEqCase{#1}{{GW190403B_o3afin}{--}{GW190408H_o3afin}{0.00}{GW190412B_o3afin}{0.00}{GW190413A_o3afin}{--}{GW190413E_o3afin}{0.01}{GW190421I_o3afin}{0.01}{GW190425B_o3afin}{--}{GW190426N_o3afin}{--}{GW190503E_o3afin}{0.00}{GW190512G_o3afin}{0.01}{GW190513E_o3afin}{0.01}{GW190514E_o3afin}{--}{GW190517B_o3afin}{0.00}{GW190519J_o3afin}{0.00}{GW190521B_o3afin}{0.04}{GW190521E_o3afin}{0.00}{GW190527H_o3afin}{--}{GW190602E_o3afin}{0.00}{GW190620B_o3afin}{--}{GW190630E_o3afin}{--}{GW190701E_o3afin}{0.13}{GW190706F_o3afin}{0.00}{GW190707E_o3afin}{0.00}{GW190708M_o3afin}{--}{GW190719H_o3afin}{--}{GW190720A_o3afin}{0.00}{GW190725F_o3afin}{0.41}{GW190727B_o3afin}{0.00}{GW190728D_o3afin}{0.00}{GW190731E_o3afin}{0.20}{GW190803B_o3afin}{0.04}{GW190805J_o3afin}{--}{GW190814H_o3afin}{0.00}{GW190828A_o3afin}{0.00}{GW190828B_o3afin}{0.04}{GW190910B_o3afin}{--}{GW190915K_o3afin}{0.00}{GW190916K_o3afin}{0.34}{GW190917B_o3afin}{--}{GW190924A_o3afin}{0.01}{GW190925J_o3afin}{0.65}{GW190926C_o3afin}{--}{GW190929B_o3afin}{0.36}{GW190930C_o3afin}{0.13}}}

\DeclareRobustCommand{\MBTAALLSKYPASTRO}[1]{\IfEqCase{#1}{{GW190403B_o3afin}{--}{GW190408H_o3afin}{1.00}{GW190412B_o3afin}{1.00}{GW190413A_o3afin}{--}{GW190413E_o3afin}{0.99}{GW190421I_o3afin}{0.99}{GW190425B_o3afin}{--}{GW190426N_o3afin}{--}{GW190503E_o3afin}{1.00}{GW190512G_o3afin}{0.99}{GW190513E_o3afin}{0.99}{GW190514E_o3afin}{--}{GW190517B_o3afin}{1.00}{GW190519J_o3afin}{1.00}{GW190521B_o3afin}{0.96}{GW190521E_o3afin}{1.00}{GW190527H_o3afin}{--}{GW190602E_o3afin}{1.00}{GW190620B_o3afin}{--}{GW190630E_o3afin}{--}{GW190701E_o3afin}{0.87}{GW190706F_o3afin}{1.00}{GW190707E_o3afin}{1.00}{GW190708M_o3afin}{--}{GW190719H_o3afin}{--}{GW190720A_o3afin}{1.00}{GW190725F_o3afin}{0.59}{GW190727B_o3afin}{1.00}{GW190728D_o3afin}{1.00}{GW190731E_o3afin}{0.80}{GW190803B_o3afin}{0.96}{GW190805J_o3afin}{--}{GW190814H_o3afin}{1.00}{GW190828A_o3afin}{1.00}{GW190828B_o3afin}{0.96}{GW190910B_o3afin}{--}{GW190915K_o3afin}{1.00}{GW190916K_o3afin}{0.66}{GW190917B_o3afin}{--}{GW190924A_o3afin}{0.99}{GW190925J_o3afin}{0.35}{GW190926C_o3afin}{--}{GW190929B_o3afin}{0.64}{GW190930C_o3afin}{0.87}}}

\DeclareRobustCommand{\MBTAALLSKYMEETSPASTROTHRESH}[1]{\IfEqCase{#1}{{GW190403B_o3afin}{}{GW190408H_o3afin}{}{GW190412B_o3afin}{}{GW190413A_o3afin}{}{GW190413E_o3afin}{}{GW190421I_o3afin}{}{GW190425B_o3afin}{}{GW190426N_o3afin}{}{GW190503E_o3afin}{}{GW190512G_o3afin}{}{GW190513E_o3afin}{}{GW190514E_o3afin}{}{GW190517B_o3afin}{}{GW190519J_o3afin}{}{GW190521B_o3afin}{}{GW190521E_o3afin}{}{GW190527H_o3afin}{}{GW190602E_o3afin}{}{GW190620B_o3afin}{}{GW190630E_o3afin}{}{GW190701E_o3afin}{}{GW190706F_o3afin}{}{GW190707E_o3afin}{}{GW190708M_o3afin}{}{GW190719H_o3afin}{}{GW190720A_o3afin}{}{GW190725F_o3afin}{}{GW190727B_o3afin}{}{GW190728D_o3afin}{}{GW190731E_o3afin}{}{GW190803B_o3afin}{}{GW190805J_o3afin}{}{GW190814H_o3afin}{}{GW190828A_o3afin}{}{GW190828B_o3afin}{}{GW190910B_o3afin}{}{GW190915K_o3afin}{}{GW190916K_o3afin}{}{GW190917B_o3afin}{}{GW190924A_o3afin}{}{GW190925J_o3afin}{\it }{GW190926C_o3afin}{}{GW190929B_o3afin}{}{GW190930C_o3afin}{}}}

\DeclareRobustCommand{\PYCBCHIGHMASSPBBH}[1]{\IfEqCase{#1}{{GW190403B_o3afin}{0.61}{GW190408H_o3afin}{1.00}{GW190412B_o3afin}{1.00}{GW190413A_o3afin}{0.93}{GW190413E_o3afin}{0.99}{GW190421I_o3afin}{1.00}{GW190425B_o3afin}{--}{GW190426N_o3afin}{0.75}{GW190503E_o3afin}{1.00}{GW190512G_o3afin}{1.00}{GW190513E_o3afin}{1.00}{GW190514E_o3afin}{0.76}{GW190517B_o3afin}{1.00}{GW190519J_o3afin}{1.00}{GW190521B_o3afin}{1.00}{GW190521E_o3afin}{1.00}{GW190527H_o3afin}{0.33}{GW190602E_o3afin}{1.00}{GW190620B_o3afin}{--}{GW190630E_o3afin}{1.00}{GW190701E_o3afin}{1.00}{GW190706F_o3afin}{1.00}{GW190707E_o3afin}{0.93}{GW190708M_o3afin}{--}{GW190719H_o3afin}{0.92}{GW190720A_o3afin}{1.00}{GW190725F_o3afin}{0.58}{GW190727B_o3afin}{1.00}{GW190728D_o3afin}{0.97}{GW190731E_o3afin}{0.83}{GW190803B_o3afin}{0.97}{GW190805J_o3afin}{0.95}{GW190814H_o3afin}{--}{GW190828A_o3afin}{1.00}{GW190828B_o3afin}{1.00}{GW190910B_o3afin}{--}{GW190915K_o3afin}{1.00}{GW190916K_o3afin}{0.64}{GW190917B_o3afin}{--}{GW190924A_o3afin}{0.44}{GW190925J_o3afin}{0.99}{GW190926C_o3afin}{0.09}{GW190929B_o3afin}{0.41}{GW190930C_o3afin}{0.85}}}

\DeclareRobustCommand{\PYCBCHIGHMASSPBNS}[1]{\IfEqCase{#1}{{GW190403B_o3afin}{0.00}{GW190408H_o3afin}{0.00}{GW190412B_o3afin}{0.00}{GW190413A_o3afin}{0.00}{GW190413E_o3afin}{0.00}{GW190421I_o3afin}{0.00}{GW190425B_o3afin}{--}{GW190426N_o3afin}{0.00}{GW190503E_o3afin}{0.00}{GW190512G_o3afin}{0.00}{GW190513E_o3afin}{0.00}{GW190514E_o3afin}{0.00}{GW190517B_o3afin}{0.00}{GW190519J_o3afin}{0.00}{GW190521B_o3afin}{0.00}{GW190521E_o3afin}{0.00}{GW190527H_o3afin}{0.00}{GW190602E_o3afin}{0.00}{GW190620B_o3afin}{--}{GW190630E_o3afin}{0.00}{GW190701E_o3afin}{0.00}{GW190706F_o3afin}{0.00}{GW190707E_o3afin}{0.00}{GW190708M_o3afin}{--}{GW190719H_o3afin}{0.00}{GW190720A_o3afin}{0.00}{GW190725F_o3afin}{0.00}{GW190727B_o3afin}{0.00}{GW190728D_o3afin}{0.00}{GW190731E_o3afin}{0.00}{GW190803B_o3afin}{0.00}{GW190805J_o3afin}{0.00}{GW190814H_o3afin}{--}{GW190828A_o3afin}{0.00}{GW190828B_o3afin}{0.00}{GW190910B_o3afin}{--}{GW190915K_o3afin}{0.00}{GW190916K_o3afin}{0.00}{GW190917B_o3afin}{--}{GW190924A_o3afin}{0.00}{GW190925J_o3afin}{0.00}{GW190926C_o3afin}{0.00}{GW190929B_o3afin}{0.00}{GW190930C_o3afin}{0.00}}}

\DeclareRobustCommand{\PYCBCHIGHMASSPNSBH}[1]{\IfEqCase{#1}{{GW190403B_o3afin}{0.00}{GW190408H_o3afin}{0.00}{GW190412B_o3afin}{0.00}{GW190413A_o3afin}{0.00}{GW190413E_o3afin}{0.00}{GW190421I_o3afin}{0.00}{GW190425B_o3afin}{--}{GW190426N_o3afin}{0.00}{GW190503E_o3afin}{0.00}{GW190512G_o3afin}{0.00}{GW190513E_o3afin}{0.00}{GW190514E_o3afin}{0.00}{GW190517B_o3afin}{0.00}{GW190519J_o3afin}{0.00}{GW190521B_o3afin}{0.00}{GW190521E_o3afin}{0.00}{GW190527H_o3afin}{0.00}{GW190602E_o3afin}{0.00}{GW190620B_o3afin}{--}{GW190630E_o3afin}{0.00}{GW190701E_o3afin}{0.00}{GW190706F_o3afin}{0.00}{GW190707E_o3afin}{0.07}{GW190708M_o3afin}{--}{GW190719H_o3afin}{0.00}{GW190720A_o3afin}{0.00}{GW190725F_o3afin}{0.24}{GW190727B_o3afin}{0.00}{GW190728D_o3afin}{0.03}{GW190731E_o3afin}{0.00}{GW190803B_o3afin}{0.00}{GW190805J_o3afin}{0.00}{GW190814H_o3afin}{--}{GW190828A_o3afin}{0.00}{GW190828B_o3afin}{0.00}{GW190910B_o3afin}{--}{GW190915K_o3afin}{0.00}{GW190916K_o3afin}{0.00}{GW190917B_o3afin}{--}{GW190924A_o3afin}{0.56}{GW190925J_o3afin}{0.00}{GW190926C_o3afin}{0.00}{GW190929B_o3afin}{0.00}{GW190930C_o3afin}{0.15}}}

\DeclareRobustCommand{\PYCBCHIGHMASSPMASSGAP}[1]{\IfEqCase{#1}{{GW190403B_o3afin}{--}{GW190408H_o3afin}{--}{GW190412B_o3afin}{--}{GW190413A_o3afin}{--}{GW190413E_o3afin}{--}{GW190421I_o3afin}{--}{GW190425B_o3afin}{--}{GW190426N_o3afin}{--}{GW190503E_o3afin}{--}{GW190512G_o3afin}{--}{GW190513E_o3afin}{--}{GW190514E_o3afin}{--}{GW190517B_o3afin}{--}{GW190519J_o3afin}{--}{GW190521B_o3afin}{--}{GW190521E_o3afin}{--}{GW190527H_o3afin}{--}{GW190602E_o3afin}{--}{GW190620B_o3afin}{--}{GW190630E_o3afin}{--}{GW190701E_o3afin}{--}{GW190706F_o3afin}{--}{GW190707E_o3afin}{--}{GW190708M_o3afin}{--}{GW190719H_o3afin}{--}{GW190720A_o3afin}{--}{GW190725F_o3afin}{--}{GW190727B_o3afin}{--}{GW190728D_o3afin}{--}{GW190731E_o3afin}{--}{GW190803B_o3afin}{--}{GW190805J_o3afin}{--}{GW190814H_o3afin}{--}{GW190828A_o3afin}{--}{GW190828B_o3afin}{--}{GW190910B_o3afin}{--}{GW190915K_o3afin}{--}{GW190916K_o3afin}{--}{GW190917B_o3afin}{--}{GW190924A_o3afin}{--}{GW190925J_o3afin}{--}{GW190926C_o3afin}{--}{GW190929B_o3afin}{--}{GW190930C_o3afin}{--}}}

\DeclareRobustCommand{\PYCBCHIGHMASSPTERRES}[1]{\IfEqCase{#1}{{GW190403B_o3afin}{0.39}{GW190408H_o3afin}{0.00}{GW190412B_o3afin}{0.00}{GW190413A_o3afin}{0.07}{GW190413E_o3afin}{0.01}{GW190421I_o3afin}{0.00}{GW190425B_o3afin}{--}{GW190426N_o3afin}{0.25}{GW190503E_o3afin}{0.00}{GW190512G_o3afin}{0.00}{GW190513E_o3afin}{0.00}{GW190514E_o3afin}{0.24}{GW190517B_o3afin}{0.00}{GW190519J_o3afin}{0.00}{GW190521B_o3afin}{0.00}{GW190521E_o3afin}{0.00}{GW190527H_o3afin}{0.67}{GW190602E_o3afin}{0.00}{GW190620B_o3afin}{--}{GW190630E_o3afin}{0.00}{GW190701E_o3afin}{0.00}{GW190706F_o3afin}{0.00}{GW190707E_o3afin}{0.00}{GW190708M_o3afin}{--}{GW190719H_o3afin}{0.08}{GW190720A_o3afin}{0.00}{GW190725F_o3afin}{0.18}{GW190727B_o3afin}{0.00}{GW190728D_o3afin}{0.00}{GW190731E_o3afin}{0.17}{GW190803B_o3afin}{0.03}{GW190805J_o3afin}{0.05}{GW190814H_o3afin}{--}{GW190828A_o3afin}{0.00}{GW190828B_o3afin}{0.00}{GW190910B_o3afin}{--}{GW190915K_o3afin}{0.00}{GW190916K_o3afin}{0.36}{GW190917B_o3afin}{--}{GW190924A_o3afin}{0.00}{GW190925J_o3afin}{0.01}{GW190926C_o3afin}{0.91}{GW190929B_o3afin}{0.59}{GW190930C_o3afin}{0.00}}}

\DeclareRobustCommand{\PYCBCHIGHMASSPASTRO}[1]{\IfEqCase{#1}{{GW190403B_o3afin}{0.61}{GW190408H_o3afin}{1.00}{GW190412B_o3afin}{1.00}{GW190413A_o3afin}{0.93}{GW190413E_o3afin}{0.99}{GW190421I_o3afin}{1.00}{GW190425B_o3afin}{--}{GW190426N_o3afin}{0.75}{GW190503E_o3afin}{1.00}{GW190512G_o3afin}{1.00}{GW190513E_o3afin}{1.00}{GW190514E_o3afin}{0.76}{GW190517B_o3afin}{1.00}{GW190519J_o3afin}{1.00}{GW190521B_o3afin}{1.00}{GW190521E_o3afin}{1.00}{GW190527H_o3afin}{0.33}{GW190602E_o3afin}{1.00}{GW190620B_o3afin}{--}{GW190630E_o3afin}{1.00}{GW190701E_o3afin}{1.00}{GW190706F_o3afin}{1.00}{GW190707E_o3afin}{1.00}{GW190708M_o3afin}{--}{GW190719H_o3afin}{0.92}{GW190720A_o3afin}{1.00}{GW190725F_o3afin}{0.82}{GW190727B_o3afin}{1.00}{GW190728D_o3afin}{1.00}{GW190731E_o3afin}{0.83}{GW190803B_o3afin}{0.97}{GW190805J_o3afin}{0.95}{GW190814H_o3afin}{--}{GW190828A_o3afin}{1.00}{GW190828B_o3afin}{1.00}{GW190910B_o3afin}{--}{GW190915K_o3afin}{1.00}{GW190916K_o3afin}{0.64}{GW190917B_o3afin}{--}{GW190924A_o3afin}{1.00}{GW190925J_o3afin}{0.99}{GW190926C_o3afin}{0.09}{GW190929B_o3afin}{0.41}{GW190930C_o3afin}{1.00}}}

\DeclareRobustCommand{\PYCBCHIGHMASSMEETSPASTROTHRESH}[1]{\IfEqCase{#1}{{GW190403B_o3afin}{}{GW190408H_o3afin}{}{GW190412B_o3afin}{}{GW190413A_o3afin}{}{GW190413E_o3afin}{}{GW190421I_o3afin}{}{GW190425B_o3afin}{}{GW190426N_o3afin}{}{GW190503E_o3afin}{}{GW190512G_o3afin}{}{GW190513E_o3afin}{}{GW190514E_o3afin}{}{GW190517B_o3afin}{}{GW190519J_o3afin}{}{GW190521B_o3afin}{}{GW190521E_o3afin}{}{GW190527H_o3afin}{\it }{GW190602E_o3afin}{}{GW190620B_o3afin}{}{GW190630E_o3afin}{}{GW190701E_o3afin}{}{GW190706F_o3afin}{}{GW190707E_o3afin}{}{GW190708M_o3afin}{}{GW190719H_o3afin}{}{GW190720A_o3afin}{}{GW190725F_o3afin}{}{GW190727B_o3afin}{}{GW190728D_o3afin}{}{GW190731E_o3afin}{}{GW190803B_o3afin}{}{GW190805J_o3afin}{}{GW190814H_o3afin}{}{GW190828A_o3afin}{}{GW190828B_o3afin}{}{GW190910B_o3afin}{}{GW190915K_o3afin}{}{GW190916K_o3afin}{}{GW190917B_o3afin}{}{GW190924A_o3afin}{}{GW190925J_o3afin}{}{GW190926C_o3afin}{\it }{GW190929B_o3afin}{\it }{GW190930C_o3afin}{}}}

\DeclareRobustCommand{\PYCBCALLSKYPBBH}[1]{\IfEqCase{#1}{{GW190403B_o3afin}{--}{GW190408H_o3afin}{1.00}{GW190412B_o3afin}{1.00}{GW190413A_o3afin}{0.13}{GW190413E_o3afin}{0.48}{GW190421I_o3afin}{0.75}{GW190425B_o3afin}{--}{GW190426N_o3afin}{--}{GW190503E_o3afin}{1.00}{GW190512G_o3afin}{1.00}{GW190513E_o3afin}{0.49}{GW190514E_o3afin}{--}{GW190517B_o3afin}{1.00}{GW190519J_o3afin}{1.00}{GW190521B_o3afin}{0.96}{GW190521E_o3afin}{1.00}{GW190527H_o3afin}{--}{GW190602E_o3afin}{0.98}{GW190620B_o3afin}{--}{GW190630E_o3afin}{--}{GW190701E_o3afin}{0.99}{GW190706F_o3afin}{1.00}{GW190707E_o3afin}{0.93}{GW190708M_o3afin}{--}{GW190719H_o3afin}{--}{GW190720A_o3afin}{0.95}{GW190725F_o3afin}{0.79}{GW190727B_o3afin}{1.00}{GW190728D_o3afin}{0.97}{GW190731E_o3afin}{--}{GW190803B_o3afin}{0.17}{GW190805J_o3afin}{--}{GW190814H_o3afin}{0.54}{GW190828A_o3afin}{1.00}{GW190828B_o3afin}{1.00}{GW190910B_o3afin}{--}{GW190915K_o3afin}{1.00}{GW190916K_o3afin}{--}{GW190917B_o3afin}{--}{GW190924A_o3afin}{0.44}{GW190925J_o3afin}{0.02}{GW190926C_o3afin}{--}{GW190929B_o3afin}{0.14}{GW190930C_o3afin}{0.93}}}

\DeclareRobustCommand{\PYCBCALLSKYPBNS}[1]{\IfEqCase{#1}{{GW190403B_o3afin}{--}{GW190408H_o3afin}{0.00}{GW190412B_o3afin}{0.00}{GW190413A_o3afin}{0.00}{GW190413E_o3afin}{0.00}{GW190421I_o3afin}{0.00}{GW190425B_o3afin}{--}{GW190426N_o3afin}{--}{GW190503E_o3afin}{0.00}{GW190512G_o3afin}{0.00}{GW190513E_o3afin}{0.00}{GW190514E_o3afin}{--}{GW190517B_o3afin}{0.00}{GW190519J_o3afin}{0.00}{GW190521B_o3afin}{0.00}{GW190521E_o3afin}{0.00}{GW190527H_o3afin}{--}{GW190602E_o3afin}{0.00}{GW190620B_o3afin}{--}{GW190630E_o3afin}{--}{GW190701E_o3afin}{0.00}{GW190706F_o3afin}{0.00}{GW190707E_o3afin}{0.00}{GW190708M_o3afin}{--}{GW190719H_o3afin}{--}{GW190720A_o3afin}{0.00}{GW190725F_o3afin}{0.00}{GW190727B_o3afin}{0.00}{GW190728D_o3afin}{0.00}{GW190731E_o3afin}{--}{GW190803B_o3afin}{0.00}{GW190805J_o3afin}{--}{GW190814H_o3afin}{0.00}{GW190828A_o3afin}{0.00}{GW190828B_o3afin}{0.00}{GW190910B_o3afin}{--}{GW190915K_o3afin}{0.00}{GW190916K_o3afin}{--}{GW190917B_o3afin}{--}{GW190924A_o3afin}{0.00}{GW190925J_o3afin}{0.00}{GW190926C_o3afin}{--}{GW190929B_o3afin}{0.00}{GW190930C_o3afin}{0.00}}}

\DeclareRobustCommand{\PYCBCALLSKYPNSBH}[1]{\IfEqCase{#1}{{GW190403B_o3afin}{--}{GW190408H_o3afin}{0.00}{GW190412B_o3afin}{0.00}{GW190413A_o3afin}{0.00}{GW190413E_o3afin}{0.00}{GW190421I_o3afin}{0.00}{GW190425B_o3afin}{--}{GW190426N_o3afin}{--}{GW190503E_o3afin}{0.00}{GW190512G_o3afin}{0.00}{GW190513E_o3afin}{0.00}{GW190514E_o3afin}{--}{GW190517B_o3afin}{0.00}{GW190519J_o3afin}{0.00}{GW190521B_o3afin}{0.00}{GW190521E_o3afin}{0.00}{GW190527H_o3afin}{--}{GW190602E_o3afin}{0.00}{GW190620B_o3afin}{--}{GW190630E_o3afin}{--}{GW190701E_o3afin}{0.00}{GW190706F_o3afin}{0.00}{GW190707E_o3afin}{0.07}{GW190708M_o3afin}{--}{GW190719H_o3afin}{--}{GW190720A_o3afin}{0.05}{GW190725F_o3afin}{0.17}{GW190727B_o3afin}{0.00}{GW190728D_o3afin}{0.03}{GW190731E_o3afin}{--}{GW190803B_o3afin}{0.00}{GW190805J_o3afin}{--}{GW190814H_o3afin}{0.46}{GW190828A_o3afin}{0.00}{GW190828B_o3afin}{0.00}{GW190910B_o3afin}{--}{GW190915K_o3afin}{0.00}{GW190916K_o3afin}{--}{GW190917B_o3afin}{--}{GW190924A_o3afin}{0.56}{GW190925J_o3afin}{0.00}{GW190926C_o3afin}{--}{GW190929B_o3afin}{0.00}{GW190930C_o3afin}{0.07}}}

\DeclareRobustCommand{\PYCBCALLSKYPMASSGAP}[1]{\IfEqCase{#1}{{GW190403B_o3afin}{--}{GW190408H_o3afin}{--}{GW190412B_o3afin}{--}{GW190413A_o3afin}{--}{GW190413E_o3afin}{--}{GW190421I_o3afin}{--}{GW190425B_o3afin}{--}{GW190426N_o3afin}{--}{GW190503E_o3afin}{--}{GW190512G_o3afin}{--}{GW190513E_o3afin}{--}{GW190514E_o3afin}{--}{GW190517B_o3afin}{--}{GW190519J_o3afin}{--}{GW190521B_o3afin}{--}{GW190521E_o3afin}{--}{GW190527H_o3afin}{--}{GW190602E_o3afin}{--}{GW190620B_o3afin}{--}{GW190630E_o3afin}{--}{GW190701E_o3afin}{--}{GW190706F_o3afin}{--}{GW190707E_o3afin}{--}{GW190708M_o3afin}{--}{GW190719H_o3afin}{--}{GW190720A_o3afin}{--}{GW190725F_o3afin}{--}{GW190727B_o3afin}{--}{GW190728D_o3afin}{--}{GW190731E_o3afin}{--}{GW190803B_o3afin}{--}{GW190805J_o3afin}{--}{GW190814H_o3afin}{--}{GW190828A_o3afin}{--}{GW190828B_o3afin}{--}{GW190910B_o3afin}{--}{GW190915K_o3afin}{--}{GW190916K_o3afin}{--}{GW190917B_o3afin}{--}{GW190924A_o3afin}{--}{GW190925J_o3afin}{--}{GW190926C_o3afin}{--}{GW190929B_o3afin}{--}{GW190930C_o3afin}{--}}}

\DeclareRobustCommand{\PYCBCALLSKYPTERRES}[1]{\IfEqCase{#1}{{GW190403B_o3afin}{--}{GW190408H_o3afin}{0.00}{GW190412B_o3afin}{0.00}{GW190413A_o3afin}{0.87}{GW190413E_o3afin}{0.52}{GW190421I_o3afin}{0.25}{GW190425B_o3afin}{--}{GW190426N_o3afin}{--}{GW190503E_o3afin}{0.00}{GW190512G_o3afin}{0.00}{GW190513E_o3afin}{0.51}{GW190514E_o3afin}{--}{GW190517B_o3afin}{0.00}{GW190519J_o3afin}{0.00}{GW190521B_o3afin}{0.04}{GW190521E_o3afin}{0.00}{GW190527H_o3afin}{--}{GW190602E_o3afin}{0.02}{GW190620B_o3afin}{--}{GW190630E_o3afin}{--}{GW190701E_o3afin}{0.01}{GW190706F_o3afin}{0.00}{GW190707E_o3afin}{0.00}{GW190708M_o3afin}{--}{GW190719H_o3afin}{--}{GW190720A_o3afin}{0.00}{GW190725F_o3afin}{0.04}{GW190727B_o3afin}{0.00}{GW190728D_o3afin}{0.00}{GW190731E_o3afin}{--}{GW190803B_o3afin}{0.83}{GW190805J_o3afin}{--}{GW190814H_o3afin}{0.00}{GW190828A_o3afin}{0.00}{GW190828B_o3afin}{0.00}{GW190910B_o3afin}{--}{GW190915K_o3afin}{0.00}{GW190916K_o3afin}{--}{GW190917B_o3afin}{--}{GW190924A_o3afin}{0.00}{GW190925J_o3afin}{0.98}{GW190926C_o3afin}{--}{GW190929B_o3afin}{0.86}{GW190930C_o3afin}{0.00}}}

\DeclareRobustCommand{\PYCBCALLSKYPASTRO}[1]{\IfEqCase{#1}{{GW190403B_o3afin}{--}{GW190408H_o3afin}{1.00}{GW190412B_o3afin}{1.00}{GW190413A_o3afin}{0.13}{GW190413E_o3afin}{0.48}{GW190421I_o3afin}{0.75}{GW190425B_o3afin}{--}{GW190426N_o3afin}{--}{GW190503E_o3afin}{1.00}{GW190512G_o3afin}{1.00}{GW190513E_o3afin}{0.49}{GW190514E_o3afin}{--}{GW190517B_o3afin}{1.00}{GW190519J_o3afin}{1.00}{GW190521B_o3afin}{0.96}{GW190521E_o3afin}{1.00}{GW190527H_o3afin}{--}{GW190602E_o3afin}{0.98}{GW190620B_o3afin}{--}{GW190630E_o3afin}{--}{GW190701E_o3afin}{0.99}{GW190706F_o3afin}{1.00}{GW190707E_o3afin}{1.00}{GW190708M_o3afin}{--}{GW190719H_o3afin}{--}{GW190720A_o3afin}{1.00}{GW190725F_o3afin}{0.96}{GW190727B_o3afin}{1.00}{GW190728D_o3afin}{1.00}{GW190731E_o3afin}{--}{GW190803B_o3afin}{0.17}{GW190805J_o3afin}{--}{GW190814H_o3afin}{1.00}{GW190828A_o3afin}{1.00}{GW190828B_o3afin}{1.00}{GW190910B_o3afin}{--}{GW190915K_o3afin}{1.00}{GW190916K_o3afin}{--}{GW190917B_o3afin}{--}{GW190924A_o3afin}{1.00}{GW190925J_o3afin}{0.02}{GW190926C_o3afin}{--}{GW190929B_o3afin}{0.14}{GW190930C_o3afin}{1.00}}}

\DeclareRobustCommand{\PYCBCALLSKYMEETSPASTROTHRESH}[1]{\IfEqCase{#1}{{GW190403B_o3afin}{}{GW190408H_o3afin}{}{GW190412B_o3afin}{}{GW190413A_o3afin}{\it }{GW190413E_o3afin}{\it }{GW190421I_o3afin}{}{GW190425B_o3afin}{}{GW190426N_o3afin}{}{GW190503E_o3afin}{}{GW190512G_o3afin}{}{GW190513E_o3afin}{\it }{GW190514E_o3afin}{}{GW190517B_o3afin}{}{GW190519J_o3afin}{}{GW190521B_o3afin}{}{GW190521E_o3afin}{}{GW190527H_o3afin}{}{GW190602E_o3afin}{}{GW190620B_o3afin}{}{GW190630E_o3afin}{}{GW190701E_o3afin}{}{GW190706F_o3afin}{}{GW190707E_o3afin}{}{GW190708M_o3afin}{}{GW190719H_o3afin}{}{GW190720A_o3afin}{}{GW190725F_o3afin}{}{GW190727B_o3afin}{}{GW190728D_o3afin}{}{GW190731E_o3afin}{}{GW190803B_o3afin}{\it }{GW190805J_o3afin}{}{GW190814H_o3afin}{}{GW190828A_o3afin}{}{GW190828B_o3afin}{}{GW190910B_o3afin}{}{GW190915K_o3afin}{}{GW190916K_o3afin}{}{GW190917B_o3afin}{}{GW190924A_o3afin}{}{GW190925J_o3afin}{\it }{GW190926C_o3afin}{}{GW190929B_o3afin}{\it }{GW190930C_o3afin}{}}}

\DeclareRobustCommand{\GSTLALALLSKYPBBH}[1]{\IfEqCase{#1}{{GW190403B_o3afin}{--}{GW190408H_o3afin}{1.00}{GW190412B_o3afin}{1.00}{GW190413A_o3afin}{--}{GW190413E_o3afin}{0.04}{GW190421I_o3afin}{1.00}{GW190425B_o3afin}{0.00}{GW190426N_o3afin}{--}{GW190503E_o3afin}{1.00}{GW190512G_o3afin}{1.00}{GW190513E_o3afin}{1.00}{GW190514E_o3afin}{0.00}{GW190517B_o3afin}{1.00}{GW190519J_o3afin}{1.00}{GW190521B_o3afin}{0.79}{GW190521E_o3afin}{1.00}{GW190527H_o3afin}{0.85}{GW190602E_o3afin}{1.00}{GW190620B_o3afin}{0.99}{GW190630E_o3afin}{1.00}{GW190701E_o3afin}{0.99}{GW190706F_o3afin}{1.00}{GW190707E_o3afin}{1.00}{GW190708M_o3afin}{1.00}{GW190719H_o3afin}{--}{GW190720A_o3afin}{1.00}{GW190725F_o3afin}{--}{GW190727B_o3afin}{1.00}{GW190728D_o3afin}{1.00}{GW190731E_o3afin}{0.78}{GW190803B_o3afin}{0.94}{GW190805J_o3afin}{--}{GW190814H_o3afin}{0.19}{GW190828A_o3afin}{1.00}{GW190828B_o3afin}{1.00}{GW190910B_o3afin}{1.00}{GW190915K_o3afin}{1.00}{GW190916K_o3afin}{0.09}{GW190917B_o3afin}{0.77}{GW190924A_o3afin}{1.00}{GW190925J_o3afin}{--}{GW190926C_o3afin}{0.54}{GW190929B_o3afin}{0.87}{GW190930C_o3afin}{0.76}}}

\DeclareRobustCommand{\GSTLALALLSKYPBNS}[1]{\IfEqCase{#1}{{GW190403B_o3afin}{--}{GW190408H_o3afin}{0.00}{GW190412B_o3afin}{0.00}{GW190413A_o3afin}{--}{GW190413E_o3afin}{0.00}{GW190421I_o3afin}{0.00}{GW190425B_o3afin}{0.78}{GW190426N_o3afin}{--}{GW190503E_o3afin}{0.00}{GW190512G_o3afin}{0.00}{GW190513E_o3afin}{0.00}{GW190514E_o3afin}{0.00}{GW190517B_o3afin}{0.00}{GW190519J_o3afin}{0.00}{GW190521B_o3afin}{0.00}{GW190521E_o3afin}{0.00}{GW190527H_o3afin}{0.00}{GW190602E_o3afin}{0.00}{GW190620B_o3afin}{0.00}{GW190630E_o3afin}{0.00}{GW190701E_o3afin}{0.00}{GW190706F_o3afin}{0.00}{GW190707E_o3afin}{0.00}{GW190708M_o3afin}{0.00}{GW190719H_o3afin}{--}{GW190720A_o3afin}{0.00}{GW190725F_o3afin}{--}{GW190727B_o3afin}{0.00}{GW190728D_o3afin}{0.00}{GW190731E_o3afin}{0.00}{GW190803B_o3afin}{0.00}{GW190805J_o3afin}{--}{GW190814H_o3afin}{0.00}{GW190828A_o3afin}{0.00}{GW190828B_o3afin}{0.00}{GW190910B_o3afin}{0.00}{GW190915K_o3afin}{0.00}{GW190916K_o3afin}{0.00}{GW190917B_o3afin}{0.00}{GW190924A_o3afin}{0.00}{GW190925J_o3afin}{--}{GW190926C_o3afin}{0.00}{GW190929B_o3afin}{0.00}{GW190930C_o3afin}{0.00}}}

\DeclareRobustCommand{\GSTLALALLSKYPNSBH}[1]{\IfEqCase{#1}{{GW190403B_o3afin}{--}{GW190408H_o3afin}{0.00}{GW190412B_o3afin}{0.00}{GW190413A_o3afin}{--}{GW190413E_o3afin}{0.00}{GW190421I_o3afin}{0.00}{GW190425B_o3afin}{0.00}{GW190426N_o3afin}{--}{GW190503E_o3afin}{0.00}{GW190512G_o3afin}{0.00}{GW190513E_o3afin}{0.00}{GW190514E_o3afin}{0.00}{GW190517B_o3afin}{0.00}{GW190519J_o3afin}{0.00}{GW190521B_o3afin}{0.00}{GW190521E_o3afin}{0.00}{GW190527H_o3afin}{0.00}{GW190602E_o3afin}{0.00}{GW190620B_o3afin}{0.00}{GW190630E_o3afin}{0.00}{GW190701E_o3afin}{0.00}{GW190706F_o3afin}{0.00}{GW190707E_o3afin}{0.00}{GW190708M_o3afin}{0.00}{GW190719H_o3afin}{--}{GW190720A_o3afin}{0.00}{GW190725F_o3afin}{--}{GW190727B_o3afin}{0.00}{GW190728D_o3afin}{0.00}{GW190731E_o3afin}{0.00}{GW190803B_o3afin}{0.00}{GW190805J_o3afin}{--}{GW190814H_o3afin}{0.81}{GW190828A_o3afin}{0.00}{GW190828B_o3afin}{0.00}{GW190910B_o3afin}{0.00}{GW190915K_o3afin}{0.00}{GW190916K_o3afin}{0.00}{GW190917B_o3afin}{0.00}{GW190924A_o3afin}{0.00}{GW190925J_o3afin}{--}{GW190926C_o3afin}{0.00}{GW190929B_o3afin}{0.00}{GW190930C_o3afin}{0.00}}}

\DeclareRobustCommand{\GSTLALALLSKYPMASSGAP}[1]{\IfEqCase{#1}{{GW190403B_o3afin}{--}{GW190408H_o3afin}{--}{GW190412B_o3afin}{--}{GW190413A_o3afin}{--}{GW190413E_o3afin}{--}{GW190421I_o3afin}{--}{GW190425B_o3afin}{--}{GW190426N_o3afin}{--}{GW190503E_o3afin}{--}{GW190512G_o3afin}{--}{GW190513E_o3afin}{--}{GW190514E_o3afin}{--}{GW190517B_o3afin}{--}{GW190519J_o3afin}{--}{GW190521B_o3afin}{--}{GW190521E_o3afin}{--}{GW190527H_o3afin}{--}{GW190602E_o3afin}{--}{GW190620B_o3afin}{--}{GW190630E_o3afin}{--}{GW190701E_o3afin}{--}{GW190706F_o3afin}{--}{GW190707E_o3afin}{--}{GW190708M_o3afin}{--}{GW190719H_o3afin}{--}{GW190720A_o3afin}{--}{GW190725F_o3afin}{--}{GW190727B_o3afin}{--}{GW190728D_o3afin}{--}{GW190731E_o3afin}{--}{GW190803B_o3afin}{--}{GW190805J_o3afin}{--}{GW190814H_o3afin}{--}{GW190828A_o3afin}{--}{GW190828B_o3afin}{--}{GW190910B_o3afin}{--}{GW190915K_o3afin}{--}{GW190916K_o3afin}{--}{GW190917B_o3afin}{--}{GW190924A_o3afin}{--}{GW190925J_o3afin}{--}{GW190926C_o3afin}{--}{GW190929B_o3afin}{--}{GW190930C_o3afin}{--}}}

\DeclareRobustCommand{\GSTLALALLSKYPTERRES}[1]{\IfEqCase{#1}{{GW190403B_o3afin}{--}{GW190408H_o3afin}{0.00}{GW190412B_o3afin}{0.00}{GW190413A_o3afin}{--}{GW190413E_o3afin}{0.96}{GW190421I_o3afin}{0.00}{GW190425B_o3afin}{0.22}{GW190426N_o3afin}{--}{GW190503E_o3afin}{0.00}{GW190512G_o3afin}{0.00}{GW190513E_o3afin}{0.00}{GW190514E_o3afin}{1.00}{GW190517B_o3afin}{0.00}{GW190519J_o3afin}{0.00}{GW190521B_o3afin}{0.21}{GW190521E_o3afin}{0.00}{GW190527H_o3afin}{0.15}{GW190602E_o3afin}{0.00}{GW190620B_o3afin}{0.01}{GW190630E_o3afin}{0.00}{GW190701E_o3afin}{0.01}{GW190706F_o3afin}{0.00}{GW190707E_o3afin}{0.00}{GW190708M_o3afin}{0.00}{GW190719H_o3afin}{--}{GW190720A_o3afin}{0.00}{GW190725F_o3afin}{--}{GW190727B_o3afin}{0.00}{GW190728D_o3afin}{0.00}{GW190731E_o3afin}{0.22}{GW190803B_o3afin}{0.06}{GW190805J_o3afin}{--}{GW190814H_o3afin}{0.00}{GW190828A_o3afin}{0.00}{GW190828B_o3afin}{0.00}{GW190910B_o3afin}{0.00}{GW190915K_o3afin}{0.00}{GW190916K_o3afin}{0.91}{GW190917B_o3afin}{0.23}{GW190924A_o3afin}{0.00}{GW190925J_o3afin}{--}{GW190926C_o3afin}{0.46}{GW190929B_o3afin}{0.13}{GW190930C_o3afin}{0.24}}}

\DeclareRobustCommand{\GSTLALALLSKYPASTRO}[1]{\IfEqCase{#1}{{GW190403B_o3afin}{--}{GW190408H_o3afin}{1.00}{GW190412B_o3afin}{1.00}{GW190413A_o3afin}{--}{GW190413E_o3afin}{0.04}{GW190421I_o3afin}{1.00}{GW190425B_o3afin}{0.78}{GW190426N_o3afin}{--}{GW190503E_o3afin}{1.00}{GW190512G_o3afin}{1.00}{GW190513E_o3afin}{1.00}{GW190514E_o3afin}{0.00}{GW190517B_o3afin}{1.00}{GW190519J_o3afin}{1.00}{GW190521B_o3afin}{0.79}{GW190521E_o3afin}{1.00}{GW190527H_o3afin}{0.85}{GW190602E_o3afin}{1.00}{GW190620B_o3afin}{0.99}{GW190630E_o3afin}{1.00}{GW190701E_o3afin}{0.99}{GW190706F_o3afin}{1.00}{GW190707E_o3afin}{1.00}{GW190708M_o3afin}{1.00}{GW190719H_o3afin}{--}{GW190720A_o3afin}{1.00}{GW190725F_o3afin}{--}{GW190727B_o3afin}{1.00}{GW190728D_o3afin}{1.00}{GW190731E_o3afin}{0.78}{GW190803B_o3afin}{0.94}{GW190805J_o3afin}{--}{GW190814H_o3afin}{1.00}{GW190828A_o3afin}{1.00}{GW190828B_o3afin}{1.00}{GW190910B_o3afin}{1.00}{GW190915K_o3afin}{1.00}{GW190916K_o3afin}{0.09}{GW190917B_o3afin}{0.77}{GW190924A_o3afin}{1.00}{GW190925J_o3afin}{--}{GW190926C_o3afin}{0.54}{GW190929B_o3afin}{0.87}{GW190930C_o3afin}{0.76}}}

\DeclareRobustCommand{\GSTLALALLSKYMEETSPASTROTHRESH}[1]{\IfEqCase{#1}{{GW190403B_o3afin}{}{GW190408H_o3afin}{}{GW190412B_o3afin}{}{GW190413A_o3afin}{}{GW190413E_o3afin}{\it }{GW190421I_o3afin}{}{GW190425B_o3afin}{}{GW190426N_o3afin}{}{GW190503E_o3afin}{}{GW190512G_o3afin}{}{GW190513E_o3afin}{}{GW190514E_o3afin}{\it }{GW190517B_o3afin}{}{GW190519J_o3afin}{}{GW190521B_o3afin}{}{GW190521E_o3afin}{}{GW190527H_o3afin}{}{GW190602E_o3afin}{}{GW190620B_o3afin}{}{GW190630E_o3afin}{}{GW190701E_o3afin}{}{GW190706F_o3afin}{}{GW190707E_o3afin}{}{GW190708M_o3afin}{}{GW190719H_o3afin}{}{GW190720A_o3afin}{}{GW190725F_o3afin}{}{GW190727B_o3afin}{}{GW190728D_o3afin}{}{GW190731E_o3afin}{}{GW190803B_o3afin}{}{GW190805J_o3afin}{}{GW190814H_o3afin}{}{GW190828A_o3afin}{}{GW190828B_o3afin}{}{GW190910B_o3afin}{}{GW190915K_o3afin}{}{GW190916K_o3afin}{\it }{GW190917B_o3afin}{}{GW190924A_o3afin}{}{GW190925J_o3afin}{}{GW190926C_o3afin}{}{GW190929B_o3afin}{}{GW190930C_o3afin}{}}}


\DeclareRobustCommand{\GSTLALSNRTHRESH}{4.0}

\DeclareRobustCommand{\NUMGSTLAL}{34}

\DeclareRobustCommand{\NUMMBTA}{29}

\DeclareRobustCommand{\NUMPYCBC}{36}

\DeclareRobustCommand{\NUMPYCBCBBH}{35}

\DeclareRobustCommand{\NUMPYCBCHYPERBANK}{23}

\DeclareRobustCommand{\NUMUNIQUEGSTLALEVENTS}{7}

\DeclareRobustCommand{\NUMUNIQUEMBTAEVENTS}{0}

\DeclareRobustCommand{\NUMUNIQUEPYCBCEVENTS}{7}

\DeclareRobustCommand{\TWOPIPES}{30}

\DeclareRobustCommand{\THREEPIPES}{25}

\DeclareRobustCommand{\ALLCBCPIPES}{25}

\DeclareRobustCommand{\ALLPIPES}{25}


\DeclareRobustCommand{\MINFAR}[1]{\IfEqCase{#1}{{GW190403B_o3afin}{\ensuremath{\ensuremath{7.7} \mathrm{yr}^{-1}}}{GW190408H_o3afin}{\ensuremath{\ensuremath{8.7 \times 10^{-5}} \mathrm{yr}^{-1}}}{GW190412B_o3afin}{\ensuremath{\ensuremath{1.0 \times 10^{-5}} \mathrm{yr}^{-1}}}{GW190413A_o3afin}{\ensuremath{\ensuremath{0.82} \mathrm{yr}^{-1}}}{GW190413E_o3afin}{\ensuremath{\ensuremath{0.34} \mathrm{yr}^{-1}}}{GW190421I_o3afin}{\ensuremath{\ensuremath{0.014} \mathrm{yr}^{-1}}}{GW190425B_o3afin}{\ensuremath{\ensuremath{0.034} \mathrm{yr}^{-1}}}{GW190426N_o3afin}{\ensuremath{\ensuremath{4.1} \mathrm{yr}^{-1}}}{GW190503E_o3afin}{\ensuremath{\ensuremath{1.0 \times 10^{-5}} \mathrm{yr}^{-1}}}{GW190512G_o3afin}{\ensuremath{\ensuremath{1.0 \times 10^{-5}} \mathrm{yr}^{-1}}}{GW190513E_o3afin}{\ensuremath{\ensuremath{1.3 \times 10^{-5}} \mathrm{yr}^{-1}}}{GW190514E_o3afin}{\ensuremath{\ensuremath{2.8} \mathrm{yr}^{-1}}}{GW190517B_o3afin}{\ensuremath{\ensuremath{0.11} \mathrm{yr}^{-1}}}{GW190519J_o3afin}{\ensuremath{\ensuremath{7.0 \times 10^{-5}} \mathrm{yr}^{-1}}}{GW190521B_o3afin}{\ensuremath{\ensuremath{0.0013} \mathrm{yr}^{-1}}}{GW190521E_o3afin}{\ensuremath{\ensuremath{1.0 \times 10^{-5}} \mathrm{yr}^{-1}}}{GW190527H_o3afin}{\ensuremath{\ensuremath{0.23} \mathrm{yr}^{-1}}}{GW190602E_o3afin}{\ensuremath{\ensuremath{1.0 \times 10^{-5}} \mathrm{yr}^{-1}}}{GW190620B_o3afin}{\ensuremath{\ensuremath{0.011} \mathrm{yr}^{-1}}}{GW190630E_o3afin}{\ensuremath{\ensuremath{1.0 \times 10^{-5}} \mathrm{yr}^{-1}}}{GW190701E_o3afin}{\ensuremath{\ensuremath{0.56} \mathrm{yr}^{-1}}}{GW190706F_o3afin}{\ensuremath{\ensuremath{0.34} \mathrm{yr}^{-1}}}{GW190707E_o3afin}{\ensuremath{\ensuremath{1.0 \times 10^{-5}} \mathrm{yr}^{-1}}}{GW190708M_o3afin}{\ensuremath{\ensuremath{3.1 \times 10^{-4}} \mathrm{yr}^{-1}}}{GW190719H_o3afin}{\ensuremath{\ensuremath{0.63} \mathrm{yr}^{-1}}}{GW190720A_o3afin}{\ensuremath{\ensuremath{0.094} \mathrm{yr}^{-1}}}{GW190725F_o3afin}{\ensuremath{\ensuremath{0.46} \mathrm{yr}^{-1}}}{GW190727B_o3afin}{\ensuremath{\ensuremath{1.0 \times 10^{-5}} \mathrm{yr}^{-1}}}{GW190728D_o3afin}{\ensuremath{\ensuremath{7.5 \times 10^{-4}} \mathrm{yr}^{-1}}}{GW190731E_o3afin}{\ensuremath{\ensuremath{1.9} \mathrm{yr}^{-1}}}{GW190803B_o3afin}{\ensuremath{\ensuremath{0.39} \mathrm{yr}^{-1}}}{GW190805J_o3afin}{\ensuremath{\ensuremath{0.63} \mathrm{yr}^{-1}}}{GW190814H_o3afin}{\ensuremath{\ensuremath{2.0 \times 10^{-4}} \mathrm{yr}^{-1}}}{GW190828A_o3afin}{\ensuremath{\ensuremath{1.0 \times 10^{-5}} \mathrm{yr}^{-1}}}{GW190828B_o3afin}{\ensuremath{\ensuremath{2.8 \times 10^{-4}} \mathrm{yr}^{-1}}}{GW190910B_o3afin}{\ensuremath{\ensuremath{0.0029} \mathrm{yr}^{-1}}}{GW190915K_o3afin}{\ensuremath{\ensuremath{7.0 \times 10^{-5}} \mathrm{yr}^{-1}}}{GW190916K_o3afin}{\ensuremath{\ensuremath{6.9 \times 10^{3}} \mathrm{yr}^{-1}}}{GW190917B_o3afin}{\ensuremath{\ensuremath{0.66} \mathrm{yr}^{-1}}}{GW190924A_o3afin}{\ensuremath{\ensuremath{1.0 \times 10^{-5}} \mathrm{yr}^{-1}}}{GW190925J_o3afin}{\ensuremath{\ensuremath{0.0072} \mathrm{yr}^{-1}}}{GW190926C_o3afin}{\ensuremath{\ensuremath{1.1} \mathrm{yr}^{-1}}}{GW190929B_o3afin}{\ensuremath{\ensuremath{0.16} \mathrm{yr}^{-1}}}{GW190930C_o3afin}{\ensuremath{\ensuremath{0.018} \mathrm{yr}^{-1}}}}}

\DeclareRobustCommand{\MBTAALLSKYFAR}[1]{\IfEqCase{#1}{{GW190403B_o3afin}{--}{GW190408H_o3afin}{\ensuremath{8.7 \times 10^{-5}}}{GW190412B_o3afin}{\ensuremath{< \ensuremath{1.0 \times 10^{-5}}}}{GW190413A_o3afin}{--}{GW190413E_o3afin}{\ensuremath{0.34}}{GW190421I_o3afin}{\ensuremath{1.2}}{GW190425B_o3afin}{--}{GW190426N_o3afin}{--}{GW190503E_o3afin}{\ensuremath{0.013}}{GW190512G_o3afin}{\ensuremath{0.038}}{GW190513E_o3afin}{\ensuremath{0.11}}{GW190514E_o3afin}{--}{GW190517B_o3afin}{\ensuremath{0.11}}{GW190519J_o3afin}{\ensuremath{7.0 \times 10^{-5}}}{GW190521B_o3afin}{\ensuremath{0.042}}{GW190521E_o3afin}{\ensuremath{< \ensuremath{1.0 \times 10^{-5}}}}{GW190527H_o3afin}{--}{GW190602E_o3afin}{\ensuremath{3.0 \times 10^{-4}}}{GW190620B_o3afin}{--}{GW190630E_o3afin}{--}{GW190701E_o3afin}{\ensuremath{35}}{GW190706F_o3afin}{\ensuremath{0.0015}}{GW190707E_o3afin}{\ensuremath{0.032}}{GW190708M_o3afin}{--}{GW190719H_o3afin}{--}{GW190720A_o3afin}{\ensuremath{0.094}}{GW190725F_o3afin}{\ensuremath{3.1}}{GW190727B_o3afin}{\ensuremath{0.023}}{GW190728D_o3afin}{\ensuremath{7.5 \times 10^{-4}}}{GW190731E_o3afin}{\ensuremath{6.1}}{GW190803B_o3afin}{\ensuremath{77}}{GW190805J_o3afin}{--}{GW190814H_o3afin}{\ensuremath{< \ensuremath{2.0 \times 10^{-4}}}}{GW190828A_o3afin}{\ensuremath{< \ensuremath{1.0 \times 10^{-5}}}}{GW190828B_o3afin}{\ensuremath{0.16}}{GW190910B_o3afin}{--}{GW190915K_o3afin}{\ensuremath{0.0055}}{GW190916K_o3afin}{\ensuremath{6.9 \times 10^{3}}}{GW190917B_o3afin}{--}{GW190924A_o3afin}{\ensuremath{0.0049}}{GW190925J_o3afin}{\ensuremath{100}}{GW190926C_o3afin}{--}{GW190929B_o3afin}{\ensuremath{2.9}}{GW190930C_o3afin}{\ensuremath{0.34}}}}

\DeclareRobustCommand{\MBTAALLSKYIFAR}[1]{\IfEqCase{#1}{{GW190403B_o3afin}{--}{GW190408H_o3afin}{4.1}{GW190412B_o3afin}{5.0}{GW190413A_o3afin}{--}{GW190413E_o3afin}{0.5}{GW190421I_o3afin}{-0.1}{GW190425B_o3afin}{--}{GW190426N_o3afin}{--}{GW190503E_o3afin}{1.9}{GW190512G_o3afin}{1.4}{GW190513E_o3afin}{1.0}{GW190514E_o3afin}{--}{GW190517B_o3afin}{1.0}{GW190519J_o3afin}{4.2}{GW190521B_o3afin}{1.4}{GW190521E_o3afin}{5.0}{GW190527H_o3afin}{--}{GW190602E_o3afin}{3.5}{GW190620B_o3afin}{--}{GW190630E_o3afin}{--}{GW190701E_o3afin}{-1.5}{GW190706F_o3afin}{2.8}{GW190707E_o3afin}{1.5}{GW190708M_o3afin}{--}{GW190719H_o3afin}{--}{GW190720A_o3afin}{1.0}{GW190725F_o3afin}{-0.5}{GW190727B_o3afin}{1.6}{GW190728D_o3afin}{3.1}{GW190731E_o3afin}{-0.8}{GW190803B_o3afin}{-1.9}{GW190805J_o3afin}{--}{GW190814H_o3afin}{3.7}{GW190828A_o3afin}{5.0}{GW190828B_o3afin}{0.8}{GW190910B_o3afin}{--}{GW190915K_o3afin}{2.3}{GW190916K_o3afin}{-3.8}{GW190917B_o3afin}{--}{GW190924A_o3afin}{2.3}{GW190925J_o3afin}{-2.0}{GW190926C_o3afin}{--}{GW190929B_o3afin}{-0.5}{GW190930C_o3afin}{0.5}}}

\DeclareRobustCommand{\MBTAALLSKYSNR}[1]{\IfEqCase{#1}{{GW190403B_o3afin}{--}{GW190408H_o3afin}{14.4}{GW190412B_o3afin}{18.2}{GW190413A_o3afin}{--}{GW190413E_o3afin}{10.3}{GW190421I_o3afin}{9.7}{GW190425B_o3afin}{--}{GW190426N_o3afin}{--}{GW190503E_o3afin}{12.8}{GW190512G_o3afin}{11.7}{GW190513E_o3afin}{13.0}{GW190514E_o3afin}{--}{GW190517B_o3afin}{11.3}{GW190519J_o3afin}{13.7}{GW190521B_o3afin}{13.0}{GW190521E_o3afin}{22.2}{GW190527H_o3afin}{--}{GW190602E_o3afin}{12.6}{GW190620B_o3afin}{--}{GW190630E_o3afin}{--}{GW190701E_o3afin}{11.3}{GW190706F_o3afin}{11.9}{GW190707E_o3afin}{12.6}{GW190708M_o3afin}{--}{GW190719H_o3afin}{--}{GW190720A_o3afin}{11.6}{GW190725F_o3afin}{9.8}{GW190727B_o3afin}{12.0}{GW190728D_o3afin}{13.1}{GW190731E_o3afin}{9.1}{GW190803B_o3afin}{9.0}{GW190805J_o3afin}{--}{GW190814H_o3afin}{20.4}{GW190828A_o3afin}{15.2}{GW190828B_o3afin}{10.8}{GW190910B_o3afin}{--}{GW190915K_o3afin}{12.7}{GW190916K_o3afin}{8.2}{GW190917B_o3afin}{--}{GW190924A_o3afin}{11.9}{GW190925J_o3afin}{9.4}{GW190926C_o3afin}{--}{GW190929B_o3afin}{10.3}{GW190930C_o3afin}{10.0}}}

\DeclareRobustCommand{\PYCBCHIGHMASSFAR}[1]{\IfEqCase{#1}{{GW190403B_o3afin}{\ensuremath{7.7}}{GW190408H_o3afin}{\ensuremath{< \ensuremath{1.2 \times 10^{-4}}}}{GW190412B_o3afin}{\ensuremath{< \ensuremath{1.2 \times 10^{-4}}}}{GW190413A_o3afin}{\ensuremath{0.82}}{GW190413E_o3afin}{\ensuremath{0.18}}{GW190421I_o3afin}{\ensuremath{0.014}}{GW190425B_o3afin}{--}{GW190426N_o3afin}{\ensuremath{4.1}}{GW190503E_o3afin}{\ensuremath{0.0026}}{GW190512G_o3afin}{\ensuremath{< \ensuremath{1.1 \times 10^{-4}}}}{GW190513E_o3afin}{\ensuremath{0.044}}{GW190514E_o3afin}{\ensuremath{2.8}}{GW190517B_o3afin}{\ensuremath{3.5 \times 10^{-4}}}{GW190519J_o3afin}{\ensuremath{< \ensuremath{1.1 \times 10^{-4}}}}{GW190521B_o3afin}{\ensuremath{0.0013}}{GW190521E_o3afin}{\ensuremath{< \ensuremath{2.3 \times 10^{-5}}}}{GW190527H_o3afin}{\ensuremath{19}}{GW190602E_o3afin}{\ensuremath{0.013}}{GW190620B_o3afin}{--}{GW190630E_o3afin}{\ensuremath{0.24}}{GW190701E_o3afin}{\ensuremath{0.56}}{GW190706F_o3afin}{\ensuremath{0.34}}{GW190707E_o3afin}{\ensuremath{< \ensuremath{1.9 \times 10^{-5}}}}{GW190708M_o3afin}{--}{GW190719H_o3afin}{\ensuremath{0.63}}{GW190720A_o3afin}{\ensuremath{< \ensuremath{7.8 \times 10^{-5}}}}{GW190725F_o3afin}{\ensuremath{2.9}}{GW190727B_o3afin}{\ensuremath{2.0 \times 10^{-4}}}{GW190728D_o3afin}{\ensuremath{< \ensuremath{7.8 \times 10^{-5}}}}{GW190731E_o3afin}{\ensuremath{1.9}}{GW190803B_o3afin}{\ensuremath{0.39}}{GW190805J_o3afin}{\ensuremath{0.63}}{GW190814H_o3afin}{--}{GW190828A_o3afin}{\ensuremath{< \ensuremath{7.0 \times 10^{-5}}}}{GW190828B_o3afin}{\ensuremath{1.1 \times 10^{-4}}}{GW190910B_o3afin}{--}{GW190915K_o3afin}{\ensuremath{< \ensuremath{7.0 \times 10^{-5}}}}{GW190916K_o3afin}{\ensuremath{4.7}}{GW190917B_o3afin}{--}{GW190924A_o3afin}{\ensuremath{8.3 \times 10^{-5}}}{GW190925J_o3afin}{\ensuremath{0.0072}}{GW190926C_o3afin}{\ensuremath{87}}{GW190929B_o3afin}{\ensuremath{14}}{GW190930C_o3afin}{\ensuremath{0.012}}}}

\DeclareRobustCommand{\PYCBCHIGHMASSIFAR}[1]{\IfEqCase{#1}{{GW190403B_o3afin}{-0.9}{GW190408H_o3afin}{3.9}{GW190412B_o3afin}{3.9}{GW190413A_o3afin}{0.1}{GW190413E_o3afin}{0.7}{GW190421I_o3afin}{1.8}{GW190425B_o3afin}{--}{GW190426N_o3afin}{-0.6}{GW190503E_o3afin}{2.6}{GW190512G_o3afin}{4.0}{GW190513E_o3afin}{1.4}{GW190514E_o3afin}{-0.4}{GW190517B_o3afin}{3.5}{GW190519J_o3afin}{4.0}{GW190521B_o3afin}{2.9}{GW190521E_o3afin}{4.6}{GW190527H_o3afin}{-1.3}{GW190602E_o3afin}{1.9}{GW190620B_o3afin}{--}{GW190630E_o3afin}{0.6}{GW190701E_o3afin}{0.2}{GW190706F_o3afin}{0.5}{GW190707E_o3afin}{4.7}{GW190708M_o3afin}{--}{GW190719H_o3afin}{0.2}{GW190720A_o3afin}{4.1}{GW190725F_o3afin}{-0.5}{GW190727B_o3afin}{3.7}{GW190728D_o3afin}{4.1}{GW190731E_o3afin}{-0.3}{GW190803B_o3afin}{0.4}{GW190805J_o3afin}{0.2}{GW190814H_o3afin}{--}{GW190828A_o3afin}{4.2}{GW190828B_o3afin}{4.0}{GW190910B_o3afin}{--}{GW190915K_o3afin}{4.2}{GW190916K_o3afin}{-0.7}{GW190917B_o3afin}{--}{GW190924A_o3afin}{4.1}{GW190925J_o3afin}{2.1}{GW190926C_o3afin}{-1.9}{GW190929B_o3afin}{-1.1}{GW190930C_o3afin}{1.9}}}

\DeclareRobustCommand{\PYCBCHIGHMASSSNR}[1]{\IfEqCase{#1}{{GW190403B_o3afin}{8.0}{GW190408H_o3afin}{13.7\ensuremath{{}^{\ddag}}}{GW190412B_o3afin}{17.9\ensuremath{{}^{\ddag}}}{GW190413A_o3afin}{8.5}{GW190413E_o3afin}{8.9\ensuremath{{}^{\ddag}}}{GW190421I_o3afin}{10.1}{GW190425B_o3afin}{--}{GW190426N_o3afin}{9.6}{GW190503E_o3afin}{12.2\ensuremath{{}^{\ddag}}}{GW190512G_o3afin}{12.4\ensuremath{{}^{\ddag}}}{GW190513E_o3afin}{11.8\ensuremath{{}^{\ddag}}}{GW190514E_o3afin}{8.4}{GW190517B_o3afin}{10.3\ensuremath{{}^{\ddag}}}{GW190519J_o3afin}{13.2\ensuremath{{}^{\ddag}}}{GW190521B_o3afin}{13.6\ensuremath{{}^{\ddag}}}{GW190521E_o3afin}{24.0}{GW190527H_o3afin}{8.4}{GW190602E_o3afin}{11.9\ensuremath{{}^{\ddag}}}{GW190620B_o3afin}{--}{GW190630E_o3afin}{15.1}{GW190701E_o3afin}{11.7}{GW190706F_o3afin}{12.6\ensuremath{{}^{\ddag}}}{GW190707E_o3afin}{13.0}{GW190708M_o3afin}{--}{GW190719H_o3afin}{8.0}{GW190720A_o3afin}{11.4}{GW190725F_o3afin}{8.8\ensuremath{{}^{\ddag}}}{GW190727B_o3afin}{11.1\ensuremath{{}^{\ddag}}}{GW190728D_o3afin}{13.0\ensuremath{{}^{\ddag}}}{GW190731E_o3afin}{7.8}{GW190803B_o3afin}{8.7\ensuremath{{}^{\ddag}}}{GW190805J_o3afin}{8.3}{GW190814H_o3afin}{--}{GW190828A_o3afin}{15.9\ensuremath{{}^{\ddag}}}{GW190828B_o3afin}{10.5\ensuremath{{}^{\ddag}}}{GW190910B_o3afin}{--}{GW190915K_o3afin}{13.1\ensuremath{{}^{\ddag}}}{GW190916K_o3afin}{7.9\ensuremath{{}^{\ddag}}}{GW190917B_o3afin}{--}{GW190924A_o3afin}{12.5\ensuremath{{}^{\ddag}}}{GW190925J_o3afin}{9.9}{GW190926C_o3afin}{7.8\ensuremath{{}^{\ddag}}}{GW190929B_o3afin}{8.5\ensuremath{{}^{\ddag}}}{GW190930C_o3afin}{10.0}}}

\DeclareRobustCommand{\PYCBCALLSKYFAR}[1]{\IfEqCase{#1}{{GW190403B_o3afin}{--}{GW190408H_o3afin}{\ensuremath{2.5 \times 10^{-4}}}{GW190412B_o3afin}{\ensuremath{< \ensuremath{1.1 \times 10^{-4}}}}{GW190413A_o3afin}{\ensuremath{170}}{GW190413E_o3afin}{\ensuremath{21}}{GW190421I_o3afin}{\ensuremath{5.9}}{GW190425B_o3afin}{--}{GW190426N_o3afin}{--}{GW190503E_o3afin}{\ensuremath{0.038}}{GW190512G_o3afin}{\ensuremath{1.1 \times 10^{-4}}}{GW190513E_o3afin}{\ensuremath{19}}{GW190514E_o3afin}{--}{GW190517B_o3afin}{\ensuremath{0.0095}}{GW190519J_o3afin}{\ensuremath{< \ensuremath{1.0 \times 10^{-4}}}}{GW190521B_o3afin}{\ensuremath{0.44}}{GW190521E_o3afin}{\ensuremath{< \ensuremath{1.8 \times 10^{-5}}}}{GW190527H_o3afin}{--}{GW190602E_o3afin}{\ensuremath{0.29}}{GW190620B_o3afin}{--}{GW190630E_o3afin}{--}{GW190701E_o3afin}{\ensuremath{0.064}}{GW190706F_o3afin}{\ensuremath{3.7 \times 10^{-4}}}{GW190707E_o3afin}{\ensuremath{< \ensuremath{1.0 \times 10^{-5}}}}{GW190708M_o3afin}{--}{GW190719H_o3afin}{--}{GW190720A_o3afin}{\ensuremath{1.4 \times 10^{-4}}}{GW190725F_o3afin}{\ensuremath{0.46}}{GW190727B_o3afin}{\ensuremath{0.0056}}{GW190728D_o3afin}{\ensuremath{< \ensuremath{8.2 \times 10^{-5}}}}{GW190731E_o3afin}{--}{GW190803B_o3afin}{\ensuremath{81}}{GW190805J_o3afin}{--}{GW190814H_o3afin}{\ensuremath{0.17}}{GW190828A_o3afin}{\ensuremath{< \ensuremath{8.5 \times 10^{-5}}}}{GW190828B_o3afin}{\ensuremath{2.8 \times 10^{-4}}}{GW190910B_o3afin}{--}{GW190915K_o3afin}{\ensuremath{6.8 \times 10^{-4}}}{GW190916K_o3afin}{--}{GW190917B_o3afin}{--}{GW190924A_o3afin}{\ensuremath{< \ensuremath{8.2 \times 10^{-5}}}}{GW190925J_o3afin}{\ensuremath{73}}{GW190926C_o3afin}{--}{GW190929B_o3afin}{\ensuremath{120}}{GW190930C_o3afin}{\ensuremath{0.018}}}}

\DeclareRobustCommand{\PYCBCALLSKYIFAR}[1]{\IfEqCase{#1}{{GW190403B_o3afin}{--}{GW190408H_o3afin}{3.6}{GW190412B_o3afin}{3.9}{GW190413A_o3afin}{-2.2}{GW190413E_o3afin}{-1.3}{GW190421I_o3afin}{-0.8}{GW190425B_o3afin}{--}{GW190426N_o3afin}{--}{GW190503E_o3afin}{1.4}{GW190512G_o3afin}{4.0}{GW190513E_o3afin}{-1.3}{GW190514E_o3afin}{--}{GW190517B_o3afin}{2.0}{GW190519J_o3afin}{4.0}{GW190521B_o3afin}{0.4}{GW190521E_o3afin}{4.8}{GW190527H_o3afin}{--}{GW190602E_o3afin}{0.5}{GW190620B_o3afin}{--}{GW190630E_o3afin}{--}{GW190701E_o3afin}{1.2}{GW190706F_o3afin}{3.4}{GW190707E_o3afin}{5.0}{GW190708M_o3afin}{--}{GW190719H_o3afin}{--}{GW190720A_o3afin}{3.9}{GW190725F_o3afin}{0.3}{GW190727B_o3afin}{2.2}{GW190728D_o3afin}{4.1}{GW190731E_o3afin}{--}{GW190803B_o3afin}{-1.9}{GW190805J_o3afin}{--}{GW190814H_o3afin}{0.8}{GW190828A_o3afin}{4.1}{GW190828B_o3afin}{3.6}{GW190910B_o3afin}{--}{GW190915K_o3afin}{3.2}{GW190916K_o3afin}{--}{GW190917B_o3afin}{--}{GW190924A_o3afin}{4.1}{GW190925J_o3afin}{-1.9}{GW190926C_o3afin}{--}{GW190929B_o3afin}{-2.1}{GW190930C_o3afin}{1.7}}}

\DeclareRobustCommand{\PYCBCALLSKYSNR}[1]{\IfEqCase{#1}{{GW190403B_o3afin}{--}{GW190408H_o3afin}{13.1\ensuremath{{}^{\ddag}}}{GW190412B_o3afin}{17.4\ensuremath{{}^{\ddag}}}{GW190413A_o3afin}{8.5}{GW190413E_o3afin}{9.3\ensuremath{{}^{\ddag}}}{GW190421I_o3afin}{10.1}{GW190425B_o3afin}{--}{GW190426N_o3afin}{--}{GW190503E_o3afin}{12.2\ensuremath{{}^{\ddag}}}{GW190512G_o3afin}{12.4\ensuremath{{}^{\ddag}}}{GW190513E_o3afin}{11.6\ensuremath{{}^{\ddag}}}{GW190514E_o3afin}{--}{GW190517B_o3afin}{10.4\ensuremath{{}^{\ddag}}}{GW190519J_o3afin}{13.2\ensuremath{{}^{\ddag}}}{GW190521B_o3afin}{13.7\ensuremath{{}^{\ddag}}}{GW190521E_o3afin}{24.0}{GW190527H_o3afin}{--}{GW190602E_o3afin}{11.9\ensuremath{{}^{\ddag}}}{GW190620B_o3afin}{--}{GW190630E_o3afin}{--}{GW190701E_o3afin}{11.9}{GW190706F_o3afin}{11.7\ensuremath{{}^{\ddag}}}{GW190707E_o3afin}{13.0}{GW190708M_o3afin}{--}{GW190719H_o3afin}{--}{GW190720A_o3afin}{10.6\ensuremath{{}^{\ddag}}}{GW190725F_o3afin}{9.1\ensuremath{{}^{\ddag}}}{GW190727B_o3afin}{11.4\ensuremath{{}^{\ddag}}}{GW190728D_o3afin}{13.0\ensuremath{{}^{\ddag}}}{GW190731E_o3afin}{--}{GW190803B_o3afin}{8.7\ensuremath{{}^{\ddag}}}{GW190805J_o3afin}{--}{GW190814H_o3afin}{19.5}{GW190828A_o3afin}{13.9\ensuremath{{}^{\ddag}}}{GW190828B_o3afin}{10.5\ensuremath{{}^{\ddag}}}{GW190910B_o3afin}{--}{GW190915K_o3afin}{13.0\ensuremath{{}^{\ddag}}}{GW190916K_o3afin}{--}{GW190917B_o3afin}{--}{GW190924A_o3afin}{12.4\ensuremath{{}^{\ddag}}}{GW190925J_o3afin}{9.0}{GW190926C_o3afin}{--}{GW190929B_o3afin}{9.4\ensuremath{{}^{\ddag}}}{GW190930C_o3afin}{9.8}}}

\DeclareRobustCommand{\GSTLALALLSKYFAR}[1]{\IfEqCase{#1}{{GW190403B_o3afin}{--}{GW190408H_o3afin}{\ensuremath{< \ensuremath{1.0 \times 10^{-5}}}}{GW190412B_o3afin}{\ensuremath{< \ensuremath{1.0 \times 10^{-5}}}}{GW190413A_o3afin}{--}{GW190413E_o3afin}{\ensuremath{39}}{GW190421I_o3afin}{\ensuremath{0.0028}}{GW190425B_o3afin}{\ensuremath{0.034}\ensuremath{{}^\dagger}}{GW190426N_o3afin}{--}{GW190503E_o3afin}{\ensuremath{< \ensuremath{1.0 \times 10^{-5}}}}{GW190512G_o3afin}{\ensuremath{< \ensuremath{1.0 \times 10^{-5}}}}{GW190513E_o3afin}{\ensuremath{1.3 \times 10^{-5}}}{GW190514E_o3afin}{\ensuremath{450}}{GW190517B_o3afin}{\ensuremath{0.0045}}{GW190519J_o3afin}{\ensuremath{< \ensuremath{1.0 \times 10^{-5}}}}{GW190521B_o3afin}{\ensuremath{0.20}}{GW190521E_o3afin}{\ensuremath{< \ensuremath{1.0 \times 10^{-5}}}}{GW190527H_o3afin}{\ensuremath{0.23}}{GW190602E_o3afin}{\ensuremath{< \ensuremath{1.0 \times 10^{-5}}}}{GW190620B_o3afin}{\ensuremath{0.011}\ensuremath{{}^\dagger}}{GW190630E_o3afin}{\ensuremath{< \ensuremath{1.0 \times 10^{-5}}}}{GW190701E_o3afin}{\ensuremath{0.0057}}{GW190706F_o3afin}{\ensuremath{5.0 \times 10^{-5}}}{GW190707E_o3afin}{\ensuremath{< \ensuremath{1.0 \times 10^{-5}}}}{GW190708M_o3afin}{\ensuremath{3.1 \times 10^{-4}}\ensuremath{{}^\dagger}}{GW190719H_o3afin}{--}{GW190720A_o3afin}{\ensuremath{< \ensuremath{1.0 \times 10^{-5}}}}{GW190725F_o3afin}{--}{GW190727B_o3afin}{\ensuremath{< \ensuremath{1.0 \times 10^{-5}}}}{GW190728D_o3afin}{\ensuremath{< \ensuremath{1.0 \times 10^{-5}}}}{GW190731E_o3afin}{\ensuremath{0.33}}{GW190803B_o3afin}{\ensuremath{0.073}}{GW190805J_o3afin}{--}{GW190814H_o3afin}{\ensuremath{< \ensuremath{1.0 \times 10^{-5}}}}{GW190828A_o3afin}{\ensuremath{< \ensuremath{1.0 \times 10^{-5}}}}{GW190828B_o3afin}{\ensuremath{3.5 \times 10^{-5}}}{GW190910B_o3afin}{\ensuremath{0.0029}\ensuremath{{}^\dagger}}{GW190915K_o3afin}{\ensuremath{< \ensuremath{1.0 \times 10^{-5}}}}{GW190916K_o3afin}{\ensuremath{12}}{GW190917B_o3afin}{\ensuremath{0.66}}{GW190924A_o3afin}{\ensuremath{< \ensuremath{1.0 \times 10^{-5}}}}{GW190925J_o3afin}{--}{GW190926C_o3afin}{\ensuremath{1.1}}{GW190929B_o3afin}{\ensuremath{0.16}}{GW190930C_o3afin}{\ensuremath{0.43}}}}

\DeclareRobustCommand{\GSTLALALLSKYIFAR}[1]{\IfEqCase{#1}{{GW190403B_o3afin}{--}{GW190408H_o3afin}{5.0}{GW190412B_o3afin}{5.0}{GW190413A_o3afin}{--}{GW190413E_o3afin}{-1.6}{GW190421I_o3afin}{2.5}{GW190425B_o3afin}{1.5\ensuremath{{}^\dagger}}{GW190426N_o3afin}{--}{GW190503E_o3afin}{5.0}{GW190512G_o3afin}{5.0}{GW190513E_o3afin}{4.9}{GW190514E_o3afin}{-2.6}{GW190517B_o3afin}{2.3}{GW190519J_o3afin}{5.0}{GW190521B_o3afin}{0.7}{GW190521E_o3afin}{5.0}{GW190527H_o3afin}{0.6}{GW190602E_o3afin}{5.0}{GW190620B_o3afin}{2.0\ensuremath{{}^\dagger}}{GW190630E_o3afin}{5.0}{GW190701E_o3afin}{2.2}{GW190706F_o3afin}{4.3}{GW190707E_o3afin}{5.0}{GW190708M_o3afin}{3.5\ensuremath{{}^\dagger}}{GW190719H_o3afin}{--}{GW190720A_o3afin}{5.0}{GW190725F_o3afin}{--}{GW190727B_o3afin}{5.0}{GW190728D_o3afin}{5.0}{GW190731E_o3afin}{0.5}{GW190803B_o3afin}{1.1}{GW190805J_o3afin}{--}{GW190814H_o3afin}{5.0}{GW190828A_o3afin}{5.0}{GW190828B_o3afin}{4.5}{GW190910B_o3afin}{2.5\ensuremath{{}^\dagger}}{GW190915K_o3afin}{5.0}{GW190916K_o3afin}{-1.1}{GW190917B_o3afin}{0.2}{GW190924A_o3afin}{5.0}{GW190925J_o3afin}{--}{GW190926C_o3afin}{-0.1}{GW190929B_o3afin}{0.8}{GW190930C_o3afin}{0.4}}}

\DeclareRobustCommand{\GSTLALALLSKYSNR}[1]{\IfEqCase{#1}{{GW190403B_o3afin}{--}{GW190408H_o3afin}{14.7}{GW190412B_o3afin}{19.0}{GW190413A_o3afin}{--}{GW190413E_o3afin}{10.1}{GW190421I_o3afin}{10.5}{GW190425B_o3afin}{12.9}{GW190426N_o3afin}{--}{GW190503E_o3afin}{12.0}{GW190512G_o3afin}{12.2}{GW190513E_o3afin}{12.3}{GW190514E_o3afin}{8.3}{GW190517B_o3afin}{10.8}{GW190519J_o3afin}{12.4}{GW190521B_o3afin}{13.3}{GW190521E_o3afin}{24.4}{GW190527H_o3afin}{8.7}{GW190602E_o3afin}{12.3}{GW190620B_o3afin}{10.9}{GW190630E_o3afin}{15.2}{GW190701E_o3afin}{11.7}{GW190706F_o3afin}{12.5}{GW190707E_o3afin}{13.2}{GW190708M_o3afin}{13.1}{GW190719H_o3afin}{--}{GW190720A_o3afin}{11.5}{GW190725F_o3afin}{--}{GW190727B_o3afin}{12.1}{GW190728D_o3afin}{13.4}{GW190731E_o3afin}{8.5}{GW190803B_o3afin}{9.1}{GW190805J_o3afin}{--}{GW190814H_o3afin}{22.2}{GW190828A_o3afin}{16.3}{GW190828B_o3afin}{11.1}{GW190910B_o3afin}{13.4}{GW190915K_o3afin}{13.0}{GW190916K_o3afin}{8.2}{GW190917B_o3afin}{9.5}{GW190924A_o3afin}{13.0}{GW190925J_o3afin}{--}{GW190926C_o3afin}{9.0}{GW190929B_o3afin}{10.1}{GW190930C_o3afin}{10.1}}}

\newcommand{\Mc}{\ensuremath{\mathcal{M}}\xspace}
\newcommand{\Mtot}{\ensuremath{M}\xspace}
\newcommand{\Msun}{\ensuremath{\text{M}_\odot}\xspace}
\newcommand{\massone}{\ensuremath{m_1}\xspace}
\newcommand{\masstwo}{\ensuremath{m_2}\xspace}
\newcommand{\Mf}{\ensuremath{M_\mathrm{f}}\xspace}
\newcommand{\massratio}{\ensuremath{q}\xspace}
\newcommand{\masscomponent}{\ensuremath{m_i}\xspace}

\newcommand{\chieff}{\ensuremath{\chi_\mathrm{eff}}\xspace}
\newcommand{\chip}{\ensuremath{\chi_\mathrm{p}}\xspace}
\newcommand{\chif}{\ensuremath{\chi_\mathrm{f}}\xspace}
\newcommand{\spintilt}{\ensuremath{\theta_{{JS}}}\xspace}
\newcommand{\spinone}{\ensuremath{\chi_1}\xspace}
\newcommand{\vecspinone}{\ensuremath{\vec\chi_1}\xspace}
\newcommand{\spintwo}{\ensuremath{\chi_2}\xspace}
\newcommand{\vecspintwo}{\ensuremath{\vec\chi_2}\xspace}

\newcommand{\DL}{\ensuremath{D_\mathrm{L}}\xspace}
\newcommand{\DC}{\ensuremath{D_\mathrm{c}}\xspace}
\newcommand{\redshift}{\ensuremath{z}\xspace}

\newcommand{\skyareasymbol}{\ensuremath{\Delta \Omega}\xspace}

\newcommand\perMpcyr{\ensuremath{\text{Mpc}^{-3}\,\text{yr}^{-1}}}
\newcommand\Mpcyr{\ensuremath{\text{Mpc}^{3}\,\text{yr}}}
\newcommand\perGpcyr{\ensuremath{\text{Gpc}^{-3}\,\text{yr}^{-1}}}
\newcommand\Gpcyr{\ensuremath{\text{Gpc}^{3}\,\text{yr}}}
\newcommand{\DKLchip}{D_\mathrm{KL}^{\chi_\text{p}}}
\newcommand{\DKLchieff}{D_\mathrm{KL}^{\chi_\text{eff}}}
\newcommand{\VT}{\ensuremath{\langle VT \rangle}}
\newcommand{\pastro}{\ensuremath{p_{\mathrm{astro}}}}
\newcommand{\pterr}{\ensuremath{p_{\mathrm{terr}}}}
\newcommand{\pbbh}{\ensuremath{p_{\mathrm{BBH}}}}
\newcommand{\pbns}{\ensuremath{p_{\mathrm{BNS}}}}
\newcommand{\pnsbh}{\ensuremath{p_{\mathrm{NSBH}}}}
\newcommand{\comovingvt}{\ensuremath{\langle VT_\mathrm{c} \rangle}}
\newcommand{\Gpc}{\ensuremath{\text{Gpc}}\xspace}

\newcommand{\RUNSTART}{\fixme{1 April 2019 15:00 UTC}}
\newcommand{\ARUNEND}{\fixme{1 October 2019 15:00 UTC}}

\newcommand{\RUNEND}{\fixme{27 March 2020 17:00 UTC}}

\newcommand{\CONTAMINATION}{\fixme{0\%}}

\newcommand{\DETECTIONRATE}{\fixme{$\sim$2 per week}}
\newcommand{\NUMOPA}{\fixme{24}}
\newcommand{\NUMOPANOTFOUND}{\fixme{24}}

\newcommand{\VIRGORANGE}{\fixme{45 Mpc}}
\newcommand{\HANFORDRANGE}{\fixme{108 Mpc}}
\newcommand{\LIVINGSTONRANGE}{\fixme{135 Mpc}}

\newcommand{\VIRGORANGEOTWO}{\fixme{26 Mpc}}
\newcommand{\HANFORDRANGEOTWO}{\fixme{66 Mpc}}
\newcommand{\LIVINGSTONRANGEOTWO}{\fixme{88 Mpc}}

\newcommand{\VIRGORANGEINCREASE}{\fixme{1.73}}
\newcommand{\HANFORDRANGEINCREASE}{\fixme{1.64}}
\newcommand{\LIVINGSTONRANGEINCREASE}{\fixme{1.53}}

\newcommand{\VIRGODUTYCYCLE}{\fixme{76}}
\newcommand{\HANFORDDUTYCYCLE}{\fixme{71}}
\newcommand{\LIVINGSTONDUTYCYCLE}{\fixme{76}}
\newcommand{\ONEDETECTORDUTYCYCLE}{\fixme{96.9}}
\newcommand{\TWODETECTORSDUTYCYCLE}{\fixme{81.9}}
\newcommand{\THREEDETECTORSDUTYCYCLE}{\fixme{44.5}}

\newcommand{\VIRGODAYS}{\fixme{139.5}}
\newcommand{\HANFORDDAYS}{\fixme{130.3}}
\newcommand{\LIVINGSTONDAYS}{\fixme{138.5}}
\newcommand{\ONEDETECTORDAYS}{\fixme{177.3}}
\newcommand{\TWODETECTORSDAYS}{\fixme{149.9}}
\newcommand{\THREEDETECTORSDAYS}{\fixme{81.4}}

\newcommand{\HANFORDPOWER}{\fixme{37}}
\newcommand{\HANFORDPOWEROTWO}{\fixme{30}}
\newcommand{\LIVINGSTONPOWER}{\fixme{40}}
\newcommand{\LIVINGSTONPOWEROTWO}{\fixme{25}}
\newcommand{\VIRGOPOWER}{\fixme{19}}
\newcommand{\VIRGOPOWEROTWO}{\fixme{10}}
\newcommand{\VIRGOPOWERMAX}{\fixme{50}}

\newcommand{\VIRGORANGEPLOT}{\fixme{50 Mpc}}
\newcommand{\HANFORDRANGEPLOT}{\fixme{109 Mpc}}
\newcommand{\LIVINGSTONRANGEPLOT}{\fixme{136 Mpc}}

\newcommand{\HANFORDIMPROVEMENTHF}{\fixme{1.68}}
\newcommand{\LIVINGSTONIMPROVEMENTHF}{\fixme{1.96}}

\newcommand{\LLOOTWOGLITCHRATE}{\fixme{0.2 per minute}}
\newcommand{\LLOOTHREEGLITCHRATE}{\fixme{0.8 per minute}}

\newcommand{\BLIPRATE}{\fixme{1.4 per hour}}

\newcommand{\SCATTERINGEVENTS}{\fixme{7}}

\newcommand{\LHOCATONE}{\fixme{0.27}}
\newcommand{\LLOCATONE}{\fixme{0.08}}
\newcommand{\VIRGOCATONE}{\fixme{0.15}}

\newcommand{\LHOCATTWOCBC}{\fixme{0.37}}
\newcommand{\LLOCATTWOCBC}{\fixme{0.10}}
\newcommand{\VIRGOCATTWOCBC}{\fixme{--}}

\newcommand{\LHOCATTWOBURST}{\fixme{0.83}}
\newcommand{\LLOCATTWOBURST}{\fixme{0.64}}
\newcommand{\VIRGOCATTWOBURST}{\fixme{--}}

\newcommand{\LHOCATTHREEBURST}{\fixme{0.19}}
\newcommand{\LLOCATTHREEBURST}{\fixme{0.15}}
\newcommand{\VIRGOCATTHREEBURST}{\fixme{--}}


\newcommand{\VIRGONOISESUBRANGEINCREASE}{\fixme{3 Mpc}}

\newcommand{\GWTCTWOEVENTS}{\fixme{36}}
\newcommand{\GWTCTWOEVENTSTOT}{\fixme{39}}
\newcommand{\GWTCTHREEEVENTS}{\fixme{35}}

\DeclareRobustCommand{\NUMSUBBBHGSTLAL}[1]{4.13}
\DeclareRobustCommand{\NUMSUBNSBHGSTLAL}[1]{0.73}
\DeclareRobustCommand{\NUMSUBBNSGSTLAL}[1]{0.02}
\DeclareRobustCommand{\NUMALLBBHGSTLAL}[1]{34.59}
\DeclareRobustCommand{\NUMALLNSBHGSTLAL}[1]{1.54}
\DeclareRobustCommand{\NUMALLBNSGSTLAL}[1]{0.81}
\DeclareRobustCommand{\NUMSUBBBHMBTA}[1]{2.55}
\DeclareRobustCommand{\NUMSUBNSBHMBTA}[1]{0.51}
\DeclareRobustCommand{\NUMSUBBNSMBTA}[1]{0.49}
\DeclareRobustCommand{\NUMALLBBHMBTA}[1]{29.66}
\DeclareRobustCommand{\NUMALLNSBHMBTA}[1]{0.66}
\DeclareRobustCommand{\NUMALLBNSMBTA}[1]{0.49}
\DeclareRobustCommand{\NUMSUBBBHPYCBC}[1]{3.73}
\DeclareRobustCommand{\NUMSUBNSBHPYCBC}[1]{2.39}
\DeclareRobustCommand{\NUMSUBBNSPYCBC}[1]{0.22}
\DeclareRobustCommand{\NUMALLBBHPYCBC}[1]{24.95}
\DeclareRobustCommand{\NUMALLNSBHPYCBC}[1]{3.80}
\DeclareRobustCommand{\NUMALLBNSPYCBC}[1]{0.22}
\DeclareRobustCommand{\NUMSUBBBHPYCBCBBH}[1]{12.40}
\DeclareRobustCommand{\NUMSUBNSBHPYCBCBBH}[1]{0.36}
\DeclareRobustCommand{\NUMSUBBNSPYCBCBBH}[1]{0.00}
\DeclareRobustCommand{\NUMALLBBHPYCBCBBH}[1]{44.50}
\DeclareRobustCommand{\NUMALLNSBHPYCBCBBH}[1]{1.40}
\DeclareRobustCommand{\NUMALLBNSPYCBCBBH}[1]{0.00}

\newcommand{\chirpmassdetminus}[1]{\IfEqCase{#1}{{GW190926C_o3afin}{8.4}{GW190925J_o3afin}{0.7}{GW190917B_o3afin}{0.03}{GW190916K_o3afin}{12.9}{GW190805J_o3afin}{12.4}{GW190725F_o3afin}{0.3}{GW190426N_o3afin}{27.5}{GW190403B_o3afin}{31.0}{GW150914_o3afin}{1.6}{GW151012_o3afin}{1.5}{GW151226_o3afin}{0.06}{GW170104_o3afin}{1.7}{GW170608_o3afin}{0.05}{GW170729_o3afin}{12.3}{GW170809_o3afin}{2.0}{GW170814_o3afin}{1.3}{GW170818_o3afin}{2.5}{GW170823_o3afin}{4.8}{GW190930C_o3afin}{0.2}{GW190929B_o3afin}{13.9}{GW190924A_o3afin}{0.02}{GW190915K_o3afin}{3.2}{GW190910B_o3afin}{3.9}{GW190828B_o3afin}{0.7}{GW190828A_o3afin}{2.5}{GW190814H_o3afin}{0.02}{GW190803B_o3afin}{6.2}{GW190731E_o3afin}{8.7}{GW190728D_o3afin}{0.08}{GW190727B_o3afin}{5.7}{GW190720A_o3afin}{0.1}{GW190719H_o3afin}{7.2}{GW190708M_o3afin}{0.2}{GW190707E_o3afin}{0.10}{GW190706F_o3afin}{19.0}{GW190701E_o3afin}{8.1}{GW190630E_o3afin}{1.7}{GW190620B_o3afin}{14.2}{GW190602E_o3afin}{17.9}{GW190527H_o3afin}{5.5}{GW190521E_o3afin}{2.6}{GW190521B_o3afin}{33.7}{GW190519J_o3afin}{10.9}{GW190517B_o3afin}{5.6}{GW190514E_o3afin}{10.2}{GW190513E_o3afin}{3.2}{GW190512G_o3afin}{0.7}{GW190503E_o3afin}{6.7}{GW190425B_o3afin}{0.0006}{GW190421I_o3afin}{6.2}{GW190413E_o3afin}{14.4}{GW190413A_o3afin}{6.1}{GW190412B_o3afin}{0.2}{GW190408H_o3afin}{1.5}}}
\newcommand{\chirpmassdetmed}[1]{\IfEqCase{#1}{{GW190926C_o3afin}{37.9}{GW190925J_o3afin}{18.5}{GW190917B_o3afin}{4.21}{GW190916K_o3afin}{48.3}{GW190805J_o3afin}{62.1}{GW190725F_o3afin}{9.0}{GW190426N_o3afin}{131.9}{GW190403B_o3afin}{77.3}{GW150914_o3afin}{30.7}{GW151012_o3afin}{18.6}{GW151226_o3afin}{9.71}{GW170104_o3afin}{25.7}{GW170608_o3afin}{8.50}{GW170729_o3afin}{50.9}{GW170809_o3afin}{29.9}{GW170814_o3afin}{27.0}{GW170818_o3afin}{32.5}{GW170823_o3afin}{38.8}{GW190930C_o3afin}{9.9}{GW190929B_o3afin}{55.5}{GW190924A_o3afin}{6.44}{GW190915K_o3afin}{32.4}{GW190910B_o3afin}{43.3}{GW190828B_o3afin}{17.3}{GW190828A_o3afin}{33.8}{GW190814H_o3afin}{6.42}{GW190803B_o3afin}{42.8}{GW190731E_o3afin}{46.8}{GW190728D_o3afin}{10.1}{GW190727B_o3afin}{44.9}{GW190720A_o3afin}{10.4}{GW190719H_o3afin}{36.9}{GW190708M_o3afin}{15.5}{GW190707E_o3afin}{9.90}{GW190706F_o3afin}{74.7}{GW190701E_o3afin}{55.5}{GW190630E_o3afin}{29.5}{GW190620B_o3afin}{58.4}{GW190602E_o3afin}{72.8}{GW190527H_o3afin}{34.6}{GW190521E_o3afin}{39.7}{GW190521B_o3afin}{101.0}{GW190519J_o3afin}{65.1}{GW190517B_o3afin}{35.7}{GW190514E_o3afin}{48.6}{GW190513E_o3afin}{30.4}{GW190512G_o3afin}{18.6}{GW190503E_o3afin}{37.8}{GW190425B_o3afin}{1.49}{GW190421I_o3afin}{45.9}{GW190413E_o3afin}{55.3}{GW190413A_o3afin}{38.5}{GW190412B_o3afin}{15.2}{GW190408H_o3afin}{23.8}}}
\newcommand{\chirpmassdetplus}[1]{\IfEqCase{#1}{{GW190926C_o3afin}{17.2}{GW190925J_o3afin}{0.8}{GW190917B_o3afin}{0.03}{GW190916K_o3afin}{17.3}{GW190805J_o3afin}{9.5}{GW190725F_o3afin}{0.1}{GW190426N_o3afin}{23.6}{GW190403B_o3afin}{30.2}{GW150914_o3afin}{1.8}{GW151012_o3afin}{2.9}{GW151226_o3afin}{0.08}{GW170104_o3afin}{1.7}{GW170608_o3afin}{0.05}{GW170729_o3afin}{9.7}{GW170809_o3afin}{2.4}{GW170814_o3afin}{1.5}{GW170818_o3afin}{2.8}{GW170823_o3afin}{5.2}{GW190930C_o3afin}{0.2}{GW190929B_o3afin}{15.2}{GW190924A_o3afin}{0.03}{GW190915K_o3afin}{3.1}{GW190910B_o3afin}{4.3}{GW190828B_o3afin}{0.6}{GW190828A_o3afin}{2.7}{GW190814H_o3afin}{0.02}{GW190803B_o3afin}{5.6}{GW190731E_o3afin}{8.4}{GW190728D_o3afin}{0.1}{GW190727B_o3afin}{5.4}{GW190720A_o3afin}{0.1}{GW190719H_o3afin}{12.0}{GW190708M_o3afin}{0.2}{GW190707E_o3afin}{0.1}{GW190706F_o3afin}{13.5}{GW190701E_o3afin}{7.3}{GW190630E_o3afin}{2.1}{GW190620B_o3afin}{9.0}{GW190602E_o3afin}{12.3}{GW190527H_o3afin}{13.5}{GW190521E_o3afin}{3.0}{GW190521B_o3afin}{29.0}{GW190519J_o3afin}{8.7}{GW190517B_o3afin}{4.3}{GW190514E_o3afin}{8.0}{GW190513E_o3afin}{5.9}{GW190512G_o3afin}{0.7}{GW190503E_o3afin}{6.1}{GW190425B_o3afin}{0.0007}{GW190421I_o3afin}{5.7}{GW190413E_o3afin}{10.2}{GW190413A_o3afin}{6.7}{GW190412B_o3afin}{0.4}{GW190408H_o3afin}{1.2}}}
\newcommand{\massratiominus}[1]{\IfEqCase{#1}{{GW190926C_o3afin}{0.27}{GW190925J_o3afin}{0.30}{GW190917B_o3afin}{0.09}{GW190916K_o3afin}{0.32}{GW190805J_o3afin}{0.33}{GW190725F_o3afin}{0.36}{GW190426N_o3afin}{0.48}{GW190403B_o3afin}{0.12}{GW150914_o3afin}{0.22}{GW151012_o3afin}{0.32}{GW151226_o3afin}{0.34}{GW170104_o3afin}{0.26}{GW170608_o3afin}{0.33}{GW170729_o3afin}{0.22}{GW170809_o3afin}{0.25}{GW170814_o3afin}{0.23}{GW170818_o3afin}{0.24}{GW170823_o3afin}{0.30}{GW190930C_o3afin}{0.27}{GW190929B_o3afin}{0.20}{GW190924A_o3afin}{0.30}{GW190915K_o3afin}{0.29}{GW190910B_o3afin}{0.23}{GW190828B_o3afin}{0.16}{GW190828A_o3afin}{0.24}{GW190814H_o3afin}{0.01}{GW190803B_o3afin}{0.31}{GW190731E_o3afin}{0.32}{GW190728D_o3afin}{0.36}{GW190727B_o3afin}{0.30}{GW190720A_o3afin}{0.24}{GW190719H_o3afin}{0.41}{GW190708M_o3afin}{0.18}{GW190707E_o3afin}{0.20}{GW190706F_o3afin}{0.25}{GW190701E_o3afin}{0.31}{GW190630E_o3afin}{0.22}{GW190620B_o3afin}{0.32}{GW190602E_o3afin}{0.34}{GW190527H_o3afin}{0.36}{GW190521E_o3afin}{0.21}{GW190521B_o3afin}{0.38}{GW190519J_o3afin}{0.22}{GW190517B_o3afin}{0.30}{GW190514E_o3afin}{0.35}{GW190513E_o3afin}{0.19}{GW190512G_o3afin}{0.18}{GW190503E_o3afin}{0.29}{GW190425B_o3afin}{0.20}{GW190421I_o3afin}{0.32}{GW190413E_o3afin}{0.32}{GW190413A_o3afin}{0.31}{GW190412B_o3afin}{0.10}{GW190408H_o3afin}{0.26}}}
\newcommand{\massratiomed}[1]{\IfEqCase{#1}{{GW190926C_o3afin}{0.50}{GW190925J_o3afin}{0.75}{GW190917B_o3afin}{0.21}{GW190916K_o3afin}{0.55}{GW190805J_o3afin}{0.68}{GW190725F_o3afin}{0.53}{GW190426N_o3afin}{0.76}{GW190403B_o3afin}{0.23}{GW150914_o3afin}{0.88}{GW151012_o3afin}{0.55}{GW151226_o3afin}{0.53}{GW170104_o3afin}{0.73}{GW170608_o3afin}{0.74}{GW170729_o3afin}{0.55}{GW170809_o3afin}{0.71}{GW170814_o3afin}{0.81}{GW170818_o3afin}{0.80}{GW170823_o3afin}{0.78}{GW190930C_o3afin}{0.49}{GW190929B_o3afin}{0.40}{GW190924A_o3afin}{0.58}{GW190915K_o3afin}{0.76}{GW190910B_o3afin}{0.79}{GW190828B_o3afin}{0.44}{GW190828A_o3afin}{0.82}{GW190814H_o3afin}{0.11}{GW190803B_o3afin}{0.75}{GW190731E_o3afin}{0.72}{GW190728D_o3afin}{0.64}{GW190727B_o3afin}{0.79}{GW190720A_o3afin}{0.53}{GW190719H_o3afin}{0.56}{GW190708M_o3afin}{0.58}{GW190707E_o3afin}{0.66}{GW190706F_o3afin}{0.54}{GW190701E_o3afin}{0.76}{GW190630E_o3afin}{0.68}{GW190620B_o3afin}{0.61}{GW190602E_o3afin}{0.63}{GW190527H_o3afin}{0.64}{GW190521E_o3afin}{0.77}{GW190521B_o3afin}{0.59}{GW190519J_o3afin}{0.63}{GW190517B_o3afin}{0.62}{GW190514E_o3afin}{0.71}{GW190513E_o3afin}{0.51}{GW190512G_o3afin}{0.54}{GW190503E_o3afin}{0.69}{GW190425B_o3afin}{0.64}{GW190421I_o3afin}{0.78}{GW190413E_o3afin}{0.60}{GW190413A_o3afin}{0.73}{GW190412B_o3afin}{0.33}{GW190408H_o3afin}{0.75}}}
\newcommand{\massratioplus}[1]{\IfEqCase{#1}{{GW190926C_o3afin}{0.42}{GW190925J_o3afin}{0.22}{GW190917B_o3afin}{0.32}{GW190916K_o3afin}{0.40}{GW190805J_o3afin}{0.28}{GW190725F_o3afin}{0.41}{GW190426N_o3afin}{0.22}{GW190403B_o3afin}{0.57}{GW150914_o3afin}{0.11}{GW151012_o3afin}{0.38}{GW151226_o3afin}{0.41}{GW170104_o3afin}{0.23}{GW170608_o3afin}{0.23}{GW170729_o3afin}{0.37}{GW170809_o3afin}{0.25}{GW170814_o3afin}{0.17}{GW170818_o3afin}{0.18}{GW170823_o3afin}{0.20}{GW190930C_o3afin}{0.43}{GW190929B_o3afin}{0.34}{GW190924A_o3afin}{0.32}{GW190915K_o3afin}{0.21}{GW190910B_o3afin}{0.18}{GW190828B_o3afin}{0.38}{GW190828A_o3afin}{0.16}{GW190814H_o3afin}{0.01}{GW190803B_o3afin}{0.22}{GW190731E_o3afin}{0.25}{GW190728D_o3afin}{0.30}{GW190727B_o3afin}{0.19}{GW190720A_o3afin}{0.36}{GW190719H_o3afin}{0.39}{GW190708M_o3afin}{0.36}{GW190707E_o3afin}{0.27}{GW190706F_o3afin}{0.34}{GW190701E_o3afin}{0.21}{GW190630E_o3afin}{0.28}{GW190620B_o3afin}{0.34}{GW190602E_o3afin}{0.32}{GW190527H_o3afin}{0.31}{GW190521E_o3afin}{0.18}{GW190521B_o3afin}{0.33}{GW190519J_o3afin}{0.26}{GW190517B_o3afin}{0.32}{GW190514E_o3afin}{0.26}{GW190513E_o3afin}{0.41}{GW190512G_o3afin}{0.36}{GW190503E_o3afin}{0.27}{GW190425B_o3afin}{0.31}{GW190421I_o3afin}{0.20}{GW190413E_o3afin}{0.35}{GW190413A_o3afin}{0.24}{GW190412B_o3afin}{0.17}{GW190408H_o3afin}{0.21}}}
\newcommand{\spinoneminus}[1]{\IfEqCase{#1}{{GW190926C_o3afin}{0.36}{GW190925J_o3afin}{0.32}{GW190917B_o3afin}{0.21}{GW190916K_o3afin}{0.43}{GW190805J_o3afin}{0.59}{GW190725F_o3afin}{0.32}{GW190426N_o3afin}{0.58}{GW190403B_o3afin}{0.31}{GW150914_o3afin}{0.41}{GW151012_o3afin}{0.34}{GW151226_o3afin}{0.46}{GW170104_o3afin}{0.32}{GW170608_o3afin}{0.26}{GW170729_o3afin}{0.53}{GW170809_o3afin}{0.32}{GW170814_o3afin}{0.37}{GW170818_o3afin}{0.47}{GW170823_o3afin}{0.39}{GW190930C_o3afin}{0.34}{GW190929B_o3afin}{0.31}{GW190924A_o3afin}{0.21}{GW190915K_o3afin}{0.49}{GW190910B_o3afin}{0.29}{GW190828B_o3afin}{0.24}{GW190828A_o3afin}{0.39}{GW190814H_o3afin}{0.03}{GW190803B_o3afin}{0.38}{GW190731E_o3afin}{0.35}{GW190728D_o3afin}{0.29}{GW190727B_o3afin}{0.45}{GW190720A_o3afin}{0.31}{GW190719H_o3afin}{0.52}{GW190708M_o3afin}{0.20}{GW190707E_o3afin}{0.18}{GW190706F_o3afin}{0.51}{GW190701E_o3afin}{0.40}{GW190630E_o3afin}{0.24}{GW190620B_o3afin}{0.55}{GW190602E_o3afin}{0.42}{GW190527H_o3afin}{0.33}{GW190521E_o3afin}{0.30}{GW190521B_o3afin}{0.63}{GW190519J_o3afin}{0.46}{GW190517B_o3afin}{0.30}{GW190514E_o3afin}{0.42}{GW190513E_o3afin}{0.37}{GW190512G_o3afin}{0.18}{GW190503E_o3afin}{0.39}{GW190425B_o3afin}{0.23}{GW190421I_o3afin}{0.39}{GW190413E_o3afin}{0.54}{GW190413A_o3afin}{0.38}{GW190412B_o3afin}{0.20}{GW190408H_o3afin}{0.28}}}
\newcommand{\spinonemed}[1]{\IfEqCase{#1}{{GW190926C_o3afin}{0.39}{GW190925J_o3afin}{0.35}{GW190917B_o3afin}{0.23}{GW190916K_o3afin}{0.48}{GW190805J_o3afin}{0.75}{GW190725F_o3afin}{0.35}{GW190426N_o3afin}{0.67}{GW190403B_o3afin}{0.89}{GW150914_o3afin}{0.45}{GW151012_o3afin}{0.39}{GW151226_o3afin}{0.61}{GW170104_o3afin}{0.35}{GW170608_o3afin}{0.29}{GW170729_o3afin}{0.60}{GW170809_o3afin}{0.35}{GW170814_o3afin}{0.41}{GW170818_o3afin}{0.52}{GW170823_o3afin}{0.44}{GW190930C_o3afin}{0.39}{GW190929B_o3afin}{0.34}{GW190924A_o3afin}{0.23}{GW190915K_o3afin}{0.55}{GW190910B_o3afin}{0.32}{GW190828B_o3afin}{0.27}{GW190828A_o3afin}{0.44}{GW190814H_o3afin}{0.03}{GW190803B_o3afin}{0.42}{GW190731E_o3afin}{0.39}{GW190728D_o3afin}{0.33}{GW190727B_o3afin}{0.50}{GW190720A_o3afin}{0.36}{GW190719H_o3afin}{0.60}{GW190708M_o3afin}{0.23}{GW190707E_o3afin}{0.20}{GW190706F_o3afin}{0.64}{GW190701E_o3afin}{0.44}{GW190630E_o3afin}{0.27}{GW190620B_o3afin}{0.70}{GW190602E_o3afin}{0.46}{GW190527H_o3afin}{0.37}{GW190521E_o3afin}{0.33}{GW190521B_o3afin}{0.71}{GW190519J_o3afin}{0.61}{GW190517B_o3afin}{0.90}{GW190514E_o3afin}{0.47}{GW190513E_o3afin}{0.42}{GW190512G_o3afin}{0.20}{GW190503E_o3afin}{0.43}{GW190425B_o3afin}{0.26}{GW190421I_o3afin}{0.43}{GW190413E_o3afin}{0.60}{GW190413A_o3afin}{0.42}{GW190412B_o3afin}{0.31}{GW190408H_o3afin}{0.31}}}
\newcommand{\spinoneplus}[1]{\IfEqCase{#1}{{GW190926C_o3afin}{0.52}{GW190925J_o3afin}{0.50}{GW190917B_o3afin}{0.63}{GW190916K_o3afin}{0.46}{GW190805J_o3afin}{0.22}{GW190725F_o3afin}{0.52}{GW190426N_o3afin}{0.30}{GW190403B_o3afin}{0.09}{GW150914_o3afin}{0.43}{GW151012_o3afin}{0.49}{GW151226_o3afin}{0.34}{GW170104_o3afin}{0.49}{GW170608_o3afin}{0.44}{GW170729_o3afin}{0.35}{GW170809_o3afin}{0.52}{GW170814_o3afin}{0.49}{GW170818_o3afin}{0.41}{GW170823_o3afin}{0.48}{GW190930C_o3afin}{0.39}{GW190929B_o3afin}{0.54}{GW190924A_o3afin}{0.46}{GW190915K_o3afin}{0.39}{GW190910B_o3afin}{0.52}{GW190828B_o3afin}{0.45}{GW190828A_o3afin}{0.47}{GW190814H_o3afin}{0.07}{GW190803B_o3afin}{0.50}{GW190731E_o3afin}{0.52}{GW190728D_o3afin}{0.38}{GW190727B_o3afin}{0.44}{GW190720A_o3afin}{0.37}{GW190719H_o3afin}{0.36}{GW190708M_o3afin}{0.47}{GW190707E_o3afin}{0.47}{GW190706F_o3afin}{0.32}{GW190701E_o3afin}{0.48}{GW190630E_o3afin}{0.44}{GW190620B_o3afin}{0.27}{GW190602E_o3afin}{0.47}{GW190527H_o3afin}{0.53}{GW190521E_o3afin}{0.48}{GW190521B_o3afin}{0.26}{GW190519J_o3afin}{0.33}{GW190517B_o3afin}{0.09}{GW190514E_o3afin}{0.47}{GW190513E_o3afin}{0.46}{GW190512G_o3afin}{0.49}{GW190503E_o3afin}{0.48}{GW190425B_o3afin}{0.47}{GW190421I_o3afin}{0.49}{GW190413E_o3afin}{0.36}{GW190413A_o3afin}{0.50}{GW190412B_o3afin}{0.22}{GW190408H_o3afin}{0.51}}}
\newcommand{\spintwominus}[1]{\IfEqCase{#1}{{GW190926C_o3afin}{0.43}{GW190925J_o3afin}{0.37}{GW190917B_o3afin}{0.43}{GW190916K_o3afin}{0.47}{GW190805J_o3afin}{0.52}{GW190725F_o3afin}{0.44}{GW190426N_o3afin}{0.50}{GW190403B_o3afin}{0.47}{GW150914_o3afin}{0.37}{GW151012_o3afin}{0.41}{GW151226_o3afin}{0.41}{GW170104_o3afin}{0.34}{GW170608_o3afin}{0.28}{GW170729_o3afin}{0.44}{GW170809_o3afin}{0.36}{GW170814_o3afin}{0.38}{GW170818_o3afin}{0.42}{GW170823_o3afin}{0.39}{GW190930C_o3afin}{0.41}{GW190929B_o3afin}{0.43}{GW190924A_o3afin}{0.30}{GW190915K_o3afin}{0.40}{GW190910B_o3afin}{0.33}{GW190828B_o3afin}{0.39}{GW190828A_o3afin}{0.37}{GW190814H_o3afin}{0.43}{GW190803B_o3afin}{0.40}{GW190731E_o3afin}{0.42}{GW190728D_o3afin}{0.34}{GW190727B_o3afin}{0.42}{GW190720A_o3afin}{0.42}{GW190719H_o3afin}{0.46}{GW190708M_o3afin}{0.30}{GW190707E_o3afin}{0.31}{GW190706F_o3afin}{0.46}{GW190701E_o3afin}{0.41}{GW190630E_o3afin}{0.38}{GW190620B_o3afin}{0.51}{GW190602E_o3afin}{0.46}{GW190527H_o3afin}{0.36}{GW190521E_o3afin}{0.38}{GW190521B_o3afin}{0.48}{GW190519J_o3afin}{0.52}{GW190517B_o3afin}{0.54}{GW190514E_o3afin}{0.43}{GW190513E_o3afin}{0.41}{GW190512G_o3afin}{0.37}{GW190503E_o3afin}{0.40}{GW190425B_o3afin}{0.28}{GW190421I_o3afin}{0.41}{GW190413E_o3afin}{0.44}{GW190413A_o3afin}{0.41}{GW190412B_o3afin}{0.37}{GW190408H_o3afin}{0.34}}}
\newcommand{\spintwomed}[1]{\IfEqCase{#1}{{GW190926C_o3afin}{0.48}{GW190925J_o3afin}{0.41}{GW190917B_o3afin}{0.48}{GW190916K_o3afin}{0.52}{GW190805J_o3afin}{0.59}{GW190725F_o3afin}{0.49}{GW190426N_o3afin}{0.55}{GW190403B_o3afin}{0.53}{GW150914_o3afin}{0.42}{GW151012_o3afin}{0.46}{GW151226_o3afin}{0.45}{GW170104_o3afin}{0.38}{GW170608_o3afin}{0.31}{GW170729_o3afin}{0.48}{GW170809_o3afin}{0.40}{GW170814_o3afin}{0.42}{GW170818_o3afin}{0.48}{GW170823_o3afin}{0.43}{GW190930C_o3afin}{0.46}{GW190929B_o3afin}{0.47}{GW190924A_o3afin}{0.33}{GW190915K_o3afin}{0.45}{GW190910B_o3afin}{0.36}{GW190828B_o3afin}{0.44}{GW190828A_o3afin}{0.41}{GW190814H_o3afin}{0.47}{GW190803B_o3afin}{0.44}{GW190731E_o3afin}{0.46}{GW190728D_o3afin}{0.37}{GW190727B_o3afin}{0.47}{GW190720A_o3afin}{0.47}{GW190719H_o3afin}{0.51}{GW190708M_o3afin}{0.34}{GW190707E_o3afin}{0.33}{GW190706F_o3afin}{0.51}{GW190701E_o3afin}{0.45}{GW190630E_o3afin}{0.43}{GW190620B_o3afin}{0.57}{GW190602E_o3afin}{0.52}{GW190527H_o3afin}{0.39}{GW190521E_o3afin}{0.43}{GW190521B_o3afin}{0.53}{GW190519J_o3afin}{0.59}{GW190517B_o3afin}{0.62}{GW190514E_o3afin}{0.48}{GW190513E_o3afin}{0.45}{GW190512G_o3afin}{0.40}{GW190503E_o3afin}{0.44}{GW190425B_o3afin}{0.32}{GW190421I_o3afin}{0.46}{GW190413E_o3afin}{0.49}{GW190413A_o3afin}{0.46}{GW190412B_o3afin}{0.41}{GW190408H_o3afin}{0.37}}}
\newcommand{\spintwoplus}[1]{\IfEqCase{#1}{{GW190926C_o3afin}{0.45}{GW190925J_o3afin}{0.50}{GW190917B_o3afin}{0.45}{GW190916K_o3afin}{0.42}{GW190805J_o3afin}{0.37}{GW190725F_o3afin}{0.45}{GW190426N_o3afin}{0.40}{GW190403B_o3afin}{0.43}{GW150914_o3afin}{0.49}{GW151012_o3afin}{0.46}{GW151226_o3afin}{0.47}{GW170104_o3afin}{0.52}{GW170608_o3afin}{0.52}{GW170729_o3afin}{0.46}{GW170809_o3afin}{0.51}{GW170814_o3afin}{0.50}{GW170818_o3afin}{0.46}{GW170823_o3afin}{0.49}{GW190930C_o3afin}{0.46}{GW190929B_o3afin}{0.46}{GW190924A_o3afin}{0.53}{GW190915K_o3afin}{0.48}{GW190910B_o3afin}{0.53}{GW190828B_o3afin}{0.49}{GW190828A_o3afin}{0.49}{GW190814H_o3afin}{0.46}{GW190803B_o3afin}{0.49}{GW190731E_o3afin}{0.47}{GW190728D_o3afin}{0.51}{GW190727B_o3afin}{0.47}{GW190720A_o3afin}{0.45}{GW190719H_o3afin}{0.43}{GW190708M_o3afin}{0.53}{GW190707E_o3afin}{0.54}{GW190706F_o3afin}{0.43}{GW190701E_o3afin}{0.48}{GW190630E_o3afin}{0.48}{GW190620B_o3afin}{0.39}{GW190602E_o3afin}{0.43}{GW190527H_o3afin}{0.52}{GW190521E_o3afin}{0.47}{GW190521B_o3afin}{0.42}{GW190519J_o3afin}{0.36}{GW190517B_o3afin}{0.34}{GW190514E_o3afin}{0.47}{GW190513E_o3afin}{0.47}{GW190512G_o3afin}{0.51}{GW190503E_o3afin}{0.49}{GW190425B_o3afin}{0.44}{GW190421I_o3afin}{0.47}{GW190413E_o3afin}{0.44}{GW190413A_o3afin}{0.47}{GW190412B_o3afin}{0.50}{GW190408H_o3afin}{0.52}}}
\newcommand{\tiltoneminus}[1]{\IfEqCase{#1}{{GW190926C_o3afin}{1.02}{GW190925J_o3afin}{0.87}{GW190917B_o3afin}{1.34}{GW190916K_o3afin}{0.75}{GW190805J_o3afin}{0.56}{GW190725F_o3afin}{1.10}{GW190426N_o3afin}{0.77}{GW190403B_o3afin}{0.29}{GW150914_o3afin}{0.99}{GW151012_o3afin}{0.82}{GW151226_o3afin}{0.63}{GW170104_o3afin}{1.02}{GW170608_o3afin}{0.88}{GW170729_o3afin}{0.60}{GW170809_o3afin}{0.90}{GW170814_o3afin}{0.84}{GW170818_o3afin}{0.96}{GW170823_o3afin}{0.96}{GW190930C_o3afin}{0.71}{GW190929B_o3afin}{1.00}{GW190924A_o3afin}{0.98}{GW190915K_o3afin}{0.91}{GW190910B_o3afin}{1.00}{GW190828B_o3afin}{0.90}{GW190828A_o3afin}{0.74}{GW190814H_o3afin}{1.23}{GW190803B_o3afin}{1.03}{GW190731E_o3afin}{0.92}{GW190728D_o3afin}{0.75}{GW190727B_o3afin}{0.90}{GW190720A_o3afin}{0.71}{GW190719H_o3afin}{0.67}{GW190708M_o3afin}{0.90}{GW190707E_o3afin}{1.07}{GW190706F_o3afin}{0.65}{GW190701E_o3afin}{1.10}{GW190630E_o3afin}{0.88}{GW190620B_o3afin}{0.63}{GW190602E_o3afin}{0.87}{GW190527H_o3afin}{0.85}{GW190521E_o3afin}{0.92}{GW190521B_o3afin}{1.51}{GW190519J_o3afin}{0.61}{GW190517B_o3afin}{0.47}{GW190514E_o3afin}{1.14}{GW190513E_o3afin}{0.74}{GW190512G_o3afin}{0.99}{GW190503E_o3afin}{1.10}{GW190425B_o3afin}{0.78}{GW190421I_o3afin}{1.08}{GW190413E_o3afin}{0.99}{GW190413A_o3afin}{1.08}{GW190412B_o3afin}{0.43}{GW190408H_o3afin}{1.05}}}
\newcommand{\tiltonemed}[1]{\IfEqCase{#1}{{GW190926C_o3afin}{1.63}{GW190925J_o3afin}{1.31}{GW190917B_o3afin}{1.98}{GW190916K_o3afin}{1.01}{GW190805J_o3afin}{0.81}{GW190725F_o3afin}{1.57}{GW190426N_o3afin}{1.04}{GW190403B_o3afin}{0.38}{GW150914_o3afin}{1.73}{GW151012_o3afin}{1.18}{GW151226_o3afin}{1.02}{GW170104_o3afin}{1.76}{GW170608_o3afin}{1.36}{GW170729_o3afin}{0.83}{GW170809_o3afin}{1.37}{GW170814_o3afin}{1.28}{GW170818_o3afin}{1.68}{GW170823_o3afin}{1.45}{GW190930C_o3afin}{0.97}{GW190929B_o3afin}{1.69}{GW190924A_o3afin}{1.47}{GW190915K_o3afin}{1.61}{GW190910B_o3afin}{1.54}{GW190828B_o3afin}{1.36}{GW190828A_o3afin}{1.09}{GW190814H_o3afin}{1.57}{GW190803B_o3afin}{1.62}{GW190731E_o3afin}{1.36}{GW190728D_o3afin}{1.05}{GW190727B_o3afin}{1.33}{GW190720A_o3afin}{1.00}{GW190719H_o3afin}{0.96}{GW190708M_o3afin}{1.35}{GW190707E_o3afin}{1.73}{GW190706F_o3afin}{0.92}{GW190701E_o3afin}{1.81}{GW190630E_o3afin}{1.29}{GW190620B_o3afin}{0.87}{GW190602E_o3afin}{1.30}{GW190527H_o3afin}{1.26}{GW190521E_o3afin}{1.44}{GW190521B_o3afin}{2.10}{GW190519J_o3afin}{0.86}{GW190517B_o3afin}{0.66}{GW190514E_o3afin}{1.80}{GW190513E_o3afin}{1.06}{GW190512G_o3afin}{1.49}{GW190503E_o3afin}{1.78}{GW190425B_o3afin}{1.26}{GW190421I_o3afin}{1.83}{GW190413E_o3afin}{1.62}{GW190413A_o3afin}{1.65}{GW190412B_o3afin}{0.65}{GW190408H_o3afin}{1.71}}}
\newcommand{\tiltoneplus}[1]{\IfEqCase{#1}{{GW190926C_o3afin}{1.00}{GW190925J_o3afin}{1.06}{GW190917B_o3afin}{0.86}{GW190916K_o3afin}{1.27}{GW190805J_o3afin}{1.01}{GW190725F_o3afin}{0.95}{GW190426N_o3afin}{1.28}{GW190403B_o3afin}{0.70}{GW150914_o3afin}{0.91}{GW151012_o3afin}{1.16}{GW151226_o3afin}{0.64}{GW170104_o3afin}{0.90}{GW170608_o3afin}{0.97}{GW170729_o3afin}{1.12}{GW170809_o3afin}{1.10}{GW170814_o3afin}{1.01}{GW170818_o3afin}{0.90}{GW170823_o3afin}{1.06}{GW190930C_o3afin}{1.08}{GW190929B_o3afin}{0.98}{GW190924A_o3afin}{0.95}{GW190915K_o3afin}{0.86}{GW190910B_o3afin}{1.06}{GW190828B_o3afin}{1.03}{GW190828A_o3afin}{1.03}{GW190814H_o3afin}{1.20}{GW190803B_o3afin}{1.02}{GW190731E_o3afin}{1.17}{GW190728D_o3afin}{1.13}{GW190727B_o3afin}{1.09}{GW190720A_o3afin}{1.05}{GW190719H_o3afin}{1.07}{GW190708M_o3afin}{0.99}{GW190707E_o3afin}{0.93}{GW190706F_o3afin}{0.93}{GW190701E_o3afin}{0.92}{GW190630E_o3afin}{1.10}{GW190620B_o3afin}{0.86}{GW190602E_o3afin}{1.06}{GW190527H_o3afin}{1.11}{GW190521E_o3afin}{1.10}{GW190521B_o3afin}{0.71}{GW190519J_o3afin}{0.75}{GW190517B_o3afin}{0.58}{GW190514E_o3afin}{0.94}{GW190513E_o3afin}{1.15}{GW190512G_o3afin}{1.06}{GW190503E_o3afin}{0.93}{GW190425B_o3afin}{0.74}{GW190421I_o3afin}{0.90}{GW190413E_o3afin}{0.94}{GW190413A_o3afin}{1.00}{GW190412B_o3afin}{0.66}{GW190408H_o3afin}{0.93}}}
\newcommand{\tilttwominus}[1]{\IfEqCase{#1}{{GW190926C_o3afin}{1.14}{GW190925J_o3afin}{0.91}{GW190917B_o3afin}{1.19}{GW190916K_o3afin}{0.91}{GW190805J_o3afin}{0.83}{GW190725F_o3afin}{1.16}{GW190426N_o3afin}{0.91}{GW190403B_o3afin}{0.91}{GW150914_o3afin}{1.05}{GW151012_o3afin}{0.93}{GW151226_o3afin}{0.98}{GW170104_o3afin}{1.11}{GW170608_o3afin}{0.93}{GW170729_o3afin}{0.96}{GW170809_o3afin}{0.93}{GW170814_o3afin}{0.94}{GW170818_o3afin}{1.07}{GW170823_o3afin}{0.99}{GW190930C_o3afin}{0.83}{GW190929B_o3afin}{1.16}{GW190924A_o3afin}{0.97}{GW190915K_o3afin}{1.10}{GW190910B_o3afin}{1.09}{GW190828B_o3afin}{0.99}{GW190828A_o3afin}{0.88}{GW190814H_o3afin}{1.03}{GW190803B_o3afin}{1.11}{GW190731E_o3afin}{0.98}{GW190728D_o3afin}{0.81}{GW190727B_o3afin}{0.96}{GW190720A_o3afin}{0.74}{GW190719H_o3afin}{0.91}{GW190708M_o3afin}{0.96}{GW190707E_o3afin}{1.13}{GW190706F_o3afin}{0.90}{GW190701E_o3afin}{1.14}{GW190630E_o3afin}{0.82}{GW190620B_o3afin}{0.77}{GW190602E_o3afin}{0.95}{GW190527H_o3afin}{0.93}{GW190521E_o3afin}{0.82}{GW190521B_o3afin}{0.99}{GW190519J_o3afin}{0.73}{GW190517B_o3afin}{0.84}{GW190514E_o3afin}{1.17}{GW190513E_o3afin}{0.91}{GW190512G_o3afin}{1.00}{GW190503E_o3afin}{1.07}{GW190425B_o3afin}{0.78}{GW190421I_o3afin}{1.16}{GW190413E_o3afin}{1.07}{GW190413A_o3afin}{1.16}{GW190412B_o3afin}{0.88}{GW190408H_o3afin}{1.07}}}
\newcommand{\tilttwomed}[1]{\IfEqCase{#1}{{GW190926C_o3afin}{1.66}{GW190925J_o3afin}{1.32}{GW190917B_o3afin}{1.78}{GW190916K_o3afin}{1.24}{GW190805J_o3afin}{1.13}{GW190725F_o3afin}{1.71}{GW190426N_o3afin}{1.26}{GW190403B_o3afin}{1.21}{GW150914_o3afin}{1.61}{GW151012_o3afin}{1.33}{GW151226_o3afin}{1.38}{GW170104_o3afin}{1.62}{GW170608_o3afin}{1.36}{GW170729_o3afin}{1.30}{GW170809_o3afin}{1.35}{GW170814_o3afin}{1.46}{GW170818_o3afin}{1.71}{GW170823_o3afin}{1.43}{GW190930C_o3afin}{1.16}{GW190929B_o3afin}{1.65}{GW190924A_o3afin}{1.40}{GW190915K_o3afin}{1.66}{GW190910B_o3afin}{1.64}{GW190828B_o3afin}{1.41}{GW190828A_o3afin}{1.28}{GW190814H_o3afin}{1.46}{GW190803B_o3afin}{1.62}{GW190731E_o3afin}{1.42}{GW190728D_o3afin}{1.12}{GW190727B_o3afin}{1.40}{GW190720A_o3afin}{1.02}{GW190719H_o3afin}{1.25}{GW190708M_o3afin}{1.38}{GW190707E_o3afin}{1.68}{GW190706F_o3afin}{1.21}{GW190701E_o3afin}{1.74}{GW190630E_o3afin}{1.19}{GW190620B_o3afin}{1.04}{GW190602E_o3afin}{1.33}{GW190527H_o3afin}{1.34}{GW190521E_o3afin}{1.14}{GW190521B_o3afin}{1.43}{GW190519J_o3afin}{1.02}{GW190517B_o3afin}{1.14}{GW190514E_o3afin}{1.72}{GW190513E_o3afin}{1.28}{GW190512G_o3afin}{1.44}{GW190503E_o3afin}{1.58}{GW190425B_o3afin}{1.35}{GW190421I_o3afin}{1.82}{GW190413E_o3afin}{1.53}{GW190413A_o3afin}{1.69}{GW190412B_o3afin}{1.24}{GW190408H_o3afin}{1.60}}}
\newcommand{\tilttwoplus}[1]{\IfEqCase{#1}{{GW190926C_o3afin}{1.05}{GW190925J_o3afin}{1.13}{GW190917B_o3afin}{0.97}{GW190916K_o3afin}{1.27}{GW190805J_o3afin}{1.29}{GW190725F_o3afin}{0.95}{GW190426N_o3afin}{1.25}{GW190403B_o3afin}{1.29}{GW150914_o3afin}{1.01}{GW151012_o3afin}{1.20}{GW151226_o3afin}{1.24}{GW170104_o3afin}{1.02}{GW170608_o3afin}{1.15}{GW170729_o3afin}{1.26}{GW170809_o3afin}{1.18}{GW170814_o3afin}{1.09}{GW170818_o3afin}{0.97}{GW170823_o3afin}{1.11}{GW190930C_o3afin}{1.25}{GW190929B_o3afin}{1.06}{GW190924A_o3afin}{1.08}{GW190915K_o3afin}{1.02}{GW190910B_o3afin}{1.02}{GW190828B_o3afin}{1.19}{GW190828A_o3afin}{1.20}{GW190814H_o3afin}{1.13}{GW190803B_o3afin}{1.04}{GW190731E_o3afin}{1.17}{GW190728D_o3afin}{1.26}{GW190727B_o3afin}{1.15}{GW190720A_o3afin}{1.31}{GW190719H_o3afin}{1.26}{GW190708M_o3afin}{1.13}{GW190707E_o3afin}{0.93}{GW190706F_o3afin}{1.27}{GW190701E_o3afin}{0.99}{GW190630E_o3afin}{1.15}{GW190620B_o3afin}{1.36}{GW190602E_o3afin}{1.20}{GW190527H_o3afin}{1.19}{GW190521E_o3afin}{1.19}{GW190521B_o3afin}{1.19}{GW190519J_o3afin}{1.24}{GW190517B_o3afin}{1.34}{GW190514E_o3afin}{0.99}{GW190513E_o3afin}{1.24}{GW190512G_o3afin}{1.08}{GW190503E_o3afin}{1.08}{GW190425B_o3afin}{0.99}{GW190421I_o3afin}{0.93}{GW190413E_o3afin}{1.11}{GW190413A_o3afin}{1.02}{GW190412B_o3afin}{1.20}{GW190408H_o3afin}{1.02}}}
\newcommand{\phionetwominus}[1]{\IfEqCase{#1}{{GW190926C_o3afin}{2.81}{GW190925J_o3afin}{2.73}{GW190917B_o3afin}{2.84}{GW190916K_o3afin}{2.79}{GW190805J_o3afin}{2.79}{GW190725F_o3afin}{2.85}{GW190426N_o3afin}{2.90}{GW190403B_o3afin}{2.77}{GW150914_o3afin}{2.67}{GW151012_o3afin}{2.76}{GW151226_o3afin}{2.81}{GW170104_o3afin}{2.80}{GW170608_o3afin}{2.84}{GW170729_o3afin}{2.58}{GW170809_o3afin}{2.88}{GW170814_o3afin}{2.78}{GW170818_o3afin}{2.82}{GW170823_o3afin}{2.80}{GW190930C_o3afin}{2.89}{GW190929B_o3afin}{2.84}{GW190924A_o3afin}{2.82}{GW190915K_o3afin}{2.93}{GW190910B_o3afin}{2.67}{GW190828B_o3afin}{2.82}{GW190828A_o3afin}{2.60}{GW190814H_o3afin}{2.82}{GW190803B_o3afin}{2.78}{GW190731E_o3afin}{2.82}{GW190728D_o3afin}{2.77}{GW190727B_o3afin}{2.88}{GW190720A_o3afin}{2.80}{GW190719H_o3afin}{2.81}{GW190708M_o3afin}{2.79}{GW190707E_o3afin}{2.80}{GW190706F_o3afin}{2.82}{GW190701E_o3afin}{2.83}{GW190630E_o3afin}{2.69}{GW190620B_o3afin}{2.80}{GW190602E_o3afin}{2.88}{GW190527H_o3afin}{2.94}{GW190521E_o3afin}{2.62}{GW190521B_o3afin}{2.97}{GW190519J_o3afin}{2.67}{GW190517B_o3afin}{2.78}{GW190514E_o3afin}{2.83}{GW190513E_o3afin}{2.80}{GW190512G_o3afin}{2.81}{GW190503E_o3afin}{2.84}{GW190425B_o3afin}{2.42}{GW190421I_o3afin}{2.83}{GW190413E_o3afin}{2.84}{GW190413A_o3afin}{2.82}{GW190412B_o3afin}{2.90}{GW190408H_o3afin}{2.78}}}
\newcommand{\phionetwomed}[1]{\IfEqCase{#1}{{GW190926C_o3afin}{3.11}{GW190925J_o3afin}{3.08}{GW190917B_o3afin}{3.14}{GW190916K_o3afin}{3.13}{GW190805J_o3afin}{3.08}{GW190725F_o3afin}{3.15}{GW190426N_o3afin}{3.18}{GW190403B_o3afin}{3.06}{GW150914_o3afin}{3.01}{GW151012_o3afin}{3.09}{GW151226_o3afin}{3.12}{GW170104_o3afin}{3.15}{GW170608_o3afin}{3.14}{GW170729_o3afin}{2.92}{GW170809_o3afin}{3.19}{GW170814_o3afin}{3.11}{GW170818_o3afin}{3.10}{GW170823_o3afin}{3.11}{GW190930C_o3afin}{3.20}{GW190929B_o3afin}{3.16}{GW190924A_o3afin}{3.14}{GW190915K_o3afin}{3.24}{GW190910B_o3afin}{3.00}{GW190828B_o3afin}{3.15}{GW190828A_o3afin}{2.93}{GW190814H_o3afin}{3.12}{GW190803B_o3afin}{3.07}{GW190731E_o3afin}{3.15}{GW190728D_o3afin}{3.13}{GW190727B_o3afin}{3.20}{GW190720A_o3afin}{3.11}{GW190719H_o3afin}{3.14}{GW190708M_o3afin}{3.12}{GW190707E_o3afin}{3.14}{GW190706F_o3afin}{3.16}{GW190701E_o3afin}{3.14}{GW190630E_o3afin}{3.06}{GW190620B_o3afin}{3.13}{GW190602E_o3afin}{3.18}{GW190527H_o3afin}{3.20}{GW190521E_o3afin}{3.02}{GW190521B_o3afin}{3.25}{GW190519J_o3afin}{3.02}{GW190517B_o3afin}{3.02}{GW190514E_o3afin}{3.14}{GW190513E_o3afin}{3.11}{GW190512G_o3afin}{3.13}{GW190503E_o3afin}{3.17}{GW190425B_o3afin}{2.78}{GW190421I_o3afin}{3.13}{GW190413E_o3afin}{3.13}{GW190413A_o3afin}{3.15}{GW190412B_o3afin}{3.30}{GW190408H_o3afin}{3.11}}}
\newcommand{\phionetwoplus}[1]{\IfEqCase{#1}{{GW190926C_o3afin}{2.84}{GW190925J_o3afin}{2.89}{GW190917B_o3afin}{2.83}{GW190916K_o3afin}{2.82}{GW190805J_o3afin}{2.90}{GW190725F_o3afin}{2.83}{GW190426N_o3afin}{2.81}{GW190403B_o3afin}{2.95}{GW150914_o3afin}{2.94}{GW151012_o3afin}{2.86}{GW151226_o3afin}{2.85}{GW170104_o3afin}{2.78}{GW170608_o3afin}{2.78}{GW170729_o3afin}{3.02}{GW170809_o3afin}{2.77}{GW170814_o3afin}{2.84}{GW170818_o3afin}{2.89}{GW170823_o3afin}{2.87}{GW190930C_o3afin}{2.75}{GW190929B_o3afin}{2.82}{GW190924A_o3afin}{2.81}{GW190915K_o3afin}{2.76}{GW190910B_o3afin}{2.90}{GW190828B_o3afin}{2.79}{GW190828A_o3afin}{3.03}{GW190814H_o3afin}{2.85}{GW190803B_o3afin}{2.90}{GW190731E_o3afin}{2.80}{GW190728D_o3afin}{2.79}{GW190727B_o3afin}{2.80}{GW190720A_o3afin}{2.85}{GW190719H_o3afin}{2.84}{GW190708M_o3afin}{2.86}{GW190707E_o3afin}{2.77}{GW190706F_o3afin}{2.85}{GW190701E_o3afin}{2.84}{GW190630E_o3afin}{2.87}{GW190620B_o3afin}{2.82}{GW190602E_o3afin}{2.79}{GW190527H_o3afin}{2.75}{GW190521E_o3afin}{2.82}{GW190521B_o3afin}{2.75}{GW190519J_o3afin}{2.91}{GW190517B_o3afin}{3.01}{GW190514E_o3afin}{2.83}{GW190513E_o3afin}{2.85}{GW190512G_o3afin}{2.85}{GW190503E_o3afin}{2.80}{GW190425B_o3afin}{3.06}{GW190421I_o3afin}{2.84}{GW190413E_o3afin}{2.86}{GW190413A_o3afin}{2.83}{GW190412B_o3afin}{2.61}{GW190408H_o3afin}{2.84}}}
\newcommand{\phijlminus}[1]{\IfEqCase{#1}{{GW190926C_o3afin}{2.82}{GW190925J_o3afin}{2.85}{GW190917B_o3afin}{2.81}{GW190916K_o3afin}{2.79}{GW190805J_o3afin}{2.81}{GW190725F_o3afin}{2.90}{GW190426N_o3afin}{2.90}{GW190403B_o3afin}{2.77}{GW150914_o3afin}{2.02}{GW151012_o3afin}{2.84}{GW151226_o3afin}{2.66}{GW170104_o3afin}{2.71}{GW170608_o3afin}{2.84}{GW170729_o3afin}{2.80}{GW170809_o3afin}{3.44}{GW170814_o3afin}{2.85}{GW170818_o3afin}{3.13}{GW170823_o3afin}{2.81}{GW190930C_o3afin}{2.94}{GW190929B_o3afin}{2.88}{GW190924A_o3afin}{2.79}{GW190915K_o3afin}{2.60}{GW190910B_o3afin}{2.80}{GW190828B_o3afin}{2.88}{GW190828A_o3afin}{2.72}{GW190814H_o3afin}{2.03}{GW190803B_o3afin}{2.99}{GW190731E_o3afin}{2.71}{GW190728D_o3afin}{2.92}{GW190727B_o3afin}{2.63}{GW190720A_o3afin}{2.64}{GW190719H_o3afin}{2.81}{GW190708M_o3afin}{2.82}{GW190707E_o3afin}{2.83}{GW190706F_o3afin}{2.77}{GW190701E_o3afin}{2.29}{GW190630E_o3afin}{2.80}{GW190620B_o3afin}{2.90}{GW190602E_o3afin}{2.86}{GW190527H_o3afin}{2.75}{GW190521E_o3afin}{2.82}{GW190521B_o3afin}{2.88}{GW190519J_o3afin}{2.90}{GW190517B_o3afin}{2.24}{GW190514E_o3afin}{2.81}{GW190513E_o3afin}{2.82}{GW190512G_o3afin}{2.86}{GW190503E_o3afin}{3.47}{GW190425B_o3afin}{2.63}{GW190421I_o3afin}{2.51}{GW190413E_o3afin}{2.94}{GW190413A_o3afin}{2.93}{GW190412B_o3afin}{1.61}{GW190408H_o3afin}{2.97}}}
\newcommand{\phijlmed}[1]{\IfEqCase{#1}{{GW190926C_o3afin}{3.13}{GW190925J_o3afin}{3.20}{GW190917B_o3afin}{3.11}{GW190916K_o3afin}{3.10}{GW190805J_o3afin}{3.08}{GW190725F_o3afin}{3.22}{GW190426N_o3afin}{3.22}{GW190403B_o3afin}{3.08}{GW150914_o3afin}{2.19}{GW151012_o3afin}{3.15}{GW151226_o3afin}{2.99}{GW170104_o3afin}{3.04}{GW170608_o3afin}{3.16}{GW170729_o3afin}{3.12}{GW170809_o3afin}{3.78}{GW170814_o3afin}{3.10}{GW170818_o3afin}{3.80}{GW170823_o3afin}{3.11}{GW190930C_o3afin}{3.24}{GW190929B_o3afin}{3.22}{GW190924A_o3afin}{3.13}{GW190915K_o3afin}{3.04}{GW190910B_o3afin}{3.13}{GW190828B_o3afin}{3.16}{GW190828A_o3afin}{2.97}{GW190814H_o3afin}{2.43}{GW190803B_o3afin}{3.35}{GW190731E_o3afin}{3.04}{GW190728D_o3afin}{3.22}{GW190727B_o3afin}{2.97}{GW190720A_o3afin}{2.96}{GW190719H_o3afin}{3.13}{GW190708M_o3afin}{3.17}{GW190707E_o3afin}{3.16}{GW190706F_o3afin}{3.13}{GW190701E_o3afin}{2.66}{GW190630E_o3afin}{3.12}{GW190620B_o3afin}{3.19}{GW190602E_o3afin}{3.15}{GW190527H_o3afin}{3.08}{GW190521E_o3afin}{3.19}{GW190521B_o3afin}{3.21}{GW190519J_o3afin}{3.20}{GW190517B_o3afin}{2.62}{GW190514E_o3afin}{3.12}{GW190513E_o3afin}{3.15}{GW190512G_o3afin}{3.16}{GW190503E_o3afin}{3.78}{GW190425B_o3afin}{2.97}{GW190421I_o3afin}{2.83}{GW190413E_o3afin}{3.24}{GW190413A_o3afin}{3.24}{GW190412B_o3afin}{1.77}{GW190408H_o3afin}{3.27}}}
\newcommand{\phijlplus}[1]{\IfEqCase{#1}{{GW190926C_o3afin}{2.81}{GW190925J_o3afin}{2.77}{GW190917B_o3afin}{2.86}{GW190916K_o3afin}{2.90}{GW190805J_o3afin}{2.90}{GW190725F_o3afin}{2.74}{GW190426N_o3afin}{2.78}{GW190403B_o3afin}{2.88}{GW150914_o3afin}{3.90}{GW151012_o3afin}{2.82}{GW151226_o3afin}{2.97}{GW170104_o3afin}{2.90}{GW170608_o3afin}{2.81}{GW170729_o3afin}{2.84}{GW170809_o3afin}{2.23}{GW170814_o3afin}{2.96}{GW170818_o3afin}{2.03}{GW170823_o3afin}{2.84}{GW190930C_o3afin}{2.75}{GW190929B_o3afin}{2.73}{GW190924A_o3afin}{2.80}{GW190915K_o3afin}{2.86}{GW190910B_o3afin}{2.85}{GW190828B_o3afin}{2.84}{GW190828A_o3afin}{3.07}{GW190814H_o3afin}{3.37}{GW190803B_o3afin}{2.57}{GW190731E_o3afin}{2.91}{GW190728D_o3afin}{2.75}{GW190727B_o3afin}{2.98}{GW190720A_o3afin}{3.03}{GW190719H_o3afin}{2.80}{GW190708M_o3afin}{2.79}{GW190707E_o3afin}{2.82}{GW190706F_o3afin}{2.85}{GW190701E_o3afin}{3.22}{GW190630E_o3afin}{2.83}{GW190620B_o3afin}{2.81}{GW190602E_o3afin}{2.85}{GW190527H_o3afin}{2.89}{GW190521E_o3afin}{2.75}{GW190521B_o3afin}{2.72}{GW190519J_o3afin}{2.79}{GW190517B_o3afin}{3.17}{GW190514E_o3afin}{2.81}{GW190513E_o3afin}{2.81}{GW190512G_o3afin}{2.81}{GW190503E_o3afin}{2.22}{GW190425B_o3afin}{2.92}{GW190421I_o3afin}{3.09}{GW190413E_o3afin}{2.72}{GW190413A_o3afin}{2.75}{GW190412B_o3afin}{4.34}{GW190408H_o3afin}{2.72}}}
\newcommand{\thetajnminus}[1]{\IfEqCase{#1}{{GW190926C_o3afin}{1.19}{GW190925J_o3afin}{0.58}{GW190917B_o3afin}{1.16}{GW190916K_o3afin}{1.35}{GW190805J_o3afin}{0.74}{GW190725F_o3afin}{0.74}{GW190426N_o3afin}{1.72}{GW190403B_o3afin}{1.66}{GW150914_o3afin}{0.71}{GW151012_o3afin}{1.40}{GW151226_o3afin}{0.66}{GW170104_o3afin}{0.86}{GW170608_o3afin}{2.06}{GW170729_o3afin}{1.03}{GW170809_o3afin}{0.59}{GW170814_o3afin}{0.48}{GW170818_o3afin}{0.50}{GW170823_o3afin}{1.48}{GW190930C_o3afin}{0.55}{GW190929B_o3afin}{0.97}{GW190924A_o3afin}{0.64}{GW190915K_o3afin}{1.48}{GW190910B_o3afin}{1.25}{GW190828B_o3afin}{1.52}{GW190828A_o3afin}{2.02}{GW190814H_o3afin}{0.24}{GW190803B_o3afin}{0.70}{GW190731E_o3afin}{0.98}{GW190728D_o3afin}{0.90}{GW190727B_o3afin}{1.27}{GW190720A_o3afin}{1.99}{GW190719H_o3afin}{1.33}{GW190708M_o3afin}{1.18}{GW190707E_o3afin}{1.89}{GW190706F_o3afin}{1.50}{GW190701E_o3afin}{0.42}{GW190630E_o3afin}{1.16}{GW190620B_o3afin}{1.74}{GW190602E_o3afin}{1.87}{GW190527H_o3afin}{0.87}{GW190521E_o3afin}{1.10}{GW190521B_o3afin}{1.07}{GW190519J_o3afin}{1.02}{GW190517B_o3afin}{1.18}{GW190514E_o3afin}{1.20}{GW190513E_o3afin}{0.58}{GW190512G_o3afin}{1.50}{GW190503E_o3afin}{0.62}{GW190425B_o3afin}{1.38}{GW190421I_o3afin}{1.70}{GW190413E_o3afin}{1.52}{GW190413A_o3afin}{0.58}{GW190412B_o3afin}{0.40}{GW190408H_o3afin}{0.79}}}
\newcommand{\thetajnmed}[1]{\IfEqCase{#1}{{GW190926C_o3afin}{1.67}{GW190925J_o3afin}{0.77}{GW190917B_o3afin}{1.35}{GW190916K_o3afin}{1.61}{GW190805J_o3afin}{1.00}{GW190725F_o3afin}{1.00}{GW190426N_o3afin}{2.08}{GW190403B_o3afin}{1.84}{GW150914_o3afin}{2.70}{GW151012_o3afin}{1.70}{GW151226_o3afin}{0.88}{GW170104_o3afin}{1.10}{GW170608_o3afin}{2.37}{GW170729_o3afin}{1.35}{GW170809_o3afin}{2.61}{GW170814_o3afin}{0.69}{GW170818_o3afin}{2.46}{GW170823_o3afin}{1.73}{GW190930C_o3afin}{0.72}{GW190929B_o3afin}{1.45}{GW190924A_o3afin}{0.84}{GW190915K_o3afin}{1.84}{GW190910B_o3afin}{1.62}{GW190828B_o3afin}{1.86}{GW190828A_o3afin}{2.38}{GW190814H_o3afin}{0.90}{GW190803B_o3afin}{0.91}{GW190731E_o3afin}{1.23}{GW190728D_o3afin}{1.11}{GW190727B_o3afin}{1.54}{GW190720A_o3afin}{2.59}{GW190719H_o3afin}{1.61}{GW190708M_o3afin}{1.39}{GW190707E_o3afin}{2.12}{GW190706F_o3afin}{1.86}{GW190701E_o3afin}{0.58}{GW190630E_o3afin}{1.41}{GW190620B_o3afin}{2.07}{GW190602E_o3afin}{2.13}{GW190527H_o3afin}{1.15}{GW190521E_o3afin}{1.48}{GW190521B_o3afin}{1.38}{GW190519J_o3afin}{1.61}{GW190517B_o3afin}{2.12}{GW190514E_o3afin}{1.47}{GW190513E_o3afin}{0.79}{GW190512G_o3afin}{1.79}{GW190503E_o3afin}{2.50}{GW190425B_o3afin}{1.70}{GW190421I_o3afin}{2.03}{GW190413E_o3afin}{1.85}{GW190413A_o3afin}{0.79}{GW190412B_o3afin}{0.92}{GW190408H_o3afin}{1.01}}}
\newcommand{\thetajnplus}[1]{\IfEqCase{#1}{{GW190926C_o3afin}{1.03}{GW190925J_o3afin}{2.07}{GW190917B_o3afin}{1.60}{GW190916K_o3afin}{1.27}{GW190805J_o3afin}{1.75}{GW190725F_o3afin}{1.79}{GW190426N_o3afin}{0.80}{GW190403B_o3afin}{1.12}{GW150914_o3afin}{0.32}{GW151012_o3afin}{1.13}{GW151226_o3afin}{2.00}{GW170104_o3afin}{1.79}{GW170608_o3afin}{0.58}{GW170729_o3afin}{1.44}{GW170809_o3afin}{0.39}{GW170814_o3afin}{1.92}{GW170818_o3afin}{0.47}{GW170823_o3afin}{1.16}{GW190930C_o3afin}{2.06}{GW190929B_o3afin}{1.18}{GW190924A_o3afin}{1.95}{GW190915K_o3afin}{0.99}{GW190910B_o3afin}{1.16}{GW190828B_o3afin}{0.96}{GW190828A_o3afin}{0.57}{GW190814H_o3afin}{1.40}{GW190803B_o3afin}{1.93}{GW190731E_o3afin}{1.63}{GW190728D_o3afin}{1.77}{GW190727B_o3afin}{1.35}{GW190720A_o3afin}{0.41}{GW190719H_o3afin}{1.26}{GW190708M_o3afin}{1.54}{GW190707E_o3afin}{0.81}{GW190706F_o3afin}{0.96}{GW190701E_o3afin}{0.55}{GW190630E_o3afin}{1.43}{GW190620B_o3afin}{0.82}{GW190602E_o3afin}{0.79}{GW190527H_o3afin}{1.65}{GW190521E_o3afin}{1.27}{GW190521B_o3afin}{1.40}{GW190519J_o3afin}{0.95}{GW190517B_o3afin}{0.70}{GW190514E_o3afin}{1.38}{GW190513E_o3afin}{2.01}{GW190512G_o3afin}{1.07}{GW190503E_o3afin}{0.46}{GW190425B_o3afin}{1.14}{GW190421I_o3afin}{0.85}{GW190413E_o3afin}{1.01}{GW190413A_o3afin}{2.01}{GW190412B_o3afin}{1.69}{GW190408H_o3afin}{1.85}}}
\newcommand{\psiminus}[1]{\IfEqCase{#1}{{GW190926C_o3afin}{1.36}{GW190925J_o3afin}{1.23}{GW190917B_o3afin}{1.35}{GW190916K_o3afin}{1.91}{GW190805J_o3afin}{2.08}{GW190725F_o3afin}{1.50}{GW190426N_o3afin}{1.22}{GW190403B_o3afin}{1.40}{GW150914_o3afin}{1.28}{GW151012_o3afin}{1.26}{GW151226_o3afin}{1.38}{GW170104_o3afin}{1.33}{GW170608_o3afin}{1.37}{GW170729_o3afin}{1.60}{GW170809_o3afin}{1.39}{GW170814_o3afin}{1.23}{GW170818_o3afin}{1.33}{GW170823_o3afin}{1.38}{GW190930C_o3afin}{1.47}{GW190929B_o3afin}{1.61}{GW190924A_o3afin}{1.07}{GW190915K_o3afin}{1.27}{GW190910B_o3afin}{1.09}{GW190828B_o3afin}{1.13}{GW190828A_o3afin}{1.05}{GW190814H_o3afin}{0.33}{GW190803B_o3afin}{1.43}{GW190731E_o3afin}{1.85}{GW190728D_o3afin}{1.59}{GW190727B_o3afin}{1.92}{GW190720A_o3afin}{1.45}{GW190719H_o3afin}{1.44}{GW190708M_o3afin}{1.50}{GW190707E_o3afin}{1.40}{GW190706F_o3afin}{1.47}{GW190701E_o3afin}{1.90}{GW190630E_o3afin}{1.19}{GW190620B_o3afin}{1.71}{GW190602E_o3afin}{1.59}{GW190527H_o3afin}{1.77}{GW190521E_o3afin}{0.90}{GW190521B_o3afin}{1.27}{GW190519J_o3afin}{1.13}{GW190517B_o3afin}{1.19}{GW190514E_o3afin}{2.00}{GW190513E_o3afin}{1.68}{GW190512G_o3afin}{1.48}{GW190503E_o3afin}{1.28}{GW190425B_o3afin}{1.41}{GW190421I_o3afin}{1.47}{GW190413E_o3afin}{1.30}{GW190413A_o3afin}{1.33}{GW190412B_o3afin}{2.36}{GW190408H_o3afin}{1.35}}}
\newcommand{\psimed}[1]{\IfEqCase{#1}{{GW190926C_o3afin}{1.63}{GW190925J_o3afin}{1.39}{GW190917B_o3afin}{1.56}{GW190916K_o3afin}{2.12}{GW190805J_o3afin}{2.26}{GW190725F_o3afin}{1.69}{GW190426N_o3afin}{1.48}{GW190403B_o3afin}{1.58}{GW150914_o3afin}{1.45}{GW151012_o3afin}{1.52}{GW151226_o3afin}{1.54}{GW170104_o3afin}{1.51}{GW170608_o3afin}{1.56}{GW170729_o3afin}{1.83}{GW170809_o3afin}{1.60}{GW170814_o3afin}{1.41}{GW170818_o3afin}{1.53}{GW170823_o3afin}{1.53}{GW190930C_o3afin}{1.66}{GW190929B_o3afin}{1.70}{GW190924A_o3afin}{1.20}{GW190915K_o3afin}{1.41}{GW190910B_o3afin}{1.59}{GW190828B_o3afin}{1.27}{GW190828A_o3afin}{1.17}{GW190814H_o3afin}{0.46}{GW190803B_o3afin}{1.59}{GW190731E_o3afin}{2.11}{GW190728D_o3afin}{1.71}{GW190727B_o3afin}{2.08}{GW190720A_o3afin}{1.63}{GW190719H_o3afin}{1.60}{GW190708M_o3afin}{1.65}{GW190707E_o3afin}{1.56}{GW190706F_o3afin}{1.60}{GW190701E_o3afin}{2.09}{GW190630E_o3afin}{1.61}{GW190620B_o3afin}{1.96}{GW190602E_o3afin}{1.73}{GW190527H_o3afin}{2.03}{GW190521E_o3afin}{1.45}{GW190521B_o3afin}{1.61}{GW190519J_o3afin}{1.69}{GW190517B_o3afin}{1.38}{GW190514E_o3afin}{2.24}{GW190513E_o3afin}{1.87}{GW190512G_o3afin}{1.66}{GW190503E_o3afin}{1.47}{GW190425B_o3afin}{1.57}{GW190421I_o3afin}{1.64}{GW190413E_o3afin}{1.49}{GW190413A_o3afin}{1.52}{GW190412B_o3afin}{2.49}{GW190408H_o3afin}{1.52}}}
\newcommand{\psiplus}[1]{\IfEqCase{#1}{{GW190926C_o3afin}{1.23}{GW190925J_o3afin}{1.58}{GW190917B_o3afin}{1.35}{GW190916K_o3afin}{3.61}{GW190805J_o3afin}{3.49}{GW190725F_o3afin}{1.26}{GW190426N_o3afin}{1.38}{GW190403B_o3afin}{1.40}{GW150914_o3afin}{1.49}{GW151012_o3afin}{1.35}{GW151226_o3afin}{1.44}{GW170104_o3afin}{1.46}{GW170608_o3afin}{1.38}{GW170729_o3afin}{1.11}{GW170809_o3afin}{1.33}{GW170814_o3afin}{1.56}{GW170818_o3afin}{1.40}{GW170823_o3afin}{1.45}{GW190930C_o3afin}{1.28}{GW190929B_o3afin}{1.35}{GW190924A_o3afin}{1.78}{GW190915K_o3afin}{1.57}{GW190910B_o3afin}{1.09}{GW190828B_o3afin}{1.72}{GW190828A_o3afin}{1.85}{GW190814H_o3afin}{2.57}{GW190803B_o3afin}{1.39}{GW190731E_o3afin}{3.53}{GW190728D_o3afin}{1.32}{GW190727B_o3afin}{3.59}{GW190720A_o3afin}{1.31}{GW190719H_o3afin}{1.38}{GW190708M_o3afin}{1.35}{GW190707E_o3afin}{1.42}{GW190706F_o3afin}{1.41}{GW190701E_o3afin}{3.55}{GW190630E_o3afin}{1.11}{GW190620B_o3afin}{3.51}{GW190602E_o3afin}{1.28}{GW190527H_o3afin}{3.57}{GW190521E_o3afin}{0.96}{GW190521B_o3afin}{1.22}{GW190519J_o3afin}{3.41}{GW190517B_o3afin}{1.56}{GW190514E_o3afin}{3.51}{GW190513E_o3afin}{1.11}{GW190512G_o3afin}{1.27}{GW190503E_o3afin}{1.48}{GW190425B_o3afin}{1.40}{GW190421I_o3afin}{1.33}{GW190413E_o3afin}{1.45}{GW190413A_o3afin}{1.40}{GW190412B_o3afin}{0.55}{GW190408H_o3afin}{1.43}}}
\newcommand{\phaseminus}[1]{\IfEqCase{#1}{{GW190926C_o3afin}{2.74}{GW190925J_o3afin}{2.99}{GW190917B_o3afin}{2.39}{GW190916K_o3afin}{2.89}{GW190805J_o3afin}{2.84}{GW190725F_o3afin}{3.30}{GW190426N_o3afin}{2.82}{GW190403B_o3afin}{2.83}{GW150914_o3afin}{2.94}{GW151012_o3afin}{2.73}{GW151226_o3afin}{1.71}{GW170104_o3afin}{2.52}{GW170608_o3afin}{2.46}{GW170729_o3afin}{2.46}{GW170809_o3afin}{2.90}{GW170814_o3afin}{2.08}{GW170818_o3afin}{2.27}{GW170823_o3afin}{2.51}{GW190930C_o3afin}{2.80}{GW190929B_o3afin}{3.05}{GW190924A_o3afin}{2.10}{GW190915K_o3afin}{3.16}{GW190910B_o3afin}{1.98}{GW190828B_o3afin}{1.97}{GW190828A_o3afin}{3.22}{GW190814H_o3afin}{0.33}{GW190803B_o3afin}{3.07}{GW190731E_o3afin}{2.85}{GW190728D_o3afin}{3.10}{GW190727B_o3afin}{2.75}{GW190720A_o3afin}{3.02}{GW190719H_o3afin}{2.81}{GW190708M_o3afin}{2.70}{GW190707E_o3afin}{2.79}{GW190706F_o3afin}{2.43}{GW190701E_o3afin}{2.81}{GW190630E_o3afin}{4.11}{GW190620B_o3afin}{2.78}{GW190602E_o3afin}{2.80}{GW190527H_o3afin}{2.83}{GW190521E_o3afin}{2.05}{GW190521B_o3afin}{3.04}{GW190519J_o3afin}{2.42}{GW190517B_o3afin}{2.40}{GW190514E_o3afin}{2.83}{GW190513E_o3afin}{3.11}{GW190512G_o3afin}{3.66}{GW190503E_o3afin}{2.40}{GW190425B_o3afin}{3.35}{GW190421I_o3afin}{2.57}{GW190413E_o3afin}{2.83}{GW190413A_o3afin}{2.77}{GW190412B_o3afin}{1.50}{GW190408H_o3afin}{2.73}}}
\newcommand{\phasemed}[1]{\IfEqCase{#1}{{GW190926C_o3afin}{3.07}{GW190925J_o3afin}{3.27}{GW190917B_o3afin}{2.55}{GW190916K_o3afin}{3.22}{GW190805J_o3afin}{3.16}{GW190725F_o3afin}{3.47}{GW190426N_o3afin}{3.15}{GW190403B_o3afin}{3.15}{GW150914_o3afin}{3.34}{GW151012_o3afin}{3.11}{GW151226_o3afin}{1.95}{GW170104_o3afin}{2.86}{GW170608_o3afin}{2.79}{GW170729_o3afin}{3.05}{GW170809_o3afin}{3.22}{GW170814_o3afin}{2.40}{GW170818_o3afin}{2.55}{GW170823_o3afin}{2.78}{GW190930C_o3afin}{3.17}{GW190929B_o3afin}{3.40}{GW190924A_o3afin}{2.59}{GW190915K_o3afin}{3.57}{GW190910B_o3afin}{2.50}{GW190828B_o3afin}{2.80}{GW190828A_o3afin}{3.58}{GW190814H_o3afin}{2.94}{GW190803B_o3afin}{3.35}{GW190731E_o3afin}{3.17}{GW190728D_o3afin}{3.56}{GW190727B_o3afin}{3.08}{GW190720A_o3afin}{3.43}{GW190719H_o3afin}{3.12}{GW190708M_o3afin}{3.04}{GW190707E_o3afin}{3.11}{GW190706F_o3afin}{2.75}{GW190701E_o3afin}{3.14}{GW190630E_o3afin}{4.32}{GW190620B_o3afin}{3.09}{GW190602E_o3afin}{3.21}{GW190527H_o3afin}{3.15}{GW190521E_o3afin}{2.68}{GW190521B_o3afin}{3.98}{GW190519J_o3afin}{2.75}{GW190517B_o3afin}{2.92}{GW190514E_o3afin}{3.14}{GW190513E_o3afin}{3.53}{GW190512G_o3afin}{3.86}{GW190503E_o3afin}{2.76}{GW190425B_o3afin}{3.67}{GW190421I_o3afin}{2.89}{GW190413E_o3afin}{3.13}{GW190413A_o3afin}{3.08}{GW190412B_o3afin}{2.17}{GW190408H_o3afin}{3.16}}}
\newcommand{\phaseplus}[1]{\IfEqCase{#1}{{GW190926C_o3afin}{2.91}{GW190925J_o3afin}{2.73}{GW190917B_o3afin}{3.56}{GW190916K_o3afin}{2.74}{GW190805J_o3afin}{2.81}{GW190725F_o3afin}{2.64}{GW190426N_o3afin}{2.80}{GW190403B_o3afin}{2.76}{GW150914_o3afin}{2.53}{GW151012_o3afin}{2.80}{GW151226_o3afin}{4.08}{GW170104_o3afin}{3.10}{GW170608_o3afin}{3.14}{GW170729_o3afin}{2.66}{GW170809_o3afin}{2.69}{GW170814_o3afin}{3.51}{GW170818_o3afin}{3.44}{GW170823_o3afin}{3.18}{GW190930C_o3afin}{2.76}{GW190929B_o3afin}{2.55}{GW190924A_o3afin}{3.12}{GW190915K_o3afin}{2.32}{GW190910B_o3afin}{3.32}{GW190828B_o3afin}{1.92}{GW190828A_o3afin}{2.30}{GW190814H_o3afin}{0.40}{GW190803B_o3afin}{2.64}{GW190731E_o3afin}{2.79}{GW190728D_o3afin}{2.29}{GW190727B_o3afin}{2.86}{GW190720A_o3afin}{2.46}{GW190719H_o3afin}{2.84}{GW190708M_o3afin}{2.88}{GW190707E_o3afin}{2.82}{GW190706F_o3afin}{3.19}{GW190701E_o3afin}{2.81}{GW190630E_o3afin}{1.79}{GW190620B_o3afin}{2.88}{GW190602E_o3afin}{2.66}{GW190527H_o3afin}{2.82}{GW190521E_o3afin}{2.72}{GW190521B_o3afin}{1.62}{GW190519J_o3afin}{3.21}{GW190517B_o3afin}{2.87}{GW190514E_o3afin}{2.82}{GW190513E_o3afin}{2.42}{GW190512G_o3afin}{2.24}{GW190503E_o3afin}{3.14}{GW190425B_o3afin}{2.28}{GW190421I_o3afin}{3.09}{GW190413E_o3afin}{2.85}{GW190413A_o3afin}{2.86}{GW190412B_o3afin}{3.42}{GW190408H_o3afin}{2.73}}}
\newcommand{\luminositydistanceminus}[1]{\IfEqCase{#1}{{GW190926C_o3afin}{1.73}{GW190925J_o3afin}{0.35}{GW190917B_o3afin}{0.31}{GW190916K_o3afin}{2.38}{GW190805J_o3afin}{3.08}{GW190725F_o3afin}{0.43}{GW190426N_o3afin}{2.28}{GW190403B_o3afin}{4.29}{GW150914_o3afin}{0.16}{GW151012_o3afin}{0.49}{GW151226_o3afin}{0.20}{GW170104_o3afin}{0.48}{GW170608_o3afin}{0.13}{GW170729_o3afin}{1.23}{GW170809_o3afin}{0.38}{GW170814_o3afin}{0.23}{GW170818_o3afin}{0.41}{GW170823_o3afin}{0.93}{GW190930C_o3afin}{0.32}{GW190929B_o3afin}{1.37}{GW190924A_o3afin}{0.22}{GW190915K_o3afin}{0.65}{GW190910B_o3afin}{0.63}{GW190828B_o3afin}{0.65}{GW190828A_o3afin}{0.92}{GW190814H_o3afin}{0.05}{GW190803B_o3afin}{1.47}{GW190731E_o3afin}{1.77}{GW190728D_o3afin}{0.38}{GW190727B_o3afin}{1.23}{GW190720A_o3afin}{0.26}{GW190719H_o3afin}{2.07}{GW190708M_o3afin}{0.39}{GW190707E_o3afin}{0.40}{GW190706F_o3afin}{2.00}{GW190701E_o3afin}{0.74}{GW190630E_o3afin}{0.36}{GW190620B_o3afin}{1.32}{GW190602E_o3afin}{1.28}{GW190527H_o3afin}{1.23}{GW190521E_o3afin}{0.53}{GW190521B_o3afin}{1.80}{GW190519J_o3afin}{0.96}{GW190517B_o3afin}{0.88}{GW190514E_o3afin}{2.07}{GW190513E_o3afin}{0.81}{GW190512G_o3afin}{0.59}{GW190503E_o3afin}{0.60}{GW190425B_o3afin}{0.06}{GW190421I_o3afin}{1.24}{GW190413E_o3afin}{1.83}{GW190413A_o3afin}{1.40}{GW190412B_o3afin}{0.22}{GW190408H_o3afin}{0.62}}}
\newcommand{\luminositydistancemed}[1]{\IfEqCase{#1}{{GW190926C_o3afin}{3.28}{GW190925J_o3afin}{0.93}{GW190917B_o3afin}{0.72}{GW190916K_o3afin}{4.94}{GW190805J_o3afin}{6.13}{GW190725F_o3afin}{1.03}{GW190426N_o3afin}{4.58}{GW190403B_o3afin}{8.28}{GW150914_o3afin}{0.47}{GW151012_o3afin}{1.00}{GW151226_o3afin}{0.46}{GW170104_o3afin}{1.11}{GW170608_o3afin}{0.34}{GW170729_o3afin}{2.49}{GW170809_o3afin}{1.07}{GW170814_o3afin}{0.61}{GW170818_o3afin}{1.08}{GW170823_o3afin}{1.97}{GW190930C_o3afin}{0.77}{GW190929B_o3afin}{3.13}{GW190924A_o3afin}{0.55}{GW190915K_o3afin}{1.75}{GW190910B_o3afin}{1.52}{GW190828B_o3afin}{1.54}{GW190828A_o3afin}{2.07}{GW190814H_o3afin}{0.23}{GW190803B_o3afin}{3.19}{GW190731E_o3afin}{3.33}{GW190728D_o3afin}{0.88}{GW190727B_o3afin}{3.07}{GW190720A_o3afin}{0.77}{GW190719H_o3afin}{3.73}{GW190708M_o3afin}{0.93}{GW190707E_o3afin}{0.85}{GW190706F_o3afin}{3.63}{GW190701E_o3afin}{2.09}{GW190630E_o3afin}{0.87}{GW190620B_o3afin}{2.91}{GW190602E_o3afin}{2.84}{GW190527H_o3afin}{2.52}{GW190521E_o3afin}{1.08}{GW190521B_o3afin}{3.31}{GW190519J_o3afin}{2.60}{GW190517B_o3afin}{1.79}{GW190514E_o3afin}{3.89}{GW190513E_o3afin}{2.21}{GW190512G_o3afin}{1.46}{GW190503E_o3afin}{1.52}{GW190425B_o3afin}{0.15}{GW190421I_o3afin}{2.59}{GW190413E_o3afin}{3.80}{GW190413A_o3afin}{3.32}{GW190412B_o3afin}{0.72}{GW190408H_o3afin}{1.54}}}
\newcommand{\luminositydistanceplus}[1]{\IfEqCase{#1}{{GW190926C_o3afin}{3.40}{GW190925J_o3afin}{0.46}{GW190917B_o3afin}{0.30}{GW190916K_o3afin}{3.71}{GW190805J_o3afin}{3.72}{GW190725F_o3afin}{0.52}{GW190426N_o3afin}{3.40}{GW190403B_o3afin}{6.72}{GW150914_o3afin}{0.14}{GW151012_o3afin}{0.64}{GW151226_o3afin}{0.16}{GW170104_o3afin}{0.39}{GW170608_o3afin}{0.12}{GW170729_o3afin}{1.69}{GW170809_o3afin}{0.31}{GW170814_o3afin}{0.16}{GW170818_o3afin}{0.43}{GW170823_o3afin}{0.84}{GW190930C_o3afin}{0.32}{GW190929B_o3afin}{2.51}{GW190924A_o3afin}{0.22}{GW190915K_o3afin}{0.71}{GW190910B_o3afin}{1.09}{GW190828B_o3afin}{0.69}{GW190828A_o3afin}{0.65}{GW190814H_o3afin}{0.04}{GW190803B_o3afin}{1.63}{GW190731E_o3afin}{2.35}{GW190728D_o3afin}{0.26}{GW190727B_o3afin}{1.30}{GW190720A_o3afin}{0.65}{GW190719H_o3afin}{3.12}{GW190708M_o3afin}{0.31}{GW190707E_o3afin}{0.34}{GW190706F_o3afin}{2.60}{GW190701E_o3afin}{0.77}{GW190630E_o3afin}{0.53}{GW190620B_o3afin}{1.71}{GW190602E_o3afin}{1.93}{GW190527H_o3afin}{2.08}{GW190521E_o3afin}{0.58}{GW190521B_o3afin}{2.79}{GW190519J_o3afin}{1.72}{GW190517B_o3afin}{1.75}{GW190514E_o3afin}{2.61}{GW190513E_o3afin}{0.99}{GW190512G_o3afin}{0.51}{GW190503E_o3afin}{0.63}{GW190425B_o3afin}{0.08}{GW190421I_o3afin}{1.49}{GW190413E_o3afin}{2.48}{GW190413A_o3afin}{1.91}{GW190412B_o3afin}{0.24}{GW190408H_o3afin}{0.44}}}
\newcommand{\geocenttimeminus}[1]{\IfEqCase{#1}{{GW190926C_o3afin}{0.01}{GW190925J_o3afin}{0.002}{GW190917B_o3afin}{0.008}{GW190916K_o3afin}{0.03}{GW190805J_o3afin}{0.05}{GW190725F_o3afin}{0.005}{GW190426N_o3afin}{0.03}{GW190403B_o3afin}{0.05}{GW150914_o3afin}{0.01}{GW151012_o3afin}{0.02}{GW151226_o3afin}{0.01}{GW170104_o3afin}{0.02}{GW170608_o3afin}{0.0007}{GW170729_o3afin}{0.03}{GW170809_o3afin}{0.0}{GW170814_o3afin}{0.01}{GW170818_o3afin}{0.0}{GW170823_o3afin}{0.03}{GW190930C_o3afin}{0.02}{GW190929B_o3afin}{0.01}{GW190924A_o3afin}{0.005}{GW190915K_o3afin}{0.005}{GW190910B_o3afin}{0.03}{GW190828B_o3afin}{0.002}{GW190828A_o3afin}{0.002}{GW190814H_o3afin}{0.0003}{GW190803B_o3afin}{0.02}{GW190731E_o3afin}{0.02}{GW190728D_o3afin}{0.03}{GW190727B_o3afin}{0.04}{GW190720A_o3afin}{0.0}{GW190719H_o3afin}{0.03}{GW190708M_o3afin}{0.02}{GW190707E_o3afin}{0.03}{GW190706F_o3afin}{0.04}{GW190701E_o3afin}{0.01}{GW190630E_o3afin}{0.02}{GW190620B_o3afin}{0.04}{GW190602E_o3afin}{0.008}{GW190527H_o3afin}{0.02}{GW190521E_o3afin}{0.005}{GW190521B_o3afin}{0.04}{GW190519J_o3afin}{0.04}{GW190517B_o3afin}{0.02}{GW190514E_o3afin}{0.04}{GW190513E_o3afin}{0.01}{GW190512G_o3afin}{0.02}{GW190503E_o3afin}{0.005}{GW190425B_o3afin}{0.03}{GW190421I_o3afin}{0.003}{GW190413E_o3afin}{0.004}{GW190413A_o3afin}{0.006}{GW190412B_o3afin}{0.009}{GW190408H_o3afin}{0.002}}}
\newcommand{\geocenttimemed}[1]{\IfEqCase{#1}{{GW190926C_o3afin}{1253509434.1}{GW190925J_o3afin}{1253489343.1}{GW190917B_o3afin}{1252756008.0}{GW190916K_o3afin}{1252699636.9}{GW190805J_o3afin}{1249074715.4}{GW190725F_o3afin}{1248112066.4}{GW190426N_o3afin}{1240340820.6}{GW190403B_o3afin}{1238303737.2}{GW150914_o3afin}{1126259462.4}{GW151012_o3afin}{1128678900.4}{GW151226_o3afin}{1135136350.6}{GW170104_o3afin}{1167559936.6}{GW170608_o3afin}{1180922494.5}{GW170729_o3afin}{1185389807.3}{GW170809_o3afin}{1186302519.7}{GW170814_o3afin}{1186741861.5}{GW170818_o3afin}{1187058327.1}{GW170823_o3afin}{1187529256.5}{GW190930C_o3afin}{1253885759.2}{GW190929B_o3afin}{1253755327.5}{GW190924A_o3afin}{1253326744.8}{GW190915K_o3afin}{1252627040.7}{GW190910B_o3afin}{1252150105.3}{GW190828B_o3afin}{1251010527.9}{GW190828A_o3afin}{1251009263.7}{GW190814H_o3afin}{1249852257.0}{GW190803B_o3afin}{1248834439.9}{GW190731E_o3afin}{1248617394.6}{GW190728D_o3afin}{1248331528.5}{GW190727B_o3afin}{1248242632.0}{GW190720A_o3afin}{1247616534.7}{GW190719H_o3afin}{1247608532.9}{GW190708M_o3afin}{1246663515.4}{GW190707E_o3afin}{1246527224.2}{GW190706F_o3afin}{1246487219.3}{GW190701E_o3afin}{1246048404.6}{GW190630E_o3afin}{1245955943.2}{GW190620B_o3afin}{1245035079.3}{GW190602E_o3afin}{1243533585.1}{GW190527H_o3afin}{1242984073.8}{GW190521E_o3afin}{1242459857.5}{GW190521B_o3afin}{1242442967.4}{GW190519J_o3afin}{1242315362.4}{GW190517B_o3afin}{1242107479.8}{GW190514E_o3afin}{1241852074.9}{GW190513E_o3afin}{1241816086.7}{GW190512G_o3afin}{1241719652.4}{GW190503E_o3afin}{1240944862.3}{GW190425B_o3afin}{1240215503.0}{GW190421I_o3afin}{1239917954.2}{GW190413E_o3afin}{1239198206.7}{GW190413A_o3afin}{1239168612.5}{GW190412B_o3afin}{1239082262.2}{GW190408H_o3afin}{1238782700.3}}}
\newcommand{\geocenttimeplus}[1]{\IfEqCase{#1}{{GW190926C_o3afin}{0.04}{GW190925J_o3afin}{0.009}{GW190917B_o3afin}{0.04}{GW190916K_o3afin}{0.01}{GW190805J_o3afin}{0.0}{GW190725F_o3afin}{0.04}{GW190426N_o3afin}{0.01}{GW190403B_o3afin}{0.01}{GW150914_o3afin}{0.007}{GW151012_o3afin}{0.02}{GW151226_o3afin}{0.03}{GW170104_o3afin}{0.02}{GW170608_o3afin}{0.010}{GW170729_o3afin}{0.005}{GW170809_o3afin}{0.007}{GW170814_o3afin}{0.003}{GW170818_o3afin}{0.01}{GW170823_o3afin}{0.01}{GW190930C_o3afin}{0.004}{GW190929B_o3afin}{0.04}{GW190924A_o3afin}{0.007}{GW190915K_o3afin}{0.0}{GW190910B_o3afin}{0.01}{GW190828B_o3afin}{0.04}{GW190828A_o3afin}{0.04}{GW190814H_o3afin}{0.0004}{GW190803B_o3afin}{0.01}{GW190731E_o3afin}{0.02}{GW190728D_o3afin}{0.00009}{GW190727B_o3afin}{0.0008}{GW190720A_o3afin}{0.005}{GW190719H_o3afin}{0.01}{GW190708M_o3afin}{0.02}{GW190707E_o3afin}{0.005}{GW190706F_o3afin}{0.003}{GW190701E_o3afin}{0.0}{GW190630E_o3afin}{0.01}{GW190620B_o3afin}{0.0}{GW190602E_o3afin}{0.008}{GW190527H_o3afin}{0.01}{GW190521E_o3afin}{0.007}{GW190521B_o3afin}{0.007}{GW190519J_o3afin}{0.005}{GW190517B_o3afin}{0.01}{GW190514E_o3afin}{0.02}{GW190513E_o3afin}{0.03}{GW190512G_o3afin}{0.0}{GW190503E_o3afin}{0.006}{GW190425B_o3afin}{0.02}{GW190421I_o3afin}{0.009}{GW190413E_o3afin}{0.03}{GW190413A_o3afin}{0.03}{GW190412B_o3afin}{0.001}{GW190408H_o3afin}{0.002}}}
\newcommand{\loglikelihoodminus}[1]{\IfEqCase{#1}{{GW190926C_o3afin}{5.1}{GW190925J_o3afin}{5.6}{GW190917B_o3afin}{6.2}{GW190916K_o3afin}{4.7}{GW190805J_o3afin}{5.6}{GW190725F_o3afin}{6.4}{GW190426N_o3afin}{4.8}{GW190403B_o3afin}{8.5}{GW150914_o3afin}{4.8}{GW151012_o3afin}{4.9}{GW151226_o3afin}{4.7}{GW170104_o3afin}{4.7}{GW170608_o3afin}{7.8}{GW170729_o3afin}{5.0}{GW170809_o3afin}{5.0}{GW170814_o3afin}{6.1}{GW170818_o3afin}{4.9}{GW170823_o3afin}{4.5}{GW190930C_o3afin}{5.3}{GW190929B_o3afin}{4.7}{GW190924A_o3afin}{8.4}{GW190915K_o3afin}{4.8}{GW190910B_o3afin}{4.6}{GW190828B_o3afin}{5.1}{GW190828A_o3afin}{4.9}{GW190814H_o3afin}{5.3}{GW190803B_o3afin}{4.4}{GW190731E_o3afin}{4.2}{GW190728D_o3afin}{10.3}{GW190727B_o3afin}{7.3}{GW190720A_o3afin}{10.0}{GW190719H_o3afin}{5.0}{GW190708M_o3afin}{5.1}{GW190707E_o3afin}{7.1}{GW190706F_o3afin}{5.4}{GW190701E_o3afin}{4.1}{GW190630E_o3afin}{5.7}{GW190620B_o3afin}{5.2}{GW190602E_o3afin}{4.3}{GW190527H_o3afin}{5.6}{GW190521E_o3afin}{5.1}{GW190521B_o3afin}{5.0}{GW190519J_o3afin}{5.4}{GW190517B_o3afin}{6.5}{GW190514E_o3afin}{4.8}{GW190513E_o3afin}{4.7}{GW190512G_o3afin}{5.3}{GW190503E_o3afin}{4.8}{GW190425B_o3afin}{5.2}{GW190421I_o3afin}{4.3}{GW190413E_o3afin}{5.7}{GW190413A_o3afin}{6.1}{GW190412B_o3afin}{26.6}{GW190408H_o3afin}{5.0}}}
\newcommand{\loglikelihoodmed}[1]{\IfEqCase{#1}{{GW190926C_o3afin}{21.7}{GW190925J_o3afin}{33.6}{GW190917B_o3afin}{-339908.3}{GW190916K_o3afin}{22.6}{GW190805J_o3afin}{23.9}{GW190725F_o3afin}{-108191.8}{GW190426N_o3afin}{28.9}{GW190403B_o3afin}{20.2}{GW150914_o3afin}{323.7}{GW151012_o3afin}{31.1}{GW151226_o3afin}{64.8}{GW170104_o3afin}{83.9}{GW170608_o3afin}{-56203.2}{GW170729_o3afin}{47.8}{GW170809_o3afin}{70.7}{GW170814_o3afin}{144.5}{GW170818_o3afin}{61.0}{GW170823_o3afin}{64.1}{GW190930C_o3afin}{34.3}{GW190929B_o3afin}{38.3}{GW190924A_o3afin}{-340445.9}{GW190915K_o3afin}{75.1}{GW190910B_o3afin}{94.7}{GW190828B_o3afin}{40.6}{GW190828A_o3afin}{124.9}{GW190814H_o3afin}{-169891.5}{GW190803B_o3afin}{33.9}{GW190731E_o3afin}{30.1}{GW190728D_o3afin}{-84372.9}{GW190727B_o3afin}{58.1}{GW190720A_o3afin}{-84773.0}{GW190719H_o3afin}{20.8}{GW190708M_o3afin}{77.3}{GW190707E_o3afin}{-56508.8}{GW190706F_o3afin}{80.5}{GW190701E_o3afin}{54.0}{GW190630E_o3afin}{122.1}{GW190620B_o3afin}{63.8}{GW190602E_o3afin}{78.2}{GW190527H_o3afin}{23.7}{GW190521E_o3afin}{323.5}{GW190521B_o3afin}{95.2}{GW190519J_o3afin}{117.1}{GW190517B_o3afin}{45.8}{GW190514E_o3afin}{24.2}{GW190513E_o3afin}{68.3}{GW190512G_o3afin}{69.5}{GW190503E_o3afin}{63.6}{GW190425B_o3afin}{72.3}{GW190421I_o3afin}{48.5}{GW190413E_o3afin}{47.9}{GW190413A_o3afin}{31.7}{GW190412B_o3afin}{173.5}{GW190408H_o3afin}{96.2}}}
\newcommand{\loglikelihoodplus}[1]{\IfEqCase{#1}{{GW190926C_o3afin}{3.9}{GW190925J_o3afin}{3.9}{GW190917B_o3afin}{4.2}{GW190916K_o3afin}{3.8}{GW190805J_o3afin}{3.9}{GW190725F_o3afin}{4.4}{GW190426N_o3afin}{4.0}{GW190403B_o3afin}{5.2}{GW150914_o3afin}{5.0}{GW151012_o3afin}{4.7}{GW151226_o3afin}{4.2}{GW170104_o3afin}{3.9}{GW170608_o3afin}{56314.0}{GW170729_o3afin}{3.8}{GW170809_o3afin}{4.2}{GW170814_o3afin}{7.3}{GW170818_o3afin}{4.7}{GW170823_o3afin}{3.0}{GW190930C_o3afin}{5.9}{GW190929B_o3afin}{3.5}{GW190924A_o3afin}{340511.4}{GW190915K_o3afin}{3.7}{GW190910B_o3afin}{4.7}{GW190828B_o3afin}{3.9}{GW190828A_o3afin}{3.3}{GW190814H_o3afin}{3.4}{GW190803B_o3afin}{2.6}{GW190731E_o3afin}{2.8}{GW190728D_o3afin}{84451.8}{GW190727B_o3afin}{3.6}{GW190720A_o3afin}{84825.9}{GW190719H_o3afin}{3.1}{GW190708M_o3afin}{4.6}{GW190707E_o3afin}{56586.2}{GW190706F_o3afin}{3.7}{GW190701E_o3afin}{2.9}{GW190630E_o3afin}{5.6}{GW190620B_o3afin}{3.9}{GW190602E_o3afin}{2.8}{GW190527H_o3afin}{3.2}{GW190521E_o3afin}{4.1}{GW190521B_o3afin}{5.0}{GW190519J_o3afin}{3.6}{GW190517B_o3afin}{4.8}{GW190514E_o3afin}{2.8}{GW190513E_o3afin}{3.6}{GW190512G_o3afin}{4.4}{GW190503E_o3afin}{3.7}{GW190425B_o3afin}{4.3}{GW190421I_o3afin}{2.5}{GW190413E_o3afin}{3.2}{GW190413A_o3afin}{3.3}{GW190412B_o3afin}{10.0}{GW190408H_o3afin}{4.2}}}
\newcommand{\logpriorminus}[1]{\IfEqCase{#1}{{GW190926C_o3afin}{10.8}{GW190917B_o3afin}{10.1}{GW190725F_o3afin}{10.1}{GW190426N_o3afin}{10.5}{GW151226_o3afin}{9.2}{GW190929B_o3afin}{10.8}{GW190828B_o3afin}{10.8}{GW190828A_o3afin}{11.0}{GW190814H_o3afin}{10.1}{GW190803B_o3afin}{11.0}{GW190719H_o3afin}{9.0}{GW190602E_o3afin}{11.0}{GW190521B_o3afin}{11.4}{GW190425B_o3afin}{12.7}{GW190421I_o3afin}{8.8}{GW190413E_o3afin}{11.1}{GW190413A_o3afin}{11.2}}}
\newcommand{\logpriormed}[1]{\IfEqCase{#1}{{GW190926C_o3afin}{139.8}{GW190917B_o3afin}{171.8}{GW190725F_o3afin}{170.8}{GW190426N_o3afin}{131.4}{GW151226_o3afin}{81.6}{GW190929B_o3afin}{141.1}{GW190828B_o3afin}{145.8}{GW190828A_o3afin}{144.1}{GW190814H_o3afin}{172.6}{GW190803B_o3afin}{143.0}{GW190719H_o3afin}{108.3}{GW190602E_o3afin}{132.0}{GW190521B_o3afin}{124.4}{GW190425B_o3afin}{65.6}{GW190421I_o3afin}{98.5}{GW190413E_o3afin}{134.1}{GW190413A_o3afin}{135.2}}}
\newcommand{\logpriorplus}[1]{\IfEqCase{#1}{{GW190926C_o3afin}{8.8}{GW190917B_o3afin}{8.4}{GW190725F_o3afin}{8.3}{GW190426N_o3afin}{8.7}{GW151226_o3afin}{7.0}{GW190929B_o3afin}{8.8}{GW190828B_o3afin}{8.7}{GW190828A_o3afin}{8.7}{GW190814H_o3afin}{8.2}{GW190803B_o3afin}{8.7}{GW190719H_o3afin}{6.9}{GW190602E_o3afin}{8.8}{GW190521B_o3afin}{8.9}{GW190425B_o3afin}{7.8}{GW190421I_o3afin}{6.8}{GW190413E_o3afin}{8.9}{GW190413A_o3afin}{8.9}}}
\newcommand{\totalmassdetminus}[1]{\IfEqCase{#1}{{GW190926C_o3afin}{13.0}{GW190925J_o3afin}{1.9}{GW190917B_o3afin}{3.3}{GW190916K_o3afin}{21.9}{GW190805J_o3afin}{21.2}{GW190725F_o3afin}{1.3}{GW190426N_o3afin}{44.8}{GW190403B_o3afin}{54.0}{GW150914_o3afin}{3.6}{GW151012_o3afin}{4.5}{GW151226_o3afin}{1.4}{GW170104_o3afin}{3.9}{GW170608_o3afin}{0.3}{GW170729_o3afin}{21.0}{GW170809_o3afin}{4.6}{GW170814_o3afin}{3.0}{GW170818_o3afin}{5.7}{GW170823_o3afin}{9.9}{GW190930C_o3afin}{1.9}{GW190929B_o3afin}{19.3}{GW190924A_o3afin}{0.7}{GW190915K_o3afin}{6.8}{GW190910B_o3afin}{7.9}{GW190828B_o3afin}{3.7}{GW190828A_o3afin}{5.3}{GW190814H_o3afin}{1.3}{GW190803B_o3afin}{11.5}{GW190731E_o3afin}{14.7}{GW190728D_o3afin}{0.7}{GW190727B_o3afin}{10.8}{GW190720A_o3afin}{1.5}{GW190719H_o3afin}{14.8}{GW190708M_o3afin}{1.7}{GW190707E_o3afin}{0.7}{GW190706F_o3afin}{27.9}{GW190701E_o3afin}{15.2}{GW190630E_o3afin}{3.6}{GW190620B_o3afin}{22.1}{GW190602E_o3afin}{24.4}{GW190527H_o3afin}{6.9}{GW190521E_o3afin}{5.1}{GW190521B_o3afin}{36.0}{GW190519J_o3afin}{18.2}{GW190517B_o3afin}{8.5}{GW190514E_o3afin}{19.2}{GW190513E_o3afin}{8.2}{GW190512G_o3afin}{2.7}{GW190503E_o3afin}{13.1}{GW190425B_o3afin}{0.1}{GW190421I_o3afin}{11.6}{GW190413E_o3afin}{20.5}{GW190413A_o3afin}{13.0}{GW190412B_o3afin}{4.5}{GW190408H_o3afin}{3.2}}}
\newcommand{\totalmassdetmed}[1]{\IfEqCase{#1}{{GW190926C_o3afin}{95.0}{GW190925J_o3afin}{43.4}{GW190917B_o3afin}{13.5}{GW190916K_o3afin}{119.7}{GW190805J_o3afin}{147.4}{GW190725F_o3afin}{21.8}{GW190426N_o3afin}{312.3}{GW190403B_o3afin}{236.2}{GW150914_o3afin}{70.9}{GW151012_o3afin}{46.4}{GW151226_o3afin}{23.7}{GW170104_o3afin}{60.3}{GW170608_o3afin}{19.8}{GW170729_o3afin}{122.9}{GW170809_o3afin}{70.6}{GW170814_o3afin}{62.8}{GW170818_o3afin}{75.8}{GW170823_o3afin}{90.8}{GW190930C_o3afin}{24.4}{GW190929B_o3afin}{144.0}{GW190924A_o3afin}{15.5}{GW190915K_o3afin}{75.9}{GW190910B_o3afin}{100.8}{GW190828B_o3afin}{43.9}{GW190828A_o3afin}{78.6}{GW190814H_o3afin}{27.2}{GW190803B_o3afin}{100.3}{GW190731E_o3afin}{110.3}{GW190728D_o3afin}{23.9}{GW190727B_o3afin}{104.6}{GW190720A_o3afin}{25.3}{GW190719H_o3afin}{91.3}{GW190708M_o3afin}{37.1}{GW190707E_o3afin}{23.4}{GW190706F_o3afin}{182.7}{GW190701E_o3afin}{130.2}{GW190630E_o3afin}{69.7}{GW190620B_o3afin}{140.5}{GW190602E_o3afin}{173.8}{GW190527H_o3afin}{82.6}{GW190521E_o3afin}{92.4}{GW190521B_o3afin}{243.3}{GW190519J_o3afin}{154.8}{GW190517B_o3afin}{85.8}{GW190514E_o3afin}{114.6}{GW190513E_o3afin}{75.5}{GW190512G_o3afin}{45.3}{GW190503E_o3afin}{89.6}{GW190425B_o3afin}{3.5}{GW190421I_o3afin}{107.3}{GW190413E_o3afin}{133.3}{GW190413A_o3afin}{90.9}{GW190412B_o3afin}{42.1}{GW190408H_o3afin}{55.9}}}
\newcommand{\totalmassdetplus}[1]{\IfEqCase{#1}{{GW190926C_o3afin}{43.2}{GW190925J_o3afin}{3.7}{GW190917B_o3afin}{3.2}{GW190916K_o3afin}{39.5}{GW190805J_o3afin}{21.6}{GW190725F_o3afin}{9.2}{GW190426N_o3afin}{61.4}{GW190403B_o3afin}{40.2}{GW150914_o3afin}{4.1}{GW151012_o3afin}{12.4}{GW151226_o3afin}{9.2}{GW170104_o3afin}{4.1}{GW170608_o3afin}{2.3}{GW170729_o3afin}{21.6}{GW170809_o3afin}{6.1}{GW170814_o3afin}{3.6}{GW170818_o3afin}{6.5}{GW170823_o3afin}{11.7}{GW190930C_o3afin}{6.9}{GW190929B_o3afin}{31.6}{GW190924A_o3afin}{3.2}{GW190915K_o3afin}{7.6}{GW190910B_o3afin}{9.4}{GW190828B_o3afin}{5.8}{GW190828A_o3afin}{6.1}{GW190814H_o3afin}{1.4}{GW190803B_o3afin}{12.6}{GW190731E_o3afin}{19.1}{GW190728D_o3afin}{5.3}{GW190727B_o3afin}{12.8}{GW190720A_o3afin}{4.3}{GW190719H_o3afin}{75.2}{GW190708M_o3afin}{2.8}{GW190707E_o3afin}{1.5}{GW190706F_o3afin}{25.9}{GW190701E_o3afin}{16.1}{GW190630E_o3afin}{4.8}{GW190620B_o3afin}{20.5}{GW190602E_o3afin}{25.7}{GW190527H_o3afin}{39.8}{GW190521E_o3afin}{6.9}{GW190521B_o3afin}{58.3}{GW190519J_o3afin}{18.7}{GW190517B_o3afin}{10.0}{GW190514E_o3afin}{20.1}{GW190513E_o3afin}{14.1}{GW190512G_o3afin}{4.2}{GW190503E_o3afin}{13.2}{GW190425B_o3afin}{0.3}{GW190421I_o3afin}{12.6}{GW190413E_o3afin}{20.5}{GW190413A_o3afin}{15.3}{GW190412B_o3afin}{5.4}{GW190408H_o3afin}{3.2}}}
\newcommand{\massonedetminus}[1]{\IfEqCase{#1}{{GW190926C_o3afin}{14.0}{GW190925J_o3afin}{3.2}{GW190917B_o3afin}{4.5}{GW190916K_o3afin}{21.5}{GW190805J_o3afin}{15.3}{GW190725F_o3afin}{3.6}{GW190426N_o3afin}{31.5}{GW190403B_o3afin}{58.1}{GW150914_o3afin}{2.9}{GW151012_o3afin}{6.9}{GW151226_o3afin}{4.0}{GW170104_o3afin}{4.9}{GW170608_o3afin}{1.5}{GW170729_o3afin}{14.5}{GW170809_o3afin}{6.3}{GW170814_o3afin}{3.5}{GW170818_o3afin}{5.0}{GW170823_o3afin}{7.9}{GW190930C_o3afin}{4.6}{GW190929B_o3afin}{18.3}{GW190924A_o3afin}{2.0}{GW190915K_o3afin}{6.0}{GW190910B_o3afin}{6.5}{GW190828B_o3afin}{8.3}{GW190828A_o3afin}{4.6}{GW190814H_o3afin}{1.4}{GW190803B_o3afin}{8.9}{GW190731E_o3afin}{10.8}{GW190728D_o3afin}{2.6}{GW190727B_o3afin}{8.0}{GW190720A_o3afin}{3.9}{GW190719H_o3afin}{16.3}{GW190708M_o3afin}{5.1}{GW190707E_o3afin}{2.3}{GW190706F_o3afin}{18.6}{GW190701E_o3afin}{10.8}{GW190630E_o3afin}{6.5}{GW190620B_o3afin}{17.3}{GW190602E_o3afin}{18.0}{GW190527H_o3afin}{9.3}{GW190521E_o3afin}{5.4}{GW190521B_o3afin}{17.6}{GW190519J_o3afin}{12.4}{GW190517B_o3afin}{10.0}{GW190514E_o3afin}{12.6}{GW190513E_o3afin}{12.8}{GW190512G_o3afin}{6.7}{GW190503E_o3afin}{10.3}{GW190425B_o3afin}{0.4}{GW190421I_o3afin}{8.7}{GW190413E_o3afin}{15.6}{GW190413A_o3afin}{10.0}{GW190412B_o3afin}{6.6}{GW190408H_o3afin}{3.9}}}
\newcommand{\massonedetmed}[1]{\IfEqCase{#1}{{GW190926C_o3afin}{63.7}{GW190925J_o3afin}{24.7}{GW190917B_o3afin}{11.1}{GW190916K_o3afin}{78.3}{GW190805J_o3afin}{87.2}{GW190725F_o3afin}{14.2}{GW190426N_o3afin}{178.5}{GW190403B_o3afin}{186.0}{GW150914_o3afin}{38.0}{GW151012_o3afin}{29.6}{GW151226_o3afin}{15.5}{GW170104_o3afin}{34.8}{GW170608_o3afin}{11.4}{GW170729_o3afin}{77.9}{GW170809_o3afin}{41.2}{GW170814_o3afin}{34.7}{GW170818_o3afin}{42.1}{GW170823_o3afin}{51.7}{GW190930C_o3afin}{16.4}{GW190929B_o3afin}{101.7}{GW190924A_o3afin}{9.8}{GW190915K_o3afin}{43.0}{GW190910B_o3afin}{56.5}{GW190828B_o3afin}{30.4}{GW190828A_o3afin}{43.3}{GW190814H_o3afin}{24.5}{GW190803B_o3afin}{57.7}{GW190731E_o3afin}{64.9}{GW190728D_o3afin}{14.6}{GW190727B_o3afin}{58.9}{GW190720A_o3afin}{16.6}{GW190719H_o3afin}{58.8}{GW190708M_o3afin}{23.4}{GW190707E_o3afin}{14.1}{GW190706F_o3afin}{117.5}{GW190701E_o3afin}{74.6}{GW190630E_o3afin}{41.4}{GW190620B_o3afin}{87.1}{GW190602E_o3afin}{106.8}{GW190527H_o3afin}{51.1}{GW190521E_o3afin}{52.2}{GW190521B_o3afin}{152.4}{GW190519J_o3afin}{94.8}{GW190517B_o3afin}{52.9}{GW190514E_o3afin}{67.0}{GW190513E_o3afin}{50.0}{GW190512G_o3afin}{29.3}{GW190503E_o3afin}{53.3}{GW190425B_o3afin}{2.2}{GW190421I_o3afin}{60.8}{GW190413E_o3afin}{83.1}{GW190413A_o3afin}{52.8}{GW190412B_o3afin}{31.8}{GW190408H_o3afin}{31.7}}}
\newcommand{\massonedetplus}[1]{\IfEqCase{#1}{{GW190926C_o3afin}{31.5}{GW190925J_o3afin}{7.7}{GW190917B_o3afin}{3.7}{GW190916K_o3afin}{32.3}{GW190805J_o3afin}{24.9}{GW190725F_o3afin}{12.2}{GW190426N_o3afin}{87.8}{GW190403B_o3afin}{37.5}{GW150914_o3afin}{5.1}{GW151012_o3afin}{17.1}{GW151226_o3afin}{12.2}{GW170104_o3afin}{7.7}{GW170608_o3afin}{4.4}{GW170729_o3afin}{17.9}{GW170809_o3afin}{9.6}{GW170814_o3afin}{6.0}{GW170818_o3afin}{8.0}{GW170823_o3afin}{11.9}{GW190930C_o3afin}{9.3}{GW190929B_o3afin}{26.0}{GW190924A_o3afin}{4.8}{GW190915K_o3afin}{11.1}{GW190910B_o3afin}{8.7}{GW190828B_o3afin}{8.3}{GW190828A_o3afin}{7.4}{GW190814H_o3afin}{1.5}{GW190803B_o3afin}{13.2}{GW190731E_o3afin}{15.6}{GW190728D_o3afin}{8.2}{GW190727B_o3afin}{13.1}{GW190720A_o3afin}{6.4}{GW190719H_o3afin}{75.1}{GW190708M_o3afin}{5.0}{GW190707E_o3afin}{3.0}{GW190706F_o3afin}{22.6}{GW190701E_o3afin}{16.0}{GW190630E_o3afin}{8.1}{GW190620B_o3afin}{23.9}{GW190602E_o3afin}{24.9}{GW190527H_o3afin}{29.9}{GW190521E_o3afin}{7.6}{GW190521B_o3afin}{31.6}{GW190519J_o3afin}{15.8}{GW190517B_o3afin}{14.8}{GW190514E_o3afin}{21.4}{GW190513E_o3afin}{14.7}{GW190512G_o3afin}{6.9}{GW190503E_o3afin}{12.2}{GW190425B_o3afin}{0.5}{GW190421I_o3afin}{13.1}{GW190413E_o3afin}{19.3}{GW190413A_o3afin}{15.0}{GW190412B_o3afin}{6.8}{GW190408H_o3afin}{7.0}}}
\newcommand{\masstwodetminus}[1]{\IfEqCase{#1}{{GW190926C_o3afin}{14.2}{GW190925J_o3afin}{4.2}{GW190917B_o3afin}{0.5}{GW190916K_o3afin}{21.2}{GW190805J_o3afin}{23.3}{GW190725F_o3afin}{3.0}{GW190426N_o3afin}{63.9}{GW190403B_o3afin}{24.1}{GW150914_o3afin}{5.0}{GW151012_o3afin}{6.0}{GW151226_o3afin}{3.0}{GW170104_o3afin}{5.6}{GW170608_o3afin}{2.1}{GW170729_o3afin}{17.5}{GW170809_o3afin}{6.5}{GW170814_o3afin}{4.6}{GW170818_o3afin}{6.1}{GW170823_o3afin}{11.0}{GW190930C_o3afin}{2.3}{GW190929B_o3afin}{18.1}{GW190924A_o3afin}{1.6}{GW190915K_o3afin}{8.0}{GW190910B_o3afin}{9.2}{GW190828B_o3afin}{2.7}{GW190828A_o3afin}{7.0}{GW190814H_o3afin}{0.1}{GW190803B_o3afin}{13.8}{GW190731E_o3afin}{17.5}{GW190728D_o3afin}{2.9}{GW190727B_o3afin}{13.2}{GW190720A_o3afin}{2.1}{GW190719H_o3afin}{16.4}{GW190708M_o3afin}{2.2}{GW190707E_o3afin}{1.5}{GW190706F_o3afin}{27.4}{GW190701E_o3afin}{17.9}{GW190630E_o3afin}{5.6}{GW190620B_o3afin}{24.3}{GW190602E_o3afin}{31.6}{GW190527H_o3afin}{13.4}{GW190521E_o3afin}{7.2}{GW190521B_o3afin}{52.2}{GW190519J_o3afin}{18.9}{GW190517B_o3afin}{12.2}{GW190514E_o3afin}{18.2}{GW190513E_o3afin}{7.0}{GW190512G_o3afin}{3.0}{GW190503E_o3afin}{12.6}{GW190425B_o3afin}{0.2}{GW190421I_o3afin}{14.6}{GW190413E_o3afin}{23.8}{GW190413A_o3afin}{11.6}{GW190412B_o3afin}{1.6}{GW190408H_o3afin}{5.0}}}
\newcommand{\masstwodetmed}[1]{\IfEqCase{#1}{{GW190926C_o3afin}{32.0}{GW190925J_o3afin}{18.5}{GW190917B_o3afin}{2.3}{GW190916K_o3afin}{42.4}{GW190805J_o3afin}{59.6}{GW190725F_o3afin}{7.6}{GW190426N_o3afin}{132.6}{GW190403B_o3afin}{44.6}{GW150914_o3afin}{33.0}{GW151012_o3afin}{16.3}{GW151226_o3afin}{8.2}{GW170104_o3afin}{25.3}{GW170608_o3afin}{8.4}{GW170729_o3afin}{44.1}{GW170809_o3afin}{29.2}{GW170814_o3afin}{28.0}{GW170818_o3afin}{33.4}{GW170823_o3afin}{39.4}{GW190930C_o3afin}{8.0}{GW190929B_o3afin}{41.6}{GW190924A_o3afin}{5.7}{GW190915K_o3afin}{32.5}{GW190910B_o3afin}{44.5}{GW190828B_o3afin}{13.4}{GW190828A_o3afin}{35.4}{GW190814H_o3afin}{2.7}{GW190803B_o3afin}{43.0}{GW190731E_o3afin}{46.3}{GW190728D_o3afin}{9.4}{GW190727B_o3afin}{46.2}{GW190720A_o3afin}{8.8}{GW190719H_o3afin}{32.8}{GW190708M_o3afin}{13.7}{GW190707E_o3afin}{9.3}{GW190706F_o3afin}{63.7}{GW190701E_o3afin}{56.1}{GW190630E_o3afin}{28.2}{GW190620B_o3afin}{53.3}{GW190602E_o3afin}{67.6}{GW190527H_o3afin}{32.4}{GW190521E_o3afin}{40.4}{GW190521B_o3afin}{89.7}{GW190519J_o3afin}{59.9}{GW190517B_o3afin}{32.7}{GW190514E_o3afin}{47.3}{GW190513E_o3afin}{25.6}{GW190512G_o3afin}{15.8}{GW190503E_o3afin}{36.5}{GW190425B_o3afin}{1.4}{GW190421I_o3afin}{46.9}{GW190413E_o3afin}{50.2}{GW190413A_o3afin}{37.9}{GW190412B_o3afin}{10.3}{GW190408H_o3afin}{23.8}}}
\newcommand{\masstwodetplus}[1]{\IfEqCase{#1}{{GW190926C_o3afin}{21.0}{GW190925J_o3afin}{2.7}{GW190917B_o3afin}{1.2}{GW190916K_o3afin}{24.1}{GW190805J_o3afin}{16.4}{GW190725F_o3afin}{2.4}{GW190426N_o3afin}{32.3}{GW190403B_o3afin}{61.3}{GW150914_o3afin}{3.2}{GW151012_o3afin}{5.7}{GW151226_o3afin}{2.6}{GW170104_o3afin}{4.4}{GW170608_o3afin}{1.2}{GW170729_o3afin}{18.1}{GW170809_o3afin}{5.6}{GW170814_o3afin}{3.2}{GW170818_o3afin}{5.1}{GW170823_o3afin}{8.1}{GW190930C_o3afin}{2.8}{GW190929B_o3afin}{25.1}{GW190924A_o3afin}{1.4}{GW190915K_o3afin}{5.6}{GW190910B_o3afin}{7.5}{GW190828B_o3afin}{4.8}{GW190828A_o3afin}{4.7}{GW190814H_o3afin}{0.1}{GW190803B_o3afin}{9.2}{GW190731E_o3afin}{12.7}{GW190728D_o3afin}{1.9}{GW190727B_o3afin}{8.2}{GW190720A_o3afin}{2.4}{GW190719H_o3afin}{14.4}{GW190708M_o3afin}{3.6}{GW190707E_o3afin}{1.7}{GW190706F_o3afin}{27.2}{GW190701E_o3afin}{12.1}{GW190630E_o3afin}{5.8}{GW190620B_o3afin}{17.6}{GW190602E_o3afin}{23.0}{GW190527H_o3afin}{12.0}{GW190521E_o3afin}{5.9}{GW190521B_o3afin}{49.0}{GW190519J_o3afin}{16.9}{GW190517B_o3afin}{9.6}{GW190514E_o3afin}{12.1}{GW190513E_o3afin}{10.9}{GW190512G_o3afin}{4.6}{GW190503E_o3afin}{10.2}{GW190425B_o3afin}{0.3}{GW190421I_o3afin}{9.0}{GW190413E_o3afin}{19.5}{GW190413A_o3afin}{10.0}{GW190412B_o3afin}{2.2}{GW190408H_o3afin}{3.6}}}
\newcommand{\raminus}[1]{\IfEqCase{#1}{{GW190926C_o3afin}{2.13707}{GW190925J_o3afin}{0.13948}{GW190917B_o3afin}{3.35797}{GW190916K_o3afin}{3.16955}{GW190805J_o3afin}{5.95123}{GW190725F_o3afin}{3.78850}{GW190426N_o3afin}{0.41802}{GW190403B_o3afin}{1.74360}{GW150914_o3afin}{1.05857}{GW151012_o3afin}{1.87695}{GW151226_o3afin}{2.81545}{GW170104_o3afin}{1.64993}{GW170608_o3afin}{0.41868}{GW170729_o3afin}{3.14901}{GW170809_o3afin}{0.07202}{GW170814_o3afin}{0.13677}{GW170818_o3afin}{0.02351}{GW170823_o3afin}{3.57690}{GW190930C_o3afin}{5.34489}{GW190929B_o3afin}{1.27912}{GW190924A_o3afin}{0.15785}{GW190915K_o3afin}{0.15832}{GW190910B_o3afin}{1.26395}{GW190828B_o3afin}{0.26102}{GW190828A_o3afin}{0.10437}{GW190814H_o3afin}{0.03145}{GW190803B_o3afin}{0.31697}{GW190731E_o3afin}{1.55462}{GW190728D_o3afin}{3.93720}{GW190727B_o3afin}{0.87113}{GW190720A_o3afin}{4.95254}{GW190719H_o3afin}{2.57083}{GW190708M_o3afin}{2.50489}{GW190707E_o3afin}{1.11204}{GW190706F_o3afin}{2.33965}{GW190701E_o3afin}{0.03004}{GW190630E_o3afin}{3.07834}{GW190620B_o3afin}{3.73424}{GW190602E_o3afin}{0.15103}{GW190527H_o3afin}{4.86821}{GW190521E_o3afin}{0.52850}{GW190521B_o3afin}{3.36590}{GW190519J_o3afin}{3.55924}{GW190517B_o3afin}{0.15621}{GW190514E_o3afin}{2.73360}{GW190513E_o3afin}{0.17461}{GW190512G_o3afin}{0.39774}{GW190503E_o3afin}{0.07953}{GW190425B_o3afin}{2.69749}{GW190421I_o3afin}{1.89130}{GW190413E_o3afin}{0.38197}{GW190413A_o3afin}{0.87676}{GW190412B_o3afin}{0.47993}{GW190408H_o3afin}{5.85324}}}
\newcommand{\ramed}[1]{\IfEqCase{#1}{{GW190926C_o3afin}{2.91707}{GW190925J_o3afin}{3.17079}{GW190917B_o3afin}{4.13416}{GW190916K_o3afin}{3.29976}{GW190805J_o3afin}{6.06653}{GW190725F_o3afin}{4.93705}{GW190426N_o3afin}{1.16049}{GW190403B_o3afin}{3.63704}{GW150914_o3afin}{2.03438}{GW151012_o3afin}{2.54442}{GW151226_o3afin}{3.33419}{GW170104_o3afin}{2.28830}{GW170608_o3afin}{2.13424}{GW170729_o3afin}{5.17934}{GW170809_o3afin}{0.26030}{GW170814_o3afin}{0.81504}{GW170818_o3afin}{5.95667}{GW170823_o3afin}{4.21371}{GW190930C_o3afin}{5.58492}{GW190929B_o3afin}{2.95883}{GW190924A_o3afin}{2.33106}{GW190915K_o3afin}{3.40295}{GW190910B_o3afin}{1.84118}{GW190828B_o3afin}{2.46731}{GW190828A_o3afin}{2.48234}{GW190814H_o3afin}{0.22650}{GW190803B_o3afin}{1.66444}{GW190731E_o3afin}{2.54985}{GW190728D_o3afin}{5.47387}{GW190727B_o3afin}{2.51789}{GW190720A_o3afin}{5.19311}{GW190719H_o3afin}{2.84667}{GW190708M_o3afin}{2.96118}{GW190707E_o3afin}{3.28188}{GW190706F_o3afin}{2.89716}{GW190701E_o3afin}{0.66129}{GW190630E_o3afin}{5.90962}{GW190620B_o3afin}{4.28664}{GW190602E_o3afin}{1.29824}{GW190527H_o3afin}{5.17669}{GW190521E_o3afin}{4.89988}{GW190521B_o3afin}{3.46644}{GW190519J_o3afin}{3.60868}{GW190517B_o3afin}{4.12702}{GW190514E_o3afin}{3.52828}{GW190513E_o3afin}{0.90826}{GW190512G_o3afin}{4.36518}{GW190503E_o3afin}{1.65642}{GW190425B_o3afin}{3.16064}{GW190421I_o3afin}{3.46412}{GW190413E_o3afin}{2.70580}{GW190413A_o3afin}{1.16764}{GW190412B_o3afin}{3.80387}{GW190408H_o3afin}{6.05326}}}
\newcommand{\raplus}[1]{\IfEqCase{#1}{{GW190926C_o3afin}{3.13345}{GW190925J_o3afin}{1.08594}{GW190917B_o3afin}{0.91384}{GW190916K_o3afin}{2.88457}{GW190805J_o3afin}{0.16098}{GW190725F_o3afin}{0.67713}{GW190426N_o3afin}{3.40414}{GW190403B_o3afin}{2.40628}{GW150914_o3afin}{0.57524}{GW151012_o3afin}{2.73738}{GW151226_o3afin}{1.59970}{GW170104_o3afin}{3.71410}{GW170608_o3afin}{0.09869}{GW170729_o3afin}{0.67240}{GW170809_o3afin}{0.19186}{GW170814_o3afin}{0.05173}{GW170818_o3afin}{0.02275}{GW170823_o3afin}{1.04460}{GW190930C_o3afin}{0.49295}{GW190929B_o3afin}{2.53584}{GW190924A_o3afin}{0.16418}{GW190915K_o3afin}{0.08026}{GW190910B_o3afin}{2.93063}{GW190828B_o3afin}{3.36960}{GW190828A_o3afin}{3.31431}{GW190814H_o3afin}{0.16869}{GW190803B_o3afin}{1.68254}{GW190731E_o3afin}{0.84665}{GW190728D_o3afin}{0.22765}{GW190727B_o3afin}{3.67834}{GW190720A_o3afin}{0.92118}{GW190719H_o3afin}{3.24957}{GW190708M_o3afin}{2.98342}{GW190707E_o3afin}{2.45791}{GW190706F_o3afin}{3.13187}{GW190701E_o3afin}{0.03016}{GW190630E_o3afin}{0.07448}{GW190620B_o3afin}{0.29440}{GW190602E_o3afin}{0.31398}{GW190527H_o3afin}{0.75047}{GW190521E_o3afin}{0.67394}{GW190521B_o3afin}{2.73350}{GW190519J_o3afin}{2.64215}{GW190517B_o3afin}{1.69291}{GW190514E_o3afin}{1.92904}{GW190513E_o3afin}{4.12350}{GW190512G_o3afin}{0.24082}{GW190503E_o3afin}{0.08693}{GW190425B_o3afin}{1.46077}{GW190421I_o3afin}{0.19087}{GW190413E_o3afin}{1.96969}{GW190413A_o3afin}{2.29177}{GW190412B_o3afin}{0.22594}{GW190408H_o3afin}{0.11085}}}
\newcommand{\decminus}[1]{\IfEqCase{#1}{{GW190926C_o3afin}{0.93134}{GW190925J_o3afin}{0.31084}{GW190917B_o3afin}{0.68809}{GW190916K_o3afin}{1.26250}{GW190805J_o3afin}{0.52907}{GW190725F_o3afin}{0.63636}{GW190426N_o3afin}{0.57250}{GW190403B_o3afin}{1.68509}{GW150914_o3afin}{0.08773}{GW151012_o3afin}{0.98915}{GW151226_o3afin}{1.13915}{GW170104_o3afin}{1.49935}{GW170608_o3afin}{0.33749}{GW170729_o3afin}{0.43187}{GW170809_o3afin}{0.17568}{GW170814_o3afin}{0.08492}{GW170818_o3afin}{0.08519}{GW170823_o3afin}{0.56251}{GW190930C_o3afin}{0.69379}{GW190929B_o3afin}{0.80299}{GW190924A_o3afin}{0.27642}{GW190915K_o3afin}{0.47852}{GW190910B_o3afin}{0.94217}{GW190828B_o3afin}{0.41747}{GW190828A_o3afin}{0.43983}{GW190814H_o3afin}{0.12461}{GW190803B_o3afin}{0.84758}{GW190731E_o3afin}{0.71307}{GW190728D_o3afin}{1.43066}{GW190727B_o3afin}{0.51881}{GW190720A_o3afin}{1.52792}{GW190719H_o3afin}{1.20597}{GW190708M_o3afin}{1.08391}{GW190707E_o3afin}{0.65083}{GW190706F_o3afin}{1.57562}{GW190701E_o3afin}{0.08516}{GW190630E_o3afin}{0.61778}{GW190620B_o3afin}{1.10303}{GW190602E_o3afin}{0.31914}{GW190527H_o3afin}{0.77761}{GW190521E_o3afin}{0.67057}{GW190521B_o3afin}{1.61088}{GW190519J_o3afin}{1.15082}{GW190517B_o3afin}{0.23687}{GW190514E_o3afin}{1.38536}{GW190513E_o3afin}{1.28393}{GW190512G_o3afin}{0.09850}{GW190503E_o3afin}{0.08574}{GW190425B_o3afin}{0.83681}{GW190421I_o3afin}{0.64377}{GW190413E_o3afin}{0.10403}{GW190413A_o3afin}{0.08378}{GW190412B_o3afin}{0.47567}{GW190408H_o3afin}{0.68310}}}
\newcommand{\decmed}[1]{\IfEqCase{#1}{{GW190926C_o3afin}{-0.20359}{GW190925J_o3afin}{-0.16077}{GW190917B_o3afin}{-0.16457}{GW190916K_o3afin}{0.26309}{GW190805J_o3afin}{-0.17800}{GW190725F_o3afin}{-0.10213}{GW190426N_o3afin}{-0.13262}{GW190403B_o3afin}{0.42013}{GW150914_o3afin}{-1.19327}{GW151012_o3afin}{0.00049}{GW151226_o3afin}{-0.03701}{GW170104_o3afin}{0.74589}{GW170608_o3afin}{0.58790}{GW170729_o3afin}{-0.81926}{GW170809_o3afin}{-0.49337}{GW170814_o3afin}{-0.79056}{GW170818_o3afin}{0.37393}{GW170823_o3afin}{-0.35183}{GW190930C_o3afin}{0.71892}{GW190929B_o3afin}{-0.13588}{GW190924A_o3afin}{0.12720}{GW190915K_o3afin}{0.67186}{GW190910B_o3afin}{0.00534}{GW190828B_o3afin}{-0.70310}{GW190828A_o3afin}{-0.40131}{GW190814H_o3afin}{-0.43858}{GW190803B_o3afin}{0.66114}{GW190731E_o3afin}{-0.67919}{GW190728D_o3afin}{0.13886}{GW190727B_o3afin}{-0.72380}{GW190720A_o3afin}{0.63008}{GW190719H_o3afin}{0.19323}{GW190708M_o3afin}{0.29891}{GW190707E_o3afin}{-0.26844}{GW190706F_o3afin}{0.46920}{GW190701E_o3afin}{-0.12043}{GW190630E_o3afin}{-0.13613}{GW190620B_o3afin}{0.39043}{GW190602E_o3afin}{-0.47785}{GW190527H_o3afin}{-0.51399}{GW190521E_o3afin}{0.32807}{GW190521B_o3afin}{0.45673}{GW190519J_o3afin}{0.65743}{GW190517B_o3afin}{-0.75282}{GW190514E_o3afin}{0.67271}{GW190513E_o3afin}{0.74847}{GW190512G_o3afin}{-0.45822}{GW190503E_o3afin}{-0.89249}{GW190425B_o3afin}{-0.07833}{GW190421I_o3afin}{-0.74627}{GW190413E_o3afin}{-0.53417}{GW190413A_o3afin}{-0.74623}{GW190412B_o3afin}{0.61792}{GW190408H_o3afin}{0.91208}}}
\newcommand{\decplus}[1]{\IfEqCase{#1}{{GW190926C_o3afin}{1.01898}{GW190925J_o3afin}{0.52540}{GW190917B_o3afin}{1.23785}{GW190916K_o3afin}{0.68074}{GW190805J_o3afin}{0.88905}{GW190725F_o3afin}{1.02712}{GW190426N_o3afin}{0.66782}{GW190403B_o3afin}{0.42801}{GW150914_o3afin}{0.21040}{GW151012_o3afin}{1.08858}{GW151226_o3afin}{1.01571}{GW170104_o3afin}{0.58815}{GW170608_o3afin}{0.64565}{GW170729_o3afin}{1.64349}{GW170809_o3afin}{0.26044}{GW170814_o3afin}{0.18052}{GW170818_o3afin}{0.07543}{GW170823_o3afin}{1.43880}{GW190930C_o3afin}{0.36627}{GW190929B_o3afin}{1.22727}{GW190924A_o3afin}{0.27665}{GW190915K_o3afin}{0.54839}{GW190910B_o3afin}{0.92435}{GW190828B_o3afin}{1.29796}{GW190828A_o3afin}{1.23223}{GW190814H_o3afin}{0.03517}{GW190803B_o3afin}{0.52024}{GW190731E_o3afin}{1.07697}{GW190728D_o3afin}{0.33902}{GW190727B_o3afin}{1.65712}{GW190720A_o3afin}{0.04431}{GW190719H_o3afin}{0.97021}{GW190708M_o3afin}{0.88697}{GW190707E_o3afin}{1.38516}{GW190706F_o3afin}{0.59996}{GW190701E_o3afin}{0.08883}{GW190630E_o3afin}{0.74477}{GW190620B_o3afin}{0.73118}{GW190602E_o3afin}{0.50968}{GW190527H_o3afin}{0.85373}{GW190521E_o3afin}{0.26182}{GW190521B_o3afin}{0.30010}{GW190519J_o3afin}{0.34635}{GW190517B_o3afin}{0.85516}{GW190514E_o3afin}{0.69678}{GW190513E_o3afin}{0.32930}{GW190512G_o3afin}{0.41949}{GW190503E_o3afin}{0.11551}{GW190425B_o3afin}{0.98096}{GW190421I_o3afin}{0.60435}{GW190413E_o3afin}{1.16715}{GW190413A_o3afin}{2.11832}{GW190412B_o3afin}{0.04122}{GW190408H_o3afin}{0.32646}}}
\newcommand{\symmetricmassratiominus}[1]{\IfEqCase{#1}{{GW190926C_o3afin}{0.07}{GW190925J_o3afin}{0.03}{GW190917B_o3afin}{0.05}{GW190916K_o3afin}{0.08}{GW190805J_o3afin}{0.05}{GW190725F_o3afin}{0.10}{GW190426N_o3afin}{0.07}{GW190403B_o3afin}{0.06}{GW150914_o3afin}{0.009}{GW151012_o3afin}{0.08}{GW151226_o3afin}{0.09}{GW170104_o3afin}{0.03}{GW170608_o3afin}{0.04}{GW170729_o3afin}{0.04}{GW170809_o3afin}{0.03}{GW170814_o3afin}{0.01}{GW170818_o3afin}{0.02}{GW170823_o3afin}{0.03}{GW190930C_o3afin}{0.07}{GW190929B_o3afin}{0.06}{GW190924A_o3afin}{0.06}{GW190915K_o3afin}{0.03}{GW190910B_o3afin}{0.02}{GW190828B_o3afin}{0.04}{GW190828A_o3afin}{0.02}{GW190814H_o3afin}{0.007}{GW190803B_o3afin}{0.03}{GW190731E_o3afin}{0.04}{GW190728D_o3afin}{0.07}{GW190727B_o3afin}{0.03}{GW190720A_o3afin}{0.05}{GW190719H_o3afin}{0.12}{GW190708M_o3afin}{0.03}{GW190707E_o3afin}{0.02}{GW190706F_o3afin}{0.05}{GW190701E_o3afin}{0.03}{GW190630E_o3afin}{0.02}{GW190620B_o3afin}{0.06}{GW190602E_o3afin}{0.06}{GW190527H_o3afin}{0.07}{GW190521E_o3afin}{0.01}{GW190521B_o3afin}{0.09}{GW190519J_o3afin}{0.03}{GW190517B_o3afin}{0.05}{GW190514E_o3afin}{0.05}{GW190513E_o3afin}{0.04}{GW190512G_o3afin}{0.03}{GW190503E_o3afin}{0.04}{GW190425B_o3afin}{0.03}{GW190421I_o3afin}{0.03}{GW190413E_o3afin}{0.06}{GW190413A_o3afin}{0.03}{GW190412B_o3afin}{0.03}{GW190408H_o3afin}{0.02}}}
\newcommand{\symmetricmassratiomed}[1]{\IfEqCase{#1}{{GW190926C_o3afin}{0.22}{GW190925J_o3afin}{0.245}{GW190917B_o3afin}{0.14}{GW190916K_o3afin}{0.23}{GW190805J_o3afin}{0.241}{GW190725F_o3afin}{0.23}{GW190426N_o3afin}{0.245}{GW190403B_o3afin}{0.15}{GW150914_o3afin}{0.249}{GW151012_o3afin}{0.23}{GW151226_o3afin}{0.23}{GW170104_o3afin}{0.244}{GW170608_o3afin}{0.244}{GW170729_o3afin}{0.23}{GW170809_o3afin}{0.243}{GW170814_o3afin}{0.247}{GW170818_o3afin}{0.247}{GW170823_o3afin}{0.246}{GW190930C_o3afin}{0.22}{GW190929B_o3afin}{0.20}{GW190924A_o3afin}{0.23}{GW190915K_o3afin}{0.245}{GW190910B_o3afin}{0.247}{GW190828B_o3afin}{0.21}{GW190828A_o3afin}{0.248}{GW190814H_o3afin}{0.090}{GW190803B_o3afin}{0.245}{GW190731E_o3afin}{0.243}{GW190728D_o3afin}{0.24}{GW190727B_o3afin}{0.247}{GW190720A_o3afin}{0.23}{GW190719H_o3afin}{0.23}{GW190708M_o3afin}{0.23}{GW190707E_o3afin}{0.24}{GW190706F_o3afin}{0.23}{GW190701E_o3afin}{0.245}{GW190630E_o3afin}{0.241}{GW190620B_o3afin}{0.23}{GW190602E_o3afin}{0.24}{GW190527H_o3afin}{0.24}{GW190521E_o3afin}{0.246}{GW190521B_o3afin}{0.23}{GW190519J_o3afin}{0.24}{GW190517B_o3afin}{0.24}{GW190514E_o3afin}{0.243}{GW190513E_o3afin}{0.22}{GW190512G_o3afin}{0.23}{GW190503E_o3afin}{0.242}{GW190425B_o3afin}{0.24}{GW190421I_o3afin}{0.246}{GW190413E_o3afin}{0.23}{GW190413A_o3afin}{0.244}{GW190412B_o3afin}{0.19}{GW190408H_o3afin}{0.245}}}
\newcommand{\symmetricmassratioplus}[1]{\IfEqCase{#1}{{GW190926C_o3afin}{0.03}{GW190925J_o3afin}{0.005}{GW190917B_o3afin}{0.08}{GW190916K_o3afin}{0.02}{GW190805J_o3afin}{0.009}{GW190725F_o3afin}{0.02}{GW190426N_o3afin}{0.005}{GW190403B_o3afin}{0.09}{GW150914_o3afin}{0.001}{GW151012_o3afin}{0.02}{GW151226_o3afin}{0.02}{GW170104_o3afin}{0.006}{GW170608_o3afin}{0.006}{GW170729_o3afin}{0.02}{GW170809_o3afin}{0.007}{GW170814_o3afin}{0.003}{GW170818_o3afin}{0.003}{GW170823_o3afin}{0.004}{GW190930C_o3afin}{0.03}{GW190929B_o3afin}{0.04}{GW190924A_o3afin}{0.02}{GW190915K_o3afin}{0.005}{GW190910B_o3afin}{0.003}{GW190828B_o3afin}{0.04}{GW190828A_o3afin}{0.002}{GW190814H_o3afin}{0.007}{GW190803B_o3afin}{0.005}{GW190731E_o3afin}{0.007}{GW190728D_o3afin}{0.01}{GW190727B_o3afin}{0.003}{GW190720A_o3afin}{0.02}{GW190719H_o3afin}{0.02}{GW190708M_o3afin}{0.02}{GW190707E_o3afin}{0.01}{GW190706F_o3afin}{0.02}{GW190701E_o3afin}{0.005}{GW190630E_o3afin}{0.009}{GW190620B_o3afin}{0.01}{GW190602E_o3afin}{0.01}{GW190527H_o3afin}{0.01}{GW190521E_o3afin}{0.004}{GW190521B_o3afin}{0.02}{GW190519J_o3afin}{0.01}{GW190517B_o3afin}{0.01}{GW190514E_o3afin}{0.007}{GW190513E_o3afin}{0.03}{GW190512G_o3afin}{0.02}{GW190503E_o3afin}{0.008}{GW190425B_o3afin}{0.01}{GW190421I_o3afin}{0.004}{GW190413E_o3afin}{0.02}{GW190413A_o3afin}{0.006}{GW190412B_o3afin}{0.04}{GW190408H_o3afin}{0.005}}}
\newcommand{\iotaminus}[1]{\IfEqCase{#1}{{GW190926C_o3afin}{1.19}{GW190925J_o3afin}{0.60}{GW190917B_o3afin}{1.09}{GW190916K_o3afin}{1.34}{GW190805J_o3afin}{0.77}{GW190725F_o3afin}{0.78}{GW190426N_o3afin}{1.71}{GW190403B_o3afin}{1.61}{GW150914_o3afin}{0.66}{GW151012_o3afin}{1.37}{GW151226_o3afin}{0.72}{GW170104_o3afin}{0.92}{GW170608_o3afin}{2.05}{GW170729_o3afin}{1.05}{GW170809_o3afin}{0.60}{GW170814_o3afin}{0.50}{GW170818_o3afin}{0.55}{GW170823_o3afin}{1.48}{GW190930C_o3afin}{0.56}{GW190929B_o3afin}{1.01}{GW190924A_o3afin}{0.63}{GW190915K_o3afin}{1.54}{GW190910B_o3afin}{1.22}{GW190828B_o3afin}{1.55}{GW190828A_o3afin}{1.96}{GW190814H_o3afin}{0.26}{GW190803B_o3afin}{0.77}{GW190731E_o3afin}{0.99}{GW190728D_o3afin}{0.89}{GW190727B_o3afin}{1.27}{GW190720A_o3afin}{1.93}{GW190719H_o3afin}{1.30}{GW190708M_o3afin}{1.17}{GW190707E_o3afin}{1.87}{GW190706F_o3afin}{1.53}{GW190701E_o3afin}{0.41}{GW190630E_o3afin}{1.15}{GW190620B_o3afin}{1.69}{GW190602E_o3afin}{1.87}{GW190527H_o3afin}{0.89}{GW190521E_o3afin}{1.15}{GW190521B_o3afin}{0.99}{GW190519J_o3afin}{0.99}{GW190517B_o3afin}{1.26}{GW190514E_o3afin}{1.20}{GW190513E_o3afin}{0.61}{GW190512G_o3afin}{1.49}{GW190503E_o3afin}{0.67}{GW190425B_o3afin}{1.38}{GW190421I_o3afin}{1.54}{GW190413E_o3afin}{1.56}{GW190413A_o3afin}{0.61}{GW190412B_o3afin}{0.42}{GW190408H_o3afin}{0.79}}}
\newcommand{\iotamed}[1]{\IfEqCase{#1}{{GW190926C_o3afin}{1.67}{GW190925J_o3afin}{0.80}{GW190917B_o3afin}{1.30}{GW190916K_o3afin}{1.61}{GW190805J_o3afin}{1.03}{GW190725F_o3afin}{1.04}{GW190426N_o3afin}{2.07}{GW190403B_o3afin}{1.81}{GW150914_o3afin}{2.64}{GW151012_o3afin}{1.69}{GW151226_o3afin}{0.89}{GW170104_o3afin}{1.14}{GW170608_o3afin}{2.36}{GW170729_o3afin}{1.37}{GW170809_o3afin}{2.62}{GW170814_o3afin}{0.70}{GW170818_o3afin}{2.49}{GW170823_o3afin}{1.74}{GW190930C_o3afin}{0.75}{GW190929B_o3afin}{1.51}{GW190924A_o3afin}{0.84}{GW190915K_o3afin}{1.85}{GW190910B_o3afin}{1.62}{GW190828B_o3afin}{1.87}{GW190828A_o3afin}{2.35}{GW190814H_o3afin}{0.88}{GW190803B_o3afin}{1.00}{GW190731E_o3afin}{1.24}{GW190728D_o3afin}{1.12}{GW190727B_o3afin}{1.53}{GW190720A_o3afin}{2.56}{GW190719H_o3afin}{1.59}{GW190708M_o3afin}{1.38}{GW190707E_o3afin}{2.12}{GW190706F_o3afin}{1.84}{GW190701E_o3afin}{0.56}{GW190630E_o3afin}{1.41}{GW190620B_o3afin}{2.05}{GW190602E_o3afin}{2.12}{GW190527H_o3afin}{1.15}{GW190521E_o3afin}{1.49}{GW190521B_o3afin}{1.30}{GW190519J_o3afin}{1.61}{GW190517B_o3afin}{2.11}{GW190514E_o3afin}{1.47}{GW190513E_o3afin}{0.83}{GW190512G_o3afin}{1.78}{GW190503E_o3afin}{2.51}{GW190425B_o3afin}{1.70}{GW190421I_o3afin}{1.91}{GW190413E_o3afin}{1.89}{GW190413A_o3afin}{0.82}{GW190412B_o3afin}{0.91}{GW190408H_o3afin}{1.02}}}
\newcommand{\iotaplus}[1]{\IfEqCase{#1}{{GW190926C_o3afin}{1.03}{GW190925J_o3afin}{2.05}{GW190917B_o3afin}{1.61}{GW190916K_o3afin}{1.26}{GW190805J_o3afin}{1.71}{GW190725F_o3afin}{1.72}{GW190426N_o3afin}{0.83}{GW190403B_o3afin}{1.14}{GW150914_o3afin}{0.35}{GW151012_o3afin}{1.13}{GW151226_o3afin}{2.04}{GW170104_o3afin}{1.76}{GW170608_o3afin}{0.59}{GW170729_o3afin}{1.40}{GW170809_o3afin}{0.38}{GW170814_o3afin}{1.90}{GW170818_o3afin}{0.46}{GW170823_o3afin}{1.14}{GW190930C_o3afin}{2.01}{GW190929B_o3afin}{1.11}{GW190924A_o3afin}{1.93}{GW190915K_o3afin}{1.02}{GW190910B_o3afin}{1.15}{GW190828B_o3afin}{0.99}{GW190828A_o3afin}{0.58}{GW190814H_o3afin}{1.43}{GW190803B_o3afin}{1.80}{GW190731E_o3afin}{1.63}{GW190728D_o3afin}{1.74}{GW190727B_o3afin}{1.35}{GW190720A_o3afin}{0.42}{GW190719H_o3afin}{1.25}{GW190708M_o3afin}{1.56}{GW190707E_o3afin}{0.81}{GW190706F_o3afin}{1.00}{GW190701E_o3afin}{0.54}{GW190630E_o3afin}{1.42}{GW190620B_o3afin}{0.82}{GW190602E_o3afin}{0.80}{GW190527H_o3afin}{1.64}{GW190521E_o3afin}{1.30}{GW190521B_o3afin}{1.48}{GW190519J_o3afin}{0.92}{GW190517B_o3afin}{0.71}{GW190514E_o3afin}{1.40}{GW190513E_o3afin}{1.96}{GW190512G_o3afin}{1.07}{GW190503E_o3afin}{0.46}{GW190425B_o3afin}{1.13}{GW190421I_o3afin}{0.94}{GW190413E_o3afin}{0.97}{GW190413A_o3afin}{1.97}{GW190412B_o3afin}{1.67}{GW190408H_o3afin}{1.84}}}
\newcommand{\spinonexminus}[1]{\IfEqCase{#1}{{GW190926C_o3afin}{0.56}{GW190925J_o3afin}{0.50}{GW190917B_o3afin}{0.34}{GW190916K_o3afin}{0.51}{GW190805J_o3afin}{0.58}{GW190725F_o3afin}{0.52}{GW190426N_o3afin}{0.61}{GW190403B_o3afin}{0.46}{GW150914_o3afin}{0.57}{GW151012_o3afin}{0.49}{GW151226_o3afin}{0.68}{GW170104_o3afin}{0.51}{GW170608_o3afin}{0.42}{GW170729_o3afin}{0.53}{GW170809_o3afin}{0.53}{GW170814_o3afin}{0.59}{GW170818_o3afin}{0.65}{GW170823_o3afin}{0.59}{GW190930C_o3afin}{0.43}{GW190929B_o3afin}{0.52}{GW190924A_o3afin}{0.39}{GW190915K_o3afin}{0.68}{GW190910B_o3afin}{0.50}{GW190828B_o3afin}{0.40}{GW190828A_o3afin}{0.53}{GW190814H_o3afin}{0.04}{GW190803B_o3afin}{0.57}{GW190731E_o3afin}{0.54}{GW190728D_o3afin}{0.41}{GW190727B_o3afin}{0.63}{GW190720A_o3afin}{0.40}{GW190719H_o3afin}{0.58}{GW190708M_o3afin}{0.38}{GW190707E_o3afin}{0.35}{GW190706F_o3afin}{0.61}{GW190701E_o3afin}{0.55}{GW190630E_o3afin}{0.40}{GW190620B_o3afin}{0.63}{GW190602E_o3afin}{0.61}{GW190527H_o3afin}{0.52}{GW190521E_o3afin}{0.47}{GW190521B_o3afin}{0.61}{GW190519J_o3afin}{0.56}{GW190517B_o3afin}{0.61}{GW190514E_o3afin}{0.58}{GW190513E_o3afin}{0.48}{GW190512G_o3afin}{0.36}{GW190503E_o3afin}{0.53}{GW190425B_o3afin}{0.50}{GW190421I_o3afin}{0.57}{GW190413E_o3afin}{0.70}{GW190413A_o3afin}{0.58}{GW190412B_o3afin}{0.22}{GW190408H_o3afin}{0.47}}}
\newcommand{\spinonexmed}[1]{\IfEqCase{#1}{{GW190926C_o3afin}{0.00}{GW190925J_o3afin}{0.0003}{GW190917B_o3afin}{0.00}{GW190916K_o3afin}{0.00}{GW190805J_o3afin}{-0.01}{GW190725F_o3afin}{0.0002}{GW190426N_o3afin}{0.00}{GW190403B_o3afin}{0.00}{GW150914_o3afin}{0.01}{GW151012_o3afin}{0.00}{GW151226_o3afin}{-0.01}{GW170104_o3afin}{0.0006}{GW170608_o3afin}{0.00}{GW170729_o3afin}{0.00}{GW170809_o3afin}{0.00}{GW170814_o3afin}{0.00}{GW170818_o3afin}{0.003}{GW170823_o3afin}{0.0002}{GW190930C_o3afin}{0.001}{GW190929B_o3afin}{0.00}{GW190924A_o3afin}{0.00001}{GW190915K_o3afin}{0.00}{GW190910B_o3afin}{0.00}{GW190828B_o3afin}{0.002}{GW190828A_o3afin}{0.003}{GW190814H_o3afin}{0.00}{GW190803B_o3afin}{0.001}{GW190731E_o3afin}{0.002}{GW190728D_o3afin}{0.0005}{GW190727B_o3afin}{0.001}{GW190720A_o3afin}{0.002}{GW190719H_o3afin}{0.00}{GW190708M_o3afin}{0.00}{GW190707E_o3afin}{0.00}{GW190706F_o3afin}{0.00}{GW190701E_o3afin}{0.002}{GW190630E_o3afin}{0.0001}{GW190620B_o3afin}{0.00}{GW190602E_o3afin}{0.001}{GW190527H_o3afin}{0.00}{GW190521E_o3afin}{0.00}{GW190521B_o3afin}{0.007}{GW190519J_o3afin}{0.005}{GW190517B_o3afin}{-0.08}{GW190514E_o3afin}{0.001}{GW190513E_o3afin}{0.00}{GW190512G_o3afin}{0.00}{GW190503E_o3afin}{0.004}{GW190425B_o3afin}{0.00001}{GW190421I_o3afin}{0.00}{GW190413E_o3afin}{0.00}{GW190413A_o3afin}{0.00}{GW190412B_o3afin}{0.0004}{GW190408H_o3afin}{0.00}}}
\newcommand{\spinonexplus}[1]{\IfEqCase{#1}{{GW190926C_o3afin}{0.57}{GW190925J_o3afin}{0.52}{GW190917B_o3afin}{0.34}{GW190916K_o3afin}{0.52}{GW190805J_o3afin}{0.64}{GW190725F_o3afin}{0.52}{GW190426N_o3afin}{0.62}{GW190403B_o3afin}{0.46}{GW150914_o3afin}{0.59}{GW151012_o3afin}{0.49}{GW151226_o3afin}{0.67}{GW170104_o3afin}{0.51}{GW170608_o3afin}{0.43}{GW170729_o3afin}{0.54}{GW170809_o3afin}{0.48}{GW170814_o3afin}{0.56}{GW170818_o3afin}{0.65}{GW170823_o3afin}{0.60}{GW190930C_o3afin}{0.42}{GW190929B_o3afin}{0.51}{GW190924A_o3afin}{0.38}{GW190915K_o3afin}{0.65}{GW190910B_o3afin}{0.50}{GW190828B_o3afin}{0.40}{GW190828A_o3afin}{0.56}{GW190814H_o3afin}{0.03}{GW190803B_o3afin}{0.59}{GW190731E_o3afin}{0.55}{GW190728D_o3afin}{0.39}{GW190727B_o3afin}{0.61}{GW190720A_o3afin}{0.42}{GW190719H_o3afin}{0.59}{GW190708M_o3afin}{0.39}{GW190707E_o3afin}{0.35}{GW190706F_o3afin}{0.60}{GW190701E_o3afin}{0.57}{GW190630E_o3afin}{0.41}{GW190620B_o3afin}{0.63}{GW190602E_o3afin}{0.61}{GW190527H_o3afin}{0.52}{GW190521E_o3afin}{0.47}{GW190521B_o3afin}{0.62}{GW190519J_o3afin}{0.57}{GW190517B_o3afin}{0.68}{GW190514E_o3afin}{0.57}{GW190513E_o3afin}{0.48}{GW190512G_o3afin}{0.36}{GW190503E_o3afin}{0.58}{GW190425B_o3afin}{0.43}{GW190421I_o3afin}{0.56}{GW190413E_o3afin}{0.69}{GW190413A_o3afin}{0.58}{GW190412B_o3afin}{0.26}{GW190408H_o3afin}{0.48}}}
\newcommand{\spinoneyminus}[1]{\IfEqCase{#1}{{GW190926C_o3afin}{0.56}{GW190925J_o3afin}{0.51}{GW190917B_o3afin}{0.33}{GW190916K_o3afin}{0.53}{GW190805J_o3afin}{0.59}{GW190725F_o3afin}{0.52}{GW190426N_o3afin}{0.62}{GW190403B_o3afin}{0.47}{GW150914_o3afin}{0.58}{GW151012_o3afin}{0.50}{GW151226_o3afin}{0.65}{GW170104_o3afin}{0.50}{GW170608_o3afin}{0.44}{GW170729_o3afin}{0.55}{GW170809_o3afin}{0.58}{GW170814_o3afin}{0.57}{GW170818_o3afin}{0.64}{GW170823_o3afin}{0.58}{GW190930C_o3afin}{0.43}{GW190929B_o3afin}{0.52}{GW190924A_o3afin}{0.39}{GW190915K_o3afin}{0.67}{GW190910B_o3afin}{0.48}{GW190828B_o3afin}{0.40}{GW190828A_o3afin}{0.56}{GW190814H_o3afin}{0.03}{GW190803B_o3afin}{0.57}{GW190731E_o3afin}{0.56}{GW190728D_o3afin}{0.40}{GW190727B_o3afin}{0.62}{GW190720A_o3afin}{0.38}{GW190719H_o3afin}{0.59}{GW190708M_o3afin}{0.39}{GW190707E_o3afin}{0.34}{GW190706F_o3afin}{0.61}{GW190701E_o3afin}{0.57}{GW190630E_o3afin}{0.40}{GW190620B_o3afin}{0.62}{GW190602E_o3afin}{0.60}{GW190527H_o3afin}{0.54}{GW190521E_o3afin}{0.51}{GW190521B_o3afin}{0.58}{GW190519J_o3afin}{0.56}{GW190517B_o3afin}{0.64}{GW190514E_o3afin}{0.60}{GW190513E_o3afin}{0.50}{GW190512G_o3afin}{0.36}{GW190503E_o3afin}{0.56}{GW190425B_o3afin}{0.51}{GW190421I_o3afin}{0.58}{GW190413E_o3afin}{0.69}{GW190413A_o3afin}{0.57}{GW190412B_o3afin}{0.28}{GW190408H_o3afin}{0.48}}}
\newcommand{\spinoneymed}[1]{\IfEqCase{#1}{{GW190926C_o3afin}{0.00}{GW190925J_o3afin}{0.002}{GW190917B_o3afin}{0.00}{GW190916K_o3afin}{0.0005}{GW190805J_o3afin}{0.002}{GW190725F_o3afin}{0.001}{GW190426N_o3afin}{0.0003}{GW190403B_o3afin}{0.004}{GW150914_o3afin}{0.002}{GW151012_o3afin}{0.00}{GW151226_o3afin}{0.0005}{GW170104_o3afin}{0.002}{GW170608_o3afin}{0.0001}{GW170729_o3afin}{0.0002}{GW170809_o3afin}{-0.01}{GW170814_o3afin}{0.002}{GW170818_o3afin}{0.02}{GW170823_o3afin}{0.002}{GW190930C_o3afin}{0.0004}{GW190929B_o3afin}{0.0003}{GW190924A_o3afin}{0.00}{GW190915K_o3afin}{0.002}{GW190910B_o3afin}{0.0008}{GW190828B_o3afin}{0.0003}{GW190828A_o3afin}{0.00}{GW190814H_o3afin}{0.004}{GW190803B_o3afin}{0.002}{GW190731E_o3afin}{0.00}{GW190728D_o3afin}{0.00}{GW190727B_o3afin}{0.00}{GW190720A_o3afin}{0.007}{GW190719H_o3afin}{0.0003}{GW190708M_o3afin}{0.00}{GW190707E_o3afin}{0.001}{GW190706F_o3afin}{-0.01}{GW190701E_o3afin}{0.001}{GW190630E_o3afin}{0.0008}{GW190620B_o3afin}{0.00}{GW190602E_o3afin}{0.00}{GW190527H_o3afin}{0.00}{GW190521E_o3afin}{0.00}{GW190521B_o3afin}{0.00}{GW190519J_o3afin}{0.003}{GW190517B_o3afin}{0.004}{GW190514E_o3afin}{0.00}{GW190513E_o3afin}{0.002}{GW190512G_o3afin}{0.00}{GW190503E_o3afin}{0.0006}{GW190425B_o3afin}{0.001}{GW190421I_o3afin}{0.0007}{GW190413E_o3afin}{0.0005}{GW190413A_o3afin}{0.0003}{GW190412B_o3afin}{0.06}{GW190408H_o3afin}{0.0003}}}
\newcommand{\spinoneyplus}[1]{\IfEqCase{#1}{{GW190926C_o3afin}{0.55}{GW190925J_o3afin}{0.53}{GW190917B_o3afin}{0.34}{GW190916K_o3afin}{0.51}{GW190805J_o3afin}{0.61}{GW190725F_o3afin}{0.52}{GW190426N_o3afin}{0.62}{GW190403B_o3afin}{0.46}{GW150914_o3afin}{0.57}{GW151012_o3afin}{0.51}{GW151226_o3afin}{0.66}{GW170104_o3afin}{0.51}{GW170608_o3afin}{0.44}{GW170729_o3afin}{0.53}{GW170809_o3afin}{0.47}{GW170814_o3afin}{0.59}{GW170818_o3afin}{0.66}{GW170823_o3afin}{0.60}{GW190930C_o3afin}{0.43}{GW190929B_o3afin}{0.52}{GW190924A_o3afin}{0.37}{GW190915K_o3afin}{0.69}{GW190910B_o3afin}{0.48}{GW190828B_o3afin}{0.39}{GW190828A_o3afin}{0.54}{GW190814H_o3afin}{0.04}{GW190803B_o3afin}{0.59}{GW190731E_o3afin}{0.53}{GW190728D_o3afin}{0.39}{GW190727B_o3afin}{0.61}{GW190720A_o3afin}{0.43}{GW190719H_o3afin}{0.59}{GW190708M_o3afin}{0.38}{GW190707E_o3afin}{0.35}{GW190706F_o3afin}{0.61}{GW190701E_o3afin}{0.57}{GW190630E_o3afin}{0.39}{GW190620B_o3afin}{0.63}{GW190602E_o3afin}{0.62}{GW190527H_o3afin}{0.52}{GW190521E_o3afin}{0.50}{GW190521B_o3afin}{0.59}{GW190519J_o3afin}{0.57}{GW190517B_o3afin}{0.66}{GW190514E_o3afin}{0.60}{GW190513E_o3afin}{0.51}{GW190512G_o3afin}{0.36}{GW190503E_o3afin}{0.54}{GW190425B_o3afin}{0.41}{GW190421I_o3afin}{0.58}{GW190413E_o3afin}{0.70}{GW190413A_o3afin}{0.57}{GW190412B_o3afin}{0.22}{GW190408H_o3afin}{0.46}}}
\newcommand{\spinonezminus}[1]{\IfEqCase{#1}{{GW190926C_o3afin}{0.45}{GW190925J_o3afin}{0.30}{GW190917B_o3afin}{0.64}{GW190916K_o3afin}{0.38}{GW190805J_o3afin}{0.54}{GW190725F_o3afin}{0.36}{GW190426N_o3afin}{0.56}{GW190403B_o3afin}{0.50}{GW150914_o3afin}{0.41}{GW151012_o3afin}{0.31}{GW151226_o3afin}{0.32}{GW170104_o3afin}{0.41}{GW170608_o3afin}{0.24}{GW170729_o3afin}{0.42}{GW170809_o3afin}{0.31}{GW170814_o3afin}{0.28}{GW170818_o3afin}{0.45}{GW170823_o3afin}{0.38}{GW190930C_o3afin}{0.27}{GW190929B_o3afin}{0.37}{GW190924A_o3afin}{0.23}{GW190915K_o3afin}{0.42}{GW190910B_o3afin}{0.35}{GW190828B_o3afin}{0.25}{GW190828A_o3afin}{0.28}{GW190814H_o3afin}{0.06}{GW190803B_o3afin}{0.45}{GW190731E_o3afin}{0.34}{GW190728D_o3afin}{0.26}{GW190727B_o3afin}{0.40}{GW190720A_o3afin}{0.26}{GW190719H_o3afin}{0.41}{GW190708M_o3afin}{0.19}{GW190707E_o3afin}{0.27}{GW190706F_o3afin}{0.41}{GW190701E_o3afin}{0.53}{GW190630E_o3afin}{0.25}{GW190620B_o3afin}{0.44}{GW190602E_o3afin}{0.36}{GW190527H_o3afin}{0.29}{GW190521E_o3afin}{0.31}{GW190521B_o3afin}{0.57}{GW190519J_o3afin}{0.36}{GW190517B_o3afin}{0.41}{GW190514E_o3afin}{0.56}{GW190513E_o3afin}{0.29}{GW190512G_o3afin}{0.25}{GW190503E_o3afin}{0.49}{GW190425B_o3afin}{0.16}{GW190421I_o3afin}{0.47}{GW190413E_o3afin}{0.55}{GW190413A_o3afin}{0.50}{GW190412B_o3afin}{0.19}{GW190408H_o3afin}{0.38}}}
\newcommand{\spinonezmed}[1]{\IfEqCase{#1}{{GW190926C_o3afin}{-0.01}{GW190925J_o3afin}{0.07}{GW190917B_o3afin}{-0.06}{GW190916K_o3afin}{0.19}{GW190805J_o3afin}{0.47}{GW190725F_o3afin}{0.0005}{GW190426N_o3afin}{0.25}{GW190403B_o3afin}{0.80}{GW150914_o3afin}{-0.05}{GW151012_o3afin}{0.11}{GW151226_o3afin}{0.30}{GW170104_o3afin}{-0.04}{GW170608_o3afin}{0.04}{GW170729_o3afin}{0.37}{GW170809_o3afin}{0.05}{GW170814_o3afin}{0.09}{GW170818_o3afin}{-0.04}{GW170823_o3afin}{0.03}{GW190930C_o3afin}{0.19}{GW190929B_o3afin}{-0.02}{GW190924A_o3afin}{0.01}{GW190915K_o3afin}{-0.01}{GW190910B_o3afin}{0.004}{GW190828B_o3afin}{0.04}{GW190828A_o3afin}{0.17}{GW190814H_o3afin}{0.000007}{GW190803B_o3afin}{-0.01}{GW190731E_o3afin}{0.05}{GW190728D_o3afin}{0.14}{GW190727B_o3afin}{0.08}{GW190720A_o3afin}{0.18}{GW190719H_o3afin}{0.29}{GW190708M_o3afin}{0.04}{GW190707E_o3afin}{-0.02}{GW190706F_o3afin}{0.34}{GW190701E_o3afin}{-0.07}{GW190630E_o3afin}{0.05}{GW190620B_o3afin}{0.39}{GW190602E_o3afin}{0.09}{GW190527H_o3afin}{0.08}{GW190521E_o3afin}{0.02}{GW190521B_o3afin}{-0.26}{GW190519J_o3afin}{0.35}{GW190517B_o3afin}{0.66}{GW190514E_o3afin}{-0.06}{GW190513E_o3afin}{0.16}{GW190512G_o3afin}{0.008}{GW190503E_o3afin}{-0.05}{GW190425B_o3afin}{0.07}{GW190421I_o3afin}{-0.07}{GW190413E_o3afin}{-0.01}{GW190413A_o3afin}{-0.02}{GW190412B_o3afin}{0.24}{GW190408H_o3afin}{-0.02}}}
\newcommand{\spinonezplus}[1]{\IfEqCase{#1}{{GW190926C_o3afin}{0.36}{GW190925J_o3afin}{0.34}{GW190917B_o3afin}{0.21}{GW190916K_o3afin}{0.53}{GW190805J_o3afin}{0.39}{GW190725F_o3afin}{0.39}{GW190426N_o3afin}{0.57}{GW190403B_o3afin}{0.15}{GW150914_o3afin}{0.30}{GW151012_o3afin}{0.44}{GW151226_o3afin}{0.25}{GW170104_o3afin}{0.25}{GW170608_o3afin}{0.25}{GW170729_o3afin}{0.44}{GW170809_o3afin}{0.35}{GW170814_o3afin}{0.39}{GW170818_o3afin}{0.34}{GW170823_o3afin}{0.43}{GW190930C_o3afin}{0.32}{GW190929B_o3afin}{0.31}{GW190924A_o3afin}{0.28}{GW190915K_o3afin}{0.35}{GW190910B_o3afin}{0.35}{GW190828B_o3afin}{0.24}{GW190828A_o3afin}{0.41}{GW190814H_o3afin}{0.06}{GW190803B_o3afin}{0.38}{GW190731E_o3afin}{0.52}{GW190728D_o3afin}{0.28}{GW190727B_o3afin}{0.48}{GW190720A_o3afin}{0.23}{GW190719H_o3afin}{0.47}{GW190708M_o3afin}{0.21}{GW190707E_o3afin}{0.17}{GW190706F_o3afin}{0.39}{GW190701E_o3afin}{0.34}{GW190630E_o3afin}{0.30}{GW190620B_o3afin}{0.39}{GW190602E_o3afin}{0.46}{GW190527H_o3afin}{0.42}{GW190521E_o3afin}{0.39}{GW190521B_o3afin}{0.75}{GW190519J_o3afin}{0.40}{GW190517B_o3afin}{0.24}{GW190514E_o3afin}{0.41}{GW190513E_o3afin}{0.46}{GW190512G_o3afin}{0.21}{GW190503E_o3afin}{0.35}{GW190425B_o3afin}{0.19}{GW190421I_o3afin}{0.33}{GW190413E_o3afin}{0.43}{GW190413A_o3afin}{0.39}{GW190412B_o3afin}{0.17}{GW190408H_o3afin}{0.25}}}
\newcommand{\spintwoxminus}[1]{\IfEqCase{#1}{{GW190926C_o3afin}{0.58}{GW190925J_o3afin}{0.55}{GW190917B_o3afin}{0.56}{GW190916K_o3afin}{0.57}{GW190805J_o3afin}{0.58}{GW190725F_o3afin}{0.61}{GW190426N_o3afin}{0.59}{GW190403B_o3afin}{0.55}{GW150914_o3afin}{0.57}{GW151012_o3afin}{0.56}{GW151226_o3afin}{0.55}{GW170104_o3afin}{0.54}{GW170608_o3afin}{0.47}{GW170729_o3afin}{0.57}{GW170809_o3afin}{0.55}{GW170814_o3afin}{0.60}{GW170818_o3afin}{0.60}{GW170823_o3afin}{0.56}{GW190930C_o3afin}{0.54}{GW190929B_o3afin}{0.58}{GW190924A_o3afin}{0.49}{GW190915K_o3afin}{0.59}{GW190910B_o3afin}{0.53}{GW190828B_o3afin}{0.56}{GW190828A_o3afin}{0.52}{GW190814H_o3afin}{0.58}{GW190803B_o3afin}{0.56}{GW190731E_o3afin}{0.56}{GW190728D_o3afin}{0.48}{GW190727B_o3afin}{0.58}{GW190720A_o3afin}{0.52}{GW190719H_o3afin}{0.58}{GW190708M_o3afin}{0.47}{GW190707E_o3afin}{0.52}{GW190706F_o3afin}{0.56}{GW190701E_o3afin}{0.58}{GW190630E_o3afin}{0.55}{GW190620B_o3afin}{0.58}{GW190602E_o3afin}{0.60}{GW190527H_o3afin}{0.52}{GW190521E_o3afin}{0.53}{GW190521B_o3afin}{0.62}{GW190519J_o3afin}{0.58}{GW190517B_o3afin}{0.60}{GW190514E_o3afin}{0.58}{GW190513E_o3afin}{0.55}{GW190512G_o3afin}{0.56}{GW190503E_o3afin}{0.57}{GW190425B_o3afin}{0.50}{GW190421I_o3afin}{0.57}{GW190413E_o3afin}{0.60}{GW190413A_o3afin}{0.57}{GW190412B_o3afin}{0.48}{GW190408H_o3afin}{0.53}}}
\newcommand{\spintwoxmed}[1]{\IfEqCase{#1}{{GW190926C_o3afin}{0.0001}{GW190925J_o3afin}{0.00}{GW190917B_o3afin}{0.00}{GW190916K_o3afin}{0.004}{GW190805J_o3afin}{0.00}{GW190725F_o3afin}{0.00}{GW190426N_o3afin}{0.001}{GW190403B_o3afin}{0.0009}{GW150914_o3afin}{0.0009}{GW151012_o3afin}{0.00}{GW151226_o3afin}{0.0001}{GW170104_o3afin}{0.0002}{GW170608_o3afin}{0.00}{GW170729_o3afin}{0.0004}{GW170809_o3afin}{0.00}{GW170814_o3afin}{0.0004}{GW170818_o3afin}{0.003}{GW170823_o3afin}{0.002}{GW190930C_o3afin}{0.0009}{GW190929B_o3afin}{0.002}{GW190924A_o3afin}{0.001}{GW190915K_o3afin}{0.002}{GW190910B_o3afin}{0.0008}{GW190828B_o3afin}{0.0001}{GW190828A_o3afin}{0.002}{GW190814H_o3afin}{0.00}{GW190803B_o3afin}{0.001}{GW190731E_o3afin}{0.00}{GW190728D_o3afin}{0.0004}{GW190727B_o3afin}{0.00}{GW190720A_o3afin}{0.0002}{GW190719H_o3afin}{0.00}{GW190708M_o3afin}{0.002}{GW190707E_o3afin}{0.00}{GW190706F_o3afin}{0.00}{GW190701E_o3afin}{0.00}{GW190630E_o3afin}{0.00}{GW190620B_o3afin}{0.002}{GW190602E_o3afin}{0.00}{GW190527H_o3afin}{0.00}{GW190521E_o3afin}{0.00}{GW190521B_o3afin}{0.001}{GW190519J_o3afin}{0.00}{GW190517B_o3afin}{-0.01}{GW190514E_o3afin}{0.00}{GW190513E_o3afin}{0.00}{GW190512G_o3afin}{0.00}{GW190503E_o3afin}{0.00}{GW190425B_o3afin}{0.002}{GW190421I_o3afin}{0.00}{GW190413E_o3afin}{0.00}{GW190413A_o3afin}{0.00}{GW190412B_o3afin}{0.02}{GW190408H_o3afin}{0.001}}}
\newcommand{\spintwoxplus}[1]{\IfEqCase{#1}{{GW190926C_o3afin}{0.58}{GW190925J_o3afin}{0.54}{GW190917B_o3afin}{0.58}{GW190916K_o3afin}{0.58}{GW190805J_o3afin}{0.59}{GW190725F_o3afin}{0.61}{GW190426N_o3afin}{0.60}{GW190403B_o3afin}{0.55}{GW150914_o3afin}{0.59}{GW151012_o3afin}{0.56}{GW151226_o3afin}{0.55}{GW170104_o3afin}{0.54}{GW170608_o3afin}{0.47}{GW170729_o3afin}{0.56}{GW170809_o3afin}{0.51}{GW170814_o3afin}{0.57}{GW170818_o3afin}{0.63}{GW170823_o3afin}{0.57}{GW190930C_o3afin}{0.55}{GW190929B_o3afin}{0.58}{GW190924A_o3afin}{0.48}{GW190915K_o3afin}{0.59}{GW190910B_o3afin}{0.53}{GW190828B_o3afin}{0.55}{GW190828A_o3afin}{0.53}{GW190814H_o3afin}{0.56}{GW190803B_o3afin}{0.58}{GW190731E_o3afin}{0.56}{GW190728D_o3afin}{0.47}{GW190727B_o3afin}{0.57}{GW190720A_o3afin}{0.50}{GW190719H_o3afin}{0.58}{GW190708M_o3afin}{0.49}{GW190707E_o3afin}{0.51}{GW190706F_o3afin}{0.56}{GW190701E_o3afin}{0.59}{GW190630E_o3afin}{0.54}{GW190620B_o3afin}{0.57}{GW190602E_o3afin}{0.60}{GW190527H_o3afin}{0.52}{GW190521E_o3afin}{0.52}{GW190521B_o3afin}{0.61}{GW190519J_o3afin}{0.57}{GW190517B_o3afin}{0.61}{GW190514E_o3afin}{0.58}{GW190513E_o3afin}{0.55}{GW190512G_o3afin}{0.55}{GW190503E_o3afin}{0.55}{GW190425B_o3afin}{0.46}{GW190421I_o3afin}{0.57}{GW190413E_o3afin}{0.59}{GW190413A_o3afin}{0.57}{GW190412B_o3afin}{0.56}{GW190408H_o3afin}{0.56}}}
\newcommand{\spintwoyminus}[1]{\IfEqCase{#1}{{GW190926C_o3afin}{0.59}{GW190925J_o3afin}{0.54}{GW190917B_o3afin}{0.56}{GW190916K_o3afin}{0.57}{GW190805J_o3afin}{0.58}{GW190725F_o3afin}{0.61}{GW190426N_o3afin}{0.59}{GW190403B_o3afin}{0.57}{GW150914_o3afin}{0.58}{GW151012_o3afin}{0.56}{GW151226_o3afin}{0.55}{GW170104_o3afin}{0.54}{GW170608_o3afin}{0.46}{GW170729_o3afin}{0.55}{GW170809_o3afin}{0.54}{GW170814_o3afin}{0.57}{GW170818_o3afin}{0.61}{GW170823_o3afin}{0.56}{GW190930C_o3afin}{0.53}{GW190929B_o3afin}{0.58}{GW190924A_o3afin}{0.48}{GW190915K_o3afin}{0.58}{GW190910B_o3afin}{0.53}{GW190828B_o3afin}{0.56}{GW190828A_o3afin}{0.54}{GW190814H_o3afin}{0.57}{GW190803B_o3afin}{0.57}{GW190731E_o3afin}{0.59}{GW190728D_o3afin}{0.46}{GW190727B_o3afin}{0.57}{GW190720A_o3afin}{0.50}{GW190719H_o3afin}{0.58}{GW190708M_o3afin}{0.48}{GW190707E_o3afin}{0.51}{GW190706F_o3afin}{0.58}{GW190701E_o3afin}{0.56}{GW190630E_o3afin}{0.53}{GW190620B_o3afin}{0.55}{GW190602E_o3afin}{0.59}{GW190527H_o3afin}{0.54}{GW190521E_o3afin}{0.53}{GW190521B_o3afin}{0.61}{GW190519J_o3afin}{0.57}{GW190517B_o3afin}{0.58}{GW190514E_o3afin}{0.59}{GW190513E_o3afin}{0.56}{GW190512G_o3afin}{0.56}{GW190503E_o3afin}{0.58}{GW190425B_o3afin}{0.49}{GW190421I_o3afin}{0.57}{GW190413E_o3afin}{0.60}{GW190413A_o3afin}{0.58}{GW190412B_o3afin}{0.54}{GW190408H_o3afin}{0.53}}}
\newcommand{\spintwoymed}[1]{\IfEqCase{#1}{{GW190926C_o3afin}{0.0004}{GW190925J_o3afin}{0.00}{GW190917B_o3afin}{0.000005}{GW190916K_o3afin}{0.0006}{GW190805J_o3afin}{0.00005}{GW190725F_o3afin}{0.00}{GW190426N_o3afin}{0.00}{GW190403B_o3afin}{-0.01}{GW150914_o3afin}{0.00}{GW151012_o3afin}{0.00}{GW151226_o3afin}{0.0004}{GW170104_o3afin}{0.00005}{GW170608_o3afin}{0.0005}{GW170729_o3afin}{0.00005}{GW170809_o3afin}{0.00}{GW170814_o3afin}{0.003}{GW170818_o3afin}{0.003}{GW170823_o3afin}{0.000006}{GW190930C_o3afin}{0.00005}{GW190929B_o3afin}{0.00}{GW190924A_o3afin}{0.00}{GW190915K_o3afin}{0.0007}{GW190910B_o3afin}{0.00008}{GW190828B_o3afin}{0.00}{GW190828A_o3afin}{0.00}{GW190814H_o3afin}{0.00}{GW190803B_o3afin}{0.00}{GW190731E_o3afin}{0.00}{GW190728D_o3afin}{0.001}{GW190727B_o3afin}{0.001}{GW190720A_o3afin}{0.00}{GW190719H_o3afin}{0.0004}{GW190708M_o3afin}{0.00}{GW190707E_o3afin}{0.00}{GW190706F_o3afin}{0.00}{GW190701E_o3afin}{0.00}{GW190630E_o3afin}{0.00}{GW190620B_o3afin}{0.0008}{GW190602E_o3afin}{0.001}{GW190527H_o3afin}{0.002}{GW190521E_o3afin}{0.00}{GW190521B_o3afin}{0.00}{GW190519J_o3afin}{0.00}{GW190517B_o3afin}{0.01}{GW190514E_o3afin}{0.00}{GW190513E_o3afin}{0.0006}{GW190512G_o3afin}{0.0002}{GW190503E_o3afin}{0.00}{GW190425B_o3afin}{0.005}{GW190421I_o3afin}{0.001}{GW190413E_o3afin}{0.00}{GW190413A_o3afin}{0.00}{GW190412B_o3afin}{0.00}{GW190408H_o3afin}{0.0004}}}
\newcommand{\spintwoyplus}[1]{\IfEqCase{#1}{{GW190926C_o3afin}{0.58}{GW190925J_o3afin}{0.57}{GW190917B_o3afin}{0.57}{GW190916K_o3afin}{0.56}{GW190805J_o3afin}{0.59}{GW190725F_o3afin}{0.61}{GW190426N_o3afin}{0.61}{GW190403B_o3afin}{0.55}{GW150914_o3afin}{0.58}{GW151012_o3afin}{0.53}{GW151226_o3afin}{0.55}{GW170104_o3afin}{0.54}{GW170608_o3afin}{0.48}{GW170729_o3afin}{0.56}{GW170809_o3afin}{0.54}{GW170814_o3afin}{0.60}{GW170818_o3afin}{0.61}{GW170823_o3afin}{0.56}{GW190930C_o3afin}{0.53}{GW190929B_o3afin}{0.57}{GW190924A_o3afin}{0.48}{GW190915K_o3afin}{0.60}{GW190910B_o3afin}{0.52}{GW190828B_o3afin}{0.55}{GW190828A_o3afin}{0.52}{GW190814H_o3afin}{0.58}{GW190803B_o3afin}{0.57}{GW190731E_o3afin}{0.56}{GW190728D_o3afin}{0.46}{GW190727B_o3afin}{0.59}{GW190720A_o3afin}{0.52}{GW190719H_o3afin}{0.57}{GW190708M_o3afin}{0.48}{GW190707E_o3afin}{0.52}{GW190706F_o3afin}{0.56}{GW190701E_o3afin}{0.56}{GW190630E_o3afin}{0.55}{GW190620B_o3afin}{0.56}{GW190602E_o3afin}{0.60}{GW190527H_o3afin}{0.50}{GW190521E_o3afin}{0.52}{GW190521B_o3afin}{0.62}{GW190519J_o3afin}{0.57}{GW190517B_o3afin}{0.61}{GW190514E_o3afin}{0.58}{GW190513E_o3afin}{0.53}{GW190512G_o3afin}{0.55}{GW190503E_o3afin}{0.57}{GW190425B_o3afin}{0.46}{GW190421I_o3afin}{0.58}{GW190413E_o3afin}{0.60}{GW190413A_o3afin}{0.57}{GW190412B_o3afin}{0.51}{GW190408H_o3afin}{0.52}}}
\newcommand{\spintwozminus}[1]{\IfEqCase{#1}{{GW190926C_o3afin}{0.58}{GW190925J_o3afin}{0.37}{GW190917B_o3afin}{0.61}{GW190916K_o3afin}{0.47}{GW190805J_o3afin}{0.53}{GW190725F_o3afin}{0.51}{GW190426N_o3afin}{0.51}{GW190403B_o3afin}{0.50}{GW150914_o3afin}{0.41}{GW151012_o3afin}{0.43}{GW151226_o3afin}{0.51}{GW170104_o3afin}{0.44}{GW170608_o3afin}{0.31}{GW170729_o3afin}{0.49}{GW170809_o3afin}{0.38}{GW170814_o3afin}{0.42}{GW170818_o3afin}{0.53}{GW170823_o3afin}{0.43}{GW190930C_o3afin}{0.39}{GW190929B_o3afin}{0.57}{GW190924A_o3afin}{0.30}{GW190915K_o3afin}{0.50}{GW190910B_o3afin}{0.44}{GW190828B_o3afin}{0.46}{GW190828A_o3afin}{0.39}{GW190814H_o3afin}{0.49}{GW190803B_o3afin}{0.53}{GW190731E_o3afin}{0.46}{GW190728D_o3afin}{0.35}{GW190727B_o3afin}{0.45}{GW190720A_o3afin}{0.40}{GW190719H_o3afin}{0.48}{GW190708M_o3afin}{0.33}{GW190707E_o3afin}{0.34}{GW190706F_o3afin}{0.46}{GW190701E_o3afin}{0.56}{GW190630E_o3afin}{0.33}{GW190620B_o3afin}{0.49}{GW190602E_o3afin}{0.48}{GW190527H_o3afin}{0.41}{GW190521E_o3afin}{0.35}{GW190521B_o3afin}{0.55}{GW190519J_o3afin}{0.46}{GW190517B_o3afin}{0.58}{GW190514E_o3afin}{0.58}{GW190513E_o3afin}{0.42}{GW190512G_o3afin}{0.37}{GW190503E_o3afin}{0.51}{GW190425B_o3afin}{0.25}{GW190421I_o3afin}{0.56}{GW190413E_o3afin}{0.56}{GW190413A_o3afin}{0.57}{GW190412B_o3afin}{0.38}{GW190408H_o3afin}{0.41}}}
\newcommand{\spintwozmed}[1]{\IfEqCase{#1}{{GW190926C_o3afin}{-0.02}{GW190925J_o3afin}{0.06}{GW190917B_o3afin}{-0.05}{GW190916K_o3afin}{0.10}{GW190805J_o3afin}{0.17}{GW190725F_o3afin}{-0.04}{GW190426N_o3afin}{0.10}{GW190403B_o3afin}{0.11}{GW150914_o3afin}{-0.01}{GW151012_o3afin}{0.06}{GW151226_o3afin}{0.04}{GW170104_o3afin}{-0.01}{GW170608_o3afin}{0.04}{GW170729_o3afin}{0.07}{GW170809_o3afin}{0.05}{GW170814_o3afin}{0.03}{GW170818_o3afin}{-0.04}{GW170823_o3afin}{0.03}{GW190930C_o3afin}{0.13}{GW190929B_o3afin}{-0.01}{GW190924A_o3afin}{0.03}{GW190915K_o3afin}{-0.02}{GW190910B_o3afin}{-0.01}{GW190828B_o3afin}{0.03}{GW190828A_o3afin}{0.07}{GW190814H_o3afin}{0.02}{GW190803B_o3afin}{-0.01}{GW190731E_o3afin}{0.03}{GW190728D_o3afin}{0.12}{GW190727B_o3afin}{0.04}{GW190720A_o3afin}{0.19}{GW190719H_o3afin}{0.10}{GW190708M_o3afin}{0.04}{GW190707E_o3afin}{-0.02}{GW190706F_o3afin}{0.11}{GW190701E_o3afin}{-0.04}{GW190630E_o3afin}{0.12}{GW190620B_o3afin}{0.20}{GW190602E_o3afin}{0.07}{GW190527H_o3afin}{0.05}{GW190521E_o3afin}{0.13}{GW190521B_o3afin}{0.03}{GW190519J_o3afin}{0.24}{GW190517B_o3afin}{0.17}{GW190514E_o3afin}{-0.04}{GW190513E_o3afin}{0.08}{GW190512G_o3afin}{0.03}{GW190503E_o3afin}{0.00}{GW190425B_o3afin}{0.04}{GW190421I_o3afin}{-0.07}{GW190413E_o3afin}{0.006}{GW190413A_o3afin}{-0.03}{GW190412B_o3afin}{0.08}{GW190408H_o3afin}{0.00}}}
\newcommand{\spintwozplus}[1]{\IfEqCase{#1}{{GW190926C_o3afin}{0.51}{GW190925J_o3afin}{0.49}{GW190917B_o3afin}{0.47}{GW190916K_o3afin}{0.64}{GW190805J_o3afin}{0.60}{GW190725F_o3afin}{0.51}{GW190426N_o3afin}{0.63}{GW190403B_o3afin}{0.67}{GW150914_o3afin}{0.42}{GW151012_o3afin}{0.57}{GW151226_o3afin}{0.54}{GW170104_o3afin}{0.43}{GW170608_o3afin}{0.40}{GW170729_o3afin}{0.63}{GW170809_o3afin}{0.49}{GW170814_o3afin}{0.41}{GW170818_o3afin}{0.41}{GW170823_o3afin}{0.50}{GW190930C_o3afin}{0.53}{GW190929B_o3afin}{0.51}{GW190924A_o3afin}{0.43}{GW190915K_o3afin}{0.43}{GW190910B_o3afin}{0.37}{GW190828B_o3afin}{0.56}{GW190828A_o3afin}{0.48}{GW190814H_o3afin}{0.58}{GW190803B_o3afin}{0.47}{GW190731E_o3afin}{0.55}{GW190728D_o3afin}{0.49}{GW190727B_o3afin}{0.53}{GW190720A_o3afin}{0.53}{GW190719H_o3afin}{0.62}{GW190708M_o3afin}{0.44}{GW190707E_o3afin}{0.35}{GW190706F_o3afin}{0.61}{GW190701E_o3afin}{0.43}{GW190630E_o3afin}{0.45}{GW190620B_o3afin}{0.61}{GW190602E_o3afin}{0.59}{GW190527H_o3afin}{0.51}{GW190521E_o3afin}{0.48}{GW190521B_o3afin}{0.60}{GW190519J_o3afin}{0.53}{GW190517B_o3afin}{0.61}{GW190514E_o3afin}{0.47}{GW190513E_o3afin}{0.58}{GW190512G_o3afin}{0.48}{GW190503E_o3afin}{0.49}{GW190425B_o3afin}{0.28}{GW190421I_o3afin}{0.40}{GW190413E_o3afin}{0.55}{GW190413A_o3afin}{0.47}{GW190412B_o3afin}{0.52}{GW190408H_o3afin}{0.39}}}
\newcommand{\phioneminus}[1]{\IfEqCase{#1}{{GW190926C_o3afin}{2.84}{GW190925J_o3afin}{2.76}{GW190917B_o3afin}{2.88}{GW190916K_o3afin}{2.83}{GW190805J_o3afin}{2.80}{GW190725F_o3afin}{2.80}{GW190426N_o3afin}{2.82}{GW190403B_o3afin}{2.78}{GW150914_o3afin}{2.86}{GW151012_o3afin}{2.86}{GW151226_o3afin}{2.83}{GW170104_o3afin}{2.77}{GW170608_o3afin}{2.81}{GW170729_o3afin}{2.81}{GW170809_o3afin}{3.08}{GW170814_o3afin}{2.75}{GW170818_o3afin}{2.60}{GW170823_o3afin}{2.79}{GW190930C_o3afin}{2.80}{GW190929B_o3afin}{2.80}{GW190924A_o3afin}{2.86}{GW190915K_o3afin}{2.76}{GW190910B_o3afin}{2.80}{GW190828B_o3afin}{2.83}{GW190828A_o3afin}{2.85}{GW190814H_o3afin}{2.11}{GW190803B_o3afin}{2.78}{GW190731E_o3afin}{2.90}{GW190728D_o3afin}{2.94}{GW190727B_o3afin}{2.92}{GW190720A_o3afin}{2.68}{GW190719H_o3afin}{2.82}{GW190708M_o3afin}{2.85}{GW190707E_o3afin}{2.75}{GW190706F_o3afin}{2.90}{GW190701E_o3afin}{2.78}{GW190630E_o3afin}{2.84}{GW190620B_o3afin}{2.85}{GW190602E_o3afin}{2.83}{GW190527H_o3afin}{2.83}{GW190521E_o3afin}{2.81}{GW190521B_o3afin}{2.90}{GW190519J_o3afin}{2.83}{GW190517B_o3afin}{2.72}{GW190514E_o3afin}{2.94}{GW190513E_o3afin}{2.76}{GW190512G_o3afin}{2.83}{GW190503E_o3afin}{2.83}{GW190425B_o3afin}{2.78}{GW190421I_o3afin}{2.80}{GW190413E_o3afin}{2.82}{GW190413A_o3afin}{2.82}{GW190412B_o3afin}{1.87}{GW190408H_o3afin}{2.81}}}
\newcommand{\phionemed}[1]{\IfEqCase{#1}{{GW190926C_o3afin}{3.15}{GW190925J_o3afin}{3.07}{GW190917B_o3afin}{3.19}{GW190916K_o3afin}{3.13}{GW190805J_o3afin}{3.12}{GW190725F_o3afin}{3.11}{GW190426N_o3afin}{3.14}{GW190403B_o3afin}{3.11}{GW150914_o3afin}{3.09}{GW151012_o3afin}{3.18}{GW151226_o3afin}{3.14}{GW170104_o3afin}{3.07}{GW170608_o3afin}{3.14}{GW170729_o3afin}{3.14}{GW170809_o3afin}{3.42}{GW170814_o3afin}{3.09}{GW170818_o3afin}{2.89}{GW170823_o3afin}{3.10}{GW190930C_o3afin}{3.13}{GW190929B_o3afin}{3.13}{GW190924A_o3afin}{3.14}{GW190915K_o3afin}{3.10}{GW190910B_o3afin}{3.11}{GW190828B_o3afin}{3.13}{GW190828A_o3afin}{3.16}{GW190814H_o3afin}{2.54}{GW190803B_o3afin}{3.09}{GW190731E_o3afin}{3.21}{GW190728D_o3afin}{3.25}{GW190727B_o3afin}{3.24}{GW190720A_o3afin}{2.98}{GW190719H_o3afin}{3.13}{GW190708M_o3afin}{3.16}{GW190707E_o3afin}{3.08}{GW190706F_o3afin}{3.23}{GW190701E_o3afin}{3.11}{GW190630E_o3afin}{3.11}{GW190620B_o3afin}{3.18}{GW190602E_o3afin}{3.14}{GW190527H_o3afin}{3.15}{GW190521E_o3afin}{3.18}{GW190521B_o3afin}{3.20}{GW190519J_o3afin}{3.11}{GW190517B_o3afin}{3.13}{GW190514E_o3afin}{3.22}{GW190513E_o3afin}{3.10}{GW190512G_o3afin}{3.15}{GW190503E_o3afin}{3.12}{GW190425B_o3afin}{3.09}{GW190421I_o3afin}{3.12}{GW190413E_o3afin}{3.13}{GW190413A_o3afin}{3.13}{GW190412B_o3afin}{2.22}{GW190408H_o3afin}{3.12}}}
\newcommand{\phioneplus}[1]{\IfEqCase{#1}{{GW190926C_o3afin}{2.81}{GW190925J_o3afin}{2.89}{GW190917B_o3afin}{2.79}{GW190916K_o3afin}{2.86}{GW190805J_o3afin}{2.88}{GW190725F_o3afin}{2.86}{GW190426N_o3afin}{2.84}{GW190403B_o3afin}{2.84}{GW150914_o3afin}{2.89}{GW151012_o3afin}{2.78}{GW151226_o3afin}{2.82}{GW170104_o3afin}{2.90}{GW170608_o3afin}{2.83}{GW170729_o3afin}{2.80}{GW170809_o3afin}{2.53}{GW170814_o3afin}{2.86}{GW170818_o3afin}{3.09}{GW170823_o3afin}{2.87}{GW190930C_o3afin}{2.86}{GW190929B_o3afin}{2.84}{GW190924A_o3afin}{2.80}{GW190915K_o3afin}{2.85}{GW190910B_o3afin}{2.85}{GW190828B_o3afin}{2.86}{GW190828A_o3afin}{2.83}{GW190814H_o3afin}{3.26}{GW190803B_o3afin}{2.87}{GW190731E_o3afin}{2.75}{GW190728D_o3afin}{2.75}{GW190727B_o3afin}{2.74}{GW190720A_o3afin}{2.98}{GW190719H_o3afin}{2.83}{GW190708M_o3afin}{2.80}{GW190707E_o3afin}{2.88}{GW190706F_o3afin}{2.75}{GW190701E_o3afin}{2.85}{GW190630E_o3afin}{2.86}{GW190620B_o3afin}{2.77}{GW190602E_o3afin}{2.82}{GW190527H_o3afin}{2.83}{GW190521E_o3afin}{2.78}{GW190521B_o3afin}{2.79}{GW190519J_o3afin}{2.86}{GW190517B_o3afin}{2.78}{GW190514E_o3afin}{2.73}{GW190513E_o3afin}{2.87}{GW190512G_o3afin}{2.83}{GW190503E_o3afin}{2.87}{GW190425B_o3afin}{2.88}{GW190421I_o3afin}{2.85}{GW190413E_o3afin}{2.84}{GW190413A_o3afin}{2.86}{GW190412B_o3afin}{3.66}{GW190408H_o3afin}{2.83}}}
\newcommand{\phitwominus}[1]{\IfEqCase{#1}{{GW190926C_o3afin}{2.82}{GW190925J_o3afin}{2.84}{GW190917B_o3afin}{2.82}{GW190916K_o3afin}{2.80}{GW190805J_o3afin}{2.83}{GW190725F_o3afin}{2.83}{GW190426N_o3afin}{2.85}{GW190403B_o3afin}{2.89}{GW150914_o3afin}{2.86}{GW151012_o3afin}{2.87}{GW151226_o3afin}{2.82}{GW170104_o3afin}{2.84}{GW170608_o3afin}{2.81}{GW170729_o3afin}{2.83}{GW170809_o3afin}{2.84}{GW170814_o3afin}{2.74}{GW170818_o3afin}{2.76}{GW170823_o3afin}{2.82}{GW190930C_o3afin}{2.84}{GW190929B_o3afin}{2.86}{GW190924A_o3afin}{2.83}{GW190915K_o3afin}{2.78}{GW190910B_o3afin}{2.85}{GW190828B_o3afin}{2.83}{GW190828A_o3afin}{2.84}{GW190814H_o3afin}{2.83}{GW190803B_o3afin}{2.84}{GW190731E_o3afin}{2.86}{GW190728D_o3afin}{2.77}{GW190727B_o3afin}{2.79}{GW190720A_o3afin}{2.82}{GW190719H_o3afin}{2.81}{GW190708M_o3afin}{2.85}{GW190707E_o3afin}{2.89}{GW190706F_o3afin}{2.84}{GW190701E_o3afin}{2.86}{GW190630E_o3afin}{2.82}{GW190620B_o3afin}{2.82}{GW190602E_o3afin}{2.80}{GW190527H_o3afin}{2.76}{GW190521E_o3afin}{2.87}{GW190521B_o3afin}{2.85}{GW190519J_o3afin}{2.86}{GW190517B_o3afin}{2.69}{GW190514E_o3afin}{2.89}{GW190513E_o3afin}{2.83}{GW190512G_o3afin}{2.82}{GW190503E_o3afin}{2.83}{GW190425B_o3afin}{2.74}{GW190421I_o3afin}{2.79}{GW190413E_o3afin}{2.82}{GW190413A_o3afin}{2.83}{GW190412B_o3afin}{2.87}{GW190408H_o3afin}{2.83}}}
\newcommand{\phitwomed}[1]{\IfEqCase{#1}{{GW190926C_o3afin}{3.13}{GW190925J_o3afin}{3.16}{GW190917B_o3afin}{3.14}{GW190916K_o3afin}{3.13}{GW190805J_o3afin}{3.14}{GW190725F_o3afin}{3.15}{GW190426N_o3afin}{3.15}{GW190403B_o3afin}{3.26}{GW150914_o3afin}{3.19}{GW151012_o3afin}{3.17}{GW151226_o3afin}{3.13}{GW170104_o3afin}{3.14}{GW170608_o3afin}{3.12}{GW170729_o3afin}{3.14}{GW170809_o3afin}{3.15}{GW170814_o3afin}{3.06}{GW170818_o3afin}{3.07}{GW170823_o3afin}{3.14}{GW190930C_o3afin}{3.14}{GW190929B_o3afin}{3.17}{GW190924A_o3afin}{3.15}{GW190915K_o3afin}{3.11}{GW190910B_o3afin}{3.14}{GW190828B_o3afin}{3.14}{GW190828A_o3afin}{3.15}{GW190814H_o3afin}{3.15}{GW190803B_o3afin}{3.15}{GW190731E_o3afin}{3.18}{GW190728D_o3afin}{3.09}{GW190727B_o3afin}{3.11}{GW190720A_o3afin}{3.15}{GW190719H_o3afin}{3.13}{GW190708M_o3afin}{3.17}{GW190707E_o3afin}{3.18}{GW190706F_o3afin}{3.15}{GW190701E_o3afin}{3.18}{GW190630E_o3afin}{3.15}{GW190620B_o3afin}{3.12}{GW190602E_o3afin}{3.11}{GW190527H_o3afin}{3.09}{GW190521E_o3afin}{3.20}{GW190521B_o3afin}{3.15}{GW190519J_o3afin}{3.18}{GW190517B_o3afin}{3.01}{GW190514E_o3afin}{3.19}{GW190513E_o3afin}{3.13}{GW190512G_o3afin}{3.13}{GW190503E_o3afin}{3.16}{GW190425B_o3afin}{3.02}{GW190421I_o3afin}{3.10}{GW190413E_o3afin}{3.15}{GW190413A_o3afin}{3.16}{GW190412B_o3afin}{3.16}{GW190408H_o3afin}{3.12}}}
\newcommand{\phitwoplus}[1]{\IfEqCase{#1}{{GW190926C_o3afin}{2.85}{GW190925J_o3afin}{2.80}{GW190917B_o3afin}{2.83}{GW190916K_o3afin}{2.82}{GW190805J_o3afin}{2.85}{GW190725F_o3afin}{2.81}{GW190426N_o3afin}{2.84}{GW190403B_o3afin}{2.71}{GW150914_o3afin}{2.82}{GW151012_o3afin}{2.81}{GW151226_o3afin}{2.85}{GW170104_o3afin}{2.82}{GW170608_o3afin}{2.85}{GW170729_o3afin}{2.81}{GW170809_o3afin}{2.79}{GW170814_o3afin}{2.89}{GW170818_o3afin}{2.91}{GW170823_o3afin}{2.86}{GW190930C_o3afin}{2.84}{GW190929B_o3afin}{2.80}{GW190924A_o3afin}{2.82}{GW190915K_o3afin}{2.86}{GW190910B_o3afin}{2.83}{GW190828B_o3afin}{2.82}{GW190828A_o3afin}{2.82}{GW190814H_o3afin}{2.81}{GW190803B_o3afin}{2.81}{GW190731E_o3afin}{2.80}{GW190728D_o3afin}{2.88}{GW190727B_o3afin}{2.84}{GW190720A_o3afin}{2.83}{GW190719H_o3afin}{2.84}{GW190708M_o3afin}{2.78}{GW190707E_o3afin}{2.79}{GW190706F_o3afin}{2.82}{GW190701E_o3afin}{2.82}{GW190630E_o3afin}{2.83}{GW190620B_o3afin}{2.84}{GW190602E_o3afin}{2.85}{GW190527H_o3afin}{2.85}{GW190521E_o3afin}{2.76}{GW190521B_o3afin}{2.83}{GW190519J_o3afin}{2.77}{GW190517B_o3afin}{2.92}{GW190514E_o3afin}{2.80}{GW190513E_o3afin}{2.83}{GW190512G_o3afin}{2.84}{GW190503E_o3afin}{2.80}{GW190425B_o3afin}{2.96}{GW190421I_o3afin}{2.87}{GW190413E_o3afin}{2.82}{GW190413A_o3afin}{2.82}{GW190412B_o3afin}{2.89}{GW190408H_o3afin}{2.84}}}
\newcommand{\chieffminus}[1]{\IfEqCase{#1}{{GW190926C_o3afin}{0.32}{GW190925J_o3afin}{0.15}{GW190917B_o3afin}{0.43}{GW190916K_o3afin}{0.31}{GW190805J_o3afin}{0.39}{GW190725F_o3afin}{0.16}{GW190426N_o3afin}{0.41}{GW190403B_o3afin}{0.43}{GW150914_o3afin}{0.14}{GW151012_o3afin}{0.21}{GW151226_o3afin}{0.08}{GW170104_o3afin}{0.19}{GW170608_o3afin}{0.05}{GW170729_o3afin}{0.33}{GW170809_o3afin}{0.17}{GW170814_o3afin}{0.12}{GW170818_o3afin}{0.22}{GW170823_o3afin}{0.22}{GW190930C_o3afin}{0.16}{GW190929B_o3afin}{0.28}{GW190924A_o3afin}{0.08}{GW190915K_o3afin}{0.24}{GW190910B_o3afin}{0.20}{GW190828B_o3afin}{0.17}{GW190828A_o3afin}{0.16}{GW190814H_o3afin}{0.07}{GW190803B_o3afin}{0.28}{GW190731E_o3afin}{0.25}{GW190728D_o3afin}{0.07}{GW190727B_o3afin}{0.27}{GW190720A_o3afin}{0.11}{GW190719H_o3afin}{0.32}{GW190708M_o3afin}{0.10}{GW190707E_o3afin}{0.09}{GW190706F_o3afin}{0.31}{GW190701E_o3afin}{0.31}{GW190630E_o3afin}{0.13}{GW190620B_o3afin}{0.29}{GW190602E_o3afin}{0.28}{GW190527H_o3afin}{0.22}{GW190521E_o3afin}{0.13}{GW190521B_o3afin}{0.45}{GW190519J_o3afin}{0.24}{GW190517B_o3afin}{0.28}{GW190514E_o3afin}{0.35}{GW190513E_o3afin}{0.22}{GW190512G_o3afin}{0.14}{GW190503E_o3afin}{0.30}{GW190425B_o3afin}{0.05}{GW190421I_o3afin}{0.27}{GW190413E_o3afin}{0.38}{GW190413A_o3afin}{0.32}{GW190412B_o3afin}{0.13}{GW190408H_o3afin}{0.17}}}
\newcommand{\chieffmed}[1]{\IfEqCase{#1}{{GW190926C_o3afin}{-0.02}{GW190925J_o3afin}{0.09}{GW190917B_o3afin}{-0.08}{GW190916K_o3afin}{0.20}{GW190805J_o3afin}{0.37}{GW190725F_o3afin}{-0.04}{GW190426N_o3afin}{0.23}{GW190403B_o3afin}{0.68}{GW150914_o3afin}{-0.04}{GW151012_o3afin}{0.12}{GW151226_o3afin}{0.20}{GW170104_o3afin}{-0.04}{GW170608_o3afin}{0.05}{GW170729_o3afin}{0.29}{GW170809_o3afin}{0.07}{GW170814_o3afin}{0.08}{GW170818_o3afin}{-0.06}{GW170823_o3afin}{0.05}{GW190930C_o3afin}{0.19}{GW190929B_o3afin}{-0.03}{GW190924A_o3afin}{0.03}{GW190915K_o3afin}{-0.03}{GW190910B_o3afin}{0.00}{GW190828B_o3afin}{0.05}{GW190828A_o3afin}{0.15}{GW190814H_o3afin}{0.004}{GW190803B_o3afin}{-0.01}{GW190731E_o3afin}{0.07}{GW190728D_o3afin}{0.13}{GW190727B_o3afin}{0.09}{GW190720A_o3afin}{0.19}{GW190719H_o3afin}{0.25}{GW190708M_o3afin}{0.05}{GW190707E_o3afin}{-0.04}{GW190706F_o3afin}{0.28}{GW190701E_o3afin}{-0.08}{GW190630E_o3afin}{0.10}{GW190620B_o3afin}{0.34}{GW190602E_o3afin}{0.12}{GW190527H_o3afin}{0.10}{GW190521E_o3afin}{0.10}{GW190521B_o3afin}{-0.14}{GW190519J_o3afin}{0.33}{GW190517B_o3afin}{0.49}{GW190514E_o3afin}{-0.08}{GW190513E_o3afin}{0.16}{GW190512G_o3afin}{0.02}{GW190503E_o3afin}{-0.05}{GW190425B_o3afin}{0.07}{GW190421I_o3afin}{-0.10}{GW190413E_o3afin}{-0.01}{GW190413A_o3afin}{-0.04}{GW190412B_o3afin}{0.21}{GW190408H_o3afin}{-0.03}}}
\newcommand{\chieffplus}[1]{\IfEqCase{#1}{{GW190926C_o3afin}{0.25}{GW190925J_o3afin}{0.16}{GW190917B_o3afin}{0.21}{GW190916K_o3afin}{0.33}{GW190805J_o3afin}{0.29}{GW190725F_o3afin}{0.36}{GW190426N_o3afin}{0.42}{GW190403B_o3afin}{0.16}{GW150914_o3afin}{0.12}{GW151012_o3afin}{0.28}{GW151226_o3afin}{0.23}{GW170104_o3afin}{0.15}{GW170608_o3afin}{0.13}{GW170729_o3afin}{0.25}{GW170809_o3afin}{0.17}{GW170814_o3afin}{0.13}{GW170818_o3afin}{0.19}{GW170823_o3afin}{0.21}{GW190930C_o3afin}{0.22}{GW190929B_o3afin}{0.23}{GW190924A_o3afin}{0.20}{GW190915K_o3afin}{0.19}{GW190910B_o3afin}{0.17}{GW190828B_o3afin}{0.16}{GW190828A_o3afin}{0.15}{GW190814H_o3afin}{0.07}{GW190803B_o3afin}{0.23}{GW190731E_o3afin}{0.28}{GW190728D_o3afin}{0.19}{GW190727B_o3afin}{0.25}{GW190720A_o3afin}{0.14}{GW190719H_o3afin}{0.33}{GW190708M_o3afin}{0.10}{GW190707E_o3afin}{0.10}{GW190706F_o3afin}{0.25}{GW190701E_o3afin}{0.23}{GW190630E_o3afin}{0.14}{GW190620B_o3afin}{0.22}{GW190602E_o3afin}{0.25}{GW190527H_o3afin}{0.22}{GW190521E_o3afin}{0.13}{GW190521B_o3afin}{0.50}{GW190519J_o3afin}{0.20}{GW190517B_o3afin}{0.21}{GW190514E_o3afin}{0.29}{GW190513E_o3afin}{0.29}{GW190512G_o3afin}{0.13}{GW190503E_o3afin}{0.23}{GW190425B_o3afin}{0.07}{GW190421I_o3afin}{0.21}{GW190413E_o3afin}{0.28}{GW190413A_o3afin}{0.27}{GW190412B_o3afin}{0.12}{GW190408H_o3afin}{0.13}}}
\newcommand{\chipminus}[1]{\IfEqCase{#1}{{GW190926C_o3afin}{0.30}{GW190925J_o3afin}{0.30}{GW190917B_o3afin}{0.13}{GW190916K_o3afin}{0.28}{GW190805J_o3afin}{0.32}{GW190725F_o3afin}{0.28}{GW190426N_o3afin}{0.36}{GW190403B_o3afin}{0.22}{GW150914_o3afin}{0.38}{GW151012_o3afin}{0.27}{GW151226_o3afin}{0.35}{GW170104_o3afin}{0.31}{GW170608_o3afin}{0.24}{GW170729_o3afin}{0.29}{GW170809_o3afin}{0.30}{GW170814_o3afin}{0.37}{GW170818_o3afin}{0.41}{GW170823_o3afin}{0.35}{GW190930C_o3afin}{0.21}{GW190929B_o3afin}{0.25}{GW190924A_o3afin}{0.19}{GW190915K_o3afin}{0.40}{GW190910B_o3afin}{0.30}{GW190828B_o3afin}{0.20}{GW190828A_o3afin}{0.32}{GW190814H_o3afin}{0.03}{GW190803B_o3afin}{0.34}{GW190731E_o3afin}{0.32}{GW190728D_o3afin}{0.20}{GW190727B_o3afin}{0.37}{GW190720A_o3afin}{0.20}{GW190719H_o3afin}{0.32}{GW190708M_o3afin}{0.20}{GW190707E_o3afin}{0.22}{GW190706F_o3afin}{0.33}{GW190701E_o3afin}{0.33}{GW190630E_o3afin}{0.24}{GW190620B_o3afin}{0.32}{GW190602E_o3afin}{0.34}{GW190527H_o3afin}{0.28}{GW190521E_o3afin}{0.30}{GW190521B_o3afin}{0.35}{GW190519J_o3afin}{0.28}{GW190517B_o3afin}{0.32}{GW190514E_o3afin}{0.33}{GW190513E_o3afin}{0.26}{GW190512G_o3afin}{0.20}{GW190503E_o3afin}{0.33}{GW190425B_o3afin}{0.27}{GW190421I_o3afin}{0.34}{GW190413E_o3afin}{0.41}{GW190413A_o3afin}{0.33}{GW190412B_o3afin}{0.12}{GW190408H_o3afin}{0.29}}}
\newcommand{\chipmed}[1]{\IfEqCase{#1}{{GW190926C_o3afin}{0.38}{GW190925J_o3afin}{0.39}{GW190917B_o3afin}{0.17}{GW190916K_o3afin}{0.37}{GW190805J_o3afin}{0.50}{GW190725F_o3afin}{0.37}{GW190426N_o3afin}{0.51}{GW190403B_o3afin}{0.32}{GW150914_o3afin}{0.51}{GW151012_o3afin}{0.36}{GW151226_o3afin}{0.52}{GW170104_o3afin}{0.40}{GW170608_o3afin}{0.32}{GW170729_o3afin}{0.39}{GW170809_o3afin}{0.39}{GW170814_o3afin}{0.48}{GW170818_o3afin}{0.56}{GW170823_o3afin}{0.47}{GW190930C_o3afin}{0.30}{GW190929B_o3afin}{0.31}{GW190924A_o3afin}{0.25}{GW190915K_o3afin}{0.56}{GW190910B_o3afin}{0.38}{GW190828B_o3afin}{0.26}{GW190828A_o3afin}{0.43}{GW190814H_o3afin}{0.04}{GW190803B_o3afin}{0.44}{GW190731E_o3afin}{0.41}{GW190728D_o3afin}{0.29}{GW190727B_o3afin}{0.50}{GW190720A_o3afin}{0.29}{GW190719H_o3afin}{0.45}{GW190708M_o3afin}{0.26}{GW190707E_o3afin}{0.28}{GW190706F_o3afin}{0.47}{GW190701E_o3afin}{0.44}{GW190630E_o3afin}{0.33}{GW190620B_o3afin}{0.48}{GW190602E_o3afin}{0.45}{GW190527H_o3afin}{0.36}{GW190521E_o3afin}{0.39}{GW190521B_o3afin}{0.49}{GW190519J_o3afin}{0.45}{GW190517B_o3afin}{0.55}{GW190514E_o3afin}{0.45}{GW190513E_o3afin}{0.35}{GW190512G_o3afin}{0.26}{GW190503E_o3afin}{0.43}{GW190425B_o3afin}{0.34}{GW190421I_o3afin}{0.45}{GW190413E_o3afin}{0.55}{GW190413A_o3afin}{0.44}{GW190412B_o3afin}{0.19}{GW190408H_o3afin}{0.37}}}
\newcommand{\chipplus}[1]{\IfEqCase{#1}{{GW190926C_o3afin}{0.47}{GW190925J_o3afin}{0.43}{GW190917B_o3afin}{0.42}{GW190916K_o3afin}{0.43}{GW190805J_o3afin}{0.34}{GW190725F_o3afin}{0.46}{GW190426N_o3afin}{0.37}{GW190403B_o3afin}{0.38}{GW150914_o3afin}{0.35}{GW151012_o3afin}{0.43}{GW151226_o3afin}{0.36}{GW170104_o3afin}{0.40}{GW170608_o3afin}{0.41}{GW170729_o3afin}{0.40}{GW170809_o3afin}{0.44}{GW170814_o3afin}{0.38}{GW170818_o3afin}{0.34}{GW170823_o3afin}{0.41}{GW190930C_o3afin}{0.42}{GW190929B_o3afin}{0.51}{GW190924A_o3afin}{0.41}{GW190915K_o3afin}{0.34}{GW190910B_o3afin}{0.43}{GW190828B_o3afin}{0.42}{GW190828A_o3afin}{0.41}{GW190814H_o3afin}{0.04}{GW190803B_o3afin}{0.42}{GW190731E_o3afin}{0.43}{GW190728D_o3afin}{0.39}{GW190727B_o3afin}{0.38}{GW190720A_o3afin}{0.39}{GW190719H_o3afin}{0.39}{GW190708M_o3afin}{0.44}{GW190707E_o3afin}{0.39}{GW190706F_o3afin}{0.38}{GW190701E_o3afin}{0.41}{GW190630E_o3afin}{0.36}{GW190620B_o3afin}{0.38}{GW190602E_o3afin}{0.43}{GW190527H_o3afin}{0.47}{GW190521E_o3afin}{0.37}{GW190521B_o3afin}{0.33}{GW190519J_o3afin}{0.36}{GW190517B_o3afin}{0.31}{GW190514E_o3afin}{0.42}{GW190513E_o3afin}{0.43}{GW190512G_o3afin}{0.41}{GW190503E_o3afin}{0.40}{GW190425B_o3afin}{0.37}{GW190421I_o3afin}{0.41}{GW190413E_o3afin}{0.36}{GW190413A_o3afin}{0.42}{GW190412B_o3afin}{0.22}{GW190408H_o3afin}{0.41}}}
\newcommand{\costiltoneminus}[1]{\IfEqCase{#1}{{GW190926C_o3afin}{0.81}{GW190925J_o3afin}{0.98}{GW190917B_o3afin}{0.55}{GW190916K_o3afin}{1.18}{GW190805J_o3afin}{0.94}{GW190725F_o3afin}{0.82}{GW190426N_o3afin}{1.19}{GW190403B_o3afin}{0.46}{GW150914_o3afin}{0.72}{GW151012_o3afin}{1.08}{GW151226_o3afin}{0.61}{GW170104_o3afin}{0.70}{GW170608_o3afin}{0.90}{GW170729_o3afin}{1.05}{GW170809_o3afin}{0.98}{GW170814_o3afin}{0.95}{GW170818_o3afin}{0.74}{GW170823_o3afin}{0.93}{GW190930C_o3afin}{1.03}{GW190929B_o3afin}{0.77}{GW190924A_o3afin}{0.85}{GW190915K_o3afin}{0.74}{GW190910B_o3afin}{0.89}{GW190828B_o3afin}{0.94}{GW190828A_o3afin}{0.98}{GW190814H_o3afin}{0.93}{GW190803B_o3afin}{0.83}{GW190731E_o3afin}{1.03}{GW190728D_o3afin}{1.07}{GW190727B_o3afin}{0.99}{GW190720A_o3afin}{1.00}{GW190719H_o3afin}{1.02}{GW190708M_o3afin}{0.92}{GW190707E_o3afin}{0.73}{GW190706F_o3afin}{0.88}{GW190701E_o3afin}{0.68}{GW190630E_o3afin}{1.01}{GW190620B_o3afin}{0.81}{GW190602E_o3afin}{0.98}{GW190527H_o3afin}{1.02}{GW190521E_o3afin}{0.95}{GW190521B_o3afin}{0.44}{GW190519J_o3afin}{0.70}{GW190517B_o3afin}{0.46}{GW190514E_o3afin}{0.69}{GW190513E_o3afin}{1.09}{GW190512G_o3afin}{0.91}{GW190503E_o3afin}{0.70}{GW190425B_o3afin}{0.72}{GW190421I_o3afin}{0.66}{GW190413E_o3afin}{0.78}{GW190413A_o3afin}{0.80}{GW190412B_o3afin}{0.54}{GW190408H_o3afin}{0.74}}}
\newcommand{\costiltonemed}[1]{\IfEqCase{#1}{{GW190926C_o3afin}{-0.06}{GW190925J_o3afin}{0.26}{GW190917B_o3afin}{-0.40}{GW190916K_o3afin}{0.53}{GW190805J_o3afin}{0.69}{GW190725F_o3afin}{0.005}{GW190426N_o3afin}{0.50}{GW190403B_o3afin}{0.93}{GW150914_o3afin}{-0.16}{GW151012_o3afin}{0.38}{GW151226_o3afin}{0.52}{GW170104_o3afin}{-0.19}{GW170608_o3afin}{0.21}{GW170729_o3afin}{0.67}{GW170809_o3afin}{0.20}{GW170814_o3afin}{0.29}{GW170818_o3afin}{-0.11}{GW170823_o3afin}{0.12}{GW190930C_o3afin}{0.57}{GW190929B_o3afin}{-0.12}{GW190924A_o3afin}{0.10}{GW190915K_o3afin}{-0.04}{GW190910B_o3afin}{0.03}{GW190828B_o3afin}{0.21}{GW190828A_o3afin}{0.46}{GW190814H_o3afin}{0.001}{GW190803B_o3afin}{-0.05}{GW190731E_o3afin}{0.21}{GW190728D_o3afin}{0.49}{GW190727B_o3afin}{0.24}{GW190720A_o3afin}{0.54}{GW190719H_o3afin}{0.57}{GW190708M_o3afin}{0.22}{GW190707E_o3afin}{-0.16}{GW190706F_o3afin}{0.60}{GW190701E_o3afin}{-0.24}{GW190630E_o3afin}{0.28}{GW190620B_o3afin}{0.64}{GW190602E_o3afin}{0.27}{GW190527H_o3afin}{0.30}{GW190521E_o3afin}{0.13}{GW190521B_o3afin}{-0.51}{GW190519J_o3afin}{0.66}{GW190517B_o3afin}{0.79}{GW190514E_o3afin}{-0.23}{GW190513E_o3afin}{0.49}{GW190512G_o3afin}{0.08}{GW190503E_o3afin}{-0.21}{GW190425B_o3afin}{0.31}{GW190421I_o3afin}{-0.25}{GW190413E_o3afin}{-0.05}{GW190413A_o3afin}{-0.08}{GW190412B_o3afin}{0.80}{GW190408H_o3afin}{-0.14}}}
\newcommand{\costiltoneplus}[1]{\IfEqCase{#1}{{GW190926C_o3afin}{0.88}{GW190925J_o3afin}{0.65}{GW190917B_o3afin}{1.20}{GW190916K_o3afin}{0.43}{GW190805J_o3afin}{0.28}{GW190725F_o3afin}{0.89}{GW190426N_o3afin}{0.46}{GW190403B_o3afin}{0.07}{GW150914_o3afin}{0.90}{GW151012_o3afin}{0.55}{GW151226_o3afin}{0.40}{GW170104_o3afin}{0.93}{GW170608_o3afin}{0.68}{GW170729_o3afin}{0.30}{GW170809_o3afin}{0.69}{GW170814_o3afin}{0.62}{GW170818_o3afin}{0.86}{GW170823_o3afin}{0.76}{GW190930C_o3afin}{0.40}{GW190929B_o3afin}{0.89}{GW190924A_o3afin}{0.78}{GW190915K_o3afin}{0.81}{GW190910B_o3afin}{0.83}{GW190828B_o3afin}{0.68}{GW190828A_o3afin}{0.48}{GW190814H_o3afin}{0.94}{GW190803B_o3afin}{0.88}{GW190731E_o3afin}{0.69}{GW190728D_o3afin}{0.46}{GW190727B_o3afin}{0.67}{GW190720A_o3afin}{0.42}{GW190719H_o3afin}{0.38}{GW190708M_o3afin}{0.68}{GW190707E_o3afin}{0.95}{GW190706F_o3afin}{0.36}{GW190701E_o3afin}{1.00}{GW190630E_o3afin}{0.64}{GW190620B_o3afin}{0.33}{GW190602E_o3afin}{0.64}{GW190527H_o3afin}{0.61}{GW190521E_o3afin}{0.74}{GW190521B_o3afin}{1.34}{GW190519J_o3afin}{0.31}{GW190517B_o3afin}{0.19}{GW190514E_o3afin}{1.02}{GW190513E_o3afin}{0.46}{GW190512G_o3afin}{0.80}{GW190503E_o3afin}{0.99}{GW190425B_o3afin}{0.58}{GW190421I_o3afin}{0.99}{GW190413E_o3afin}{0.86}{GW190413A_o3afin}{0.92}{GW190412B_o3afin}{0.18}{GW190408H_o3afin}{0.93}}}
\newcommand{\costilttwominus}[1]{\IfEqCase{#1}{{GW190926C_o3afin}{0.82}{GW190925J_o3afin}{1.02}{GW190917B_o3afin}{0.72}{GW190916K_o3afin}{1.13}{GW190805J_o3afin}{1.18}{GW190725F_o3afin}{0.75}{GW190426N_o3afin}{1.11}{GW190403B_o3afin}{1.15}{GW150914_o3afin}{0.83}{GW151012_o3afin}{1.06}{GW151226_o3afin}{1.06}{GW170104_o3afin}{0.83}{GW170608_o3afin}{1.02}{GW170729_o3afin}{1.10}{GW170809_o3afin}{1.04}{GW170814_o3afin}{0.94}{GW170818_o3afin}{0.76}{GW170823_o3afin}{0.96}{GW190930C_o3afin}{1.15}{GW190929B_o3afin}{0.83}{GW190924A_o3afin}{0.96}{GW190915K_o3afin}{0.80}{GW190910B_o3afin}{0.82}{GW190828B_o3afin}{1.02}{GW190828A_o3afin}{1.08}{GW190814H_o3afin}{0.96}{GW190803B_o3afin}{0.84}{GW190731E_o3afin}{1.00}{GW190728D_o3afin}{1.16}{GW190727B_o3afin}{1.00}{GW190720A_o3afin}{1.21}{GW190719H_o3afin}{1.12}{GW190708M_o3afin}{0.99}{GW190707E_o3afin}{0.75}{GW190706F_o3afin}{1.15}{GW190701E_o3afin}{0.75}{GW190630E_o3afin}{1.07}{GW190620B_o3afin}{1.24}{GW190602E_o3afin}{1.06}{GW190527H_o3afin}{1.05}{GW190521E_o3afin}{1.10}{GW190521B_o3afin}{1.01}{GW190519J_o3afin}{1.16}{GW190517B_o3afin}{1.21}{GW190514E_o3afin}{0.76}{GW190513E_o3afin}{1.10}{GW190512G_o3afin}{0.94}{GW190503E_o3afin}{0.88}{GW190425B_o3afin}{0.91}{GW190421I_o3afin}{0.68}{GW190413E_o3afin}{0.92}{GW190413A_o3afin}{0.79}{GW190412B_o3afin}{1.09}{GW190408H_o3afin}{0.84}}}
\newcommand{\costilttwomed}[1]{\IfEqCase{#1}{{GW190926C_o3afin}{-0.09}{GW190925J_o3afin}{0.25}{GW190917B_o3afin}{-0.20}{GW190916K_o3afin}{0.33}{GW190805J_o3afin}{0.43}{GW190725F_o3afin}{-0.14}{GW190426N_o3afin}{0.31}{GW190403B_o3afin}{0.35}{GW150914_o3afin}{-0.04}{GW151012_o3afin}{0.24}{GW151226_o3afin}{0.19}{GW170104_o3afin}{-0.05}{GW170608_o3afin}{0.21}{GW170729_o3afin}{0.26}{GW170809_o3afin}{0.22}{GW170814_o3afin}{0.11}{GW170818_o3afin}{-0.14}{GW170823_o3afin}{0.14}{GW190930C_o3afin}{0.40}{GW190929B_o3afin}{-0.08}{GW190924A_o3afin}{0.17}{GW190915K_o3afin}{-0.09}{GW190910B_o3afin}{-0.06}{GW190828B_o3afin}{0.16}{GW190828A_o3afin}{0.29}{GW190814H_o3afin}{0.12}{GW190803B_o3afin}{-0.05}{GW190731E_o3afin}{0.15}{GW190728D_o3afin}{0.43}{GW190727B_o3afin}{0.17}{GW190720A_o3afin}{0.53}{GW190719H_o3afin}{0.31}{GW190708M_o3afin}{0.19}{GW190707E_o3afin}{-0.11}{GW190706F_o3afin}{0.35}{GW190701E_o3afin}{-0.16}{GW190630E_o3afin}{0.37}{GW190620B_o3afin}{0.50}{GW190602E_o3afin}{0.24}{GW190527H_o3afin}{0.23}{GW190521E_o3afin}{0.42}{GW190521B_o3afin}{0.14}{GW190519J_o3afin}{0.52}{GW190517B_o3afin}{0.41}{GW190514E_o3afin}{-0.15}{GW190513E_o3afin}{0.29}{GW190512G_o3afin}{0.13}{GW190503E_o3afin}{-0.01}{GW190425B_o3afin}{0.22}{GW190421I_o3afin}{-0.25}{GW190413E_o3afin}{0.04}{GW190413A_o3afin}{-0.12}{GW190412B_o3afin}{0.33}{GW190408H_o3afin}{-0.03}}}
\newcommand{\costilttwoplus}[1]{\IfEqCase{#1}{{GW190926C_o3afin}{0.96}{GW190925J_o3afin}{0.67}{GW190917B_o3afin}{1.03}{GW190916K_o3afin}{0.62}{GW190805J_o3afin}{0.53}{GW190725F_o3afin}{0.99}{GW190426N_o3afin}{0.63}{GW190403B_o3afin}{0.60}{GW150914_o3afin}{0.88}{GW151012_o3afin}{0.68}{GW151226_o3afin}{0.73}{GW170104_o3afin}{0.92}{GW170608_o3afin}{0.70}{GW170729_o3afin}{0.68}{GW170809_o3afin}{0.69}{GW170814_o3afin}{0.75}{GW170818_o3afin}{0.94}{GW170823_o3afin}{0.76}{GW190930C_o3afin}{0.54}{GW190929B_o3afin}{0.96}{GW190924A_o3afin}{0.73}{GW190915K_o3afin}{0.94}{GW190910B_o3afin}{0.92}{GW190828B_o3afin}{0.75}{GW190828A_o3afin}{0.64}{GW190814H_o3afin}{0.80}{GW190803B_o3afin}{0.92}{GW190731E_o3afin}{0.75}{GW190728D_o3afin}{0.52}{GW190727B_o3afin}{0.74}{GW190720A_o3afin}{0.44}{GW190719H_o3afin}{0.63}{GW190708M_o3afin}{0.73}{GW190707E_o3afin}{0.96}{GW190706F_o3afin}{0.60}{GW190701E_o3afin}{0.99}{GW190630E_o3afin}{0.56}{GW190620B_o3afin}{0.46}{GW190602E_o3afin}{0.69}{GW190527H_o3afin}{0.68}{GW190521E_o3afin}{0.53}{GW190521B_o3afin}{0.77}{GW190519J_o3afin}{0.43}{GW190517B_o3afin}{0.54}{GW190514E_o3afin}{1.00}{GW190513E_o3afin}{0.65}{GW190512G_o3afin}{0.78}{GW190503E_o3afin}{0.88}{GW190425B_o3afin}{0.62}{GW190421I_o3afin}{1.04}{GW190413E_o3afin}{0.86}{GW190413A_o3afin}{0.98}{GW190412B_o3afin}{0.61}{GW190408H_o3afin}{0.89}}}
\newcommand{\comovingdistminus}[1]{\IfEqCase{#1}{{GW190926C_o3afin}{914}{GW190925J_o3afin}{270}{GW190917B_o3afin}{248}{GW190916K_o3afin}{1022}{GW190805J_o3afin}{1183}{GW190725F_o3afin}{326}{GW190426N_o3afin}{1024}{GW190403B_o3afin}{1388}{GW150914_o3afin}{137}{GW151012_o3afin}{370}{GW151226_o3afin}{170}{GW170104_o3afin}{358}{GW170608_o3afin}{113}{GW170729_o3afin}{719}{GW170809_o3afin}{283}{GW170814_o3afin}{190}{GW170818_o3afin}{301}{GW170823_o3afin}{590}{GW190930C_o3afin}{255}{GW190929B_o3afin}{719}{GW190924A_o3afin}{187}{GW190915K_o3afin}{418}{GW190910B_o3afin}{426}{GW190828B_o3afin}{435}{GW190828A_o3afin}{565}{GW190814H_o3afin}{43}{GW190803B_o3afin}{770}{GW190731E_o3afin}{929}{GW190728D_o3afin}{299}{GW190727B_o3afin}{641}{GW190720A_o3afin}{206}{GW190719H_o3afin}{1047}{GW190708M_o3afin}{299}{GW190707E_o3afin}{311}{GW190706F_o3afin}{1020}{GW190701E_o3afin}{443}{GW190630E_o3afin}{282}{GW190620B_o3afin}{716}{GW190602E_o3afin}{701}{GW190527H_o3afin}{717}{GW190521E_o3afin}{398}{GW190521B_o3afin}{950}{GW190519J_o3afin}{535}{GW190517B_o3afin}{573}{GW190514E_o3afin}{1015}{GW190513E_o3afin}{478}{GW190512G_o3afin}{403}{GW190503E_o3afin}{406}{GW190425B_o3afin}{60}{GW190421I_o3afin}{712}{GW190413E_o3afin}{894}{GW190413A_o3afin}{711}{GW190412B_o3afin}{173}{GW190408H_o3afin}{413}}}
\newcommand{\comovingdistmed}[1]{\IfEqCase{#1}{{GW190926C_o3afin}{2113}{GW190925J_o3afin}{782}{GW190917B_o3afin}{626}{GW190916K_o3afin}{2787}{GW190805J_o3afin}{3191}{GW190725F_o3afin}{853}{GW190426N_o3afin}{2653}{GW190403B_o3afin}{3809}{GW150914_o3afin}{430}{GW151012_o3afin}{834}{GW151226_o3afin}{420}{GW170104_o3afin}{912}{GW170608_o3afin}{317}{GW170729_o3afin}{1733}{GW170809_o3afin}{880}{GW170814_o3afin}{538}{GW170818_o3afin}{892}{GW170823_o3afin}{1449}{GW190930C_o3afin}{665}{GW190929B_o3afin}{2045}{GW190924A_o3afin}{493}{GW190915K_o3afin}{1320}{GW190910B_o3afin}{1179}{GW190828B_o3afin}{1195}{GW190828A_o3afin}{1509}{GW190814H_o3afin}{221}{GW190803B_o3afin}{2072}{GW190731E_o3afin}{2136}{GW190728D_o3afin}{746}{GW190727B_o3afin}{2018}{GW190720A_o3afin}{668}{GW190719H_o3afin}{2310}{GW190708M_o3afin}{783}{GW190707E_o3afin}{727}{GW190706F_o3afin}{2267}{GW190701E_o3afin}{1518}{GW190630E_o3afin}{743}{GW190620B_o3afin}{1940}{GW190602E_o3afin}{1908}{GW190527H_o3afin}{1745}{GW190521E_o3afin}{891}{GW190521B_o3afin}{2129}{GW190519J_o3afin}{1789}{GW190517B_o3afin}{1346}{GW190514E_o3afin}{2379}{GW190513E_o3afin}{1583}{GW190512G_o3afin}{1141}{GW190503E_o3afin}{1183}{GW190425B_o3afin}{141}{GW190421I_o3afin}{1782}{GW190413E_o3afin}{2341}{GW190413A_o3afin}{2135}{GW190412B_o3afin}{629}{GW190408H_o3afin}{1193}}}
\newcommand{\comovingdistplus}[1]{\IfEqCase{#1}{{GW190926C_o3afin}{1249}{GW190925J_o3afin}{312}{GW190917B_o3afin}{220}{GW190916K_o3afin}{1115}{GW190805J_o3afin}{995}{GW190725F_o3afin}{345}{GW190426N_o3afin}{1076}{GW190403B_o3afin}{1358}{GW150914_o3afin}{112}{GW151012_o3afin}{421}{GW151226_o3afin}{131}{GW170104_o3afin}{258}{GW170608_o3afin}{106}{GW170729_o3afin}{767}{GW170809_o3afin}{209}{GW170814_o3afin}{123}{GW170818_o3afin}{284}{GW170823_o3afin}{445}{GW190930C_o3afin}{235}{GW190929B_o3afin}{985}{GW190924A_o3afin}{171}{GW190915K_o3afin}{395}{GW190910B_o3afin}{613}{GW190828B_o3afin}{398}{GW190828A_o3afin}{340}{GW190814H_o3afin}{39}{GW190803B_o3afin}{671}{GW190731E_o3afin}{909}{GW190728D_o3afin}{188}{GW190727B_o3afin}{556}{GW190720A_o3afin}{451}{GW190719H_o3afin}{1100}{GW190708M_o3afin}{219}{GW190707E_o3afin}{237}{GW190706F_o3afin}{956}{GW190701E_o3afin}{400}{GW190630E_o3afin}{361}{GW190620B_o3afin}{728}{GW190602E_o3afin}{816}{GW190527H_o3afin}{913}{GW190521E_o3afin}{373}{GW190521B_o3afin}{1053}{GW190519J_o3afin}{767}{GW190517B_o3afin}{887}{GW190514E_o3afin}{928}{GW190513E_o3afin}{492}{GW190512G_o3afin}{305}{GW190503E_o3afin}{368}{GW190425B_o3afin}{69}{GW190421I_o3afin}{676}{GW190413E_o3afin}{898}{GW190413A_o3afin}{759}{GW190412B_o3afin}{180}{GW190408H_o3afin}{260}}}
\newcommand{\redshiftminus}[1]{\IfEqCase{#1}{{GW190926C_o3afin}{0.26}{GW190925J_o3afin}{0.07}{GW190917B_o3afin}{0.06}{GW190916K_o3afin}{0.32}{GW190805J_o3afin}{0.40}{GW190725F_o3afin}{0.08}{GW190426N_o3afin}{0.32}{GW190403B_o3afin}{0.53}{GW150914_o3afin}{0.03}{GW151012_o3afin}{0.09}{GW151226_o3afin}{0.04}{GW170104_o3afin}{0.09}{GW170608_o3afin}{0.03}{GW170729_o3afin}{0.19}{GW170809_o3afin}{0.07}{GW170814_o3afin}{0.05}{GW170818_o3afin}{0.07}{GW170823_o3afin}{0.15}{GW190930C_o3afin}{0.06}{GW190929B_o3afin}{0.20}{GW190924A_o3afin}{0.04}{GW190915K_o3afin}{0.11}{GW190910B_o3afin}{0.11}{GW190828B_o3afin}{0.11}{GW190828A_o3afin}{0.15}{GW190814H_o3afin}{0.01}{GW190803B_o3afin}{0.22}{GW190731E_o3afin}{0.26}{GW190728D_o3afin}{0.07}{GW190727B_o3afin}{0.18}{GW190720A_o3afin}{0.05}{GW190719H_o3afin}{0.30}{GW190708M_o3afin}{0.07}{GW190707E_o3afin}{0.08}{GW190706F_o3afin}{0.29}{GW190701E_o3afin}{0.12}{GW190630E_o3afin}{0.07}{GW190620B_o3afin}{0.20}{GW190602E_o3afin}{0.20}{GW190527H_o3afin}{0.19}{GW190521E_o3afin}{0.10}{GW190521B_o3afin}{0.27}{GW190519J_o3afin}{0.15}{GW190517B_o3afin}{0.15}{GW190514E_o3afin}{0.30}{GW190513E_o3afin}{0.13}{GW190512G_o3afin}{0.10}{GW190503E_o3afin}{0.10}{GW190425B_o3afin}{0.01}{GW190421I_o3afin}{0.19}{GW190413E_o3afin}{0.26}{GW190413A_o3afin}{0.21}{GW190412B_o3afin}{0.04}{GW190408H_o3afin}{0.11}}}
\newcommand{\redshiftmed}[1]{\IfEqCase{#1}{{GW190926C_o3afin}{0.55}{GW190925J_o3afin}{0.19}{GW190917B_o3afin}{0.15}{GW190916K_o3afin}{0.77}{GW190805J_o3afin}{0.92}{GW190725F_o3afin}{0.20}{GW190426N_o3afin}{0.73}{GW190403B_o3afin}{1.18}{GW150914_o3afin}{0.10}{GW151012_o3afin}{0.20}{GW151226_o3afin}{0.10}{GW170104_o3afin}{0.22}{GW170608_o3afin}{0.07}{GW170729_o3afin}{0.44}{GW170809_o3afin}{0.21}{GW170814_o3afin}{0.13}{GW170818_o3afin}{0.21}{GW170823_o3afin}{0.36}{GW190930C_o3afin}{0.16}{GW190929B_o3afin}{0.53}{GW190924A_o3afin}{0.11}{GW190915K_o3afin}{0.32}{GW190910B_o3afin}{0.29}{GW190828B_o3afin}{0.29}{GW190828A_o3afin}{0.38}{GW190814H_o3afin}{0.05}{GW190803B_o3afin}{0.54}{GW190731E_o3afin}{0.56}{GW190728D_o3afin}{0.18}{GW190727B_o3afin}{0.52}{GW190720A_o3afin}{0.16}{GW190719H_o3afin}{0.61}{GW190708M_o3afin}{0.19}{GW190707E_o3afin}{0.17}{GW190706F_o3afin}{0.60}{GW190701E_o3afin}{0.38}{GW190630E_o3afin}{0.18}{GW190620B_o3afin}{0.50}{GW190602E_o3afin}{0.49}{GW190527H_o3afin}{0.44}{GW190521E_o3afin}{0.21}{GW190521B_o3afin}{0.56}{GW190519J_o3afin}{0.45}{GW190517B_o3afin}{0.33}{GW190514E_o3afin}{0.64}{GW190513E_o3afin}{0.40}{GW190512G_o3afin}{0.28}{GW190503E_o3afin}{0.29}{GW190425B_o3afin}{0.03}{GW190421I_o3afin}{0.45}{GW190413E_o3afin}{0.62}{GW190413A_o3afin}{0.56}{GW190412B_o3afin}{0.15}{GW190408H_o3afin}{0.29}}}
\newcommand{\redshiftplus}[1]{\IfEqCase{#1}{{GW190926C_o3afin}{0.44}{GW190925J_o3afin}{0.08}{GW190917B_o3afin}{0.05}{GW190916K_o3afin}{0.45}{GW190805J_o3afin}{0.43}{GW190725F_o3afin}{0.09}{GW190426N_o3afin}{0.41}{GW190403B_o3afin}{0.73}{GW150914_o3afin}{0.03}{GW151012_o3afin}{0.11}{GW151226_o3afin}{0.03}{GW170104_o3afin}{0.07}{GW170608_o3afin}{0.03}{GW170729_o3afin}{0.24}{GW170809_o3afin}{0.05}{GW170814_o3afin}{0.03}{GW170818_o3afin}{0.07}{GW170823_o3afin}{0.13}{GW190930C_o3afin}{0.06}{GW190929B_o3afin}{0.33}{GW190924A_o3afin}{0.04}{GW190915K_o3afin}{0.11}{GW190910B_o3afin}{0.17}{GW190828B_o3afin}{0.11}{GW190828A_o3afin}{0.10}{GW190814H_o3afin}{0.01}{GW190803B_o3afin}{0.22}{GW190731E_o3afin}{0.31}{GW190728D_o3afin}{0.05}{GW190727B_o3afin}{0.18}{GW190720A_o3afin}{0.11}{GW190719H_o3afin}{0.39}{GW190708M_o3afin}{0.06}{GW190707E_o3afin}{0.06}{GW190706F_o3afin}{0.33}{GW190701E_o3afin}{0.11}{GW190630E_o3afin}{0.09}{GW190620B_o3afin}{0.23}{GW190602E_o3afin}{0.26}{GW190527H_o3afin}{0.29}{GW190521E_o3afin}{0.10}{GW190521B_o3afin}{0.36}{GW190519J_o3afin}{0.24}{GW190517B_o3afin}{0.26}{GW190514E_o3afin}{0.33}{GW190513E_o3afin}{0.14}{GW190512G_o3afin}{0.08}{GW190503E_o3afin}{0.10}{GW190425B_o3afin}{0.02}{GW190421I_o3afin}{0.21}{GW190413E_o3afin}{0.32}{GW190413A_o3afin}{0.25}{GW190412B_o3afin}{0.04}{GW190408H_o3afin}{0.07}}}
\newcommand{\massonesourceminus}[1]{\IfEqCase{#1}{{GW190926C_o3afin}{12.5}{GW190925J_o3afin}{2.9}{GW190917B_o3afin}{3.9}{GW190916K_o3afin}{12.6}{GW190805J_o3afin}{11.2}{GW190725F_o3afin}{3.0}{GW190426N_o3afin}{24.1}{GW190403B_o3afin}{33.0}{GW150914_o3afin}{2.6}{GW151012_o3afin}{6.3}{GW151226_o3afin}{3.6}{GW170104_o3afin}{4.2}{GW170608_o3afin}{1.4}{GW170729_o3afin}{12.8}{GW170809_o3afin}{5.3}{GW170814_o3afin}{3.3}{GW170818_o3afin}{4.2}{GW170823_o3afin}{6.2}{GW190930C_o3afin}{4.0}{GW190929B_o3afin}{16.6}{GW190924A_o3afin}{1.8}{GW190915K_o3afin}{4.9}{GW190910B_o3afin}{6.8}{GW190828B_o3afin}{6.7}{GW190828A_o3afin}{4.1}{GW190814H_o3afin}{1.4}{GW190803B_o3afin}{6.7}{GW190731E_o3afin}{9.1}{GW190728D_o3afin}{2.3}{GW190727B_o3afin}{6.0}{GW190720A_o3afin}{3.3}{GW190719H_o3afin}{11.1}{GW190708M_o3afin}{4.3}{GW190707E_o3afin}{2.0}{GW190706F_o3afin}{16.9}{GW190701E_o3afin}{8.0}{GW190630E_o3afin}{5.5}{GW190620B_o3afin}{13.3}{GW190602E_o3afin}{14.6}{GW190527H_o3afin}{8.0}{GW190521E_o3afin}{5.5}{GW190521B_o3afin}{21.7}{GW190519J_o3afin}{11.0}{GW190517B_o3afin}{9.2}{GW190514E_o3afin}{9.3}{GW190513E_o3afin}{9.7}{GW190512G_o3afin}{5.6}{GW190503E_o3afin}{7.7}{GW190425B_o3afin}{0.4}{GW190421I_o3afin}{7.4}{GW190413E_o3afin}{12.6}{GW190413A_o3afin}{6.4}{GW190412B_o3afin}{6.0}{GW190408H_o3afin}{3.5}}}
\newcommand{\massonesourcemed}[1]{\IfEqCase{#1}{{GW190926C_o3afin}{41.1}{GW190925J_o3afin}{20.8}{GW190917B_o3afin}{9.7}{GW190916K_o3afin}{43.8}{GW190805J_o3afin}{46.2}{GW190725F_o3afin}{11.8}{GW190426N_o3afin}{105.5}{GW190403B_o3afin}{85.0}{GW150914_o3afin}{34.6}{GW151012_o3afin}{24.8}{GW151226_o3afin}{14.2}{GW170104_o3afin}{28.7}{GW170608_o3afin}{10.6}{GW170729_o3afin}{54.7}{GW170809_o3afin}{34.1}{GW170814_o3afin}{30.9}{GW170818_o3afin}{34.8}{GW170823_o3afin}{38.3}{GW190930C_o3afin}{14.2}{GW190929B_o3afin}{66.3}{GW190924A_o3afin}{8.8}{GW190915K_o3afin}{32.6}{GW190910B_o3afin}{43.8}{GW190828B_o3afin}{23.7}{GW190828A_o3afin}{31.9}{GW190814H_o3afin}{23.3}{GW190803B_o3afin}{37.7}{GW190731E_o3afin}{41.8}{GW190728D_o3afin}{12.5}{GW190727B_o3afin}{38.9}{GW190720A_o3afin}{14.2}{GW190719H_o3afin}{36.6}{GW190708M_o3afin}{19.8}{GW190707E_o3afin}{12.1}{GW190706F_o3afin}{74.0}{GW190701E_o3afin}{54.1}{GW190630E_o3afin}{35.1}{GW190620B_o3afin}{58.0}{GW190602E_o3afin}{71.8}{GW190527H_o3afin}{35.6}{GW190521E_o3afin}{43.4}{GW190521B_o3afin}{98.4}{GW190519J_o3afin}{65.1}{GW190517B_o3afin}{39.2}{GW190514E_o3afin}{40.9}{GW190513E_o3afin}{36.0}{GW190512G_o3afin}{23.2}{GW190503E_o3afin}{41.3}{GW190425B_o3afin}{2.1}{GW190421I_o3afin}{42.0}{GW190413E_o3afin}{51.3}{GW190413A_o3afin}{33.7}{GW190412B_o3afin}{27.7}{GW190408H_o3afin}{24.8}}}
\newcommand{\massonesourceplus}[1]{\IfEqCase{#1}{{GW190926C_o3afin}{20.8}{GW190925J_o3afin}{6.5}{GW190917B_o3afin}{3.4}{GW190916K_o3afin}{19.9}{GW190805J_o3afin}{15.4}{GW190725F_o3afin}{10.1}{GW190426N_o3afin}{45.3}{GW190403B_o3afin}{27.8}{GW150914_o3afin}{4.4}{GW151012_o3afin}{14.5}{GW151226_o3afin}{11.1}{GW170104_o3afin}{6.6}{GW170608_o3afin}{4.0}{GW170729_o3afin}{12.7}{GW170809_o3afin}{8.0}{GW170814_o3afin}{5.4}{GW170818_o3afin}{6.5}{GW170823_o3afin}{9.5}{GW190930C_o3afin}{8.0}{GW190929B_o3afin}{21.6}{GW190924A_o3afin}{4.3}{GW190915K_o3afin}{8.8}{GW190910B_o3afin}{7.6}{GW190828B_o3afin}{6.8}{GW190828A_o3afin}{5.4}{GW190814H_o3afin}{1.4}{GW190803B_o3afin}{9.8}{GW190731E_o3afin}{12.7}{GW190728D_o3afin}{6.9}{GW190727B_o3afin}{8.9}{GW190720A_o3afin}{5.6}{GW190719H_o3afin}{42.1}{GW190708M_o3afin}{4.3}{GW190707E_o3afin}{2.6}{GW190706F_o3afin}{20.1}{GW190701E_o3afin}{12.6}{GW190630E_o3afin}{6.5}{GW190620B_o3afin}{19.2}{GW190602E_o3afin}{18.1}{GW190527H_o3afin}{18.7}{GW190521E_o3afin}{5.8}{GW190521B_o3afin}{33.6}{GW190519J_o3afin}{10.8}{GW190517B_o3afin}{13.9}{GW190514E_o3afin}{17.3}{GW190513E_o3afin}{10.6}{GW190512G_o3afin}{5.6}{GW190503E_o3afin}{10.3}{GW190425B_o3afin}{0.5}{GW190421I_o3afin}{10.1}{GW190413E_o3afin}{16.6}{GW190413A_o3afin}{10.4}{GW190412B_o3afin}{6.0}{GW190408H_o3afin}{5.4}}}
\newcommand{\masstwosourceminus}[1]{\IfEqCase{#1}{{GW190926C_o3afin}{8.2}{GW190925J_o3afin}{3.6}{GW190917B_o3afin}{0.4}{GW190916K_o3afin}{10.0}{GW190805J_o3afin}{11.3}{GW190725F_o3afin}{2.5}{GW190426N_o3afin}{36.5}{GW190403B_o3afin}{8.4}{GW150914_o3afin}{4.6}{GW151012_o3afin}{4.9}{GW151226_o3afin}{2.8}{GW170104_o3afin}{4.7}{GW170608_o3afin}{1.9}{GW170729_o3afin}{10.2}{GW170809_o3afin}{5.3}{GW170814_o3afin}{4.0}{GW170818_o3afin}{5.1}{GW170823_o3afin}{7.8}{GW190930C_o3afin}{2.1}{GW190929B_o3afin}{10.6}{GW190924A_o3afin}{1.5}{GW190915K_o3afin}{5.8}{GW190910B_o3afin}{7.3}{GW190828B_o3afin}{2.2}{GW190828A_o3afin}{5.3}{GW190814H_o3afin}{0.1}{GW190803B_o3afin}{8.5}{GW190731E_o3afin}{9.9}{GW190728D_o3afin}{2.6}{GW190727B_o3afin}{8.3}{GW190720A_o3afin}{1.8}{GW190719H_o3afin}{9.3}{GW190708M_o3afin}{2.0}{GW190707E_o3afin}{1.3}{GW190706F_o3afin}{15.4}{GW190701E_o3afin}{12.1}{GW190630E_o3afin}{5.2}{GW190620B_o3afin}{14.5}{GW190602E_o3afin}{19.6}{GW190527H_o3afin}{8.7}{GW190521E_o3afin}{6.8}{GW190521B_o3afin}{30.1}{GW190519J_o3afin}{12.7}{GW190517B_o3afin}{7.9}{GW190514E_o3afin}{10.1}{GW190513E_o3afin}{4.7}{GW190512G_o3afin}{2.6}{GW190503E_o3afin}{9.2}{GW190425B_o3afin}{0.2}{GW190421I_o3afin}{9.8}{GW190413E_o3afin}{12.7}{GW190413A_o3afin}{7.0}{GW190412B_o3afin}{1.4}{GW190408H_o3afin}{4.0}}}
\newcommand{\masstwosourcemed}[1]{\IfEqCase{#1}{{GW190926C_o3afin}{20.4}{GW190925J_o3afin}{15.5}{GW190917B_o3afin}{2.1}{GW190916K_o3afin}{23.3}{GW190805J_o3afin}{30.6}{GW190725F_o3afin}{6.3}{GW190426N_o3afin}{76.0}{GW190403B_o3afin}{20.0}{GW150914_o3afin}{30.0}{GW151012_o3afin}{13.6}{GW151226_o3afin}{7.5}{GW170104_o3afin}{20.8}{GW170608_o3afin}{7.8}{GW170729_o3afin}{30.2}{GW170809_o3afin}{24.2}{GW170814_o3afin}{24.9}{GW170818_o3afin}{27.6}{GW170823_o3afin}{29.0}{GW190930C_o3afin}{6.9}{GW190929B_o3afin}{26.8}{GW190924A_o3afin}{5.1}{GW190915K_o3afin}{24.5}{GW190910B_o3afin}{34.2}{GW190828B_o3afin}{10.4}{GW190828A_o3afin}{25.8}{GW190814H_o3afin}{2.6}{GW190803B_o3afin}{27.6}{GW190731E_o3afin}{29.0}{GW190728D_o3afin}{8.0}{GW190727B_o3afin}{30.2}{GW190720A_o3afin}{7.5}{GW190719H_o3afin}{19.9}{GW190708M_o3afin}{11.6}{GW190707E_o3afin}{7.9}{GW190706F_o3afin}{39.4}{GW190701E_o3afin}{40.5}{GW190630E_o3afin}{24.0}{GW190620B_o3afin}{35.0}{GW190602E_o3afin}{44.8}{GW190527H_o3afin}{22.2}{GW190521E_o3afin}{33.4}{GW190521B_o3afin}{57.2}{GW190519J_o3afin}{40.8}{GW190517B_o3afin}{24.0}{GW190514E_o3afin}{28.4}{GW190513E_o3afin}{18.3}{GW190512G_o3afin}{12.5}{GW190503E_o3afin}{28.3}{GW190425B_o3afin}{1.3}{GW190421I_o3afin}{32.0}{GW190413E_o3afin}{30.4}{GW190413A_o3afin}{24.2}{GW190412B_o3afin}{9.0}{GW190408H_o3afin}{18.5}}}
\newcommand{\masstwosourceplus}[1]{\IfEqCase{#1}{{GW190926C_o3afin}{11.4}{GW190925J_o3afin}{2.5}{GW190917B_o3afin}{1.1}{GW190916K_o3afin}{12.5}{GW190805J_o3afin}{11.8}{GW190725F_o3afin}{2.1}{GW190426N_o3afin}{26.2}{GW190403B_o3afin}{26.3}{GW150914_o3afin}{2.9}{GW151012_o3afin}{4.5}{GW151226_o3afin}{2.4}{GW170104_o3afin}{4.1}{GW170608_o3afin}{1.2}{GW170729_o3afin}{11.9}{GW170809_o3afin}{4.8}{GW170814_o3afin}{3.0}{GW170818_o3afin}{4.1}{GW170823_o3afin}{6.5}{GW190930C_o3afin}{2.4}{GW190929B_o3afin}{14.7}{GW190924A_o3afin}{1.2}{GW190915K_o3afin}{4.9}{GW190910B_o3afin}{6.6}{GW190828B_o3afin}{3.8}{GW190828A_o3afin}{4.9}{GW190814H_o3afin}{0.1}{GW190803B_o3afin}{7.6}{GW190731E_o3afin}{10.2}{GW190728D_o3afin}{1.7}{GW190727B_o3afin}{6.5}{GW190720A_o3afin}{2.2}{GW190719H_o3afin}{10.0}{GW190708M_o3afin}{3.1}{GW190707E_o3afin}{1.6}{GW190706F_o3afin}{18.4}{GW190701E_o3afin}{8.7}{GW190630E_o3afin}{5.5}{GW190620B_o3afin}{13.1}{GW190602E_o3afin}{15.5}{GW190527H_o3afin}{9.0}{GW190521E_o3afin}{5.2}{GW190521B_o3afin}{27.1}{GW190519J_o3afin}{11.5}{GW190517B_o3afin}{7.4}{GW190514E_o3afin}{10.0}{GW190513E_o3afin}{7.4}{GW190512G_o3afin}{3.5}{GW190503E_o3afin}{7.5}{GW190425B_o3afin}{0.3}{GW190421I_o3afin}{8.3}{GW190413E_o3afin}{11.7}{GW190413A_o3afin}{6.5}{GW190412B_o3afin}{2.0}{GW190408H_o3afin}{3.3}}}
\newcommand{\totalmasssourceminus}[1]{\IfEqCase{#1}{{GW190926C_o3afin}{12.0}{GW190925J_o3afin}{2.8}{GW190917B_o3afin}{2.8}{GW190916K_o3afin}{13.1}{GW190805J_o3afin}{13.8}{GW190725F_o3afin}{1.9}{GW190426N_o3afin}{35.7}{GW190403B_o3afin}{23.6}{GW150914_o3afin}{3.2}{GW151012_o3afin}{4.7}{GW151226_o3afin}{1.6}{GW170104_o3afin}{3.6}{GW170608_o3afin}{0.6}{GW170729_o3afin}{10.9}{GW170809_o3afin}{3.9}{GW170814_o3afin}{3.0}{GW170818_o3afin}{4.6}{GW170823_o3afin}{7.2}{GW190930C_o3afin}{2.0}{GW190929B_o3afin}{15.0}{GW190924A_o3afin}{0.9}{GW190915K_o3afin}{5.3}{GW190910B_o3afin}{9.1}{GW190828B_o3afin}{4.3}{GW190828A_o3afin}{4.3}{GW190814H_o3afin}{1.3}{GW190803B_o3afin}{8.1}{GW190731E_o3afin}{11.4}{GW190728D_o3afin}{1.4}{GW190727B_o3afin}{7.8}{GW190720A_o3afin}{2.0}{GW190719H_o3afin}{11.6}{GW190708M_o3afin}{2.2}{GW190707E_o3afin}{1.2}{GW190706F_o3afin}{16.8}{GW190701E_o3afin}{9.5}{GW190630E_o3afin}{4.8}{GW190620B_o3afin}{13.2}{GW190602E_o3afin}{14.8}{GW190527H_o3afin}{8.8}{GW190521E_o3afin}{5.9}{GW190521B_o3afin}{16.2}{GW190519J_o3afin}{13.9}{GW190517B_o3afin}{9.8}{GW190514E_o3afin}{12.1}{GW190513E_o3afin}{6.7}{GW190512G_o3afin}{3.5}{GW190503E_o3afin}{8.6}{GW190425B_o3afin}{0.1}{GW190421I_o3afin}{9.5}{GW190413E_o3afin}{11.8}{GW190413A_o3afin}{7.8}{GW190412B_o3afin}{4.4}{GW190408H_o3afin}{3.0}}}
\newcommand{\totalmasssourcemed}[1]{\IfEqCase{#1}{{GW190926C_o3afin}{61.9}{GW190925J_o3afin}{36.7}{GW190917B_o3afin}{11.8}{GW190916K_o3afin}{68.0}{GW190805J_o3afin}{76.7}{GW190725F_o3afin}{18.3}{GW190426N_o3afin}{182.3}{GW190403B_o3afin}{106.6}{GW150914_o3afin}{64.5}{GW151012_o3afin}{38.8}{GW151226_o3afin}{21.7}{GW170104_o3afin}{49.6}{GW170608_o3afin}{18.5}{GW170729_o3afin}{84.4}{GW170809_o3afin}{58.5}{GW170814_o3afin}{56.0}{GW170818_o3afin}{62.5}{GW170823_o3afin}{67.0}{GW190930C_o3afin}{21.2}{GW190929B_o3afin}{93.3}{GW190924A_o3afin}{13.9}{GW190915K_o3afin}{57.2}{GW190910B_o3afin}{78.0}{GW190828B_o3afin}{34.3}{GW190828A_o3afin}{57.2}{GW190814H_o3afin}{25.9}{GW190803B_o3afin}{65.0}{GW190731E_o3afin}{70.7}{GW190728D_o3afin}{20.7}{GW190727B_o3afin}{68.8}{GW190720A_o3afin}{21.8}{GW190719H_o3afin}{57.2}{GW190708M_o3afin}{31.4}{GW190707E_o3afin}{20.1}{GW190706F_o3afin}{112.6}{GW190701E_o3afin}{94.3}{GW190630E_o3afin}{59.4}{GW190620B_o3afin}{92.7}{GW190602E_o3afin}{115.6}{GW190527H_o3afin}{58.1}{GW190521E_o3afin}{76.3}{GW190521B_o3afin}{153.1}{GW190519J_o3afin}{105.6}{GW190517B_o3afin}{64.1}{GW190514E_o3afin}{69.3}{GW190513E_o3afin}{54.4}{GW190512G_o3afin}{35.8}{GW190503E_o3afin}{69.4}{GW190425B_o3afin}{3.4}{GW190421I_o3afin}{73.6}{GW190413E_o3afin}{81.3}{GW190413A_o3afin}{58.0}{GW190412B_o3afin}{36.8}{GW190408H_o3afin}{43.4}}}
\newcommand{\totalmasssourceplus}[1]{\IfEqCase{#1}{{GW190926C_o3afin}{22.7}{GW190925J_o3afin}{3.6}{GW190917B_o3afin}{3.0}{GW190916K_o3afin}{18.3}{GW190805J_o3afin}{19.5}{GW190725F_o3afin}{7.4}{GW190426N_o3afin}{40.2}{GW190403B_o3afin}{26.7}{GW150914_o3afin}{3.7}{GW151012_o3afin}{10.3}{GW151226_o3afin}{8.3}{GW170104_o3afin}{4.7}{GW170608_o3afin}{2.0}{GW170729_o3afin}{15.0}{GW170809_o3afin}{5.3}{GW170814_o3afin}{3.5}{GW170818_o3afin}{5.3}{GW170823_o3afin}{10.3}{GW190930C_o3afin}{5.9}{GW190929B_o3afin}{23.3}{GW190924A_o3afin}{2.8}{GW190915K_o3afin}{7.1}{GW190910B_o3afin}{9.3}{GW190828B_o3afin}{5.2}{GW190828A_o3afin}{7.9}{GW190814H_o3afin}{1.3}{GW190803B_o3afin}{12.0}{GW190731E_o3afin}{16.3}{GW190728D_o3afin}{4.2}{GW190727B_o3afin}{10.2}{GW190720A_o3afin}{3.8}{GW190719H_o3afin}{38.4}{GW190708M_o3afin}{2.8}{GW190707E_o3afin}{1.7}{GW190706F_o3afin}{27.4}{GW190701E_o3afin}{12.0}{GW190630E_o3afin}{4.7}{GW190620B_o3afin}{18.5}{GW190602E_o3afin}{19.2}{GW190527H_o3afin}{18.1}{GW190521E_o3afin}{7.0}{GW190521B_o3afin}{42.2}{GW190519J_o3afin}{14.4}{GW190517B_o3afin}{9.9}{GW190514E_o3afin}{19.8}{GW190513E_o3afin}{9.3}{GW190512G_o3afin}{4.1}{GW190503E_o3afin}{10.1}{GW190425B_o3afin}{0.3}{GW190421I_o3afin}{13.2}{GW190413E_o3afin}{16.8}{GW190413A_o3afin}{10.6}{GW190412B_o3afin}{4.7}{GW190408H_o3afin}{4.2}}}
\newcommand{\chirpmasssourceminus}[1]{\IfEqCase{#1}{{GW190926C_o3afin}{4.9}{GW190925J_o3afin}{1.1}{GW190917B_o3afin}{0.2}{GW190916K_o3afin}{5.4}{GW190805J_o3afin}{6.3}{GW190725F_o3afin}{0.5}{GW190426N_o3afin}{17.4}{GW190403B_o3afin}{8.4}{GW150914_o3afin}{1.5}{GW151012_o3afin}{1.5}{GW151226_o3afin}{0.3}{GW170104_o3afin}{1.5}{GW170608_o3afin}{0.2}{GW170729_o3afin}{5.7}{GW170809_o3afin}{1.6}{GW170814_o3afin}{1.2}{GW170818_o3afin}{2.0}{GW170823_o3afin}{3.3}{GW190930C_o3afin}{0.4}{GW190929B_o3afin}{7.4}{GW190924A_o3afin}{0.2}{GW190915K_o3afin}{2.3}{GW190910B_o3afin}{4.1}{GW190828B_o3afin}{1.0}{GW190828A_o3afin}{2.0}{GW190814H_o3afin}{0.05}{GW190803B_o3afin}{3.8}{GW190731E_o3afin}{5.3}{GW190728D_o3afin}{0.3}{GW190727B_o3afin}{3.7}{GW190720A_o3afin}{0.8}{GW190719H_o3afin}{4.3}{GW190708M_o3afin}{0.6}{GW190707E_o3afin}{0.4}{GW190706F_o3afin}{9.1}{GW190701E_o3afin}{5.0}{GW190630E_o3afin}{2.1}{GW190620B_o3afin}{7.2}{GW190602E_o3afin}{9.7}{GW190527H_o3afin}{3.9}{GW190521E_o3afin}{2.8}{GW190521B_o3afin}{14.6}{GW190519J_o3afin}{7.5}{GW190517B_o3afin}{4.2}{GW190514E_o3afin}{5.4}{GW190513E_o3afin}{2.2}{GW190512G_o3afin}{0.9}{GW190503E_o3afin}{4.4}{GW190425B_o3afin}{0.02}{GW190421I_o3afin}{4.6}{GW190413E_o3afin}{6.3}{GW190413A_o3afin}{3.4}{GW190412B_o3afin}{0.5}{GW190408H_o3afin}{1.2}}}
\newcommand{\chirpmasssourcemed}[1]{\IfEqCase{#1}{{GW190926C_o3afin}{24.4}{GW190925J_o3afin}{15.6}{GW190917B_o3afin}{3.7}{GW190916K_o3afin}{26.9}{GW190805J_o3afin}{31.9}{GW190725F_o3afin}{7.4}{GW190426N_o3afin}{76.0}{GW190403B_o3afin}{34.0}{GW150914_o3afin}{27.9}{GW151012_o3afin}{15.6}{GW151226_o3afin}{8.9}{GW170104_o3afin}{21.1}{GW170608_o3afin}{7.9}{GW170729_o3afin}{34.6}{GW170809_o3afin}{24.8}{GW170814_o3afin}{24.1}{GW170818_o3afin}{26.8}{GW170823_o3afin}{28.6}{GW190930C_o3afin}{8.5}{GW190929B_o3afin}{35.6}{GW190924A_o3afin}{5.8}{GW190915K_o3afin}{24.4}{GW190910B_o3afin}{33.5}{GW190828B_o3afin}{13.4}{GW190828A_o3afin}{24.6}{GW190814H_o3afin}{6.11}{GW190803B_o3afin}{27.6}{GW190731E_o3afin}{29.7}{GW190728D_o3afin}{8.6}{GW190727B_o3afin}{29.4}{GW190720A_o3afin}{9.0}{GW190719H_o3afin}{22.8}{GW190708M_o3afin}{13.1}{GW190707E_o3afin}{8.4}{GW190706F_o3afin}{45.6}{GW190701E_o3afin}{40.2}{GW190630E_o3afin}{25.1}{GW190620B_o3afin}{38.1}{GW190602E_o3afin}{48.0}{GW190527H_o3afin}{23.9}{GW190521E_o3afin}{32.8}{GW190521B_o3afin}{63.3}{GW190519J_o3afin}{44.3}{GW190517B_o3afin}{26.5}{GW190514E_o3afin}{29.1}{GW190513E_o3afin}{21.8}{GW190512G_o3afin}{14.6}{GW190503E_o3afin}{29.3}{GW190425B_o3afin}{1.44}{GW190421I_o3afin}{31.4}{GW190413E_o3afin}{33.3}{GW190413A_o3afin}{24.5}{GW190412B_o3afin}{13.3}{GW190408H_o3afin}{18.5}}}
\newcommand{\chirpmasssourceplus}[1]{\IfEqCase{#1}{{GW190926C_o3afin}{9.0}{GW190925J_o3afin}{1.1}{GW190917B_o3afin}{0.2}{GW190916K_o3afin}{8.2}{GW190805J_o3afin}{8.8}{GW190725F_o3afin}{0.5}{GW190426N_o3afin}{19.1}{GW190403B_o3afin}{15.1}{GW150914_o3afin}{1.7}{GW151012_o3afin}{2.3}{GW151226_o3afin}{0.3}{GW170104_o3afin}{2.0}{GW170608_o3afin}{0.2}{GW170729_o3afin}{7.0}{GW170809_o3afin}{2.2}{GW170814_o3afin}{1.4}{GW170818_o3afin}{2.3}{GW170823_o3afin}{4.5}{GW190930C_o3afin}{0.5}{GW190929B_o3afin}{10.2}{GW190924A_o3afin}{0.2}{GW190915K_o3afin}{3.0}{GW190910B_o3afin}{4.2}{GW190828B_o3afin}{1.4}{GW190828A_o3afin}{3.6}{GW190814H_o3afin}{0.06}{GW190803B_o3afin}{5.4}{GW190731E_o3afin}{7.4}{GW190728D_o3afin}{0.6}{GW190727B_o3afin}{4.6}{GW190720A_o3afin}{0.4}{GW190719H_o3afin}{8.3}{GW190708M_o3afin}{0.9}{GW190707E_o3afin}{0.6}{GW190706F_o3afin}{13.0}{GW190701E_o3afin}{5.4}{GW190630E_o3afin}{2.2}{GW190620B_o3afin}{8.5}{GW190602E_o3afin}{9.5}{GW190527H_o3afin}{6.8}{GW190521E_o3afin}{3.2}{GW190521B_o3afin}{19.6}{GW190519J_o3afin}{6.8}{GW190517B_o3afin}{4.0}{GW190514E_o3afin}{8.1}{GW190513E_o3afin}{3.8}{GW190512G_o3afin}{1.4}{GW190503E_o3afin}{4.5}{GW190425B_o3afin}{0.02}{GW190421I_o3afin}{6.0}{GW190413E_o3afin}{7.8}{GW190413A_o3afin}{4.6}{GW190412B_o3afin}{0.5}{GW190408H_o3afin}{1.9}}}
\newcommand{\finalspinminus}[1]{\IfEqCase{#1}{{GW190926C_o3afin}{0.20}{GW190925J_o3afin}{0.06}{GW190917B_o3afin}{0.05}{GW190916K_o3afin}{0.24}{GW190805J_o3afin}{0.16}{GW190725F_o3afin}{0.07}{GW190426N_o3afin}{0.16}{GW190403B_o3afin}{0.17}{GW150914_o3afin}{0.05}{GW151012_o3afin}{0.13}{GW151226_o3afin}{0.05}{GW170104_o3afin}{0.08}{GW170608_o3afin}{0.03}{GW170729_o3afin}{0.22}{GW170809_o3afin}{0.08}{GW170814_o3afin}{0.06}{GW170818_o3afin}{0.08}{GW170823_o3afin}{0.10}{GW190930C_o3afin}{0.06}{GW190929B_o3afin}{0.22}{GW190924A_o3afin}{0.04}{GW190915K_o3afin}{0.09}{GW190910B_o3afin}{0.08}{GW190828B_o3afin}{0.08}{GW190828A_o3afin}{0.07}{GW190814H_o3afin}{0.03}{GW190803B_o3afin}{0.12}{GW190731E_o3afin}{0.14}{GW190728D_o3afin}{0.04}{GW190727B_o3afin}{0.11}{GW190720A_o3afin}{0.05}{GW190719H_o3afin}{0.18}{GW190708M_o3afin}{0.05}{GW190707E_o3afin}{0.03}{GW190706F_o3afin}{0.19}{GW190701E_o3afin}{0.13}{GW190630E_o3afin}{0.07}{GW190620B_o3afin}{0.15}{GW190602E_o3afin}{0.17}{GW190527H_o3afin}{0.16}{GW190521E_o3afin}{0.06}{GW190521B_o3afin}{0.23}{GW190519J_o3afin}{0.12}{GW190517B_o3afin}{0.07}{GW190514E_o3afin}{0.16}{GW190513E_o3afin}{0.13}{GW190512G_o3afin}{0.07}{GW190503E_o3afin}{0.15}{GW190421I_o3afin}{0.12}{GW190413E_o3afin}{0.18}{GW190413A_o3afin}{0.12}{GW190412B_o3afin}{0.04}{GW190408H_o3afin}{0.07}}}
\newcommand{\finalspinmed}[1]{\IfEqCase{#1}{{GW190926C_o3afin}{0.64}{GW190925J_o3afin}{0.71}{GW190917B_o3afin}{0.42}{GW190916K_o3afin}{0.74}{GW190805J_o3afin}{0.82}{GW190725F_o3afin}{0.65}{GW190426N_o3afin}{0.77}{GW190403B_o3afin}{0.91}{GW150914_o3afin}{0.68}{GW151012_o3afin}{0.69}{GW151226_o3afin}{0.75}{GW170104_o3afin}{0.67}{GW170608_o3afin}{0.69}{GW170729_o3afin}{0.78}{GW170809_o3afin}{0.71}{GW170814_o3afin}{0.72}{GW170818_o3afin}{0.68}{GW170823_o3afin}{0.71}{GW190930C_o3afin}{0.71}{GW190929B_o3afin}{0.60}{GW190924A_o3afin}{0.67}{GW190915K_o3afin}{0.69}{GW190910B_o3afin}{0.69}{GW190828B_o3afin}{0.64}{GW190828A_o3afin}{0.74}{GW190814H_o3afin}{0.28}{GW190803B_o3afin}{0.68}{GW190731E_o3afin}{0.71}{GW190728D_o3afin}{0.71}{GW190727B_o3afin}{0.73}{GW190720A_o3afin}{0.71}{GW190719H_o3afin}{0.76}{GW190708M_o3afin}{0.68}{GW190707E_o3afin}{0.66}{GW190706F_o3afin}{0.78}{GW190701E_o3afin}{0.66}{GW190630E_o3afin}{0.70}{GW190620B_o3afin}{0.80}{GW190602E_o3afin}{0.72}{GW190527H_o3afin}{0.71}{GW190521E_o3afin}{0.71}{GW190521B_o3afin}{0.62}{GW190519J_o3afin}{0.79}{GW190517B_o3afin}{0.87}{GW190514E_o3afin}{0.66}{GW190513E_o3afin}{0.71}{GW190512G_o3afin}{0.65}{GW190503E_o3afin}{0.66}{GW190421I_o3afin}{0.66}{GW190413E_o3afin}{0.68}{GW190413A_o3afin}{0.67}{GW190412B_o3afin}{0.66}{GW190408H_o3afin}{0.67}}}
\newcommand{\finalspinplus}[1]{\IfEqCase{#1}{{GW190926C_o3afin}{0.14}{GW190925J_o3afin}{0.06}{GW190917B_o3afin}{0.14}{GW190916K_o3afin}{0.13}{GW190805J_o3afin}{0.09}{GW190725F_o3afin}{0.09}{GW190426N_o3afin}{0.14}{GW190403B_o3afin}{0.05}{GW150914_o3afin}{0.05}{GW151012_o3afin}{0.13}{GW151226_o3afin}{0.12}{GW170104_o3afin}{0.06}{GW170608_o3afin}{0.03}{GW170729_o3afin}{0.09}{GW170809_o3afin}{0.08}{GW170814_o3afin}{0.07}{GW170818_o3afin}{0.08}{GW170823_o3afin}{0.08}{GW190930C_o3afin}{0.07}{GW190929B_o3afin}{0.17}{GW190924A_o3afin}{0.05}{GW190915K_o3afin}{0.08}{GW190910B_o3afin}{0.07}{GW190828B_o3afin}{0.08}{GW190828A_o3afin}{0.07}{GW190814H_o3afin}{0.03}{GW190803B_o3afin}{0.09}{GW190731E_o3afin}{0.12}{GW190728D_o3afin}{0.04}{GW190727B_o3afin}{0.09}{GW190720A_o3afin}{0.05}{GW190719H_o3afin}{0.13}{GW190708M_o3afin}{0.04}{GW190707E_o3afin}{0.03}{GW190706F_o3afin}{0.09}{GW190701E_o3afin}{0.09}{GW190630E_o3afin}{0.06}{GW190620B_o3afin}{0.07}{GW190602E_o3afin}{0.11}{GW190527H_o3afin}{0.10}{GW190521E_o3afin}{0.06}{GW190521B_o3afin}{0.21}{GW190519J_o3afin}{0.07}{GW190517B_o3afin}{0.05}{GW190514E_o3afin}{0.12}{GW190513E_o3afin}{0.13}{GW190512G_o3afin}{0.06}{GW190503E_o3afin}{0.09}{GW190421I_o3afin}{0.09}{GW190413E_o3afin}{0.12}{GW190413A_o3afin}{0.10}{GW190412B_o3afin}{0.05}{GW190408H_o3afin}{0.06}}}
\newcommand{\finalmassdetminus}[1]{\IfEqCase{#1}{{GW190926C_o3afin}{12.0}{GW190925J_o3afin}{1.7}{GW190917B_o3afin}{3.3}{GW190916K_o3afin}{20.5}{GW190805J_o3afin}{18.3}{GW190725F_o3afin}{1.3}{GW190426N_o3afin}{39.6}{GW190403B_o3afin}{47.5}{GW150914_o3afin}{3.2}{GW151012_o3afin}{4.3}{GW151226_o3afin}{1.5}{GW170104_o3afin}{3.6}{GW170608_o3afin}{0.3}{GW170729_o3afin}{18.3}{GW170809_o3afin}{4.2}{GW170814_o3afin}{2.7}{GW170818_o3afin}{5.2}{GW170823_o3afin}{9.0}{GW190930C_o3afin}{1.9}{GW190929B_o3afin}{17.7}{GW190924A_o3afin}{0.7}{GW190915K_o3afin}{6.1}{GW190910B_o3afin}{7.1}{GW190828B_o3afin}{3.9}{GW190828A_o3afin}{4.7}{GW190814H_o3afin}{1.3}{GW190803B_o3afin}{10.4}{GW190731E_o3afin}{13.4}{GW190728D_o3afin}{0.8}{GW190727B_o3afin}{9.7}{GW190720A_o3afin}{1.6}{GW190719H_o3afin}{13.7}{GW190708M_o3afin}{1.7}{GW190707E_o3afin}{0.7}{GW190706F_o3afin}{24.2}{GW190701E_o3afin}{13.8}{GW190630E_o3afin}{3.4}{GW190620B_o3afin}{19.4}{GW190602E_o3afin}{21.5}{GW190527H_o3afin}{6.6}{GW190521E_o3afin}{4.5}{GW190521B_o3afin}{31.7}{GW190519J_o3afin}{15.6}{GW190517B_o3afin}{7.4}{GW190514E_o3afin}{17.6}{GW190513E_o3afin}{7.9}{GW190512G_o3afin}{2.7}{GW190503E_o3afin}{11.9}{GW190421I_o3afin}{10.6}{GW190413E_o3afin}{18.5}{GW190413A_o3afin}{12.0}{GW190412B_o3afin}{4.6}{GW190408H_o3afin}{2.9}}}
\newcommand{\finalmassdetmed}[1]{\IfEqCase{#1}{{GW190926C_o3afin}{91.4}{GW190925J_o3afin}{41.3}{GW190917B_o3afin}{13.3}{GW190916K_o3afin}{114.4}{GW190805J_o3afin}{138.9}{GW190725F_o3afin}{21.0}{GW190426N_o3afin}{296.0}{GW190403B_o3afin}{225.5}{GW150914_o3afin}{67.6}{GW151012_o3afin}{44.3}{GW151226_o3afin}{22.6}{GW170104_o3afin}{57.7}{GW170608_o3afin}{18.9}{GW170729_o3afin}{116.7}{GW170809_o3afin}{67.3}{GW170814_o3afin}{59.7}{GW170818_o3afin}{72.3}{GW170823_o3afin}{86.5}{GW190930C_o3afin}{23.3}{GW190929B_o3afin}{139.2}{GW190924A_o3afin}{14.8}{GW190915K_o3afin}{72.5}{GW190910B_o3afin}{96.1}{GW190828B_o3afin}{42.3}{GW190828A_o3afin}{74.5}{GW190814H_o3afin}{27.0}{GW190803B_o3afin}{95.8}{GW190731E_o3afin}{105.2}{GW190728D_o3afin}{22.8}{GW190727B_o3afin}{99.5}{GW190720A_o3afin}{24.2}{GW190719H_o3afin}{86.9}{GW190708M_o3afin}{35.5}{GW190707E_o3afin}{22.3}{GW190706F_o3afin}{173.7}{GW190701E_o3afin}{124.5}{GW190630E_o3afin}{66.4}{GW190620B_o3afin}{133.0}{GW190602E_o3afin}{165.9}{GW190527H_o3afin}{79.0}{GW190521E_o3afin}{87.8}{GW190521B_o3afin}{234.1}{GW190519J_o3afin}{146.5}{GW190517B_o3afin}{80.4}{GW190514E_o3afin}{109.7}{GW190513E_o3afin}{72.3}{GW190512G_o3afin}{43.4}{GW190503E_o3afin}{85.8}{GW190421I_o3afin}{102.7}{GW190413E_o3afin}{127.8}{GW190413A_o3afin}{86.9}{GW190412B_o3afin}{40.7}{GW190408H_o3afin}{53.3}}}
\newcommand{\finalmassdetplus}[1]{\IfEqCase{#1}{{GW190926C_o3afin}{41.5}{GW190925J_o3afin}{3.8}{GW190917B_o3afin}{3.3}{GW190916K_o3afin}{35.4}{GW190805J_o3afin}{19.5}{GW190725F_o3afin}{9.4}{GW190426N_o3afin}{55.1}{GW190403B_o3afin}{36.5}{GW150914_o3afin}{3.6}{GW151012_o3afin}{12.7}{GW151226_o3afin}{9.5}{GW170104_o3afin}{4.0}{GW170608_o3afin}{2.4}{GW170729_o3afin}{18.9}{GW170809_o3afin}{5.7}{GW170814_o3afin}{3.3}{GW170818_o3afin}{5.9}{GW170823_o3afin}{10.7}{GW190930C_o3afin}{7.1}{GW190929B_o3afin}{29.2}{GW190924A_o3afin}{3.3}{GW190915K_o3afin}{7.1}{GW190910B_o3afin}{8.5}{GW190828B_o3afin}{6.0}{GW190828A_o3afin}{5.4}{GW190814H_o3afin}{1.4}{GW190803B_o3afin}{11.4}{GW190731E_o3afin}{17.1}{GW190728D_o3afin}{5.5}{GW190727B_o3afin}{11.6}{GW190720A_o3afin}{4.4}{GW190719H_o3afin}{75.1}{GW190708M_o3afin}{3.0}{GW190707E_o3afin}{1.6}{GW190706F_o3afin}{22.7}{GW190701E_o3afin}{14.9}{GW190630E_o3afin}{4.6}{GW190620B_o3afin}{18.7}{GW190602E_o3afin}{23.0}{GW190527H_o3afin}{38.1}{GW190521E_o3afin}{6.2}{GW190521B_o3afin}{50.1}{GW190519J_o3afin}{16.3}{GW190517B_o3afin}{9.1}{GW190514E_o3afin}{18.9}{GW190513E_o3afin}{13.0}{GW190512G_o3afin}{4.4}{GW190503E_o3afin}{12.1}{GW190421I_o3afin}{11.6}{GW190413E_o3afin}{18.8}{GW190413A_o3afin}{14.2}{GW190412B_o3afin}{5.5}{GW190408H_o3afin}{3.1}}}
\newcommand{\finalmasssourceminus}[1]{\IfEqCase{#1}{{GW190926C_o3afin}{11.8}{GW190925J_o3afin}{2.6}{GW190917B_o3afin}{2.9}{GW190916K_o3afin}{12.6}{GW190805J_o3afin}{13.2}{GW190725F_o3afin}{1.8}{GW190426N_o3afin}{33.6}{GW190403B_o3afin}{24.3}{GW150914_o3afin}{2.9}{GW151012_o3afin}{4.6}{GW151226_o3afin}{1.6}{GW170104_o3afin}{3.4}{GW170608_o3afin}{0.6}{GW170729_o3afin}{10.2}{GW170809_o3afin}{3.6}{GW170814_o3afin}{2.7}{GW170818_o3afin}{4.2}{GW170823_o3afin}{6.8}{GW190930C_o3afin}{2.0}{GW190929B_o3afin}{14.6}{GW190924A_o3afin}{0.9}{GW190915K_o3afin}{5.0}{GW190910B_o3afin}{8.6}{GW190828B_o3afin}{4.3}{GW190828A_o3afin}{4.0}{GW190814H_o3afin}{1.3}{GW190803B_o3afin}{7.6}{GW190731E_o3afin}{10.8}{GW190728D_o3afin}{1.4}{GW190727B_o3afin}{7.3}{GW190720A_o3afin}{2.0}{GW190719H_o3afin}{11.1}{GW190708M_o3afin}{2.1}{GW190707E_o3afin}{1.2}{GW190706F_o3afin}{15.9}{GW190701E_o3afin}{8.9}{GW190630E_o3afin}{4.5}{GW190620B_o3afin}{12.4}{GW190602E_o3afin}{13.9}{GW190527H_o3afin}{8.5}{GW190521E_o3afin}{5.4}{GW190521B_o3afin}{16.0}{GW190519J_o3afin}{12.9}{GW190517B_o3afin}{9.4}{GW190514E_o3afin}{11.5}{GW190513E_o3afin}{6.6}{GW190512G_o3afin}{3.4}{GW190503E_o3afin}{7.9}{GW190421I_o3afin}{9.0}{GW190413E_o3afin}{11.5}{GW190413A_o3afin}{7.3}{GW190412B_o3afin}{4.5}{GW190408H_o3afin}{2.9}}}
\newcommand{\finalmasssourcemed}[1]{\IfEqCase{#1}{{GW190926C_o3afin}{59.6}{GW190925J_o3afin}{34.9}{GW190917B_o3afin}{11.6}{GW190916K_o3afin}{65.0}{GW190805J_o3afin}{72.4}{GW190725F_o3afin}{17.6}{GW190426N_o3afin}{172.9}{GW190403B_o3afin}{102.2}{GW150914_o3afin}{61.5}{GW151012_o3afin}{37.1}{GW151226_o3afin}{20.7}{GW170104_o3afin}{47.5}{GW170608_o3afin}{17.7}{GW170729_o3afin}{80.3}{GW170809_o3afin}{55.7}{GW170814_o3afin}{53.2}{GW170818_o3afin}{59.7}{GW170823_o3afin}{63.9}{GW190930C_o3afin}{20.2}{GW190929B_o3afin}{90.3}{GW190924A_o3afin}{13.3}{GW190915K_o3afin}{54.7}{GW190910B_o3afin}{74.4}{GW190828B_o3afin}{33.0}{GW190828A_o3afin}{54.3}{GW190814H_o3afin}{25.7}{GW190803B_o3afin}{62.1}{GW190731E_o3afin}{67.4}{GW190728D_o3afin}{19.7}{GW190727B_o3afin}{65.4}{GW190720A_o3afin}{20.8}{GW190719H_o3afin}{54.5}{GW190708M_o3afin}{30.1}{GW190707E_o3afin}{19.2}{GW190706F_o3afin}{107.3}{GW190701E_o3afin}{90.2}{GW190630E_o3afin}{56.6}{GW190620B_o3afin}{88.0}{GW190602E_o3afin}{110.5}{GW190527H_o3afin}{55.5}{GW190521E_o3afin}{72.6}{GW190521B_o3afin}{147.4}{GW190519J_o3afin}{100.0}{GW190517B_o3afin}{60.1}{GW190514E_o3afin}{66.4}{GW190513E_o3afin}{52.1}{GW190512G_o3afin}{34.3}{GW190503E_o3afin}{66.5}{GW190421I_o3afin}{70.5}{GW190413E_o3afin}{78.0}{GW190413A_o3afin}{55.5}{GW190412B_o3afin}{35.6}{GW190408H_o3afin}{41.4}}}
\newcommand{\finalmasssourceplus}[1]{\IfEqCase{#1}{{GW190926C_o3afin}{22.1}{GW190925J_o3afin}{3.5}{GW190917B_o3afin}{3.1}{GW190916K_o3afin}{17.3}{GW190805J_o3afin}{18.2}{GW190725F_o3afin}{7.7}{GW190426N_o3afin}{37.7}{GW190403B_o3afin}{26.3}{GW150914_o3afin}{3.4}{GW151012_o3afin}{10.6}{GW151226_o3afin}{8.6}{GW170104_o3afin}{4.5}{GW170608_o3afin}{2.1}{GW170729_o3afin}{13.5}{GW170809_o3afin}{5.0}{GW170814_o3afin}{3.2}{GW170818_o3afin}{4.9}{GW170823_o3afin}{9.6}{GW190930C_o3afin}{6.1}{GW190929B_o3afin}{22.3}{GW190924A_o3afin}{3.0}{GW190915K_o3afin}{6.6}{GW190910B_o3afin}{8.5}{GW190828B_o3afin}{5.3}{GW190828A_o3afin}{7.3}{GW190814H_o3afin}{1.3}{GW190803B_o3afin}{11.2}{GW190731E_o3afin}{15.3}{GW190728D_o3afin}{4.4}{GW190727B_o3afin}{9.5}{GW190720A_o3afin}{3.9}{GW190719H_o3afin}{38.3}{GW190708M_o3afin}{2.9}{GW190707E_o3afin}{1.7}{GW190706F_o3afin}{25.2}{GW190701E_o3afin}{11.2}{GW190630E_o3afin}{4.4}{GW190620B_o3afin}{17.2}{GW190602E_o3afin}{17.9}{GW190527H_o3afin}{17.9}{GW190521E_o3afin}{6.5}{GW190521B_o3afin}{40.0}{GW190519J_o3afin}{13.0}{GW190517B_o3afin}{9.9}{GW190514E_o3afin}{19.0}{GW190513E_o3afin}{8.8}{GW190512G_o3afin}{4.1}{GW190503E_o3afin}{9.4}{GW190421I_o3afin}{12.4}{GW190413E_o3afin}{16.1}{GW190413A_o3afin}{10.1}{GW190412B_o3afin}{4.8}{GW190408H_o3afin}{3.9}}}
\newcommand{\radiatedenergyminus}[1]{\IfEqCase{#1}{{GW190926C_o3afin}{1.1}{GW190925J_o3afin}{0.3}{GW190917B_o3afin}{0.04}{GW190916K_o3afin}{1.5}{GW190805J_o3afin}{1.6}{GW190725F_o3afin}{0.2}{GW190426N_o3afin}{3.8}{GW190403B_o3afin}{2.6}{GW150914_o3afin}{0.4}{GW151012_o3afin}{0.6}{GW151226_o3afin}{0.3}{GW170104_o3afin}{0.4}{GW170608_o3afin}{0.1}{GW170729_o3afin}{1.9}{GW170809_o3afin}{0.5}{GW170814_o3afin}{0.4}{GW170818_o3afin}{0.5}{GW170823_o3afin}{0.8}{GW190930C_o3afin}{0.2}{GW190929B_o3afin}{1.5}{GW190924A_o3afin}{0.1}{GW190915K_o3afin}{0.6}{GW190910B_o3afin}{0.8}{GW190828B_o3afin}{0.3}{GW190828A_o3afin}{0.5}{GW190814H_o3afin}{0.007}{GW190803B_o3afin}{0.9}{GW190731E_o3afin}{1.1}{GW190728D_o3afin}{0.2}{GW190727B_o3afin}{0.9}{GW190720A_o3afin}{0.2}{GW190719H_o3afin}{1.3}{GW190708M_o3afin}{0.2}{GW190707E_o3afin}{0.08}{GW190706F_o3afin}{2.6}{GW190701E_o3afin}{1.2}{GW190630E_o3afin}{0.5}{GW190620B_o3afin}{2.4}{GW190602E_o3afin}{2.4}{GW190527H_o3afin}{1.1}{GW190521E_o3afin}{0.7}{GW190521B_o3afin}{3.8}{GW190519J_o3afin}{1.9}{GW190517B_o3afin}{1.6}{GW190514E_o3afin}{1.0}{GW190513E_o3afin}{0.7}{GW190512G_o3afin}{0.3}{GW190503E_o3afin}{1.1}{GW190421I_o3afin}{0.9}{GW190413E_o3afin}{1.6}{GW190413A_o3afin}{0.8}{GW190412B_o3afin}{0.2}{GW190408H_o3afin}{0.3}}}
\newcommand{\radiatedenergymed}[1]{\IfEqCase{#1}{{GW190926C_o3afin}{2.3}{GW190925J_o3afin}{1.7}{GW190917B_o3afin}{0.2}{GW190916K_o3afin}{3.0}{GW190805J_o3afin}{4.3}{GW190725F_o3afin}{0.7}{GW190426N_o3afin}{9.3}{GW190403B_o3afin}{4.5}{GW150914_o3afin}{3.0}{GW151012_o3afin}{1.6}{GW151226_o3afin}{1.0}{GW170104_o3afin}{2.2}{GW170608_o3afin}{0.9}{GW170729_o3afin}{4.3}{GW170809_o3afin}{2.7}{GW170814_o3afin}{2.7}{GW170818_o3afin}{2.8}{GW170823_o3afin}{3.2}{GW190930C_o3afin}{0.9}{GW190929B_o3afin}{3.1}{GW190924A_o3afin}{0.6}{GW190915K_o3afin}{2.6}{GW190910B_o3afin}{3.6}{GW190828B_o3afin}{1.2}{GW190828A_o3afin}{3.0}{GW190814H_o3afin}{0.2}{GW190803B_o3afin}{2.9}{GW190731E_o3afin}{3.3}{GW190728D_o3afin}{1.0}{GW190727B_o3afin}{3.4}{GW190720A_o3afin}{1.0}{GW190719H_o3afin}{2.6}{GW190708M_o3afin}{1.4}{GW190707E_o3afin}{0.9}{GW190706F_o3afin}{5.5}{GW190701E_o3afin}{4.1}{GW190630E_o3afin}{2.8}{GW190620B_o3afin}{4.9}{GW190602E_o3afin}{5.3}{GW190527H_o3afin}{2.6}{GW190521E_o3afin}{3.7}{GW190521B_o3afin}{5.9}{GW190519J_o3afin}{5.7}{GW190517B_o3afin}{3.9}{GW190514E_o3afin}{2.9}{GW190513E_o3afin}{2.3}{GW190512G_o3afin}{1.5}{GW190503E_o3afin}{3.0}{GW190421I_o3afin}{3.2}{GW190413E_o3afin}{3.4}{GW190413A_o3afin}{2.6}{GW190412B_o3afin}{1.2}{GW190408H_o3afin}{1.9}}}
\newcommand{\radiatedenergyplus}[1]{\IfEqCase{#1}{{GW190926C_o3afin}{1.3}{GW190925J_o3afin}{0.3}{GW190917B_o3afin}{0.06}{GW190916K_o3afin}{2.1}{GW190805J_o3afin}{1.9}{GW190725F_o3afin}{0.1}{GW190426N_o3afin}{4.8}{GW190403B_o3afin}{3.4}{GW150914_o3afin}{0.4}{GW151012_o3afin}{0.7}{GW151226_o3afin}{0.1}{GW170104_o3afin}{0.4}{GW170608_o3afin}{0.04}{GW170729_o3afin}{2.0}{GW170809_o3afin}{0.6}{GW170814_o3afin}{0.4}{GW170818_o3afin}{0.6}{GW170823_o3afin}{0.9}{GW190930C_o3afin}{0.1}{GW190929B_o3afin}{1.8}{GW190924A_o3afin}{0.05}{GW190915K_o3afin}{0.6}{GW190910B_o3afin}{0.8}{GW190828B_o3afin}{0.3}{GW190828A_o3afin}{0.7}{GW190814H_o3afin}{0.007}{GW190803B_o3afin}{0.9}{GW190731E_o3afin}{1.4}{GW190728D_o3afin}{0.09}{GW190727B_o3afin}{0.9}{GW190720A_o3afin}{0.1}{GW190719H_o3afin}{1.8}{GW190708M_o3afin}{0.2}{GW190707E_o3afin}{0.08}{GW190706F_o3afin}{2.8}{GW190701E_o3afin}{1.1}{GW190630E_o3afin}{0.6}{GW190620B_o3afin}{2.1}{GW190602E_o3afin}{1.9}{GW190527H_o3afin}{1.2}{GW190521E_o3afin}{0.7}{GW190521B_o3afin}{4.4}{GW190519J_o3afin}{1.8}{GW190517B_o3afin}{1.3}{GW190514E_o3afin}{1.3}{GW190513E_o3afin}{1.2}{GW190512G_o3afin}{0.3}{GW190503E_o3afin}{1.0}{GW190421I_o3afin}{0.9}{GW190413E_o3afin}{1.4}{GW190413A_o3afin}{0.9}{GW190412B_o3afin}{0.2}{GW190408H_o3afin}{0.3}}}
\newcommand{\networkoptimalsnrminus}[1]{\IfEqCase{#1}{{GW190926C_o3afin}{1.8}{GW190917B_o3afin}{1.9}{GW190725F_o3afin}{1.8}{GW190426N_o3afin}{1.8}{GW151226_o3afin}{1.7}{GW190929B_o3afin}{1.7}{GW190828B_o3afin}{1.7}{GW190828A_o3afin}{1.7}{GW190814H_o3afin}{1.7}{GW190803B_o3afin}{1.8}{GW190719H_o3afin}{1.8}{GW190602E_o3afin}{1.6}{GW190521B_o3afin}{1.7}{GW190425B_o3afin}{1.7}{GW190421I_o3afin}{1.7}{GW190413E_o3afin}{1.8}{GW190413A_o3afin}{1.8}}}
\newcommand{\networkoptimalsnrmed}[1]{\IfEqCase{#1}{{GW190926C_o3afin}{7.8}{GW190917B_o3afin}{7.8}{GW190725F_o3afin}{8.7}{GW190426N_o3afin}{8.4}{GW151226_o3afin}{12.4}{GW190929B_o3afin}{9.4}{GW190828B_o3afin}{9.9}{GW190828A_o3afin}{16.3}{GW190814H_o3afin}{25.1}{GW190803B_o3afin}{8.9}{GW190719H_o3afin}{7.5}{GW190602E_o3afin}{13.0}{GW190521B_o3afin}{14.2}{GW190425B_o3afin}{12.0}{GW190421I_o3afin}{10.4}{GW190413E_o3afin}{10.3}{GW190413A_o3afin}{8.6}}}
\newcommand{\networkoptimalsnrplus}[1]{\IfEqCase{#1}{{GW190926C_o3afin}{1.8}{GW190917B_o3afin}{1.8}{GW190725F_o3afin}{1.8}{GW190426N_o3afin}{1.7}{GW151226_o3afin}{1.7}{GW190929B_o3afin}{1.7}{GW190828B_o3afin}{1.7}{GW190828A_o3afin}{1.7}{GW190814H_o3afin}{1.7}{GW190803B_o3afin}{1.7}{GW190719H_o3afin}{1.7}{GW190602E_o3afin}{1.7}{GW190521B_o3afin}{1.7}{GW190425B_o3afin}{1.7}{GW190421I_o3afin}{1.7}{GW190413E_o3afin}{1.7}{GW190413A_o3afin}{1.8}}}
\newcommand{\networkmatchedfiltersnrIMRPminus}[1]{\IfEqCase{#1}{{GW190926C_o3afin}{0.8}{GW190925J_o3afin}{0.6}{GW190917B_o3afin}{0.8}{GW190916K_o3afin}{0.5}{GW190805J_o3afin}{0.7}{GW190725F_o3afin}{0.7}{GW190426N_o3afin}{0.6}{GW190403B_o3afin}{1.1}{GW150914_o3afin}{0.2}{GW151012_o3afin}{0.5}{GW151226_o3afin}{0.4}{GW170104_o3afin}{0.3}{GW170608_o3afin}{0.3}{GW170729_o3afin}{0.5}{GW170809_o3afin}{0.3}{GW170814_o3afin}{0.3}{GW170818_o3afin}{0.4}{GW170823_o3afin}{0.3}{GW190930C_o3afin}{0.5}{GW190929B_o3afin}{0.6}{GW190924A_o3afin}{0.4}{GW190915K_o3afin}{0.3}{GW190910B_o3afin}{0.3}{GW190828B_o3afin}{0.5}{GW190828A_o3afin}{0.3}{GW190814H_o3afin}{0.2}{GW190803B_o3afin}{0.5}{GW190731E_o3afin}{0.4}{GW190728D_o3afin}{0.4}{GW190727B_o3afin}{0.5}{GW190720A_o3afin}{0.8}{GW190719H_o3afin}{0.7}{GW190708M_o3afin}{0.3}{GW190707E_o3afin}{0.4}{GW190706F_o3afin}{0.4}{GW190701E_o3afin}{0.4}{GW190630E_o3afin}{0.3}{GW190620B_o3afin}{0.4}{GW190602E_o3afin}{0.3}{GW190527H_o3afin}{0.9}{GW190521E_o3afin}{0.2}{GW190521B_o3afin}{0.4}{GW190519J_o3afin}{0.3}{GW190517B_o3afin}{0.6}{GW190514E_o3afin}{0.6}{GW190513E_o3afin}{0.4}{GW190512G_o3afin}{0.4}{GW190503E_o3afin}{0.4}{GW190425B_o3afin}{0.4}{GW190421I_o3afin}{0.4}{GW190413E_o3afin}{0.5}{GW190413A_o3afin}{0.8}{GW190412B_o3afin}{0.3}{GW190408H_o3afin}{0.3}}}
\newcommand{\networkmatchedfiltersnrIMRPmed}[1]{\IfEqCase{#1}{{GW190926C_o3afin}{8.1}{GW190925J_o3afin}{9.7}{GW190917B_o3afin}{8.3}{GW190916K_o3afin}{8.1}{GW190805J_o3afin}{8.1}{GW190725F_o3afin}{9.1}{GW190426N_o3afin}{8.7}{GW190403B_o3afin}{7.6}{GW150914_o3afin}{26.0}{GW151012_o3afin}{9.3}{GW151226_o3afin}{12.7}{GW170104_o3afin}{13.8}{GW170608_o3afin}{15.3}{GW170729_o3afin}{10.7}{GW170809_o3afin}{12.8}{GW170814_o3afin}{17.7}{GW170818_o3afin}{12.0}{GW170823_o3afin}{12.2}{GW190930C_o3afin}{9.7}{GW190929B_o3afin}{9.7}{GW190924A_o3afin}{12.0}{GW190915K_o3afin}{13.1}{GW190910B_o3afin}{14.5}{GW190828B_o3afin}{10.2}{GW190828A_o3afin}{16.5}{GW190814H_o3afin}{25.3}{GW190803B_o3afin}{9.3}{GW190731E_o3afin}{8.8}{GW190728D_o3afin}{13.1}{GW190727B_o3afin}{11.7}{GW190720A_o3afin}{10.9}{GW190719H_o3afin}{7.9}{GW190708M_o3afin}{13.4}{GW190707E_o3afin}{13.1}{GW190706F_o3afin}{13.4}{GW190701E_o3afin}{11.2}{GW190630E_o3afin}{16.4}{GW190620B_o3afin}{12.1}{GW190602E_o3afin}{13.2}{GW190527H_o3afin}{8.0}{GW190521E_o3afin}{25.9}{GW190521B_o3afin}{14.3}{GW190519J_o3afin}{15.9}{GW190517B_o3afin}{10.8}{GW190514E_o3afin}{8.0}{GW190513E_o3afin}{12.5}{GW190512G_o3afin}{12.7}{GW190503E_o3afin}{12.2}{GW190425B_o3afin}{12.4}{GW190421I_o3afin}{10.7}{GW190413E_o3afin}{10.6}{GW190413A_o3afin}{9.0}{GW190412B_o3afin}{19.8}{GW190408H_o3afin}{14.6}}}
\newcommand{\networkmatchedfiltersnrIMRPplus}[1]{\IfEqCase{#1}{{GW190926C_o3afin}{0.6}{GW190925J_o3afin}{0.3}{GW190917B_o3afin}{0.5}{GW190916K_o3afin}{0.3}{GW190805J_o3afin}{0.5}{GW190725F_o3afin}{0.4}{GW190426N_o3afin}{0.4}{GW190403B_o3afin}{0.6}{GW150914_o3afin}{0.1}{GW151012_o3afin}{0.3}{GW151226_o3afin}{0.3}{GW170104_o3afin}{0.2}{GW170608_o3afin}{0.2}{GW170729_o3afin}{0.4}{GW170809_o3afin}{0.2}{GW170814_o3afin}{0.2}{GW170818_o3afin}{0.3}{GW170823_o3afin}{0.2}{GW190930C_o3afin}{0.3}{GW190929B_o3afin}{0.4}{GW190924A_o3afin}{0.3}{GW190915K_o3afin}{0.2}{GW190910B_o3afin}{0.2}{GW190828B_o3afin}{0.4}{GW190828A_o3afin}{0.2}{GW190814H_o3afin}{0.1}{GW190803B_o3afin}{0.3}{GW190731E_o3afin}{0.3}{GW190728D_o3afin}{0.3}{GW190727B_o3afin}{0.2}{GW190720A_o3afin}{0.3}{GW190719H_o3afin}{0.3}{GW190708M_o3afin}{0.2}{GW190707E_o3afin}{0.2}{GW190706F_o3afin}{0.2}{GW190701E_o3afin}{0.2}{GW190630E_o3afin}{0.2}{GW190620B_o3afin}{0.3}{GW190602E_o3afin}{0.2}{GW190527H_o3afin}{0.4}{GW190521E_o3afin}{0.1}{GW190521B_o3afin}{0.5}{GW190519J_o3afin}{0.2}{GW190517B_o3afin}{0.5}{GW190514E_o3afin}{0.3}{GW190513E_o3afin}{0.3}{GW190512G_o3afin}{0.3}{GW190503E_o3afin}{0.2}{GW190425B_o3afin}{0.4}{GW190421I_o3afin}{0.2}{GW190413E_o3afin}{0.4}{GW190413A_o3afin}{0.4}{GW190412B_o3afin}{0.2}{GW190408H_o3afin}{0.2}}}
\newcommand{\costhetajnminus}[1]{\IfEqCase{#1}{{GW190926C_o3afin}{0.80}{GW190925J_o3afin}{1.67}{GW190917B_o3afin}{1.20}{GW190916K_o3afin}{0.92}{GW190805J_o3afin}{1.47}{GW190725F_o3afin}{1.48}{GW190426N_o3afin}{0.48}{GW190403B_o3afin}{0.72}{GW150914_o3afin}{0.09}{GW151012_o3afin}{0.82}{GW151226_o3afin}{1.60}{GW170104_o3afin}{1.42}{GW170608_o3afin}{0.27}{GW170729_o3afin}{1.16}{GW170809_o3afin}{0.13}{GW170814_o3afin}{1.64}{GW170818_o3afin}{0.20}{GW170823_o3afin}{0.81}{GW190930C_o3afin}{1.69}{GW190929B_o3afin}{0.99}{GW190924A_o3afin}{1.61}{GW190915K_o3afin}{0.69}{GW190910B_o3afin}{0.89}{GW190828B_o3afin}{0.67}{GW190828A_o3afin}{0.26}{GW190814H_o3afin}{1.29}{GW190803B_o3afin}{1.57}{GW190731E_o3afin}{1.29}{GW190728D_o3afin}{1.41}{GW190727B_o3afin}{1.00}{GW190720A_o3afin}{0.14}{GW190719H_o3afin}{0.92}{GW190708M_o3afin}{1.16}{GW190707E_o3afin}{0.45}{GW190706F_o3afin}{0.67}{GW190701E_o3afin}{0.41}{GW190630E_o3afin}{1.11}{GW190620B_o3afin}{0.49}{GW190602E_o3afin}{0.44}{GW190527H_o3afin}{1.35}{GW190521E_o3afin}{1.02}{GW190521B_o3afin}{1.12}{GW190519J_o3afin}{0.80}{GW190517B_o3afin}{0.43}{GW190514E_o3afin}{1.06}{GW190513E_o3afin}{1.65}{GW190512G_o3afin}{0.74}{GW190503E_o3afin}{0.18}{GW190425B_o3afin}{0.82}{GW190421I_o3afin}{0.53}{GW190413E_o3afin}{0.68}{GW190413A_o3afin}{1.65}{GW190412B_o3afin}{1.47}{GW190408H_o3afin}{1.49}}}
\newcommand{\costhetajnmed}[1]{\IfEqCase{#1}{{GW190926C_o3afin}{-0.10}{GW190925J_o3afin}{0.72}{GW190917B_o3afin}{0.22}{GW190916K_o3afin}{-0.04}{GW190805J_o3afin}{0.54}{GW190725F_o3afin}{0.54}{GW190426N_o3afin}{-0.49}{GW190403B_o3afin}{-0.26}{GW150914_o3afin}{-0.90}{GW151012_o3afin}{-0.13}{GW151226_o3afin}{0.64}{GW170104_o3afin}{0.46}{GW170608_o3afin}{-0.72}{GW170729_o3afin}{0.22}{GW170809_o3afin}{-0.86}{GW170814_o3afin}{0.77}{GW170818_o3afin}{-0.77}{GW170823_o3afin}{-0.16}{GW190930C_o3afin}{0.75}{GW190929B_o3afin}{0.12}{GW190924A_o3afin}{0.67}{GW190915K_o3afin}{-0.26}{GW190910B_o3afin}{-0.05}{GW190828B_o3afin}{-0.28}{GW190828A_o3afin}{-0.72}{GW190814H_o3afin}{0.62}{GW190803B_o3afin}{0.62}{GW190731E_o3afin}{0.33}{GW190728D_o3afin}{0.44}{GW190727B_o3afin}{0.03}{GW190720A_o3afin}{-0.85}{GW190719H_o3afin}{-0.04}{GW190708M_o3afin}{0.18}{GW190707E_o3afin}{-0.53}{GW190706F_o3afin}{-0.28}{GW190701E_o3afin}{0.84}{GW190630E_o3afin}{0.16}{GW190620B_o3afin}{-0.48}{GW190602E_o3afin}{-0.53}{GW190527H_o3afin}{0.41}{GW190521E_o3afin}{0.09}{GW190521B_o3afin}{0.19}{GW190519J_o3afin}{-0.04}{GW190517B_o3afin}{-0.52}{GW190514E_o3afin}{0.10}{GW190513E_o3afin}{0.70}{GW190512G_o3afin}{-0.22}{GW190503E_o3afin}{-0.80}{GW190425B_o3afin}{-0.13}{GW190421I_o3afin}{-0.44}{GW190413E_o3afin}{-0.28}{GW190413A_o3afin}{0.70}{GW190412B_o3afin}{0.61}{GW190408H_o3afin}{0.53}}}
\newcommand{\costhetajnplus}[1]{\IfEqCase{#1}{{GW190926C_o3afin}{0.99}{GW190925J_o3afin}{0.27}{GW190917B_o3afin}{0.76}{GW190916K_o3afin}{1.01}{GW190805J_o3afin}{0.43}{GW190725F_o3afin}{0.43}{GW190426N_o3afin}{1.43}{GW190403B_o3afin}{1.25}{GW150914_o3afin}{0.50}{GW151012_o3afin}{1.09}{GW151226_o3afin}{0.34}{GW170104_o3afin}{0.52}{GW170608_o3afin}{1.67}{GW170729_o3afin}{0.73}{GW170809_o3afin}{0.43}{GW170814_o3afin}{0.21}{GW170818_o3afin}{0.39}{GW170823_o3afin}{1.13}{GW190930C_o3afin}{0.23}{GW190929B_o3afin}{0.77}{GW190924A_o3afin}{0.31}{GW190915K_o3afin}{1.20}{GW190910B_o3afin}{0.98}{GW190828B_o3afin}{1.23}{GW190828A_o3afin}{1.66}{GW190814H_o3afin}{0.17}{GW190803B_o3afin}{0.36}{GW190731E_o3afin}{0.64}{GW190728D_o3afin}{0.53}{GW190727B_o3afin}{0.93}{GW190720A_o3afin}{1.68}{GW190719H_o3afin}{1.00}{GW190708M_o3afin}{0.80}{GW190707E_o3afin}{1.50}{GW190706F_o3afin}{1.22}{GW190701E_o3afin}{0.15}{GW190630E_o3afin}{0.81}{GW190620B_o3afin}{1.42}{GW190602E_o3afin}{1.50}{GW190527H_o3afin}{0.55}{GW190521E_o3afin}{0.84}{GW190521B_o3afin}{0.76}{GW190519J_o3afin}{0.87}{GW190517B_o3afin}{1.11}{GW190514E_o3afin}{0.87}{GW190513E_o3afin}{0.28}{GW190512G_o3afin}{1.18}{GW190503E_o3afin}{0.50}{GW190425B_o3afin}{1.08}{GW190421I_o3afin}{1.39}{GW190413E_o3afin}{1.22}{GW190413A_o3afin}{0.28}{GW190412B_o3afin}{0.26}{GW190408H_o3afin}{0.44}}}
\newcommand{\cosiotaminus}[1]{\IfEqCase{#1}{{GW190926C_o3afin}{0.80}{GW190925J_o3afin}{1.66}{GW190917B_o3afin}{1.24}{GW190916K_o3afin}{0.92}{GW190805J_o3afin}{1.44}{GW190725F_o3afin}{1.43}{GW190426N_o3afin}{0.49}{GW190403B_o3afin}{0.75}{GW150914_o3afin}{0.11}{GW151012_o3afin}{0.83}{GW151226_o3afin}{1.61}{GW170104_o3afin}{1.39}{GW170608_o3afin}{0.27}{GW170729_o3afin}{1.13}{GW170809_o3afin}{0.12}{GW170814_o3afin}{1.62}{GW170818_o3afin}{0.19}{GW170823_o3afin}{0.80}{GW190930C_o3afin}{1.66}{GW190929B_o3afin}{0.93}{GW190924A_o3afin}{1.60}{GW190915K_o3afin}{0.69}{GW190910B_o3afin}{0.88}{GW190828B_o3afin}{0.67}{GW190828A_o3afin}{0.28}{GW190814H_o3afin}{1.31}{GW190803B_o3afin}{1.48}{GW190731E_o3afin}{1.29}{GW190728D_o3afin}{1.40}{GW190727B_o3afin}{1.01}{GW190720A_o3afin}{0.15}{GW190719H_o3afin}{0.94}{GW190708M_o3afin}{1.17}{GW190707E_o3afin}{0.46}{GW190706F_o3afin}{0.69}{GW190701E_o3afin}{0.40}{GW190630E_o3afin}{1.11}{GW190620B_o3afin}{0.50}{GW190602E_o3afin}{0.45}{GW190527H_o3afin}{1.35}{GW190521E_o3afin}{1.02}{GW190521B_o3afin}{1.21}{GW190519J_o3afin}{0.78}{GW190517B_o3afin}{0.44}{GW190514E_o3afin}{1.06}{GW190513E_o3afin}{1.62}{GW190512G_o3afin}{0.75}{GW190503E_o3afin}{0.18}{GW190425B_o3afin}{0.82}{GW190421I_o3afin}{0.63}{GW190413E_o3afin}{0.64}{GW190413A_o3afin}{1.62}{GW190412B_o3afin}{1.46}{GW190408H_o3afin}{1.48}}}
\newcommand{\cosiotamed}[1]{\IfEqCase{#1}{{GW190926C_o3afin}{-0.10}{GW190925J_o3afin}{0.70}{GW190917B_o3afin}{0.27}{GW190916K_o3afin}{-0.04}{GW190805J_o3afin}{0.52}{GW190725F_o3afin}{0.50}{GW190426N_o3afin}{-0.48}{GW190403B_o3afin}{-0.23}{GW150914_o3afin}{-0.88}{GW151012_o3afin}{-0.12}{GW151226_o3afin}{0.63}{GW170104_o3afin}{0.42}{GW170608_o3afin}{-0.71}{GW170729_o3afin}{0.20}{GW170809_o3afin}{-0.87}{GW170814_o3afin}{0.76}{GW170818_o3afin}{-0.80}{GW170823_o3afin}{-0.17}{GW190930C_o3afin}{0.73}{GW190929B_o3afin}{0.06}{GW190924A_o3afin}{0.67}{GW190915K_o3afin}{-0.27}{GW190910B_o3afin}{-0.05}{GW190828B_o3afin}{-0.29}{GW190828A_o3afin}{-0.70}{GW190814H_o3afin}{0.64}{GW190803B_o3afin}{0.54}{GW190731E_o3afin}{0.33}{GW190728D_o3afin}{0.43}{GW190727B_o3afin}{0.04}{GW190720A_o3afin}{-0.83}{GW190719H_o3afin}{-0.02}{GW190708M_o3afin}{0.19}{GW190707E_o3afin}{-0.52}{GW190706F_o3afin}{-0.27}{GW190701E_o3afin}{0.85}{GW190630E_o3afin}{0.16}{GW190620B_o3afin}{-0.46}{GW190602E_o3afin}{-0.53}{GW190527H_o3afin}{0.41}{GW190521E_o3afin}{0.08}{GW190521B_o3afin}{0.27}{GW190519J_o3afin}{-0.04}{GW190517B_o3afin}{-0.51}{GW190514E_o3afin}{0.10}{GW190513E_o3afin}{0.68}{GW190512G_o3afin}{-0.21}{GW190503E_o3afin}{-0.81}{GW190425B_o3afin}{-0.13}{GW190421I_o3afin}{-0.33}{GW190413E_o3afin}{-0.32}{GW190413A_o3afin}{0.68}{GW190412B_o3afin}{0.62}{GW190408H_o3afin}{0.53}}}
\newcommand{\cosiotaplus}[1]{\IfEqCase{#1}{{GW190926C_o3afin}{0.99}{GW190925J_o3afin}{0.28}{GW190917B_o3afin}{0.71}{GW190916K_o3afin}{1.01}{GW190805J_o3afin}{0.45}{GW190725F_o3afin}{0.46}{GW190426N_o3afin}{1.41}{GW190403B_o3afin}{1.22}{GW150914_o3afin}{0.48}{GW151012_o3afin}{1.07}{GW151226_o3afin}{0.36}{GW170104_o3afin}{0.56}{GW170608_o3afin}{1.66}{GW170729_o3afin}{0.75}{GW170809_o3afin}{0.44}{GW170814_o3afin}{0.22}{GW170818_o3afin}{0.43}{GW170823_o3afin}{1.14}{GW190930C_o3afin}{0.25}{GW190929B_o3afin}{0.82}{GW190924A_o3afin}{0.31}{GW190915K_o3afin}{1.23}{GW190910B_o3afin}{0.97}{GW190828B_o3afin}{1.24}{GW190828A_o3afin}{1.63}{GW190814H_o3afin}{0.18}{GW190803B_o3afin}{0.43}{GW190731E_o3afin}{0.64}{GW190728D_o3afin}{0.54}{GW190727B_o3afin}{0.92}{GW190720A_o3afin}{1.65}{GW190719H_o3afin}{0.98}{GW190708M_o3afin}{0.79}{GW190707E_o3afin}{1.49}{GW190706F_o3afin}{1.22}{GW190701E_o3afin}{0.14}{GW190630E_o3afin}{0.81}{GW190620B_o3afin}{1.40}{GW190602E_o3afin}{1.49}{GW190527H_o3afin}{0.56}{GW190521E_o3afin}{0.86}{GW190521B_o3afin}{0.68}{GW190519J_o3afin}{0.85}{GW190517B_o3afin}{1.17}{GW190514E_o3afin}{0.86}{GW190513E_o3afin}{0.30}{GW190512G_o3afin}{1.17}{GW190503E_o3afin}{0.54}{GW190425B_o3afin}{1.08}{GW190421I_o3afin}{1.26}{GW190413E_o3afin}{1.26}{GW190413A_o3afin}{0.30}{GW190412B_o3afin}{0.27}{GW190408H_o3afin}{0.45}}}
\newcommand{\chieffinfinityonlyprecavgminus}[1]{\IfEqCase{#1}{{GW190926C_o3afin}{0.32}{GW190925J_o3afin}{0.15}{GW190917B_o3afin}{0.43}{GW190916K_o3afin}{0.31}{GW190805J_o3afin}{0.39}{GW190725F_o3afin}{0.16}{GW190426N_o3afin}{0.41}{GW190403B_o3afin}{0.43}{GW150914_o3afin}{0.14}{GW151012_o3afin}{0.21}{GW151226_o3afin}{0.08}{GW170104_o3afin}{0.19}{GW170608_o3afin}{0.05}{GW170729_o3afin}{0.33}{GW170809_o3afin}{0.17}{GW170814_o3afin}{0.12}{GW170818_o3afin}{0.22}{GW170823_o3afin}{0.22}{GW190930C_o3afin}{0.16}{GW190929B_o3afin}{0.28}{GW190924A_o3afin}{0.08}{GW190915K_o3afin}{0.24}{GW190910B_o3afin}{0.20}{GW190828B_o3afin}{0.17}{GW190828A_o3afin}{0.16}{GW190814H_o3afin}{0.07}{GW190803B_o3afin}{0.28}{GW190731E_o3afin}{0.25}{GW190728D_o3afin}{0.07}{GW190727B_o3afin}{0.27}{GW190720A_o3afin}{0.11}{GW190719H_o3afin}{0.32}{GW190708M_o3afin}{0.10}{GW190707E_o3afin}{0.09}{GW190706F_o3afin}{0.31}{GW190701E_o3afin}{0.31}{GW190630E_o3afin}{0.13}{GW190620B_o3afin}{0.29}{GW190602E_o3afin}{0.28}{GW190527H_o3afin}{0.22}{GW190521E_o3afin}{0.13}{GW190521B_o3afin}{0.45}{GW190519J_o3afin}{0.24}{GW190517B_o3afin}{0.28}{GW190514E_o3afin}{0.35}{GW190513E_o3afin}{0.22}{GW190512G_o3afin}{0.14}{GW190503E_o3afin}{0.30}{GW190421I_o3afin}{0.27}{GW190413E_o3afin}{0.38}{GW190413A_o3afin}{0.32}{GW190412B_o3afin}{0.13}{GW190408H_o3afin}{0.17}}}
\newcommand{\chieffinfinityonlyprecavgmed}[1]{\IfEqCase{#1}{{GW190926C_o3afin}{-0.02}{GW190925J_o3afin}{0.09}{GW190917B_o3afin}{-0.08}{GW190916K_o3afin}{0.20}{GW190805J_o3afin}{0.37}{GW190725F_o3afin}{-0.04}{GW190426N_o3afin}{0.23}{GW190403B_o3afin}{0.68}{GW150914_o3afin}{-0.04}{GW151012_o3afin}{0.12}{GW151226_o3afin}{0.20}{GW170104_o3afin}{-0.04}{GW170608_o3afin}{0.05}{GW170729_o3afin}{0.29}{GW170809_o3afin}{0.07}{GW170814_o3afin}{0.08}{GW170818_o3afin}{-0.06}{GW170823_o3afin}{0.05}{GW190930C_o3afin}{0.19}{GW190929B_o3afin}{-0.03}{GW190924A_o3afin}{0.03}{GW190915K_o3afin}{-0.03}{GW190910B_o3afin}{0.00}{GW190828B_o3afin}{0.05}{GW190828A_o3afin}{0.15}{GW190814H_o3afin}{0.00}{GW190803B_o3afin}{-0.01}{GW190731E_o3afin}{0.07}{GW190728D_o3afin}{0.13}{GW190727B_o3afin}{0.09}{GW190720A_o3afin}{0.19}{GW190719H_o3afin}{0.25}{GW190708M_o3afin}{0.05}{GW190707E_o3afin}{-0.04}{GW190706F_o3afin}{0.28}{GW190701E_o3afin}{-0.08}{GW190630E_o3afin}{0.10}{GW190620B_o3afin}{0.34}{GW190602E_o3afin}{0.12}{GW190527H_o3afin}{0.10}{GW190521E_o3afin}{0.10}{GW190521B_o3afin}{-0.14}{GW190519J_o3afin}{0.33}{GW190517B_o3afin}{0.49}{GW190514E_o3afin}{-0.08}{GW190513E_o3afin}{0.16}{GW190512G_o3afin}{0.02}{GW190503E_o3afin}{-0.05}{GW190421I_o3afin}{-0.10}{GW190413E_o3afin}{-0.01}{GW190413A_o3afin}{-0.04}{GW190412B_o3afin}{0.21}{GW190408H_o3afin}{-0.03}}}
\newcommand{\chieffinfinityonlyprecavgplus}[1]{\IfEqCase{#1}{{GW190926C_o3afin}{0.25}{GW190925J_o3afin}{0.16}{GW190917B_o3afin}{0.21}{GW190916K_o3afin}{0.33}{GW190805J_o3afin}{0.29}{GW190725F_o3afin}{0.36}{GW190426N_o3afin}{0.42}{GW190403B_o3afin}{0.16}{GW150914_o3afin}{0.12}{GW151012_o3afin}{0.28}{GW151226_o3afin}{0.23}{GW170104_o3afin}{0.15}{GW170608_o3afin}{0.13}{GW170729_o3afin}{0.25}{GW170809_o3afin}{0.17}{GW170814_o3afin}{0.13}{GW170818_o3afin}{0.19}{GW170823_o3afin}{0.21}{GW190930C_o3afin}{0.22}{GW190929B_o3afin}{0.23}{GW190924A_o3afin}{0.20}{GW190915K_o3afin}{0.19}{GW190910B_o3afin}{0.17}{GW190828B_o3afin}{0.16}{GW190828A_o3afin}{0.15}{GW190814H_o3afin}{0.07}{GW190803B_o3afin}{0.23}{GW190731E_o3afin}{0.28}{GW190728D_o3afin}{0.19}{GW190727B_o3afin}{0.25}{GW190720A_o3afin}{0.14}{GW190719H_o3afin}{0.33}{GW190708M_o3afin}{0.10}{GW190707E_o3afin}{0.10}{GW190706F_o3afin}{0.25}{GW190701E_o3afin}{0.23}{GW190630E_o3afin}{0.14}{GW190620B_o3afin}{0.22}{GW190602E_o3afin}{0.25}{GW190527H_o3afin}{0.22}{GW190521E_o3afin}{0.13}{GW190521B_o3afin}{0.50}{GW190519J_o3afin}{0.20}{GW190517B_o3afin}{0.21}{GW190514E_o3afin}{0.29}{GW190513E_o3afin}{0.29}{GW190512G_o3afin}{0.13}{GW190503E_o3afin}{0.23}{GW190421I_o3afin}{0.21}{GW190413E_o3afin}{0.28}{GW190413A_o3afin}{0.27}{GW190412B_o3afin}{0.12}{GW190408H_o3afin}{0.13}}}
\newcommand{\chipinfinityonlyprecavgminus}[1]{\IfEqCase{#1}{{GW190926C_o3afin}{0.29}{GW190925J_o3afin}{0.30}{GW190917B_o3afin}{0.14}{GW190916K_o3afin}{0.28}{GW190805J_o3afin}{0.32}{GW190725F_o3afin}{0.28}{GW190426N_o3afin}{0.36}{GW190403B_o3afin}{0.23}{GW150914_o3afin}{0.38}{GW151012_o3afin}{0.26}{GW151226_o3afin}{0.36}{GW170104_o3afin}{0.30}{GW170608_o3afin}{0.24}{GW170729_o3afin}{0.30}{GW170809_o3afin}{0.30}{GW170814_o3afin}{0.37}{GW170818_o3afin}{0.41}{GW170823_o3afin}{0.35}{GW190930C_o3afin}{0.21}{GW190929B_o3afin}{0.25}{GW190924A_o3afin}{0.19}{GW190915K_o3afin}{0.40}{GW190910B_o3afin}{0.30}{GW190828B_o3afin}{0.20}{GW190828A_o3afin}{0.32}{GW190814H_o3afin}{0.03}{GW190803B_o3afin}{0.34}{GW190731E_o3afin}{0.32}{GW190728D_o3afin}{0.20}{GW190727B_o3afin}{0.37}{GW190720A_o3afin}{0.20}{GW190719H_o3afin}{0.32}{GW190708M_o3afin}{0.20}{GW190707E_o3afin}{0.22}{GW190706F_o3afin}{0.32}{GW190701E_o3afin}{0.33}{GW190630E_o3afin}{0.24}{GW190620B_o3afin}{0.34}{GW190602E_o3afin}{0.33}{GW190527H_o3afin}{0.28}{GW190521E_o3afin}{0.30}{GW190521B_o3afin}{0.35}{GW190519J_o3afin}{0.29}{GW190517B_o3afin}{0.31}{GW190514E_o3afin}{0.34}{GW190513E_o3afin}{0.26}{GW190512G_o3afin}{0.20}{GW190503E_o3afin}{0.32}{GW190421I_o3afin}{0.33}{GW190413E_o3afin}{0.41}{GW190413A_o3afin}{0.33}{GW190412B_o3afin}{0.12}{GW190408H_o3afin}{0.29}}}
\newcommand{\chipinfinityonlyprecavgmed}[1]{\IfEqCase{#1}{{GW190926C_o3afin}{0.38}{GW190925J_o3afin}{0.39}{GW190917B_o3afin}{0.17}{GW190916K_o3afin}{0.37}{GW190805J_o3afin}{0.50}{GW190725F_o3afin}{0.37}{GW190426N_o3afin}{0.50}{GW190403B_o3afin}{0.32}{GW150914_o3afin}{0.51}{GW151012_o3afin}{0.35}{GW151226_o3afin}{0.52}{GW170104_o3afin}{0.39}{GW170608_o3afin}{0.32}{GW170729_o3afin}{0.40}{GW170809_o3afin}{0.39}{GW170814_o3afin}{0.48}{GW170818_o3afin}{0.55}{GW170823_o3afin}{0.46}{GW190930C_o3afin}{0.30}{GW190929B_o3afin}{0.31}{GW190924A_o3afin}{0.25}{GW190915K_o3afin}{0.56}{GW190910B_o3afin}{0.38}{GW190828B_o3afin}{0.26}{GW190828A_o3afin}{0.44}{GW190814H_o3afin}{0.04}{GW190803B_o3afin}{0.44}{GW190731E_o3afin}{0.41}{GW190728D_o3afin}{0.30}{GW190727B_o3afin}{0.50}{GW190720A_o3afin}{0.29}{GW190719H_o3afin}{0.45}{GW190708M_o3afin}{0.26}{GW190707E_o3afin}{0.28}{GW190706F_o3afin}{0.46}{GW190701E_o3afin}{0.45}{GW190630E_o3afin}{0.33}{GW190620B_o3afin}{0.49}{GW190602E_o3afin}{0.44}{GW190527H_o3afin}{0.36}{GW190521E_o3afin}{0.39}{GW190521B_o3afin}{0.50}{GW190519J_o3afin}{0.45}{GW190517B_o3afin}{0.54}{GW190514E_o3afin}{0.45}{GW190513E_o3afin}{0.34}{GW190512G_o3afin}{0.26}{GW190503E_o3afin}{0.42}{GW190421I_o3afin}{0.45}{GW190413E_o3afin}{0.55}{GW190413A_o3afin}{0.44}{GW190412B_o3afin}{0.19}{GW190408H_o3afin}{0.37}}}
\newcommand{\chipinfinityonlyprecavgplus}[1]{\IfEqCase{#1}{{GW190926C_o3afin}{0.47}{GW190925J_o3afin}{0.43}{GW190917B_o3afin}{0.43}{GW190916K_o3afin}{0.44}{GW190805J_o3afin}{0.35}{GW190725F_o3afin}{0.46}{GW190426N_o3afin}{0.36}{GW190403B_o3afin}{0.37}{GW150914_o3afin}{0.34}{GW151012_o3afin}{0.43}{GW151226_o3afin}{0.36}{GW170104_o3afin}{0.40}{GW170608_o3afin}{0.41}{GW170729_o3afin}{0.39}{GW170809_o3afin}{0.44}{GW170814_o3afin}{0.38}{GW170818_o3afin}{0.35}{GW170823_o3afin}{0.41}{GW190930C_o3afin}{0.42}{GW190929B_o3afin}{0.51}{GW190924A_o3afin}{0.41}{GW190915K_o3afin}{0.35}{GW190910B_o3afin}{0.43}{GW190828B_o3afin}{0.42}{GW190828A_o3afin}{0.41}{GW190814H_o3afin}{0.04}{GW190803B_o3afin}{0.42}{GW190731E_o3afin}{0.43}{GW190728D_o3afin}{0.39}{GW190727B_o3afin}{0.38}{GW190720A_o3afin}{0.39}{GW190719H_o3afin}{0.39}{GW190708M_o3afin}{0.44}{GW190707E_o3afin}{0.39}{GW190706F_o3afin}{0.38}{GW190701E_o3afin}{0.41}{GW190630E_o3afin}{0.36}{GW190620B_o3afin}{0.38}{GW190602E_o3afin}{0.43}{GW190527H_o3afin}{0.48}{GW190521E_o3afin}{0.37}{GW190521B_o3afin}{0.32}{GW190519J_o3afin}{0.36}{GW190517B_o3afin}{0.32}{GW190514E_o3afin}{0.42}{GW190513E_o3afin}{0.43}{GW190512G_o3afin}{0.42}{GW190503E_o3afin}{0.41}{GW190421I_o3afin}{0.41}{GW190413E_o3afin}{0.36}{GW190413A_o3afin}{0.42}{GW190412B_o3afin}{0.23}{GW190408H_o3afin}{0.41}}}
\newcommand{\PEpercentBNS}[1]{\IfEqCase{#1}{{GW190926C_o3afin}{0}{GW190925J_o3afin}{0}{GW190917B_o3afin}{0}{GW190916K_o3afin}{0}{GW190805J_o3afin}{0}{GW190725F_o3afin}{0}{GW190426N_o3afin}{0}{GW190403B_o3afin}{0}{GW150914_o3afin}{0}{GW151012_o3afin}{0}{GW151226_o3afin}{0}{GW170104_o3afin}{0}{GW170608_o3afin}{0}{GW170729_o3afin}{0}{GW170809_o3afin}{0}{GW170814_o3afin}{0}{GW170818_o3afin}{0}{GW170823_o3afin}{0}{GW190930C_o3afin}{0}{GW190929B_o3afin}{0}{GW190924A_o3afin}{0}{GW190915K_o3afin}{0}{GW190910B_o3afin}{0}{GW190828B_o3afin}{0}{GW190828A_o3afin}{0}{GW190814H_o3afin}{0}{GW190803B_o3afin}{0}{GW190731E_o3afin}{0}{GW190728D_o3afin}{0}{GW190727B_o3afin}{0}{GW190720A_o3afin}{0}{GW190719H_o3afin}{0}{GW190708M_o3afin}{0}{GW190707E_o3afin}{0}{GW190706F_o3afin}{0}{GW190701E_o3afin}{0}{GW190630E_o3afin}{0}{GW190620B_o3afin}{0}{GW190602E_o3afin}{0}{GW190527H_o3afin}{0}{GW190521E_o3afin}{0}{GW190521B_o3afin}{0}{GW190519J_o3afin}{0}{GW190517B_o3afin}{0}{GW190514E_o3afin}{0}{GW190513E_o3afin}{0}{GW190512G_o3afin}{0}{GW190503E_o3afin}{0}{GW190425B_o3afin}{100}{GW190421I_o3afin}{0}{GW190413E_o3afin}{0}{GW190413A_o3afin}{0}{GW190412B_o3afin}{0}{GW190408H_o3afin}{0}}}
\newcommand{\PEpercentNSBH}[1]{\IfEqCase{#1}{{GW190926C_o3afin}{0}{GW190925J_o3afin}{0}{GW190917B_o3afin}{93}{GW190916K_o3afin}{0}{GW190805J_o3afin}{0}{GW190725F_o3afin}{2}{GW190426N_o3afin}{0}{GW190403B_o3afin}{0}{GW150914_o3afin}{0}{GW151012_o3afin}{0}{GW151226_o3afin}{0}{GW170104_o3afin}{0}{GW170608_o3afin}{0}{GW170729_o3afin}{0}{GW170809_o3afin}{0}{GW170814_o3afin}{0}{GW170818_o3afin}{0}{GW170823_o3afin}{0}{GW190930C_o3afin}{0}{GW190929B_o3afin}{0}{GW190924A_o3afin}{2}{GW190915K_o3afin}{0}{GW190910B_o3afin}{0}{GW190828B_o3afin}{0}{GW190828A_o3afin}{0}{GW190814H_o3afin}{99}{GW190803B_o3afin}{0}{GW190731E_o3afin}{0}{GW190728D_o3afin}{0}{GW190727B_o3afin}{0}{GW190720A_o3afin}{0}{GW190719H_o3afin}{0}{GW190708M_o3afin}{0}{GW190707E_o3afin}{0}{GW190706F_o3afin}{0}{GW190701E_o3afin}{0}{GW190630E_o3afin}{0}{GW190620B_o3afin}{0}{GW190602E_o3afin}{0}{GW190527H_o3afin}{0}{GW190521E_o3afin}{0}{GW190521B_o3afin}{0}{GW190519J_o3afin}{0}{GW190517B_o3afin}{0}{GW190514E_o3afin}{0}{GW190513E_o3afin}{0}{GW190512G_o3afin}{0}{GW190503E_o3afin}{0}{GW190425B_o3afin}{0}{GW190421I_o3afin}{0}{GW190413E_o3afin}{0}{GW190413A_o3afin}{0}{GW190412B_o3afin}{0}{GW190408H_o3afin}{0}}}
\newcommand{\PEpercentBBH}[1]{\IfEqCase{#1}{{GW190926C_o3afin}{100}{GW190925J_o3afin}{100}{GW190917B_o3afin}{7}{GW190916K_o3afin}{100}{GW190805J_o3afin}{100}{GW190725F_o3afin}{98}{GW190426N_o3afin}{100}{GW190403B_o3afin}{100}{GW150914_o3afin}{100}{GW151012_o3afin}{100}{GW151226_o3afin}{100}{GW170104_o3afin}{100}{GW170608_o3afin}{100}{GW170729_o3afin}{100}{GW170809_o3afin}{100}{GW170814_o3afin}{100}{GW170818_o3afin}{100}{GW170823_o3afin}{100}{GW190930C_o3afin}{100}{GW190929B_o3afin}{100}{GW190924A_o3afin}{98}{GW190915K_o3afin}{100}{GW190910B_o3afin}{100}{GW190828B_o3afin}{100}{GW190828A_o3afin}{100}{GW190814H_o3afin}{1}{GW190803B_o3afin}{100}{GW190731E_o3afin}{100}{GW190728D_o3afin}{100}{GW190727B_o3afin}{100}{GW190720A_o3afin}{100}{GW190719H_o3afin}{100}{GW190708M_o3afin}{100}{GW190707E_o3afin}{100}{GW190706F_o3afin}{100}{GW190701E_o3afin}{100}{GW190630E_o3afin}{100}{GW190620B_o3afin}{100}{GW190602E_o3afin}{100}{GW190527H_o3afin}{100}{GW190521E_o3afin}{100}{GW190521B_o3afin}{100}{GW190519J_o3afin}{100}{GW190517B_o3afin}{100}{GW190514E_o3afin}{100}{GW190513E_o3afin}{100}{GW190512G_o3afin}{100}{GW190503E_o3afin}{100}{GW190425B_o3afin}{0}{GW190421I_o3afin}{100}{GW190413E_o3afin}{100}{GW190413A_o3afin}{100}{GW190412B_o3afin}{100}{GW190408H_o3afin}{100}}}
\newcommand{\PEpercentMassGap}[1]{\IfEqCase{#1}{{W190926C_o3afin}{0}{GW190925J_o3afin}{0}{GW190917B_o3afin}{0}{GW190916K_o3afin}{0}{GW190805J_o3afin}{0}{GW190725F_o3afin}{0}{GW190426N_o3afin}{0}{GW190403B_o3afin}{0}{W150914_o3afin}{0}{GW151012_o3afin}{0}{GW151226_o3afin}{0}{GW170104_o3afin}{0}{GW170608_o3afin}{0}{GW170729_o3afin}{0}{GW170809_o3afin}{0}{GW170814_o3afin}{0}{GW170818_o3afin}{0}{GW170823_o3afin}{0}{W190930C_o3afin}{0}{GW190929B_o3afin}{0}{GW190924A_o3afin}{0}{GW190915K_o3afin}{0}{GW190910B_o3afin}{0}{GW190828B_o3afin}{0}{GW190828A_o3afin}{0}{GW190814H_o3afin}{0}{GW190803B_o3afin}{0}{GW190731E_o3afin}{0}{GW190728D_o3afin}{0}{GW190727B_o3afin}{0}{GW190720A_o3afin}{0}{GW190719H_o3afin}{0}{GW190708M_o3afin}{0}{GW190707E_o3afin}{0}{GW190706F_o3afin}{0}{GW190701E_o3afin}{0}{GW190630E_o3afin}{0}{GW190620B_o3afin}{0}{GW190602E_o3afin}{0}{GW190527H_o3afin}{0}{GW190521E_o3afin}{0}{GW190521B_o3afin}{0}{GW190519J_o3afin}{0}{GW190517B_o3afin}{0}{GW190514E_o3afin}{0}{GW190513E_o3afin}{0}{GW190512G_o3afin}{0}{GW190503E_o3afin}{0}{GW190425B_o3afin}{0}{GW190421I_o3afin}{0}{GW190413E_o3afin}{0}{GW190413A_o3afin}{0}{GW190412B_o3afin}{0}{GW190408H_o3afin}{0}}}

\newcommand{\SNAME}[1]{\IfEqCase{#1}{{GW190930A}{S190930s}{GW190929A}{S190929d}{GW190924A}{S190924h}{GW190915A}{S190915ak}{GW190910A}{S190910s}{GW190828B}{S190828l}{GW190828A}{S190828j}{GW190814A}{S190814bv}{GW190803A}{S190803e}{GW190731A}{S190731aa}{GW190728A}{S190728q}{GW190727A}{S190727h}{GW190720A}{S190720a}{GW190719A}{S190719an}{GW190708A}{S190708ap}{GW190707A}{S190707q}{GW190706A}{S190706ai}{GW190701A}{S190701ah}{GW190630A}{S190630ag}{GW190620A}{S190620e}{GW190602A}{S190602aq}{GW190527A}{S190527w}{GW190521B}{S190521r}{GW190521A}{S190521g}{GW190519A}{S190519bj}{GW190517A}{S190517h}{GW190514A}{S190514n}{GW190513A}{S190513bm}{GW190512A}{S190512at}{GW190503A}{S190503bf}{GW190425A}{S190425z}{GW190421A}{S190421ar}{GW190413B}{S190413ac}{GW190413A}{S190413i}{GW190412A}{S190412m}{GW190408A}{S190408an}{GW190926A}{S190926d}{GW190925A}{S190925ad}{GW190917A}{S190917u}{GW190916A}{S190916al}{GW190805A}{S190805bq}{GW190725A}{S190725t}{GW190426B}{S190426l}{GW190403A}{S190403cj}{GW151226A}{GW151226}{GW151012A}{GW151012}{GW150914A}{GW150914}{GW170823A}{GW170823}{GW170818A}{GW170818}{GW170814A}{GW170814}{GW170809A}{GW170809}{GW170729A}{GW170729}{GW170608A}{GW170608}{GW170104A}{GW170104}{GW190909A}{S190909w}{GW190426A}{S190426c}{GW190424A}{S190424ao}}}
\newcommand{\NNAME}[1]{\IfEqCase{#1}{{GW190930A}{GW190930\_133541}{GW190929A}{GW190929\_012149}{GW190924A}{GW190924\_021846}{GW190915A}{GW190915\_235702}{GW190910A}{GW190910\_112807}{GW190828B}{GW190828\_065509}{GW190828A}{GW190828\_063405}{GW190814A}{GW190814}{GW190803A}{GW190803\_022701}{GW190731A}{GW190731\_140936}{GW190728A}{GW190728\_064510}{GW190727A}{GW190727\_060333}{GW190720A}{GW190720\_000836}{GW190719A}{GW190719\_215514}{GW190708A}{GW190708\_232457}{GW190707A}{GW190707\_093326}{GW190706A}{GW190706\_222641}{GW190701A}{GW190701\_203306}{GW190630A}{GW190630\_185205}{GW190620A}{GW190620\_030421}{GW190602A}{GW190602\_175927}{GW190527A}{GW190527\_092055}{GW190521B}{GW190521\_074359}{GW190521A}{GW190521}{GW190519A}{GW190519\_153544}{GW190517A}{GW190517\_055101}{GW190514A}{GW190514\_065416}{GW190513A}{GW190513\_205428}{GW190512A}{GW190512\_180714}{GW190503A}{GW190503\_185404}{GW190425A}{GW190425}{GW190421A}{GW190421\_213856}{GW190413B}{GW190413\_134308}{GW190413A}{GW190413\_052954}{GW190412A}{GW190412}{GW190408A}{GW190408\_181802}{GW190926A}{GW190926\_050336}{GW190925A}{GW190925\_232845}{GW190917A}{GW190917\_114630}{GW190916A}{GW190916\_200658}{GW190805A}{GW190805\_211137}{GW190725A}{GW190725\_174728}{GW190426B}{GW190426\_190642}{GW190403A}{GW190403\_051519}{GW151226A}{GW151226}{GW151012A}{GW151012}{GW150914A}{GW150914}{GW170823A}{GW170823}{GW170818A}{GW170818}{GW170814A}{GW170814}{GW170809A}{GW170809}{GW170729A}{GW170729}{GW170608A}{GW170608}{GW170104A}{GW170104}{GW190909A}{GW190909\_114149}{GW190426A}{GW190426\_152155}{GW190424A}{GW190424\_180648}}}
\newcommand{\MINIMALNAME}[1]{\IfEqCase{#1}{{GW190930A}{GW190930}{GW190929A}{GW190929}{GW190924A}{GW190924}{GW190915A}{GW190915}{GW190910A}{GW190910}{GW190828B}{GW190828\_0655}{GW190828A}{GW190828\_0634}{GW190814A}{GW190814}{GW190803A}{GW190803}{GW190731A}{GW190731}{GW190728A}{GW190728}{GW190727A}{GW190727}{GW190720A}{GW190720}{GW190719A}{GW190719}{GW190708A}{GW190708}{GW190707A}{GW190707}{GW190706A}{GW190706}{GW190701A}{GW190701}{GW190630A}{GW190630}{GW190620A}{GW190620}{GW190602A}{GW190602}{GW190527A}{GW190527}{GW190521B}{GW190521\_07}{GW190521A}{GW190521}{GW190519A}{GW190519}{GW190517A}{GW190517}{GW190514A}{GW190514}{GW190513A}{GW190513}{GW190512A}{GW190512}{GW190503A}{GW190503}{GW190425A}{GW190425}{GW190421A}{GW190421}{GW190413B}{GW190413\_13}{GW190413A}{GW190413\_05}{GW190412A}{GW190412}{GW190408A}{GW190408}{GW190926A}{GW190926}{GW190925A}{GW190925}{GW190917A}{GW190917}{GW190916A}{GW190916}{GW190805A}{GW190805}{GW190725A}{GW190725}{GW190426B}{GW190426\_19}{GW190403A}{GW190403}{GW151226A}{GW151226}{GW151012A}{GW151012}{GW150914A}{GW150914}{GW170823A}{GW170823}{GW170818A}{GW170818}{GW170814A}{GW170814}{GW170809A}{GW170809}{GW170729A}{GW170729}{GW170608A}{GW170608}{GW170104A}{GW170104}{GW190909A}{GW190909}{GW190426A}{GW190426\_15}{GW190424A}{GW190424}}}



\newcommand{\MITIGATIONMETHOD}[1]{\IfEqCase{#1}{{GW190512G_o3afin}{None}{GW190924A_o3afin}{L1 glitch subtraction, glitch-only model}{GW190519J_o3afin}{None}{GW190517B_o3afin}{None}{GW190620B_o3afin}{None}{GW190521E_o3afin}{None}{GW190926C_o3afin}{None}{GW190910B_o3afin}{None}{GW190728D_o3afin}{None}{GW190915K_o3afin}{None}{GW190701E_o3afin}{L1 glitch subtraction, glitch+signal model}{GW190929B_o3afin}{None}{GW190413E_o3afin}{L1 glitch subtraction, glitch-only model}{GW190706F_o3afin}{None}{GW190630E_o3afin}{None}{GW190527H_o3afin}{None}{GW190719H_o3afin}{None}{GW190403B_o3afin}{None}{GW190725F_o3afin}{None}{GW190707E_o3afin}{None}{GW190720A_o3afin}{None}{GW190412B_o3afin}{None}{GW190925J_o3afin}{None}{GW190828B_o3afin}{None}{GW190413A_o3afin}{None}{GW190727B_o3afin}{\fixme{L1 $f_\text{min}$: 50 Hz}}{GW190514E_o3afin}{L1 glitch subtraction, glitch-only model}{GW190521B_o3afin}{None}{GW190916K_o3afin}{None}{GW190425B_o3afin}{L1 glitch subtraction, glitch-only model}{GW190408H_o3afin}{None}{GW190731E_o3afin}{None}{GW190426N_o3afin}{None}{GW190828A_o3afin}{None}{GW190708M_o3afin}{None}{GW190421I_o3afin}{None}{GW190803B_o3afin}{None}{GW190503E_o3afin}{L1 glitch subtraction, glitch-only model}{GW190602E_o3afin}{None}{GW190513E_o3afin}{L1 glitch subtraction, glitch-only model}{GW190930C_o3afin}{None}{GW190805J_o3afin}{None}{GW190917B_o3afin}{None}{GW190814H_o3afin}{L1 $f_\text{min}$: 30 Hz; H1 non-observing data used}}}

\newcommand{\skyarea}[1]{\IfEqCase{#1}{{GW190926C_o3afin}{2000}{GW190925J_o3afin}{2900}{GW190917B_o3afin}{1700}{GW190916K_o3afin}{2400}{GW190805J_o3afin}{1600}{GW190725F_o3afin}{2200}{GW190426N_o3afin}{4600}{GW190403B_o3afin}{3900}{GW150914_o3afin}{250}{GW151012_o3afin}{1700}{GW151226_o3afin}{950}{GW170104_o3afin}{1000}{GW170608_o3afin}{380}{GW170729_o3afin}{830}{GW170809_o3afin}{260}{GW170814_o3afin}{92}{GW170818_o3afin}{35}{GW170823_o3afin}{1800}{GW190930C_o3afin}{1600}{GW190929B_o3afin}{1700}{GW190924A_o3afin}{380}{GW190915K_o3afin}{450}{GW190910B_o3afin}{9600}{GW190828B_o3afin}{590}{GW190828A_o3afin}{340}{GW190814H_o3afin}{22}{GW190803B_o3afin}{1000}{GW190731E_o3afin}{3600}{GW190728D_o3afin}{400}{GW190727B_o3afin}{380}{GW190720A_o3afin}{260}{GW190719H_o3afin}{3600}{GW190708M_o3afin}{11000}{GW190707E_o3afin}{1200}{GW190706F_o3afin}{2600}{GW190701E_o3afin}{45}{GW190630E_o3afin}{670}{GW190620B_o3afin}{7700}{GW190602E_o3afin}{740}{GW190527H_o3afin}{3500}{GW190521E_o3afin}{470}{GW190521B_o3afin}{1000}{GW190519J_o3afin}{570}{GW190517B_o3afin}{510}{GW190514E_o3afin}{3400}{GW190513E_o3afin}{450}{GW190512G_o3afin}{230}{GW190503E_o3afin}{100}{GW190425B_o3afin}{8700}{GW190421I_o3afin}{1200}{GW190413E_o3afin}{630}{GW190413A_o3afin}{650}{GW190412B_o3afin}{240}{GW190408H_o3afin}{290}}}
\newcommand{\skyvol}[1]{\IfEqCase{#1}{{GW190926C_o3afin}{8.2}{GW190925J_o3afin}{0.45}{GW190917B_o3afin}{0.11}{GW190916K_o3afin}{15}{GW190805J_o3afin}{13}{GW190725F_o3afin}{0.38}{GW190426N_o3afin}{25}{GW190403B_o3afin}{55}{GW150914_o3afin}{0.0034}{GW151012_o3afin}{0.29}{GW151226_o3afin}{0.013}{GW170104_o3afin}{0.14}{GW170608_o3afin}{0.003}{GW170729_o3afin}{1.5}{GW170809_o3afin}{0.039}{GW170814_o3afin}{0.003}{GW170818_o3afin}{0.006}{GW170823_o3afin}{1.1}{GW190930C_o3afin}{0.13}{GW190929B_o3afin}{4.8}{GW190924A_o3afin}{0.012}{GW190915K_o3afin}{0.24}{GW190910B_o3afin}{5.2}{GW190828B_o3afin}{0.27}{GW190828A_o3afin}{0.25}{GW190814H_o3afin}{3.6\times 10^{-5}}{GW190803B_o3afin}{2.2}{GW190731E_o3afin}{9.2}{GW190728D_o3afin}{0.039}{GW190727B_o3afin}{0.76}{GW190720A_o3afin}{0.042}{GW190719H_o3afin}{13}{GW190708M_o3afin}{0.98}{GW190707E_o3afin}{0.092}{GW190706F_o3afin}{7.8}{GW190701E_o3afin}{0.035}{GW190630E_o3afin}{0.097}{GW190620B_o3afin}{13}{GW190602E_o3afin}{1.4}{GW190527H_o3afin}{6.6}{GW190521E_o3afin}{0.083}{GW190521B_o3afin}{3.4}{GW190519J_o3afin}{1.0}{GW190517B_o3afin}{0.52}{GW190514E_o3afin}{11}{GW190513E_o3afin}{0.42}{GW190512G_o3afin}{0.085}{GW190503E_o3afin}{0.044}{GW190425B_o3afin}{0.0078}{GW190421I_o3afin}{1.7}{GW190413E_o3afin}{2.3}{GW190413A_o3afin}{2.0}{GW190412B_o3afin}{0.015}{GW190408H_o3afin}{0.097}}} 
\newcommand{\SINGLEPIPELINEEVENTS}{\fixme{Four}}

\newcommand{\PROBABELOWGAP}{0.16\xspace}

\newcommand{\PROBASECONDARYBELOWGAP}{0.996\xspace}

\newcommand{\PROBBBELOWGAP}{0.00078\xspace} 

\newcommand{\PROBBSECONDARYBELOWGAP}{0.30\xspace}

\newcommand{\PROBAABOVEGAP}{0.02\xspace}

\newcommand{\PROBAASECONDARYBOVEGAP}{0.0\xspace}

\newcommand{\PROBBABOVEGAP}{0.25\xspace} 

\newcommand{\PROBBSECONDARYABOVEGAP}{0.0021\xspace} 

\newcommand{\NEWBULK}{\fixme{Four}}

\newcommand{\PROBLOWGAP}{0.18\xspace}

\newcommand{\POWERLAWPOP}{$39.7^{+20.3}_{-9.1}\,{}\Msun$\xspace}

\newcommand{\GAUSSIANPOP}{$33.1^{+4.0}_{-5.6}\,{}\Msun$\xspace}

\newcommand{\evonechione}{$\spinone = 0.89^{ +0.09}_{ -0.31}$\xspace}

\newcommand{\evonechioneOnesided}{$\spinone > 0.69$\xspace}

\newcommand{\evonechioneB}{$\spinone = 0.75^{ +0.22}_{ -0.59}$\xspace}

\newcommand{\evonechioneBOnesided}{$\spinone > 0.29$\xspace}

\newcommand{\evonechioneC}{$\spinone = 0.23^{ +0.63}_{ -0.21}$\xspace}

\newcommand{\ProbSpinEitherHighA}{$82\%$\xspace}

\newcommand{\ProbSpinEitherHighB}{$59\%$\xspace}

\newcommand{\evoneq}{$\massratio = 0.23^{ +0.60}_{ -0.12}$\xspace}

\newcommand{\evoneqC}{$\massratio = 0.21^{ +0.32}_{ -0.09}$\xspace}

\newcommand{\evoneqOnesidedValue}{0.65\xspace}
\newcommand{\evoneqOnesided}{$\massratio < \evoneqOnesidedValue$\xspace}

\newcommand{\evoneqCOnesidedValue}{0.44\xspace}
\newcommand{\evoneqCOnesided}{$\massratio < \evoneqCOnesidedValue$\xspace}

\newcommand{\evonemtotDmedian}{$\sim 182\Msun$\xspace}

\newcommand{\evonemtotEmedian}{$\sim 14\Msun$\xspace}

\newcommand{\evoneqminimum}{$\massratio > 0.063$\xspace}

\DeclareRobustCommand{\OLDEVENTSNAME}[1]{\IfEqCase{#1}{{GW150914_o3afin}{GW150914}{GW151012_o3afin}{GW151012}{GW151226_o3afin}{GW151226}{GW170104_o3afin}{GW170104}{GW170608_o3afin}{GW170608}{GW170729_o3afin}{GW170729}{GW170809_o3afin}{GW170809}{GW170814_o3afin}{GW170814}{GW170818_o3afin}{GW170818}{GW170823_o3afin}{GW170823}}}

\title{GWTC-2.1: Deep Extended Catalog of Compact Binary Coalescences Observed by LIGO and Virgo
During the First Half of the Third Observing Run}



\author{R.~Abbott}
\affiliation{LIGO Laboratory, California Institute of Technology, Pasadena, CA 91125, USA}
\author{T.~D.~Abbott}
\affiliation{Louisiana State University, Baton Rouge, LA 70803, USA}
\author{F.~Acernese}
\affiliation{Dipartimento di Farmacia, Universit\`a di Salerno, I-84084 Fisciano, Salerno, Italy}
\affiliation{INFN, Sezione di Napoli, Complesso Universitario di Monte S. Angelo, I-80126 Napoli, Italy}
\author{K.~Ackley}
\affiliation{OzGrav, School of Physics \& Astronomy, Monash University, Clayton 3800, Victoria, Australia}
\author{C.~Adams}
\affiliation{LIGO Livingston Observatory, Livingston, LA 70754, USA}
\author{N.~Adhikari}
\affiliation{University of Wisconsin-Milwaukee, Milwaukee, WI 53201, USA}
\author{R.~X.~Adhikari}
\affiliation{LIGO Laboratory, California Institute of Technology, Pasadena, CA 91125, USA}
\author{V.~B.~Adya}
\affiliation{OzGrav, Australian National University, Canberra, Australian Capital Territory 0200, Australia}
\author{C.~Affeldt}
\affiliation{Max Planck Institute for Gravitational Physics (Albert Einstein Institute), D-30167 Hannover, Germany}
\affiliation{Leibniz Universit\"at Hannover, D-30167 Hannover, Germany}
\author{D.~Agarwal}
\affiliation{Inter-University Centre for Astronomy and Astrophysics, Pune 411007, India}
% ...

\collaboration{The LIGO Scientific Collaboration and the Virgo Collaboration}




\date[\relax]{Compiled: \today}

\begin{abstract}

The second Gravitational-Wave Transient Catalog, GWTC-2, reported on \fixme{39}
compact binary coalescences observed by the Advanced LIGO and Advanced Virgo detectors
between \RUNSTART{} and \ARUNEND{}. Here, we present GWTC-2.1, which reports on
a deeper list of candidate events observed over the same period. We analyze the
final version of the strain data over this period with improved calibration and
better subtraction of excess noise, which has been publicly released. We employ three
matched-filter search pipelines for candidate identification, and estimate the
probability of astrophysical origin for each candidate event. While GWTC-2 used
a false alarm rate threshold of \fixme{2 per year}, we include in GWTC-2.1, 
\fixme{\TOTALEVENTS{}} candidates that pass a false alarm rate threshold of
\fixme{2 per day}. We calculate the source properties of a subset of
\fixme{\NUMEVENTS{}} high-significance candidates that have a probability of
astrophysical origin greater than $0.5$. Of these
candidates, \fixme{36} have been reported in GWTC-2. 
We also calculate updated source properties for all binary block hole events previously reported in GWTC-1.
If the \NUMADDCANDIDATES{}
additional high-significance candidates presented here are 
astrophysical, the mass range of events that are unambiguously identified
as binary black holes (both objects $\geq 3\Msun$) is increased compared to
GWTC-2, with total masses from \evonemtotEmedian for \NAME{GW190924A_o3afin}{}
to \evonemtotDmedian for \NAME{GW190426N_o3afin}. \fixme{The primary
components of two new candidate events (\NAME{GW190403B_o3afin}{} and
\NAME{GW190426N_o3afin}) fall in the mass gap predicted by pair-instability
supernova theory.} We also expand the population of binaries with significantly
asymmetric mass ratios reported in GWTC-2 by an additional two events
(the mass ratio is less than \evoneqOnesidedValue and \evoneqCOnesidedValue at $90\%$ probability for \NAME{GW190403B_o3afin}{}
and  \NAME{GW190917B_o3afin}{} respectively), and find
that $2$ of the \NUMADDCANDIDATES{} new events have effective inspiral spins $\chi_\mathrm{eff} >
0$ (at $90\%$ credibility), while no binary is consistent with
$\chi_\mathrm{eff} < 0$ at the same significance.  We provide updated estimates for 
rates of binary black hole and binary neutron star coalescence in the local Universe. 

\end{abstract}

\pacs{%
04.80.Nn, 
04.25.dg, 
95.85.Sz, 
97.80.-d   
04.30.Db, 
04.30.Tv  
}

\maketitle

\acrodef{LSC}[LSC]{LIGO Scientific Collaboration}
\acrodef{LVC}[LVC]{LIGO Scientific and Virgo Collaboration}
\acrodef{LVK}[LVK]{LIGO Scientific, Virgo and KAGRA Collaboration}
\acrodef{aLIGO}{Advanced Laser Interferometer Gravitational-Wave Observatory}
\acrodef{aVirgo}{Advanced Virgo}
\acrodef{LIGO}[LIGO]{Laser Interferometer Gravitational-Wave Observatory}
\acrodef{IFO}[IFO]{interferometer}
\acrodef{LHO}[LHO]{LIGO-Hanford}
\acrodef{LLO}[LLO]{LIGO-Livingston}
\acrodef{O2}[O2]{second observing run}
\acrodef{O1}[O1]{first observing run}
\acrodef{O3}[O3]{third observing run}
\acrodef{O3a}[O3a]{first half of the third observing run}
\acrodef{O3b}[O3b]{second half of the third observing run}

\acrodef{BH}[BH]{black hole}
\acrodef{BBH}[BBH]{binary black hole}
\acrodef{BNS}[BNS]{binary neutron star}
\acrodef{IMBH}[IMBH]{intermediate-mass black hole}
\acrodef{NS}[NS]{neutron star}
\acrodef{BHNS}[BHNS]{black hole--neutron star binaries}
\acrodef{NSBH}[NSBH]{neutron star--black hole binary}
\acrodef{PBH}[PBH]{primordial black hole binaries}
\acrodef{CBC}[CBC]{compact binary coalescence}
\acrodef{GW}[GW]{gravitational wave}
\acrodef{GWH}[GW]{gravitational-wave}

\acrodef{CWB}[cWB]{coherent WaveBurst}

\acrodef{SNR}[SNR]{signal-to-noise ratio}
\acrodef{FAR}[FAR]{false alarm rate}
\acrodef{IFAR}[IFAR]{inverse false alarm rate}
\acrodef{FAP}[FAP]{false alarm probability}
\acrodef{PSD}[PSD]{power spectral density}

\acrodef{GR}[GR]{general relativity}
\acrodef{NR}[NR]{numerical relativity}
\acrodef{PN}[PN]{post-Newtonian}
\acrodef{EOB}[EOB]{effective-one-body}
\acrodef{ROM}[ROM]{reduced-order model}
\acrodef{IMR}[IMR]{inspiral--merger--ringdown}

\acrodef{PDF}[PDF]{probability density function}
\acrodef{PE}[PE]{parameter estimation}
\acrodef{CL}[CL]{credible level}

\acrodef{EOS}[EoS]{equation of state}

\acrodef{LAL}[LAL]{LIGO Algorithm Library}

\acrodef{KLD}[KLD]{Kullback--Leibler divergence}
\acrodef{JSD}[JSD]{Jensen--Shannon divergence}

\newcommand{\PN}[0]{\ac{PN}\xspace}
\newcommand{\BBH}[0]{\ac{BBH}\xspace}
\newcommand{\BNS}[0]{\ac{BNS}\xspace}
\newcommand{\BH}[0]{\ac{BH}\xspace}
\newcommand{\NR}[0]{\ac{NR}\xspace}
\newcommand{\GW}[0]{\ac{GW}\xspace}
\newcommand{\SNR}[0]{\ac{SNR}\xspace}
\newcommand{\aLIGO}[0]{\ac{aLIGO}\xspace}
\newcommand{\PE}[0]{\ac{PE}\xspace}
\newcommand{\IMR}[0]{\ac{IMR}\xspace}
\newcommand{\PDF}[0]{\ac{PDF}\xspace}
\newcommand{\GR}[0]{\ac{GR}\xspace}
\newcommand{\PSD}[0]{\ac{PSD}\xspace}
\newcommand{\EOS}[0]{\ac{EOS}\xspace}
\newcommand{\LVC}[0]{\ac{LVC}\xspace}

\newcommand{\GSTLAL}{GstLAL\xspace}
\newcommand{\BAYESTAR}{BAYESTAR\xspace}
\newcommand{\CWB}{\ac{CWB}\xspace}
\newcommand{\PYCBC}{PyCBC\xspace}
\newcommand{\MBTA}{MBTA\xspace}
\newcommand{\SPIIR}{SPIIR\xspace}
\newcommand{\LALINFERENCE}{LALInference\xspace}
\newcommand{\BAYESWAVE}{BayesWave\xspace}
\newcommand{\BILBY}{Bilby\xspace}
\newcommand{\RIFT}{RIFT\xspace}
\newcommand{\LALSUITE}{LALSuite\xspace}
\newcommand{\BILBYPIPE}{{BilbyPipe}\xspace}
\newcommand{\PBILBY}{{Parallel \BILBY{}}\xspace}
\newcommand{\ASIMOV}{{Asimov}\xspace}
\newcommand{\PESUMMARY}{PESummary\xspace}

\newcommand{\PYTHON}{\texttt{Python}\xspace}
\newcommand{\NUMPY}{\texttt{NumPy}\xspace}
\newcommand{\SCIPY}{\texttt{SciPy}\xspace}
\newcommand{\PLT}{\texttt{Matplotlib}\xspace}
\newcommand{\SEABORN}{\texttt{seaborn}\xspace}
\newcommand{\GWPY}{\texttt{GWpy}\xspace}
\newcommand{\DYNESTY}{\texttt{Dynesty}\xspace}

\section{Introduction}\label{sec:intro}

We are in the era of \ac{GW} astronomy, started by the Advanced
\ac{LIGO}~\cite{TheLIGOScientific:2014jea} and the Advanced
Virgo~\cite{TheVirgo:2014hva} detectors. The \ac{O1} of the advanced detectors yielded the first detection of \acp{GW} from a \ac{BBH},
GW150914~\cite{Abbott:2016blz}. By the end of \ac{O1}, the \ac{LVC} had
reported on \fixme{three} \ac{BBH} events~\cite{TheLIGOScientific:2016pea}. The
\ac{O2} of the advanced detectors saw the first direct detection of \acp{GW}
from a \ac{BNS}, GW170817~\cite{TheLIGOScientific:2017qsa}. This event was also
detected in electromagnetic waves~\cite{GBM:2017lvd}, expanding the field of
multimessenger astronomy to include \acp{GW}. By the end of \ac{O2}, the
\ac{LVC} had reported on a total of \fixme{ten} \acp{BBH} and \fixme{one}
\ac{BNS} event, described in the first Gravitational-Wave Transient Catalog,
GWTC-1~\cite{LIGOScientific:2018mvr}. The second Gravitational-Wave Transient
Catalog, GWTC-2~\cite{Abbott:2020niy}, added \GWTCTWOEVENTSTOT{} \ac{GW} events from the \ac{O3a},
and included a total of \fixme{50} events. The \ac{GW} data
until the end of \ac{O3} have been made available to the public by the \ac{LVC}.
Since the public release of the LIGO and Virgo data, groups other than the
\ac{LVC} have also performed analyses searching for \ac{GW}
signals~\cite{Nitz:2018imz,Magee:2019vmb,Venumadhav:2019tad,Zackay:2019tzo,Venumadhav:2019lyq,Nitz:2019hdf,Zackay:2019btq,Nitz:2020bdb,Nitz:2021uxj,
nitz2021search, nitz20214ogc, Olsen:2022pin, Nitz:2022ltl} and reported additional candidate events in some cases.

\ac{GW} events between \RUNSTART{} and \ARUNEND{} (O3a) that passed a \ac{FAR}
threshold of \fixme{2 per year} were presented in GWTC-2. Here, we present
GWTC-2.1, a deep catalog that includes \TOTALEVENTS{} candidates passing a
low-significance \ac{FAR} threshold of \fixme{2 per day}. Although most of the
candidates in this catalog are noise events, they can be used for
multimessenger searches by comparing against other astronomical surveys.
Temporal and spatial coincidences between candidates in distinct astrophysical
channels could lead to multimessenger discoveries~\cite{Smith_2013,Burns_2019}.
Multimessenger observations could enhance our understanding of the physical
processes associated with such systems. Previous \ac{GW} searches, both from the \ac{LVC}~\cite{gwtc1datarelease} and independent groups~\cite{Magee:2019vmb, Nitz_2019,
gwtc1datarelease, Venumadhav:2019lyq, Nitz:2019hdf}, including 
the 3-OGC analysis of public data from \ac{O1} to \ac{O3a}~\cite{Nitz:2021uxj}, have released subthreshold candidates. 
It is computationally unfeasible to determine detailed source properties of the
large set of subthreshold \ac{GW} candidates, therefore we identify a subset
of \ac{CBC} candidates that have a probability of astrophysical origin
\pastro{}~\cite{Farr:2013yna,Abbott:2016nhf,Abbott:2016drs} greater than $0.5$, and calculate the source properties of these
events. This probability \pastro{} uses both the signal rate in addition to the noise rate
in order to determine the significance of events.
  There are \NUMEVENTS{} such
candidate events, \fixme{36} of which have already been reported in GWTC-2 and
their source properties have been described in detail~\cite{Abbott:2020niy}. 
We present the source properties with a consistent set of
state-of-the-art waveform models for all of these candidates, discussing the properties of the \NUMADDCANDIDATES{} new events that have a
\pastro{} greater than $0.5$ in detail in the body of the paper, and our results for the previously reported candidates in  Appendix~\ref{app:parameter-estimation-appendix}.
A subset of the \NUMADDCANDIDATES{} additional events have
been found in the \ac{LVC} search of \ac{O3a} data~\cite{Abbott:2021iab} for
faint gravitationally lensed counterpart images~\cite{Li:2019osa,
McIsaac:2019use}, and in the independent 3-OGC~\cite{Nitz:2021uxj} analysis. While the \NUMADDCANDIDATES{} new events presented here have
a non-negligible probability of being from noise, some
of these have astrophysically interesting source properties under the default
prior. Two of the new candidates presented here have a primary component mass
in the pair instability
gap~\cite{2017ApJ...836..244W,2019ApJ...878...49W,2019ApJ...882..121S,2019ApJ...887...53F,2020ApJ...902L..36F,2020ApJ...888...76M,2020A&A...640L..18M,2021MNRAS.501.4514C,2021MNRAS.502L..40F},
and one of those shows support for high spin and unequal masses. We also
find a new candidate whose masses are consistent with a \ac{NSBH}, although as
in the case of GW190814~\cite{GW190814A}, we cannot rule out the possibility
that the secondary component of the candidate could be a low-mass black hole.

In this work, all the analyses make use of the final version of the strain data
with improved calibration and noise subtraction, which includes non-linear
subtraction around the $60\, \rm Hz$ frequency of the US power grid~\cite{Vajente:2019ycy,
Tiwari_2015}. The data used in this work have been released to the public~\cite{o3a_public_data,gwtc2p1_detchar_data,gwtc2p1_deglitched_data}. We use three matched-filter pipelines for candidate identification:
\GSTLAL{}~\cite{Sachdev:2019vvd, Hanna:2019ezx, Messick:2016aqy},
\PYCBC{}~\cite{Allen:2005fk, Allen:2004gu, Canton:2014ena, Usman:2015kfa,
Nitz:2017svb}, and \MBTA{}~\cite{Aubin:2020goo}. 
\MBTA{} is reporting results from an archival search for the
first time. Previously, in GWTC-2, only the \GSTLAL{} matched-filter pipeline included Virgo data; now all three pipelines
analyze the data from all three detectors. For inferring the source properties, we use
waveform models that include effects of spin-induced precession of the binary
orbit, contributions from both the dominant and sub-dominant spherical harmonic
modes, and tidal effects as appropriate~\cite{Pratten:2020ceb, Pratten:2020fqn, 
Garcia-Quiros:2020qpx, Garcia-Quiros:2020qlt, Ossokine:2020kjp, Babak:2016tgq, 
Pan:2013rra, Cotesta:2018fcv, Dietrich:2017aum, Dietrich:2018uni}. 

The paper is structured as follows: Sec.~\ref{sec:data} describes
the instruments and the data that are analyzed by the searches, including
methods on calibration, data quality, and glitch mitigation.
Sec.~\ref{sec:detection} describes the methods used by the search pipelines.
Sec.~\ref{sec:search_results} describes the events in GWTC-2.1, comparison
to GWTC-2, sensitivity of the search pipelines used, and inferred rates of
\acp{BNS} and \acp{BBH}.  Sec.~\ref{sec:parameter-estimation-method}
describes the methods used for estimating the source parameters of the \ac{GW}
candidates and results, and in Sec.~\ref{sec:astro-guess}, we discuss the astrophysically
interesting events and their implications. In
Sec.~\ref{sec:conclusion} we describe the data products being released
alongside this catalog and our conclusions. Finally, in Appendix~\ref{app:parameter-estimation-appendix}, we provide the source properties of events with \pastro{} greater than 0.5 that have previously been described in GWTC-1 and GWTC-2. 
Companion results from the \ac{O3b} are presented in the third Gravitational-Wave Transient Catalog, GWTC-3~\cite{LIGOScientific:2021djp}.

\section{Instruments and Data} \label{sec:data}

The Advanced LIGO~\cite{TheLIGOScientific:2014jea} and Advanced Virgo~\cite{TheVirgo:2014hva} 
instruments are kilometer-scale laser interferometers. The two \ac{LIGO} detectors are located in Hanford, Washington
and Livingston, Louisiana in the United States, and the Virgo detector near Pisa in Italy. The advanced generation 
of interferometers began operations in 2015, and observing periods have alternated with commissioning 
periods since then~\cite{Aasi:2013wya}.
In the time between O2 and the \ac{O3}, all three detectors underwent
significant upgrades that substantially increased their sensitivity~\cite{Buikema:2020dlj,Abbott:2020niy}. 

Major instrumentation upgrades on the LIGO detectors included: replacement of main lasers to increase beam stability,
replacement of test masses to lower scattering and absorption losses, installation of acoustic mode dampers to mitigate parametric instabilities~\cite{Biscans:2019akh}, installation of a squeezed vacuum source to reduce quantum noise~\cite{Tse:2019wcy}, addressing issues with scattered light~\cite{Soni_2021}, and implementation of improved feedback control systems for the instruments. Compared to the O2 run, the Hanford \ac{BNS} range \cite{Allen:2005fk,Chen:2017wpg} increased by $64\%$ (from $66$~Mpc to $108$~Mpc), and for Livingston by $53\%$ (from $88$~Mpc to $135$~Mpc).

For Virgo, major upgrades included: replacement of the steel wire suspensions of the four test masses with fused-silica fibers~\cite{Aisa:2016rsg}, modification of the vacuum system to avoid dust contamination of the lowest suspension stage, replacement of the main laser to increase power, installation of a squeezed vacuum source to reduce quantum noise~\cite{Acernese:2019sbr}, improvements in beam stability~\cite{Blom:2015fna}, and addressing issues with scattered light. Compared to the O2 run, the Virgo \ac{BNS} range increased by $73\%$ (from $26$~Mpc to $45$~Mpc). 

The processing of the data recorded by the \ac{LIGO} and Virgo
detectors includes several steps that occur both in near-real time
to allow for the broadcasting of public alerts, and in higher latency
to shape the final data set and update the catalogs of \ac{GW} events.
Raw data calibration and the subtraction of noise from known
instrumental sources, documented in Sec.~\ref{sec:calibration},
occur first and the \ac{GW} strain data, reconstructed independently
in each detector, are then jointly processed. 
Significant \ac{GW} candidates are vetted with several data quality tests as a part of the
standard analysis procedure. This procedure is described in Sec.~\ref{sec:DQ}.

\subsection{Calibration and noise subtraction}\label{sec:calibration}

The strain data used for astrophysical analyses is derived
from the optical power variations at the output ports of the interferometers.
Calibration of the raw photodetector signal to \ac{GW} strain
requires a detailed understanding and modeling of the control system and
opto-mechanical response of the interferometers throughout an observing run.
This allows for accurate and reliable calibration of the strain and also for
quantifying its systematic and statistical uncertainty. The detailed
procedure for the calibration and the determination of the systematic and
statistical uncertainty of the \ac{LIGO} and Virgo detectors for O3 can be found
in~\cite{Sun:2020wke, virgocollaboration2021calibration, Estevez_2021}. 

There are usually two calibrations applied to the data; 
a low-latency calibration and, if needed, an offline calibration. 
The low-latency (online) estimate of the strain uses 
the best models of the detector at the time of recording. However, over the course of any observing run, 
data drop-outs due to computer failures, incomplete modeling of the detector, and unknown 
residual systematic errors are often identified. The offline calibration incorporates the necessary corrections 
and improvements, producing a better calibrated strain with better known systematic uncertainty.

\begin{figure}[t] 
\centering
\includegraphics[width=0.50\textwidth]{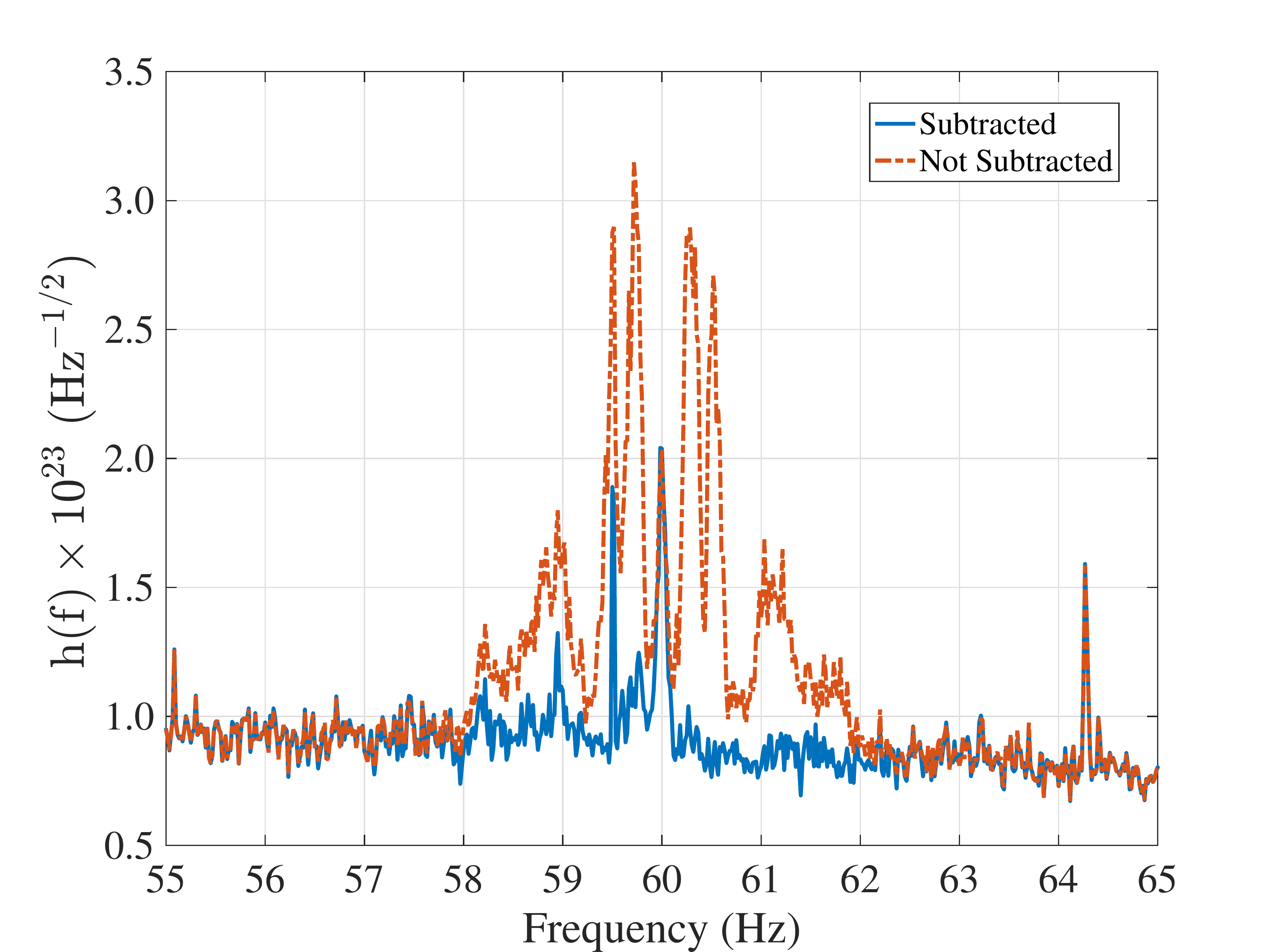} 
\caption{Comparison of the amplitude spectral density at Hanford around the $60$~Hz mains line, 
between data with subtracted non-stationary noise and data with no subtraction. 
The data correspond to a typical one-hour observation-ready data stretch during O3a.}
\label{fig:subtracted} 
\end{figure}

In addition, numerous noise sources and calibration lines that limit detectors’
sensitivity are measured and linearly subtracted from the data~\cite{Davis:2018yrz,LineSubtractionTechNote,Vajente:2019ycy,VIR-0652B-19}. 
This subtraction is performed online to generate the \ac{LIGO} and Virgo low-latency strain data, and it is also performed
when regenerating the \ac{LIGO} offline strain data. 
Additionally, noise due to non-stationary coupling of the power mains with the 
\ac{LIGO} detectors was subtracted from the offline data~\cite{Vajente:2019ycy}.
As an example of noise subtraction, Fig.~\ref{fig:subtracted} shows the improvement
in the noise levels around the $60$~Hz mains line in the Hanford detector, after non-linear noise subtraction was
applied to the strain time series. Taking as a figure of merit the \ac{BNS} range of the detectors~\cite{Allen:2005fk,Chen:2017wpg},
the subtraction results in a median range increase of $0.9$~Mpc for Hanford and $0.2$~Mpc for Livingston. 

In GWTC-2,  search pipelines and parameter estimation analyses used a mix of 
low-latency and offline calibrated frames. 
In contrast to this, all searches and analyses presented in this paper use strain 
data with the best available calibration and noise subtraction for each detector. 
For \ac{LIGO}, this corresponds to the offline recalibrated data with $60$~Hz non-linear subtraction. 
For Virgo, the online strain data stream was good enough to be used offline, 
except for the last two weeks of O3a which were reprocessed to improve subtraction of control and laser frequency noise~\cite{VIR-1201A-19}.
The strain data used in this work are publicly accessible through the Gravitational Wave Open Science Center (GWOSC)~\cite{o3a_public_data}.

In addition, the \ac{LIGO} offline data are accompanied with a much improved systematic and statistical error estimate
compared to the online data. 
 The probability distribution of the calibration uncertainty estimate
for \ac{LIGO} in O3a is characterized in~\cite{Sun:2020wke}, with the systematic error over the
detectors' bandwidth being under $3\%$ in magnitude and under $2^\circ$ in phase. 
The uncertainty in the Virgo strain data in O3a had
a maximum systematic error over the detector's bandwidth under $5\%$ in magnitude and under 2$^\circ$ in phase~\cite{virgocollaboration2021calibration}. 
Parameter estimation takes into account calibration uncertainties, as described in Sec.~\ref{sec:parameter-estimation-method}.
Given the size of calibration uncertainties in O3, there is no evidence that they have a significant impact on the 
inference of source parameters~\cite{PhysRevD.102.122004,vitale2020physical}.

\subsection{Data quality, event validation \& glitch mitigation}\label{sec:DQ}

LIGO and Virgo data quality is continuously monitored during an observing run both on site and remotely, as reported in~\cite{Davis:2021ecd,Acernese:2022jes}. 
This can include, for example, internal detector summary pages which detail the status of the detectors and interferometer subsystems~\cite{gwpy-software,VIR-0546A-16}. 
Feedback from \ac{GW} searches also gives an indication of the impact of data quality on the sensitivity of a search. 
To exclude identified instances of poor data quality from the searches and produce the results in Sec.~\ref{sec:detection}, we used the same methods and data products as reported for GWTC-2~\cite{Abbott:2020niy}. 
The data quality products used in this work are publicly available~\cite{o3a_public_data,gwtc2p1_detchar_data}.

Once a \ac{GW} event has been identified by the search pipelines, we check the quality of data around the time of the event. We followed the same procedures outlined in~\cite{Abbott:2020niy} to validate
the data quality around each new \ac{GW} candidate reported in this paper.  The
aim of these validation procedures is to identify any instrumental or
environmental noise that may impact the estimation of \ac{GW} signal
parameters.  As summarized for GWTC-2~\cite{Abbott:2020niy}, in some cases short-duration noise transients, or \textit{glitches}~\cite{TheLIGOScientific:2016zmo,Nuttall:2018xhi,Davis:2021ecd,Soni:2021cjy}, can be subtracted from the
data~\cite{Cornish:2014kda,Becsy:2016ofp,PhysRevD.98.084016,PhysRevD.103.044006}.
When this is not possible, analyses use tailored configurations, for example, a
modified low-frequency cutoff, to exclude data that could be corrupted by the
presence of a nearby glitch. 
The full list of candidate events using
candidate-specific glitch mitigation, along with the mitigation configuration,
is found in Table~\ref{tab:detchar_events}. 
These data, for the events where the glitch-mitigated data was used for the parameter estimation analysis in Sec.~\ref{sec:parameter-estimation-method}, are publicly accessible~\cite{gwtc2p1_deglitched_data}.
No candidates in this catalog have clear evidence of instrumental origin
identified through data quality validation studies. 

\begin{PE_table}
\begin{table}
{\rowcolors{1}{}{lightgray}
\begin{tabularx}{\columnwidth}{@{\extracolsep{\fill}}p{3.5cm} Z}
\textbf{Name} & \textbf{Mitigation}\\
\hline
{\EVENTNAMEBOLD{GW190413E_o3afin} \NAME{GW190413E_o3afin} } & \MITIGATIONMETHOD{GW190413E_o3afin}\\
{\EVENTNAMEBOLD{GW190425B_o3afin} \NAME{GW190425B_o3afin} } & \MITIGATIONMETHOD{GW190425B_o3afin}\\
{\EVENTNAMEBOLD{GW190503E_o3afin} \NAME{GW190503E_o3afin} } & \MITIGATIONMETHOD{GW190503E_o3afin}\\
{\EVENTNAMEBOLD{GW190513E_o3afin} \NAME{GW190513E_o3afin} } & \MITIGATIONMETHOD{GW190513E_o3afin}\\
{\EVENTNAMEBOLD{GW190514E_o3afin} \NAME{GW190514E_o3afin} } & \MITIGATIONMETHOD{GW190514E_o3afin}\\
{\EVENTNAMEBOLD{GW190701E_o3afin} \NAME{GW190701E_o3afin} } & \MITIGATIONMETHOD{GW190701E_o3afin}\\
{\EVENTNAMEBOLD{GW190727B_o3afin} \NAME{GW190727B_o3afin} } & \MITIGATIONMETHOD{GW190727B_o3afin}\\
{\EVENTNAMEBOLD{GW190814H_o3afin} \NAME{GW190814H_o3afin} } & \MITIGATIONMETHOD{GW190814H_o3afin}\\
{\EVENTNAMEBOLD{GW190924A_o3afin} \NAME{GW190924A_o3afin} } & \MITIGATIONMETHOD{GW190924A_o3afin}\\
\hline
\end{tabularx}

}
\caption{
\label{tab:detchar_events}
List of candidate-specific data usage and mitigation methods
for parameter estimates.
Only candidate events for which mitigation of instrumental artifacts was performed are listed.
The glitch subtraction methods
used for these candidate events are detailed in Sec.~\ref{sec:DQ}.
The minimum frequency is the lower limit of data
used in analyses of \ac{GW} source
properties for the listed interferometer.
}
\end{table}
\end{PE_table}

\section{Candidate identification}\label{sec:detection}

\newcommand{\PYCBCHYPERBANK}{\PYCBC{}}
\newcommand{\PYCBCBBH}{\PYCBC{}-BBH}
\newcommand{\PYCBCHYPERBANKQMIN}{1}
\newcommand{\PYCBCHYPERBANKQMAX}{1 / 100}
\newcommand{\PYCBCHYPERBANKSPINMIN}{-1}
\newcommand{\PYCBCHYPERBANKSPINMAX}{1}


\newcommand{\INJBBHMASSMIN}{\ensuremath{2}}
\newcommand{\INJBBHMASSMAX}{\ensuremath{100}}
\newcommand{\INJBBHMASSONEPOWER}{\ensuremath{-2.35}}
\newcommand{\INJBBHSPINMAX}{\ensuremath{0.998}}
\newcommand{\INJBBHMAXREDSHIFT}{\ensuremath{1.9}}

\newcommand{\INJBNSMASSMIN}{\ensuremath{1}}
\newcommand{\INJBNSMASSMAX}{\ensuremath{2.5}}
\newcommand{\INJBNSSPINMAX}{\ensuremath{0.4}}
\newcommand{\INJBNSMAXREDSHIFT}{\ensuremath{0.15}}

\newcommand{\INJNSBHMINBBHMASS}{\ensuremath{2.5}}
\newcommand{\INJNSBHMAXBBHMASS}{\ensuremath{60}}
\newcommand{\INJNSBHMINBNSMASS}{\ensuremath{1}}
\newcommand{\INJNSBHMAXBNSMASS}{\ensuremath{2.5}}
\newcommand{\INJNSBHPOWERBBHMASS}{\ensuremath{-2.35}}
\newcommand{\INJNSBHMAXBBHSPIN}{\ensuremath{0.998}}
\newcommand{\INJNSBHMAXBNSSPIN}{\ensuremath{0.4}}
\newcommand{\INJNSBHMAXREDSHIFT}{\ensuremath{0.25}}

\newcommand{\INJBBHWAVEFORM}{SEOBNRv4PHM}
\newcommand{\INJBNSWAVEFORM}{SpinTaylorT4}
\newcommand{\INJNSBHWAVEFORM}{SEOBNRv4PHM}


\newcommand{\GSTLALVTBNS}{$0.00594$}
\newcommand{\MBTAVTBNS}{$0.00631$}
\newcommand{\PYCBCVTBNS}{$0.00657$}
\newcommand{\PYCBCBBHVTBNS}{$0.00$}
\newcommand{\ANYCBCVTBNS}{$0.00781$}

\newcommand{\GSTLALVTNSBH}{$0.0174$}
\newcommand{\MBTAVTNSBH}{$0.0165$}
\newcommand{\PYCBCVTNSBH}{$0.0181$}
\newcommand{\PYCBCBBHVTNSBH}{$0.00$}
\newcommand{\ANYCBCVTNSBH}{$0.0221$}

\newcommand{\GSTLALVTBBHINJ}{$0.258$}
\newcommand{\MBTAVTBBHINJ}{$0.196$}
\newcommand{\PYCBCVTBBHINJ}{$0.194$}
\newcommand{\PYCBCBBHVTBBHINJ}{$0.234$}
\newcommand{\ANYCBCVTBBHINJ}{$0.308$}

\newcommand{\GSTLALVTBBHPOP}{$1.22$}
\newcommand{\MBTAVTBBHPOP}{$0.885$}
\newcommand{\PYCBCVTBBHPOP}{$0.914$}
\newcommand{\PYCBCBBHVTBBHPOP}{$1.20$}
\newcommand{\ANYCBCVTBBHPOP}{$1.44$}

\newcommand{\RGSTLALBBH}{\fixme{$25.0_{-6.1}^{+7.2}\,\perGpcyr$}}
\newcommand{\RPYCBCBBH}{\fixme{$26.0_{-6.8}^{+8.2}\,\perGpcyr$}}
\newcommand{\RMBTABBH}{\fixme{$25.6_{-7.8}^{+9.6}\,\perGpcyr$}}
\newcommand{\RGSTLALBNS}{\fixme{$286_{-237}^{+510}\,\perGpcyr$}}

\newcommand{\peakNoAugNoEvolutionrate}{\ensuremath{23.9^{+14.3}_{-8.6}}}
\newcommand{\BNSrate}{\ensuremath{320^{+490}_{-240}}}

\ac{GW} data is analyzed to search for candidates in two stages: first in
low-latency in order to generate public alerts that subsequently trigger
follow-up astronomical observations, and then in higher latency in the form of an
offline analysis of the archival strain data, which is used to create \ac{GW}
catalogs. Five pipelines were used in real time to analyze \ac{O3} data: a
minimally modeled generic transient search
(coherent WaveBurst~\cite{Klimenko:2004qh,Klimenko:2005xv,Klimenko:2006rh,
Klimenko:2011hz,Klimenko:2015ypf}), and four matched-filter~\cite{Allen:2005fk,
Allen:2004gu} pipelines (\GSTLAL{}~\cite{Sachdev:2019vvd, Hanna:2019ezx,
Messick:2016aqy}, \MBTA{}~\cite{Aubin:2020goo}, \PYCBC{}~\cite{ Canton:2014ena,
Usman:2015kfa, Nitz:2017svb, Davies:2020tsx}, and \SPIIR{}~\cite{chu2017low}).  Collectively,
they identified \fixme{56} unretracted candidates during \ac{O3}, \fixme{33} of
which were found in O3a.
GWTC-2~\cite{Abbott:2020niy} presented \fixme{39} events identified by
coherent WaveBurst, \GSTLAL{}, and \PYCBC{} in the first offline search over O3a.

We present here results from a refined offline search of O3a.
The search employs three matched-filter pipelines: \GSTLAL{}, \PYCBC{}, and
\MBTA{}~\cite{Aubin:2020goo}, marking the first time that \MBTA{} results from
archival data are presented and included in a \ac{GW} catalog. All three
pipelines analyze the data from all three detectors.  
While GWTC-2 imposed a \ac{FAR} ceiling of 2 per year on candidates, here we
release a deep list of \ac{GW} candidates with a \ac{FAR} smaller than \fixme{2
per day}~\cite{gwtc2p1_search_data}.  In addition, we identify the
\fixme{\NUMEVENTS} \ac{CBC} candidates with an estimated \pastro{} greater than $0.5$
(Table~\ref{tab:events}). There are also \NUMEVENTSMAR{} candidates with \pastro{} below $0.5$ that do meet the \ac{FAR} criterion used in 
GWTC-2; these are presented as marginal candidates. This \ac{GW} catalog contains the largest
number of candidates with \pastro{} greater than $0.5$ to date.

In Sec.~\ref{subsec:searches}, we first lay out a general description of
matched filter searches and in Sec.~\ref{pipelines}, we describe the
methods employed by the three \ac{CBC} searches used in this work. We describe the search results in the following Sec.~\ref{sec:search_results}.

\subsection{Matched-filter searches}
\label{subsec:searches}

The matched-filter method relies on having a model of the signal, as a function
of the physical parameters. The parameters include those that are intrinsic to
the source: two individual component masses $m_1,\,m_2$ and two dimensionless
spin vectors $\vec{\chi}_1,\,\vec{\chi}_2$ (related
to each component's spin angular momentum $\vec{S}_i$ by $\vec{\chi}_i =
c\vec{S}_i/(Gm_i^2)$), and seven extrinsic parameters that provide the
orientation and position of the source in relation to the Earth: the luminosity
distance $\DL$, two-dimensional sky position (right ascension $\alpha$ and
declination $\delta$), inclination between total angular momentum and
line-of-sight $\theta_{JN}$, time of merger $t_\mathrm{c}$, a reference phase $\phi$,
and polarization angle $\psi$. The search pipelines create a template
bank~\citep{Owen:1998dk, Harry:2009ea, Privitera:2013xza} of \ac{GW} waveforms
covering the desired intrinsic parameter space, 
and use these to filter against the data and produce \ac{SNR} time series.  
The component masses describing 
template waveforms are affected by source redshift $z$ as $m_i^{\text{det}}=(1+z)m_i$.

For each set of intrinsic parameters, extrinsic parameters affecting the signal's
amplitude and phase may be maximized over analytically~\cite{Allen:2005fk}, 
if the signal can be approximated as a pure quadrupole mode, i.e.\
$(\ell,\,|m|) = (2,\,2)$.  In particular, for this search, the templates use only the
dominant quadrupole mode and assume quasi-circular orbits with component spins
aligned with the total orbital angular momentum.  Peaks in the resulting \ac{SNR}
time series are stored as triggers. \ac{GW} candidates are formed by imposing consistency 
in time and in template intrinsic parameters between triggers in different detectors; 
in addition, \GSTLAL{} also considers non-coincident triggers as candidates~\cite{Sachdev:2019vvd}. 

When considering a single template in a single detector with stationary, Gaussian 
noise, the matched filter \ac{SNR} is an optimal statistic for ranking
candidates.  However, additional terms are needed to optimize sensitivity in  
searches of real data covering a wide signal parameter space.  To account for the 
multi-detector network, the distribution of signals over relative times, phases
and amplitudes between detectors is considered~\cite{Nitz:2017svb,Hanna:2019ezx}. 
Since detector noise is not stationary or Gaussian, signal-consistency tests
such as chi-squared~\cite{Allen:2004gu} are calculated and used to rank 
candidates. 

The distribution of noise triggers may vary strongly over the template 
masses and spins; we then model its variation empirically, as a function of
combinations of parameters that are typically well-constrained 
by \ac{GW} measurements.  The binary's chirp mass~\cite{Blanchet:1995ez},
\begin{equation}
    \Mc = \frac{(m_1 m_2)^{3/5}}{(m_1 + m_2)^{1/5}},
\end{equation}
determines to lowest order the phase evolution during the inspiral, and is
typically better constrained than the component masses.  At higher
orders, the binary phase evolution is affected by the mass ratio $q = {m_2}/{m_1}$ 
(where $m_2 \leq m_1$) and by the effective inspiral spin $\chieff$, defined
as~\cite{Ajith:2009bn}
\begin{equation}
    \chieff = \frac{(m_1 \vec{\chi}_1 + m_2 \vec{\chi}_2)\cdot\hat{L}_\mathrm{N}}{M},
\label{eq:chiEff}
\end{equation}
where $M = m_1+m_2$ is the total mass and $\hat{L}_\mathrm{N}$ is the unit vector
along the Newtonian orbital angular momentum. 
Finally, the ranking of events by the search pipelines may account for an 
assumed prior distribution of signals over masses and spins~\cite{Dent:2013cva,Fong:2018elx}. 

The significance of each candidate event is quantified by its \ac{FAR}, the 
estimated rate of events due to noise with equal or higher ranking statistic value. 
The \ac{FAR} is calculated by each search pipeline by constructing a set of
background samples designed to have the same distribution over ranking statistic as
search events in the absence of binary merger \ac{GW} signals. 

By considering also the expected distribution of \ac{GW} signal events recovered
by a given search, we may derive an estimate of the relative probabilities of noise 
(terrestrial) origin \pterr{}, and signal (astrophysical) origin 
\pastro{}~\cite{Farr:2013yna,Abbott:2016nhf,Abbott:2016drs}. 
For the bulk of released events, detailed
estimates of source 
parameters are not calculated.  Therefore, based only on the matched-filter search
results we also estimate the probability for each event to belong to three possible
astrophysical binary source classes, labeled BNS, NSBH and BBH.  
The classes are defined by binary component masses: BNS corresponds to $\{\massone,
\masstwo\} <3\,\Msun$, NSBH to $\massone >3\,\Msun$, $\masstwo <3\,\Msun$, 
and BBH to $\{\massone, \masstwo\} >3\,\Msun$. 
For MBTA, a $2.5\,\Msun$ cut is used instead of $3\,\Msun$, with a gap to $5\,\Msun$ for BBH.
These definitions are 
chosen for simplicity:
they \emph{do not}  
imply that every binary component within a given mass range is necessarily a \ac{NS} 
or a \ac{BH}. 
Such inference would ultimately require measurement of the effects of \ac{NS} matter on
observed signals, which is beyond the capabilities of the search pipelines. 
The probabilities for an event to belong to each class (\pbns{}, \pnsbh{}, \pbbh, and 
\pterr{}) are calculated from the template masses and spins recovered by the
searches, under the assumption that events from each class occur as independent 
Poisson processes. 
The calculation also requires the choice of a prior on the event counts in each category~\cite{Abbott:2016drs}. 
\GSTLAL{} used a uniform prior for the \ac{BNS} and \ac{NSBH} categories, and a 
Poisson--Jeffreys prior for the \ac{BBH} category; \MBTA used a uniform prior for 
the \ac{BNS} category, and a Poisson--Jeffreys prior for the \ac{NSBH} and \ac{BBH} 
categories; and \PYCBC{} used a Poisson--Jeffreys prior for all three categories. 
Given the number of candidates, the prior choice does not significantly impact the 
\ac{BBH} results.
Implementation details differ between pipelines, as 
summarized below; the resulting probability estimates 
are listed in Tables~\ref{tab:events} and~\ref{tab:p-astro}. 

While the \pastro{} values given here represent our best estimates of the origin  
of candidates using the information available from search pipelines, they are 
subject to statistical (random) and systematic errors, as well as in some cases 
clearly differing for a given candidate between different pipelines.  One such 
uncertainty arises from methods used to rank events between pipelines, including 
tests for noise artifacts: such tests, such as chi-squared statistics, will in 
general add (different) random variations to the ranking of a given event, 
in addition to their differing power in distinguishing signals from artifacts.  
For single-detector candidates,
there is an additional inherent uncertainty in estimating the rate of comparable 
noise events, which may only be bounded to (less than) 1 per observing time. 
An inherent source of potential systematic error also lies in the search ranking
statistic used 
in the calculation of \pastro{}: such statistics are optimized to detect a specific 
(usually broad) distribution of signals over binary intrinsic parameters.  The
resulting \pastro{} estimates may be biased if this distribution deviates significantly from
the (unknown) true signal distribution.  The risk of such bias is largest for 
regions of parameter space containing few, or zero, confirmed detections.  
For all these reasons, our current \pastro{} values may be revised in the future, 
particularly as and when current uncertainties in the true signal rate and
distributions are eventually reduced.

We next review specific methods used by individual matched-filter pipelines.
\subsection{Search pipelines}\label{pipelines}
In this section we describe the pipelines that were used to identify the candidates presented in GWTC-2.1.
\subsubsection{\GSTLAL{}}\label{sec:gstlal}

The \GSTLAL{} analysis used in this search is largely similar to the one used
in the previous analysis~\cite{Abbott:2020niy} and uses the same log-likelihood
ratio $\mathcal{L}$ as the ranking statistic. Improvements have been made to
the input data products generated by iDQ, the statistical inference framework
to autonomously detect non-Gaussian noise artifacts in strain data based on
auxiliary witness sensors~\cite{Essick:2020qpo,idq-in-gstlal}. This iDQ
timeseries is used to compute one of the terms in the log-likelihood ratio
within the \GSTLAL{} analysis, that informs the search of the presence of
non-Gaussian noise in close proximity to a GW candidate. Compared to GWTC-2,
the timeseries generated by iDQ was reprocessed offline, having
access to an expanded set of auxiliary witness sensors and trained with an
acausal binning scheme~\cite{Essick:2020qpo}.  As a result, the generated iDQ
timeseries performs better in identifying noise artifacts in strain data. In addition, for GWTC-2 the iDQ term was only used when ranking single-detector triggers, whereas now it is used for both coincident and single-detector triggers. Because of changes
in the iDQ term, the empirically determined penalty for single-detector candidates had to be retuned compared to
GWTC-2, and was increased to a penalty of $\Delta \mathcal{L} = -12$ from
$\Delta \mathcal{L} = -10$. 
The single-detector
event penalty is determined by
comparing the recovery of simulated signals in single detector versus
combinations of detectors and the sensitive volume--time for each
configuration. 

For the \GSTLAL{} analysis, \pterr{} and \pastro{} shown in
Tables~\ref{tab:events} and~\ref{tab:p-astro} are estimated following the
multicomponent population analysis \cite{Kapadia:2019uut, Farr:2013yna}. The
response of each \GSTLAL{} template to each astrophysical source class,
computed semi-analytically \cite{Fong:2018elx}, is used in estimating these
probabilities. The volume--time sensitivity of the pipeline used in this
calculation is estimated based on simulated sources injected into the pipeline
and is rescaled to the astrophysical distribution \cite{Tiwari:2018bgh}. The
volume--time ratios are used to combine triggers from various observation runs
and perform a multicomponent analysis yielding \pastro{} and merger
rates~\cite{Farr:2013yna,Kapadia:2019uut} inferred from \ac{O1} to O3a.
The astrophysical distribution assumed in this analysis uses a log-uniform
distribution for the source component masses, the component spins
aligned with the orbital angular momentum, and a uniform distribution for the
component spin magnitudes. The
\ac{BH} masses in \acp{BBH} and
\acp{NSBH} are distributed between $3\,\Msun$ and $300\,\Msun$ with aligned
component spins distributed in the range $[-0.99,0.99]$.  The \ac{NS} masses in
\acp{NSBH} and \acp{BNS} are distributed between $1\,\Msun$ and $3\,\Msun$.  In
\acp{NSBH}, the \ac{NS} spins are assumed to be aligned and distributed in the
range $[-0.4,0.4]$, whereas, in \acp{BNS} the \acp{NS} are assumed to have 
small spins in the range $[-0.05,0.05]$. These choices match previous analyses~\cite{Abbott:2020niy}.

\subsubsection{\MBTA{}}\label{sec:mbta}
The Multi-Band Template Analysis (\MBTA{}) pipeline~\cite{Aubin:2020goo} is based on matched filtering, relying on coincidences between triggers
observed in different detectors.  The version used for the offline search is
close to the online version which contributed to the LVC public
alerts~\cite{LIGOPublicGCNs}.  The archival-search version benefits from
offline-specific improvements, with a background estimate made over a longer
duration, and with a reranking of the candidates using 
information collected not just before but also after the candidate.

The parameter space covered by this analysis ranges from $1\,\Msun$ to $195\,\Msun$
for the primary (more massive) component, with total masses up to
$200\,\Msun$; or from $1\,\Msun$ to $100\,\Msun$ for the primary when the mass
of the secondary is between $1\,\Msun$ and $2\,\Msun$. Component spins are
aligned with the total angular momentum and are limited to 0.05 for objects
below 2 \Msun, and going up to 0.997 for objects above $2\,\Msun$.  The waveform
used for the search is SpinTaylorT4~\cite{Buonanno:2002fy, PhysRevD.74.029902, Buonanno:2009zt}
if both binary masses are lighter than $2\,\Msun$, and
SEOBNRv4~\cite{Bohe:2016gbl} if the mass of one of the components is above $2\,\Msun$.
The total number of templates in the bank used is 727,992.  The SNR threshold
for recording triggers in each detector is 4.5, or 4.8 if one of the
components is above $2\,\Msun$.

The \ac{FAR} is calculated for each coincident event by forming random coincidences
among single detector background triggers.
This computation is performed independently for 
three large regions of the parameter space bounded by a $2\,\Msun$ limit for the mass of each component.
These three regions are allowed to contribute equally to the background, 
while within each of them we sum the background contributions from all the templates.

The \pbns{}, \pnsbh, \pbbh, and derived \pastro{} quantities are computed as
the fraction of recovered simulated events, representative of an astrophysical population, 
to this foreground plus background estimate provided by the pipeline~\cite{MBTA_p_astro}.
The parameterizations of the populations are described in Sec.~\ref{sec:vt},
with the \textsc{Power Law + Peak} model used for BBH~\cite{o3apop}.
The rate of each type of source is adjusted
using a multicomponent population analysis~\cite{Farr:2013yna}.
To follow the population and background evolution across the parameter space, 165 subregions are used.
This finer resolution has the benefit of revealing events in population-rich areas,  
even if the overall background rate for their ranking statistic value is larger than few per year, 
as in the case of the high mass BBH event $\mathrm{GW190916\_200658}$
presented in Table~\ref{tab:events}.

\subsubsection{\PYCBC{}}\label{sec:pycbc}

In previous LVC searches~\cite{TheLIGOScientific:2016pea, Abbott:2016ezn, LIGOScientific:2018mvr, Abbott:2020niy}, 
the offline \PYCBC{}~\cite{Usman:2015kfa, pycbc-software} pipeline has analyzed data only from the two LIGO detectors. 
In this analysis, \PYCBC{} was extended to search data from the three-detector LIGO--Virgo network,
along with updates to the event ranking statistic~\cite{Davies:2020tsx} and the \pastro{} calculation
and a new method to estimate source class probability~\cite{DalCanton:2020vpm}. 

The \PYCBC{} search uses the same template bank as in GWTC-2~\cite{Abbott:2020niy},
constructed using a hybrid geometric-random algorithm outlined in \cite{Roy:2017qgg, Roy:2017oul}. 
Peaks in SNR time series exceeding a threshold of 4 constitute single-detector triggers.
Two-detector coincident events are formed from triggers with the same component masses and spins
with a physically allowed time difference between detectors, allowing for timing errors. 
Three-detector triple coincidences require triggers in all pairs of detectors to pass this
consistency test.

The detection statistic is given by the logarithm of the ratio of estimated signal event rate
density to noise event rate density.  We model the noise distribution in each detector as a
decreasing exponential of the matched-filter \ac{SNR}, reweighted based on a chi-squared
signal--glitch discriminator~\cite{Allen:2004gu, Nitz:2017lco}, with parameters that depend on the template
intrinsic
parameters.  
The signal distribution includes terms accounting for dependence on relative times of
arrival, phases and amplitudes between detectors,  
as well as relative sensitivities of the participating detectors~\cite{Nitz:2017svb}.
We estimate the \ac{FAR} separately for each combination of detectors via time-shifted 
analyses~\cite{Usman:2015kfa,Capano:2017fia}. The significance for each candidate event 
is then found through addition of the \acp{FAR} at the candidate's ranking statistic 
value over all active detector combinations~\cite{Davies:2020tsx}.

In addition to the generic \PYCBC{} search, which covers the full parameter space~\cite{Abbott:2020niy} including a 
range of possible signal types, we also conduct a focused \PYCBC{} \ac{BBH} 
search~\cite{Abbott:2020niy, Nitz:2019hdf}, 
capable of uncovering fainter \ac{BBH} mergers by imposing a prior form for the signal distribution
over the template bank~\cite{Dent:2013cva}. 
This search is targeted at systems with mass ratios from $1$ to $1/3$, primary component masses 
from $5\,\Msun$ to $350\,\Msun$, and aligned, equal component spins from $\chi=-0.998$ to $0.998$.

The inference of \pastro{} and \pterr{} for each candidate event employs a Poisson mixture 
model of signal and noise events~\cite{Farr:2013yna, Abbott:2016nhf, Abbott:2016drs}.  
Here, the distribution of signal events is estimated via a set of
simulated signals analyzed by the pipeline, and the 
rate and distribution of noise events are estimated from time-shifted 
analyses~\cite{Usman:2015kfa}.  In GWTC-2 the calculation was only performed on potential \ac{BBH} 
events with template chirp mass above $4.35\,\Msun$ (which corresponds to equal $5\,\Msun$ 
component masses).  Here, we include potential \ac{BNS} and \ac{NSBH} events by performing 
independent calculations over ranges of template chirp mass below $2.18\,\Msun$ (corresponding to
equal $2.5\,\Msun$ components), and between $2.18\,\Msun$--$4.35\,\Msun$, respectively.  
Although the implied
signal distribution over template chirp mass does not correspond to any specific astrophysical model,
it is adequate for assignment of \pastro{}
given the current knowledge of \ac{BNS} and \ac{NSBH} merger populations.  
Systematic biases in \pastro{} calculation may
arise if the (unknown) true mass distribution is different from that assumed. 
The calculation is also extended relative to 
previous analyses to account for 
different possible coincident combinations of detectors~\cite{PycbcPastro}. 
The results given here are obtained from events occurring during O3a only, except for 
the \ac{BNS} region where prior information of 1 highly significant detection was applied
to represent GW170817~\cite{TheLIGOScientific:2017qsa}. 

The estimation method for binary source class probabilities \cite{DalCanton:2020vpm} 
uses the binary chirp mass as input, and assumes a uniform density of candidate signals over
the plane of component masses $\{m_1, m_2\}$.  Here we take the classes to be defined by 
boundaries between different types of binary component at $3\,\Msun$.  
To estimate source chirp mass,
we correct the search template masses for cosmological redshift, using an estimate of the
luminosity distance
derived from the search \acp{SNR} and the corresponding templates' sensitivity. 
We then derive the relative probabilities of each source class and enforce that 
the sum of astrophysical source probabilities is
equal to \pastro{}. 

\section{Search results}
\label{sec:search_results}
\begin{figure*}
\centering
\subfloat[\GSTLAL{}]{
  \includegraphics[width=0.95\columnwidth]{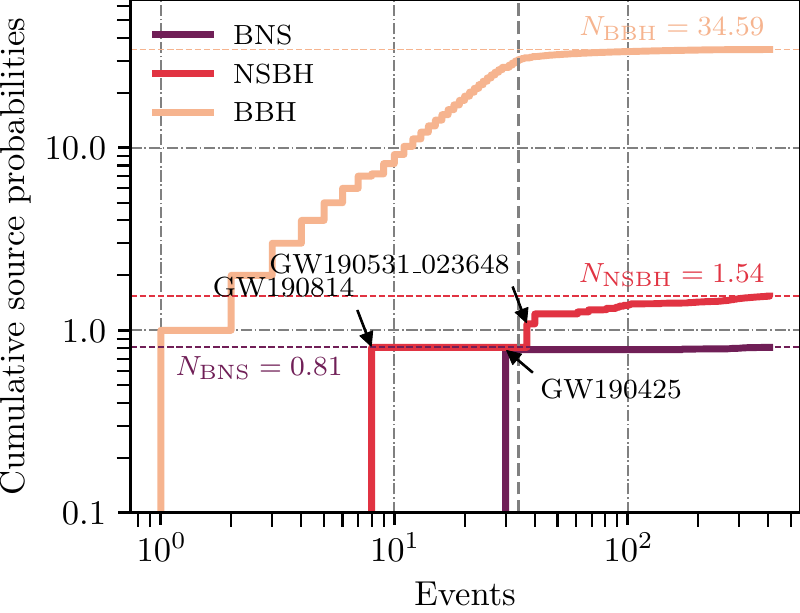}
  \label{plot:gstlal}
}\qquad
\subfloat[\MBTA{}]{
  \includegraphics[width=0.95\columnwidth]{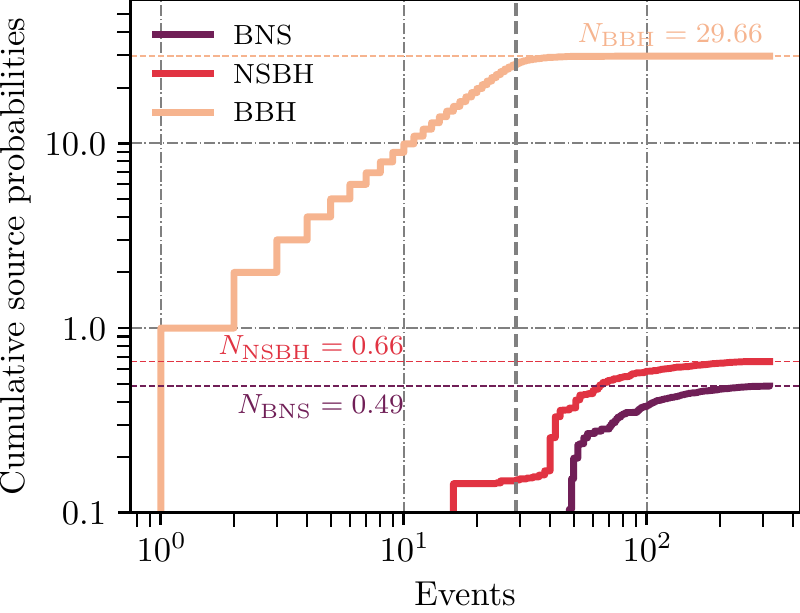}
  \label{plot:mbta}
}
\\
\subfloat[\PYCBC{}]{
  \includegraphics[width=0.95\columnwidth]{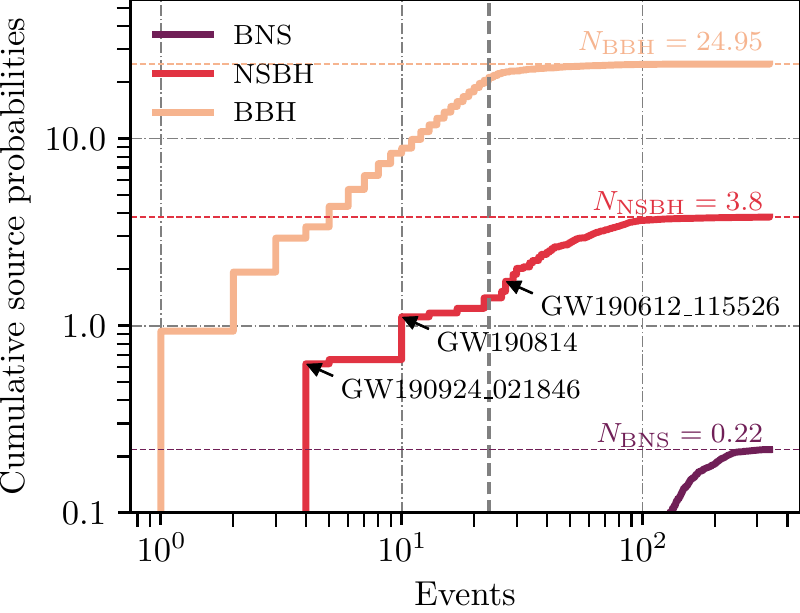}
  \label{plot:pycbc}
}\qquad
\subfloat[\PYCBCBBH{}]{
  \includegraphics[width=0.95\columnwidth]{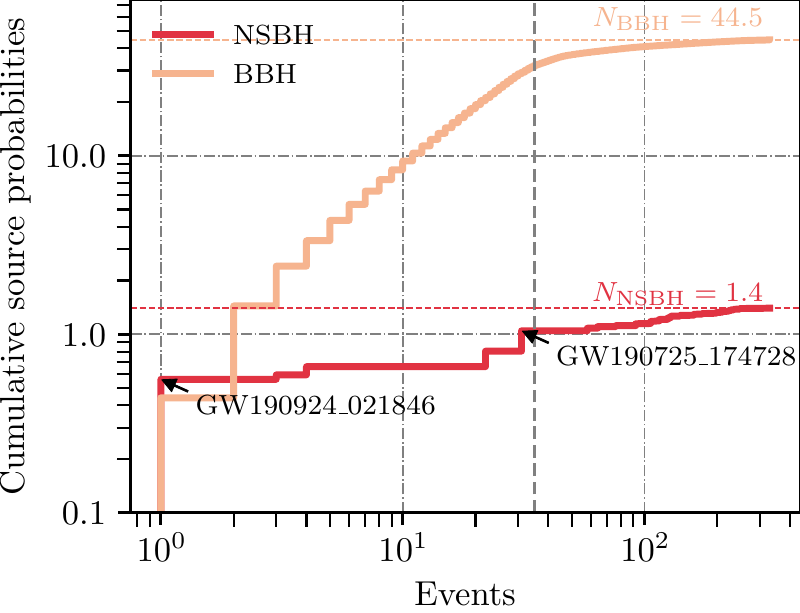}
  \label{plot:pycbc-hm}
}
\caption{\label{fig:subthreshold-pastro}Cumulative sum of \pbns{}, \pnsbh{},
\pbbh{} as a function of the candidates that pass a \ac{FAR} threshold of 2 per
day. The events are shown in decreasing order of \pastro{}. The sum of the
source probabilities shown here represents the estimated Poisson rate of sources
in each source class per \ac{O3a} experiment by the different search pipelines.  An
estimate of the rate of sources in the subthreshold candidate list is obtained
by the contribution to the sum from events with \pastro{} less than 0.5. This
estimate yields between \NUMSUBBBHMBTA{}--\NUMSUBBBHPYCBCBBH{} signals in the \ac{BBH} class,
\NUMSUBNSBHPYCBCBBH{}--\NUMSUBNSBHPYCBC{} signals in the \ac{NSBH} class, and
\NUMSUBBNSGSTLAL{}--\NUMSUBBNSMBTA{} signals in the \ac{BNS} class in the subthreshold candidates
in \ac{O3a}. The dashed vertical grey line shows where this threshold is for
each pipeline.  Names are marked for the candidate events with \pbns{} or
\pnsbh{} greater than 20\%, since these are of particular interest for
cross-correlation studies.}
\end{figure*}

We recover \TOTALEVENTS{} candidates that have \ac{FAR} less than 2 per day in
any of the search pipelines. These events and their estimated source
probabilities are shown in Fig.~\ref{fig:subthreshold-pastro}. The candidates
are shown in decreasing order of \pastro{}. The total sum of \pastro{}
represents the Poisson rate of sources that pass the \ac{FAR} threshold of 2
per day in each source class per \ac{O3a} experiment, as estimated by the search
pipelines. We find that this corresponds to between
\NUMALLBBHPYCBC{}--\NUMALLBBHPYCBCBBH{} signals in the \ac{BBH} class,
\NUMALLNSBHMBTA{}--\NUMALLNSBHPYCBC{} signals in the \ac{NSBH} class, and
\NUMALLBNSPYCBC{}--\NUMALLBNSGSTLAL{} signals in the \ac{BNS} class in
\ac{O3a}. The range represents the difference in the search pipelines. We do not
consider the \PYCBCBBH{} analysis in the estimate of the number of signals in
the \ac{BNS} class provided here, as the analysis does not search over the \ac{BNS} parameter
space. Names are marked for the
candidate events with \pbns{} or \pnsbh{} greater than 20\%. The dashed
vertical line shows the least significant event with \pastro{} greater than
0.5. An estimate of the rate of sources in the subthreshold candidate list per
\ac{O3a} experiment is obtained by the contribution to the sum from events with
\pastro{} less than 0.5. This corresponds to between
\NUMSUBBBHMBTA{}--\NUMSUBBBHPYCBCBBH{} signals in the \ac{BBH} class,
\NUMSUBNSBHPYCBCBBH{}--\NUMSUBNSBHPYCBC{} signals in the \ac{NSBH} class, and
\NUMSUBBNSGSTLAL{}--\NUMSUBBNSMBTA{} signals in the \ac{BNS} class in the subthreshold candidates
in \ac{O3a}.

We find \NUMEVENTS{} high probability \ac{CBC} candidates that have \pastro{}
greater than $0.5$. These events are listed in
Table~\ref{tab:events}. This list includes \NUMADDCANDIDATES{} new candidates that were
not present in GWTC-2~\cite{Abbott:2020niy}. These are marked in bold in Table~\ref{tab:events}. Out
of the \fixme{\NUMEVENTS} candidates, \fixme{4} were found with significant
\ac{SNR} only in one of the detectors by the \GSTLAL{} search, which is the
only pipeline that looked for \ac{GW} signals in single-detector data.
These are listed with a dagger ($^{\dagger}$) next to the \ac{FAR} in 
Table~\ref{tab:events}. For the majority of events listed in
Table~\ref{tab:events}, $\pastro{}\approx\pbbh{}$; the exceptions are listed in
Table~\ref{tab:p-astro}, which provides the list of candidates that have
\pbns{} or \pnsbh{} greater than 0.01.
\begin{event_table}
\begin{table*}
\begin{tabularx}{\textwidth}{l@{\extracolsep{\fill}}c|ccc|ccc|ccc|ccc}
\multicolumn{1}{c}{\textbf{Name}} & \textbf{Inst.} & \multicolumn{3}{c}{\textbf{\MBTA{}}} & \multicolumn{3}{c}{\textbf{\GSTLAL{}}} & \multicolumn{3}{c}{\textbf{\PYCBCHYPERBANK{}}} & \multicolumn{3}{c}{\textbf{\PYCBCBBH{}}} \\
                & & FAR \small{($\mathrm{yr}^{-1}$)} & SNR & \pastro{} & FAR \small{($\mathrm{yr}^{-1}$)} & SNR & \pastro{} & FAR \small{($\mathrm{yr}^{-1}$)} & SNR & \pastro{} & FAR \small{($\mathrm{yr}^{-1}$)} & SNR & \pastro{} \\
\hline
 {\EVENTNAMEBOLD{GW190403B_o3afin} \NAME{GW190403B_o3afin} \EVENTNAMEASTERIX{GW190403B_o3afin}}& \INSTRUMENTS{GW190403B_o3afin} &
 \MBTAALLSKYMEETSPASTROTHRESH{GW190403B_o3afin} \MBTAALLSKYFAR{GW190403B_o3afin}  &
 \MBTAALLSKYMEETSPASTROTHRESH{GW190403B_o3afin} \MBTAALLSKYSNR{GW190403B_o3afin}  &
 \MBTAALLSKYMEETSPASTROTHRESH{GW190403B_o3afin} \MBTAALLSKYPASTRO{GW190403B_o3afin}  &
 \GSTLALALLSKYMEETSPASTROTHRESH{GW190403B_o3afin} \GSTLALALLSKYFAR{GW190403B_o3afin}  &
 \GSTLALALLSKYMEETSPASTROTHRESH{GW190403B_o3afin} \GSTLALALLSKYSNR{GW190403B_o3afin}  &
 \GSTLALALLSKYMEETSPASTROTHRESH{GW190403B_o3afin} \GSTLALALLSKYPASTRO{GW190403B_o3afin}  &
 \PYCBCALLSKYMEETSPASTROTHRESH{GW190403B_o3afin} \PYCBCALLSKYFAR{GW190403B_o3afin}  &
 \PYCBCALLSKYMEETSPASTROTHRESH{GW190403B_o3afin} \PYCBCALLSKYSNR{GW190403B_o3afin}  &
 \PYCBCALLSKYMEETSPASTROTHRESH{GW190403B_o3afin} \PYCBCALLSKYPASTRO{GW190403B_o3afin}  &
 \PYCBCHIGHMASSMEETSPASTROTHRESH{GW190403B_o3afin} \PYCBCHIGHMASSFAR{GW190403B_o3afin}  &
 \PYCBCHIGHMASSMEETSPASTROTHRESH{GW190403B_o3afin} \PYCBCHIGHMASSSNR{GW190403B_o3afin}  &
 \PYCBCHIGHMASSMEETSPASTROTHRESH{GW190403B_o3afin} \PYCBCHIGHMASSPASTRO{GW190403B_o3afin}  \\
\hline
\end{tabularx}

\caption{
\label{tab:events}
Above-threshold \ac{GW} candidate list. We find \fixme{\NUMEVENTS} events that
have \pastro{} in at least one of the searches as greater than 0.5. Bold-faced
names indicate the events that were not previously reported in
GWTC-2~\cite{Abbott:2020niy}. The candidates marked with an asterisk were first
published in 3-OGC~\cite{Nitz:2021uxj}. The second column denotes the observing
instruments. Candidate events in GWTC-2.1 which do not meet the \pastro{}
threshold but were at the same time as above-threshold events are given in
italics. The \PYCBC{} and \PYCBCBBH{} network \acp{SNR} do not include
detectors with \acp{SNR} below 4; these events are marked with  double dagger
($^{\ddag}$) next to their network SNR. The 4 events marked with a dagger
($^{\dagger}$) next to their \acp{FAR} were found only in one detector by the
\GSTLAL{} search. All four were detected using the data from LIGO Livingston.
For the single-detector candidate events, the \ac{FAR} estimate involves
extrapolation. All single-detector candidate events in this list according to
the \ac{FAR} assigned to them are rarer than the background data of about 6
months collected in this analysis. Therefore, a conservative bound on the FAR
for candidates denoted by $^{\dagger}$ is $\sim 2~$yr$^{-1}$. \GSTLAL{} \acp{FAR}
have been capped at $1\times10^{-5}$~yr$^{-1}$ to be consistent with the
limiting \acp{FAR} from other pipelines. Dashes indicate that a pipeline did
not find the event with a \ac{FAR} smaller than the subthreshold \ac{FAR}
threshold of 2 per day.}
\end{table*}
\end{event_table}

\begin{event_table}
\begin{table*}
\begin{tabularx}{\textwidth}{l@{\extracolsep{\fill}}|ccc|c|ccc|c|ccc|c|cc|c}
\multicolumn{1}{c|}{\textbf{Name}} & \multicolumn{4}{c|}{\textbf{\MBTA{}}} &  \multicolumn{4}{c|}{\textbf{\GSTLAL{}}} & \multicolumn{4}{c|}{\textbf{\PYCBCHYPERBANK{}}} & \multicolumn{3}{c}{\textbf{\PYCBCBBH{}}} \\
 & \pbbh{}~ & \pnsbh{}~ & \pbns{}~ & \pastro{} & \pbbh{}~ & \pnsbh{}~ & \pbns{}~ & \pastro{} & \pbbh{}~ & \pnsbh{}~ & \pbns{}~ & \pastro{} & \pbbh{}~ & \pnsbh{}~ & \pastro{}\\
\hline
\makebox[0pt][l]{\fboxsep0pt\colorbox{lightgray}{\mystrut\hspace*{1.0\linewidth}}} {\EVENTNAMEBOLD{GW190425B_o3afin} \NAME{GW190425B_o3afin} } &\MBTAALLSKYPBBH{GW190425B_o3afin} &\MBTAALLSKYPNSBH{GW190425B_o3afin} &\MBTAALLSKYPBNS{GW190425B_o3afin} & \MBTAALLSKYPASTRO{GW190425B_o3afin} & \GSTLALALLSKYPBBH{GW190425B_o3afin} &\GSTLALALLSKYPNSBH{GW190425B_o3afin} &\GSTLALALLSKYPBNS{GW190425B_o3afin} & \GSTLALALLSKYPASTRO{GW190425B_o3afin} &\PYCBCALLSKYPBBH{GW190425B_o3afin} &\PYCBCALLSKYPNSBH{GW190425B_o3afin} &\PYCBCALLSKYPBNS{GW190425B_o3afin} & \PYCBCALLSKYPASTRO{GW190425B_o3afin} &\PYCBCHIGHMASSPBBH{GW190425B_o3afin} &\PYCBCHIGHMASSPNSBH{GW190425B_o3afin} & \PYCBCHIGHMASSPASTRO{GW190425B_o3afin} \\
 {\EVENTNAMEBOLD{GW190707E_o3afin} \NAME{GW190707E_o3afin} } &\MBTAALLSKYPBBH{GW190707E_o3afin} &\MBTAALLSKYPNSBH{GW190707E_o3afin} &\MBTAALLSKYPBNS{GW190707E_o3afin} & \MBTAALLSKYPASTRO{GW190707E_o3afin} & \GSTLALALLSKYPBBH{GW190707E_o3afin} &\GSTLALALLSKYPNSBH{GW190707E_o3afin} &\GSTLALALLSKYPBNS{GW190707E_o3afin} & \GSTLALALLSKYPASTRO{GW190707E_o3afin} &\PYCBCALLSKYPBBH{GW190707E_o3afin} &\PYCBCALLSKYPNSBH{GW190707E_o3afin} &\PYCBCALLSKYPBNS{GW190707E_o3afin} & \PYCBCALLSKYPASTRO{GW190707E_o3afin} &\PYCBCHIGHMASSPBBH{GW190707E_o3afin} &\PYCBCHIGHMASSPNSBH{GW190707E_o3afin} & \PYCBCHIGHMASSPASTRO{GW190707E_o3afin} \\
\makebox[0pt][l]{\fboxsep0pt\colorbox{lightgray}{\mystrut\hspace*{1.0\linewidth}}} {\EVENTNAMEBOLD{GW190720A_o3afin} \NAME{GW190720A_o3afin} } &\MBTAALLSKYPBBH{GW190720A_o3afin} &\MBTAALLSKYPNSBH{GW190720A_o3afin} &\MBTAALLSKYPBNS{GW190720A_o3afin} & \MBTAALLSKYPASTRO{GW190720A_o3afin} & \GSTLALALLSKYPBBH{GW190720A_o3afin} &\GSTLALALLSKYPNSBH{GW190720A_o3afin} &\GSTLALALLSKYPBNS{GW190720A_o3afin} & \GSTLALALLSKYPASTRO{GW190720A_o3afin} &\PYCBCALLSKYPBBH{GW190720A_o3afin} &\PYCBCALLSKYPNSBH{GW190720A_o3afin} &\PYCBCALLSKYPBNS{GW190720A_o3afin} & \PYCBCALLSKYPASTRO{GW190720A_o3afin} &\PYCBCHIGHMASSPBBH{GW190720A_o3afin} &\PYCBCHIGHMASSPNSBH{GW190720A_o3afin} & \PYCBCHIGHMASSPASTRO{GW190720A_o3afin} \\
 {\EVENTNAMEBOLD{GW190725F_o3afin} \NAME{GW190725F_o3afin} } &\MBTAALLSKYPBBH{GW190725F_o3afin} &\MBTAALLSKYPNSBH{GW190725F_o3afin} &\MBTAALLSKYPBNS{GW190725F_o3afin} & \MBTAALLSKYPASTRO{GW190725F_o3afin} & \GSTLALALLSKYPBBH{GW190725F_o3afin} &\GSTLALALLSKYPNSBH{GW190725F_o3afin} &\GSTLALALLSKYPBNS{GW190725F_o3afin} & \GSTLALALLSKYPASTRO{GW190725F_o3afin} &\PYCBCALLSKYPBBH{GW190725F_o3afin} &\PYCBCALLSKYPNSBH{GW190725F_o3afin} &\PYCBCALLSKYPBNS{GW190725F_o3afin} & \PYCBCALLSKYPASTRO{GW190725F_o3afin} &\PYCBCHIGHMASSPBBH{GW190725F_o3afin} &\PYCBCHIGHMASSPNSBH{GW190725F_o3afin} & \PYCBCHIGHMASSPASTRO{GW190725F_o3afin} \\
\makebox[0pt][l]{\fboxsep0pt\colorbox{lightgray}{\mystrut\hspace*{1.0\linewidth}}} {\EVENTNAMEBOLD{GW190728D_o3afin} \NAME{GW190728D_o3afin} } &\MBTAALLSKYPBBH{GW190728D_o3afin} &\MBTAALLSKYPNSBH{GW190728D_o3afin} &\MBTAALLSKYPBNS{GW190728D_o3afin} & \MBTAALLSKYPASTRO{GW190728D_o3afin} & \GSTLALALLSKYPBBH{GW190728D_o3afin} &\GSTLALALLSKYPNSBH{GW190728D_o3afin} &\GSTLALALLSKYPBNS{GW190728D_o3afin} & \GSTLALALLSKYPASTRO{GW190728D_o3afin} &\PYCBCALLSKYPBBH{GW190728D_o3afin} &\PYCBCALLSKYPNSBH{GW190728D_o3afin} &\PYCBCALLSKYPBNS{GW190728D_o3afin} & \PYCBCALLSKYPASTRO{GW190728D_o3afin} &\PYCBCHIGHMASSPBBH{GW190728D_o3afin} &\PYCBCHIGHMASSPNSBH{GW190728D_o3afin} & \PYCBCHIGHMASSPASTRO{GW190728D_o3afin} \\
 {\EVENTNAMEBOLD{GW190814H_o3afin} \NAME{GW190814H_o3afin} } &\MBTAALLSKYPBBH{GW190814H_o3afin} &\MBTAALLSKYPNSBH{GW190814H_o3afin} &\MBTAALLSKYPBNS{GW190814H_o3afin} & \MBTAALLSKYPASTRO{GW190814H_o3afin} & \GSTLALALLSKYPBBH{GW190814H_o3afin} &\GSTLALALLSKYPNSBH{GW190814H_o3afin} &\GSTLALALLSKYPBNS{GW190814H_o3afin} & \GSTLALALLSKYPASTRO{GW190814H_o3afin} &\PYCBCALLSKYPBBH{GW190814H_o3afin} &\PYCBCALLSKYPNSBH{GW190814H_o3afin} &\PYCBCALLSKYPBNS{GW190814H_o3afin} & \PYCBCALLSKYPASTRO{GW190814H_o3afin} &\PYCBCHIGHMASSPBBH{GW190814H_o3afin} &\PYCBCHIGHMASSPNSBH{GW190814H_o3afin} & \PYCBCHIGHMASSPASTRO{GW190814H_o3afin} \\
\makebox[0pt][l]{\fboxsep0pt\colorbox{lightgray}{\mystrut\hspace*{1.0\linewidth}}} {\EVENTNAMEBOLD{GW190924A_o3afin} \NAME{GW190924A_o3afin} } &\MBTAALLSKYPBBH{GW190924A_o3afin} &\MBTAALLSKYPNSBH{GW190924A_o3afin} &\MBTAALLSKYPBNS{GW190924A_o3afin} & \MBTAALLSKYPASTRO{GW190924A_o3afin} & \GSTLALALLSKYPBBH{GW190924A_o3afin} &\GSTLALALLSKYPNSBH{GW190924A_o3afin} &\GSTLALALLSKYPBNS{GW190924A_o3afin} & \GSTLALALLSKYPASTRO{GW190924A_o3afin} &\PYCBCALLSKYPBBH{GW190924A_o3afin} &\PYCBCALLSKYPNSBH{GW190924A_o3afin} &\PYCBCALLSKYPBNS{GW190924A_o3afin} & \PYCBCALLSKYPASTRO{GW190924A_o3afin} &\PYCBCHIGHMASSPBBH{GW190924A_o3afin} &\PYCBCHIGHMASSPNSBH{GW190924A_o3afin} & \PYCBCHIGHMASSPASTRO{GW190924A_o3afin} \\
 {\EVENTNAMEBOLD{GW190930C_o3afin} \NAME{GW190930C_o3afin} } &\MBTAALLSKYPBBH{GW190930C_o3afin} &\MBTAALLSKYPNSBH{GW190930C_o3afin} &\MBTAALLSKYPBNS{GW190930C_o3afin} & \MBTAALLSKYPASTRO{GW190930C_o3afin} & \GSTLALALLSKYPBBH{GW190930C_o3afin} &\GSTLALALLSKYPNSBH{GW190930C_o3afin} &\GSTLALALLSKYPBNS{GW190930C_o3afin} & \GSTLALALLSKYPASTRO{GW190930C_o3afin} &\PYCBCALLSKYPBBH{GW190930C_o3afin} &\PYCBCALLSKYPNSBH{GW190930C_o3afin} &\PYCBCALLSKYPBNS{GW190930C_o3afin} & \PYCBCALLSKYPASTRO{GW190930C_o3afin} &\PYCBCHIGHMASSPBBH{GW190930C_o3afin} &\PYCBCHIGHMASSPNSBH{GW190930C_o3afin} & \PYCBCHIGHMASSPASTRO{GW190930C_o3afin} \\
\hline
\end{tabularx}

\caption{
\label{tab:p-astro}
Source probabilities (\pbbh{}, \pbns{}, \pnsbh{}) for the high significance 
\ac{GW} candidates listed in Table~\ref{tab:events} for which \pbns{} or
\pnsbh{} is greater than 1\%. For other events in 
Table~\ref{tab:events}, \pastro{} $\approx$ \pbbh{}, and therefore we do not list
them here. Results are provided from all three matched-filter pipelines.
Dashes indicate that a pipeline did not find the event with a FAR smaller than
the subthreshold \ac{FAR} threshold of 2 per day. The classification provided here assumes a boundary of $3\,\Msun$
between \acp{NS} and \acp{BH} in the case of \GSTLAL{} and \PYCBC{}, and $2.5\, \Msun$ in the case of \MBTA{}.}
\end{table*}
\end{event_table}

\subsection{New high probability candidates}\label{ssec:hs_candidates}
We recover all the events found in GWTC-2 as having \pastro{} above 0.5, with
the exception of three: GW190424\_180648, GW190426\_152155, and GW190909\_114149.
Since the rate of
\ac{BBH} events detectable by the LIGO--Virgo detectors is greater than the rate
of detectable \ac{BNS} or \ac{NSBH} events, the \pastro{} for events in the BBH
class is higher than that of the events in the BNS or NSBH class at a
fixed \ac{FAR}. Therefore, in switching to a \pastro{} threshold from a
\ac{FAR} threshold, one can expect to add BBH events while dropping some
low-mass events. \\

All the \NUMADDCANDIDATES{} new candidates with \pastro{} greater than 0.5 are
classified as \acp{BBH}, that is, \pbbh{} is greater than \pnsbh{} and \pbns{}.
Only one new candidate, GW190725\_174728, has a  non-negligible probability in
a source class other than \ac{BBH}, with non-zero \pnsbh{}
(Table~\ref{tab:p-astro}). Out of the \NUMADDCANDIDATES{} candidates, only two
(GW190725\_174728 and GW190916\_200658) are assigned $\pastro>0.5$ by more than
one pipeline.  Differences between pipelines are expected, due to the effects
of random noise fluctuations on the different ranking statistics used, and due
to different assumed signal distributions and other choices.  In principle, a
more accurate assessment of the candidates' origins could be obtained by
considering information from all pipelines; however, this is not currently
implemented as a quantitative measure. One of the events, GW190917\_114630, is
identified as a BBH by the \GSTLAL{} pipeline, with \pbbh{} = 0.77
(Table~\ref{tab:events}). However, when its source properties are inferred by
follow-up pipelines, the mass parameters are found to be consistent with
\ac{NSBH} systems. Had it been classified as an \ac{NSBH} to begin with by the
search pipeline, the resulting \pastro{} would not have made the threshold of
0.5. There is also non-stationary noise in the LIGO 
Livingston detector at the time of this event, but we have no evidence that the \ac{FAR} of the event is misestimated. Out of the \NUMADDCANDIDATES{} 
new candidates, \fixme{5} candidates (GW190426\_190642, GW190725\_174728, GW190805\_211137,
GW190916\_200658, and GW190925\_232845) were identified in the LVC search for
gravitationally lensed candidates in \ac{O3a} data~\cite{Abbott:2021iab}, while \fixme{4} candidates
(GW190725\_174728, GW190916\_200658, GW190925\_232845, and GW190926\_050336) were
also independently identified and presented in 3-OGC~\cite{Nitz:2021uxj}.  The source properties of
all \NUMADDCANDIDATES{} candidates are discussed in
Sec.~\ref{ss:parameter-estimation-results}.

\subsection{GWTC-2 candidates with $\pastro{} < 0.5$}\label{sec:compare_search}
The three events in GWTC-2 that have a \pastro{} smaller than $0.5$ in GWTC-2.1 analyses are:

\textbf{GW190424\_180648}: This event was found by \GSTLAL{} as a single detector
\ac{BBH} event in Livingston. However, the data surrounding this event recorded
periodic glitching from a camera shutter and iDQ (Sec.~\ref{sec:gstlal}) 
heavily downranked the timespan surrounding this
event~\cite{idq-in-gstlal}. Figure~4 in~\cite{idq-in-gstlal} shows both the inspiral 
track and the surrounding glitches in the time--frequency spectrogram surrounding 
this event and the response of iDQ. While the down-ranking due to iDQ for this 
particular event remains largely the same between GWTC-2 and GWTC-2.1, the retuning of 
 the singles penalty (Sec.~\ref{sec:gstlal}) in \GSTLAL{} for GWTC-2.1 caused the significance of the event to go down. Consequently, in GWTC-2.1, this event does not meet either the \ac{FAR} threshold of 2 per year or the 
\pastro{} threshold of $0.5$.

\textbf{GW190426\_152155}: This event is in the marginal-significance candidate list for GWTC-2.1
(Table~\ref{tab:marginal-events}); the \ac{FAR} is similar to the one in
GWTC-2 and still passed the threshold of 2 per year considered in the previous
catalog. However, based on the masses recovered by the pipeline, it is assigned 
to the NSBH class with $\pnsbh{} = 0.14$. The low \pastro{} in the NSBH class 
is due to the fact that the inferred rate of detectable \acp{NSBH} is lower than that of 
detectable \acp{BBH}.

\textbf{GW190909\_114149}: This candidate \ac{BBH} event was found as a coincident event in 
Hanford and Livingston detectors by \GSTLAL{}. It is recovered now with  
smaller \ac{SNR} in the Hanford detector and is therefore ranked lower.

\begin{event_table}
\begin{table*}
\begin{tabularx}{\textwidth}{l@{\extracolsep{\fill}}c|ccc|ccc|ccc}
\multicolumn{1}{c}{\textbf{Name}} & \textbf{Inst.} & \multicolumn{3}{c}{\textbf{\MBTA{}}} & \multicolumn{3}{c}{\textbf{\GSTLAL{}}} & \multicolumn{3}{c}{\textbf{\PYCBCHYPERBANK{}}} \\
                & & FAR \small{($\mathrm{yr}^{-1}$)} & SNR & max \pastro{} & FAR \small{($\mathrm{yr}^{-1}$)} & SNR & max \pastro{} & FAR \small{($\mathrm{yr}^{-1}$)} & SNR & max \pastro{} \\
\hline
 {\EVENTNAMEBOLDMAR{GW190426H_o3asub} \NAMEMAR{GW190426H_o3asub} }& \INSTRUMENTSMAR{GW190426H_o3asub} &\MBTAALLSKYFARMAR{GW190426H_o3asub} &\MBTAALLSKYSNRMAR{GW190426H_o3asub} &\MBTAALLSKYPCATMAXMAR{GW190426H_o3asub} &\GSTLALALLSKYFARMAR{GW190426H_o3asub} &\GSTLALALLSKYSNRMAR{GW190426H_o3asub} &\GSTLALALLSKYPCATMAXMAR{GW190426H_o3asub} &\PYCBCALLSKYFARMAR{GW190426H_o3asub} &\PYCBCALLSKYSNRMAR{GW190426H_o3asub} &\PYCBCALLSKYPCATMAXMAR{GW190426H_o3asub} \\
\makebox[0pt][l]{\fboxsep0pt\colorbox{lightgray}{\mystrut\hspace*{1.0\linewidth}}} {\EVENTNAMEBOLDMAR{GW190531B_o3asub} \NAMEMAR{GW190531B_o3asub} }& \INSTRUMENTSMAR{GW190531B_o3asub} &\MBTAALLSKYFARMAR{GW190531B_o3asub} &\MBTAALLSKYSNRMAR{GW190531B_o3asub} &\MBTAALLSKYPCATMAXMAR{GW190531B_o3asub} &\GSTLALALLSKYFARMAR{GW190531B_o3asub} &\GSTLALALLSKYSNRMAR{GW190531B_o3asub} &\GSTLALALLSKYPCATMAXMAR{GW190531B_o3asub} &\PYCBCALLSKYFARMAR{GW190531B_o3asub} &\PYCBCALLSKYSNRMAR{GW190531B_o3asub} &\PYCBCALLSKYPCATMAXMAR{GW190531B_o3asub} \\
\hline
\end{tabularx}

\caption{\label{tab:marginal-events}
Marginal-significance \ac{GW} event candidate list. There are \NUMEVENTSMAR{} candidates that are found in at
least one of the searches with a \ac{FAR} less than 2 per year, but with a
\pastro{} smaller than 0.5 in all searches. The candidate in bold, GW190531\_023648, is 
a new candidate identified in GWTC-2.1, not included in GWTC-2. The column max \pastro{} shows 
the astrophysical class assigned with highest probability.  Both candidates are 
detected by \GSTLAL{} with a small \ac{FAR}, and are assigned to
the \ac{NSBH} class with \pastro{} and \pnsbh{} smaller than 0.5. 
}
\end{table*}
\end{event_table}

\subsection{Marginal-significance candidates}
The two \ac{GW} candidates that satisfy the \ac{FAR} 
criteria used by GWTC-2, but do not have \pastro{} greater than 0.5 are listed as marginal candidates in Table~\ref{tab:marginal-events}. 
Both these events were detected by \GSTLAL{} with a small \ac{FAR}, and were assigned to
the \ac{NSBH} class with \pastro{} and \pnsbh{} smaller than 0.5. Since the rate of
detectable signals in the \ac{NSBH} class is smaller than that in the \ac{BBH} class, 
the \pastro{} for these are smaller than they would be in the BBH class at
the same \ac{FAR}.

\subsection{Search sensitivity}
\label{sec:vt}

As in GWTC-2~\cite{Abbott:2020niy}, we quantify the sensitivity of the search via a campaign of simulated
signals injected into the O3a data and analyzed by the search pipelines.  We use a 
\ac{BBH} signal distribution adjusted over that used for GWTC-2 to give more even 
coverage of the inferred distribution from~\cite{o3apop}, 
changing specifically the distributions over binary mass ratio and redshift.  In
addition to the \ac{BBH} set, we also inject \ac{BNS} and \ac{NSBH} sets of 
simulated signals into the data.  The sets are generated in two stages: first, points
are sampled out to the maximum redshift considered for each set, then the samples
are reduced to sets of potentially detectable signals by imposing that the expected
LIGO Hanford–LIGO Livingston network SNR, calculated using a representative noise
\ac{PSD}, be above a threshold of $6$.  Although this threshold is below the matched-filter 
\acp{SNR} of events we consider as high-significance candidates, for detection thresholds
corresponding to \acp{FAR} significantly higher than 2 per year (the value used in GWTC-2),
the cut may remove a non-negligible fraction of potentially detectable signals,
due to random fluctuations in matched-filter SNR. The results of this simulation campaign for all the search pipelines have been made available~\cite{gwtc2p1_injection_data}.

The \ac{BNS} signals are generated using the \INJBNSWAVEFORM{} waveform 
model~\cite{Buonanno:2002fy,Buonanno:2009zt}, while the \ac{BBH} and \ac{NSBH} sets are generated using the
\INJBBHWAVEFORM{} model~\cite{Ossokine:2020kjp,Babak:2016tgq,
Pan:2013rra}.
For simulated signals with redshifted total mass below $9\,\Msun$,
the SEOBNRv4P model without higher-order multipole emission was used, as higher-order multipoles
would lie above the data sampling Nyquist frequency.  
The component spin magnitudes
$|\chi|$ are distributed uniformly up to a maximum of \INJBNSSPINMAX{} for \ac{NS} components
and \INJBBHSPINMAX{} for \ac{BBH}, with isotropically distributed orientations.  

The signal distributions over sky direction and binary orientation are isotropic.  
The distributions over redshift are proportional to the comoving volume
element $\mathrm{d}V_\mathrm{c}/\mathrm{d}z$, multiplied by a factor $(1+z)^{-1}$ accounting for time dilation, 
and by a factor $(1+z)^\kappa$ modeling possible evolution of the comoving merger rate
density with redshift (as in Appendix E of \cite{o3apop}). 
A summary of the distributions of the three injection sets is given in Table~\ref{tab:vt}.  

Given the merger distribution used for each injection set, the sensitivity of each
search over the O3 data is quantified by relating the expected number of detections,
at a specified significance threshold, to the local astrophysical merger rate as 
$N_\mathrm{det} = \mathcal{V}R(z=0)$, where $\mathcal{V}$ is an effective sensitive
hypervolume with units of volume$\times$time. This effective hypervolume is estimated
by counting the number of injected signals that are detected at the given threshold, 
here a \ac{FAR} of 2 per year.

In addition to assumed merger distributions that follow those used for the injection
sets, we also provide $\mathcal{V}$ for a fiducial \ac{BBH} population model
representative of those found to have high posterior probability in our population 
analysis of GWTC-2~\cite{o3apop}.  We choose the \textsc{Power Law + Peak}
model (defined in Appendix B.2 of \cite{o3apop})
with parameters $\alpha=2.5$, $\beta=1.5$, $m_\textrm{min}=5\,\Msun$, $m_\textrm{max}
=80\,\Msun$, $\lambda_\mathrm{peak}=0.1$, $\mu_{m}=34\,\Msun$, $\sigma_{m}=5\,\Msun$,
$\delta_{m}=3.5\,\Msun$, setting the redshift evolution to $\kappa=0$. 
The sensitivity for this \ac{BBH} population is evaluated via importance 
sampling~\cite{Tiwari:2018bgh,software-inj-release} implemented via 
\textsc{GWPopulation}~\cite{Talbot:2019okv}. 
The effective hypervolume for each search and signal population is given in 
Table~\ref{tab:vt}.

\begin{event_table}
\begin{table*}
      \begin{tabularx}{\textwidth}{l@{\extracolsep{\fill}}ccccc|ccccc}
      & \multicolumn{5}{c|}{ \textbf{Injection populations} } & \multicolumn{5}{c}{ \textbf{Sensitive hypervolume $\mathcal{V}$} (\Gpcyr) } \\
      \hline
      & mass & mass range & spin & redshift & max.\ & \multirow{2}{*}{\GSTLAL{}} & \multirow{2}{*}{\MBTA{}} & \multirow{2}{*}{\PYCBC{}} & \PYCBC{} & \multirow{2}{*}{All}\\
      & distribution & (\Msun) & range & evolution & redshift & & & & BBH &  \\
      \hline \hline
      \makecell*{BBH\\(INJ)}
          & \makecell*{$p(\massone{}) \propto \massone{}^{\INJBBHMASSONEPOWER{}}$ \\ $p(\masstwo{}|\massone{}) \propto \masstwo{}$} 
          & \makecell*{$\INJBBHMASSMIN{} < \massone{} < \INJBBHMASSMAX{}$ \\ $\INJBBHMASSMIN{} < \masstwo{} < \INJBBHMASSMAX{}$}
          & $\left|\chi_{1,2}\right| < \INJBBHSPINMAX{}$
          & \makecell*{$\kappa=1$}
          & \INJBBHMAXREDSHIFT{}  
          & \GSTLALVTBBHINJ{}
          & \MBTAVTBBHINJ{}
          & \PYCBCVTBBHINJ{}
          & \PYCBCBBHVTBBHINJ{}
          & \ANYCBCVTBBHINJ{} 
          \\
      \makecell*{BBH\\(POP)}
          & \textsc{Power Law + Peak} 
          & (see text)
          & $\left|\chi_{1,2}\right| < \INJBBHSPINMAX{}$
          & \makecell*{$\kappa=0$}
          & \INJBBHMAXREDSHIFT{} 
          & \GSTLALVTBBHPOP{}
          & \MBTAVTBBHPOP{}
          & \PYCBCVTBBHPOP{}
          & \PYCBCBBHVTBBHPOP{}
          & \ANYCBCVTBBHPOP{} \\
      \makecell*{BNS}
          & uniform
          & \makecell*{$\INJBNSMASSMIN{} < \massone{} < \INJBNSMASSMAX{}$ \\ $\INJBNSMASSMIN{} < \masstwo{} < \INJBNSMASSMAX{}$}
          & $\left|\chi_{1,2}\right| < \INJBNSSPINMAX{}$
          & \makecell*{$\kappa=0$}
          & \INJBNSMAXREDSHIFT{}
          & \GSTLALVTBNS{}
          & \MBTAVTBNS{}
          & \PYCBCVTBNS{}
          & --
          & \ANYCBCVTBNS{} \\
      \makecell*{NSBH}
          & \makecell*{$p(\massone{}) \propto \massone{}^{\INJNSBHPOWERBBHMASS{}}$ \\ uniform}
          & \makecell*{$\INJNSBHMINBBHMASS{} < \massone{} < \INJNSBHMAXBBHMASS{}$ \\ $\INJNSBHMINBNSMASS{} < \masstwo{} < \INJNSBHMAXBNSMASS{}$}
          & \makecell*{$\left|\chi_1\right| < \INJNSBHMAXBBHSPIN{}$ \\ $\left|\chi_2\right| < \INJNSBHMAXBNSSPIN{}$}
          & \makecell*{$\kappa=0$}
          & \INJNSBHMAXREDSHIFT{}
          & \GSTLALVTNSBH{}
          & \MBTAVTNSBH{}
          & \PYCBCVTNSBH{}
          & --
          & \ANYCBCVTNSBH{} \\
      \end{tabularx}
      \caption{
      \label{tab:vt}
      Measures of sensitivity for the search pipelines.  We state the sensitive hypervolume $\mathcal{V}$ for each 
      of four assumed signal populations: a \ac{BBH} population following the injected distribution, a \ac{BBH}
      population given by the \textsc{Power Law + Peak} model of~\cite{o3apop},
      and \ac{BNS} and \ac{NSBH} populations following the injected distributions.  We give estimates for each 
      search pipeline independently at a \ac{FAR} threshold of $2$ per year, and for all pipelines combined, i.e.\
      counting all injections detected in at least one pipeline at the given threshold. 
      }
\end{table*}

\end{event_table}

\subsection{Rates of \ac{BBH} and \ac{BNS} events}\label{sec:rates}

The rates of \ac{BBH} and \ac{BNS} binary mergers in the local Universe were estimated in a companion 
paper~\cite{o3apop} to GWTC-2, using the count of detected events with 
\ac{FAR} below 1 per year, combined with estimates of search sensitivity to the respective populations.  The
\ac{BBH} rate estimate was marginalized over uncertainties in the parameters of the population models used, 
while the \ac{BNS} rate estimate assumed a population \fixme{uniform in component masses between 1\,\Msun and 2.5\,\Msun}. 
The merger rate of \acp{NSBH} was recently calculated following the discovery of GW200105\_162426 and GW200115\_042309~\cite{Abbott_nsbh_2021}, and we do not update it here.

Here, we present complementary \ac{BBH} and \ac{BNS} rate estimates based solely on the matched filter
search pipeline outputs, with methods that allow us to incorporate a large number of likely noise
(background) events~\cite{Farr:2013yna} and thus avoid potential bias due to an arbitrary choice of
significance threshold.  Such methods allow for both foreground (signal) and background event distributions
with a priori unknown rates, considered as independent Poisson processes.
Furthermore, for the \GSTLAL{} pipeline we employ a multicomponent mixture analysis~\cite{Kapadia:2019uut} 
to estimate the rates of events in several astrophysical classes (BNS, NSBH, and BBH) and terrestrial. 
Every trigger is assigned probabilities of
membership in each class, 
as described in Sec.~\ref{sec:gstlal}. 
For the \MBTA{} and \PYCBC{} rate estimates, only the \ac{BBH} class is considered. 

The merger rate estimate then arises from the number of search events assigned to each class, divided
by the estimated search sensitivity obtained via injection campaigns re-weighted to an astrophysical
population model~\cite{Tiwari:2018bgh}, as discussed in the previous section. 
The population models used here to quantify search sensitivity
are in general different
from those used to obtain source classification probabilities, described in Sec.~\ref{subsec:searches}.

In both the \ac{BBH} and \ac{BNS} cases, as for other rate interval estimates derived from search 
results~\cite{LIGOScientific:2018mvr}, a Poisson--Jeffreys ($\propto R^{-1/2}$) prior was used.  The choice of
prior has little influence on estimated \ac{BBH} rate due to the large count of signals, but it has
a nontrivial effect on the \ac{BNS} rate estimate as compared to, for instance, a uniform prior. 

\ac{BBH} merger rate estimates are provided by the \GSTLAL{}, \PYCBCBBH{} and \MBTA{} pipelines.  The
astrophysical population assumed for measuring search sensitivities is given by the \textsc{Power Law + Peak}
model of~\cite{o3apop} with fiducial parameters as in Sec.~\ref{sec:vt}.  
The resulting merger rates are \RGSTLALBBH{} for \GSTLAL{}, \RPYCBCBBH{} for \PYCBCBBH{} and \RMBTABBH{} for
\MBTA.  These estimates are fully consistent with the estimate of $23.9_{-8.6}^{+14.3}\,\perGpcyr$ as derived 
in~\cite{o3apop} using only significant (\ac{FAR}$<1~\mathrm{yr^{-1}}$) events, and allowing for uncertainties in 
the population model parameters. 
Following~\cite{o3apop}, we have not included the effect of calibration uncertainties in our rate estimates.  A 
full quantitative analysis of such uncertainties would require accounting for possible frequency- and 
time-dependent amplitude systematic errors~\cite{Sun:2020wke}; these are typically 
$\sim 3\%$ or less, corresponding to a $\lesssim 10\%$ sensitive volume uncertainty which remains subdominant
to the Poisson uncertainty in the signal counts~\cite{o3apop}. 

Since the only significant event consistent with \ac{BNS} merger in O3a, GW190425, was observed in a single
detector, it is present only in the \GSTLAL{} search results.  Hence, we quote a \ac{BNS} merger rate estimate
only from the \GSTLAL{} pipeline, as we expect this to be more informative than estimates from pipelines
that did not consider single-detector triggers.  For measuring the search sensitivity to \ac{BNS} mergers, 
we use the injected population described above in Sec.~\ref{sec:vt}, 
yielding an estimated merger rate \RGSTLALBNS. 
This estimate is fully consistent within uncertainties with 
the simpler estimate of $320_{-240}^{+490}\,\perGpcyr$ derived using a fixed threshold 
in expected~\ac{SNR} to determine sensitivity to simulated signals~\cite{o3apop}. 

\newcommand*{\cjh}[1]{\textsf{\color{magenta} [\textbf{CARL}: #1]}}

\section{Estimation of source parameters}
\label{sec:parameter-estimation-method}

The physical parameters $\vec{\vartheta}$ describing each \ac{GW} source binary, corresponding to individual entries from the list of events in Table~\ref{tab:events}, are inferred directly from the data $d$ and represented as a posterior probability distribution $p(\vec{\vartheta}|d)$.
This probability distribution is evaluated through Bayes' theorem as
\begin{equation}
p(\vec{\vartheta}|d) \propto p(d|\vec{\vartheta}) \pi(\vec{\vartheta})\,,
\end{equation}
with $p(d|\vec{\vartheta})$ being the likelihood of $d$ given a set of source parameters $\vec{\vartheta}$, and $\pi(\vec{\vartheta})$ being the prior probability distribution assumed for those parameters.

The likelihood itself describes the assumptions of the underlying stochastic process generating the noise present in $d$ from a given detector.
This noise is assumed to be Gaussian, stationary and uncorrelated between pairs of detectors~\cite{LIGOScientific:2019hgc,Berry:2014jja}, as further discussed in Sec.~\ref{sec:DQ}.
This yields a Gaussian likelihood~\cite{Cutler:1994ys,Veitch:2014wba}, which for the $i^{\mathrm{th}}$ detector used in a given analysis takes the form
\begin{equation}
    \label{eqn:pe_likelihood}
    p(d^i | \vec{\vartheta}) \propto \exp\left[- \frac{1}{2}  \left\langle d^i - h_\mathrm{M}^i(\vec\vartheta) \middle| d^i - h_\mathrm{M}^i(\vec\vartheta) \right\rangle \right],
\end{equation}
with $d^i$ representing the data from this instrument.
$h_\mathrm{M}^i(\vec\vartheta)$ is the binary waveform model $h(\vec\vartheta)$ calculated for $\vec{\vartheta}$ after being projected onto the detector and adjusted to account for the uncertainty present in the offline calibration (as described in Sec.~\ref{sec:data}) of $d^i$~\cite{TheLIGOScientific:2016wfe}.
The final likelihood is evaluated coherently across the network of available detectors and is obtained by multiplication of the likelihoods in each detector.

The term from Eq.~\eqref{eqn:pe_likelihood} in angle brackets, $\langle a | b \rangle$, represents a noise-weighted inner product~\cite{Finn:1992wt, Cutler:1994ys}. 
In addition to $d^i$ and $h_M^i(\vec\vartheta)$, evaluating this inner product requires specification of the bandwidth to be used in the analysis as well as the \PSD characterizing the noise process.
The low-frequency cutoff used in our analysis is set at $f_\mathrm{low}=20$~Hz.
Time-domain waveform models are generated starting at a frequency $f_\mathrm{start}$ such that the $(\ell,\, | m |) = (3,3)$ spherical harmonic mode of the binary inspiral signal, as estimated from a set of preliminary analyses~\cite{LIGOScientific:2018mvr, Abbott:2020niy}, is present at $f_\mathrm{low}$. 
The high-frequency cutoff $f_{\mathrm{high}}$ is selected for each analysis as $f_{\mathrm{high}} = \alpha^{\mathrm{roll-off}} f_{\mathrm{Nyquist}}$ such that the ringdown frequency of the $(\ell,\, | m |) = (3,3)$ spherical harmonic mode, inferred from waveforms taken from the same set of preliminary analyses as mentioned above~\cite{LIGOScientific:2018mvr, Abbott:2020niy}, occurs below $f_{\mathrm{high}}$.
The parameter $\alpha^{\mathrm{roll-off}}$ in this expression is a scale factor chosen in order to minimize the frequency roll-off effects caused by the application of a tapering window to the time-domain data~\cite{Romero-Shaw:2020owr}.
The Nyquist frequency $f_{\mathrm{Nyquist}}$ is then selected as the smallest power-of-two-valued frequency which together with $\alpha^{\mathrm{roll-off}} = 0.875$ satisfies the constraint on $f_{\mathrm{high}}$ specified above.
Similarly, the duration of data $d$ used in each analysis is determined from a requirement that the waveforms from previous analyses~\cite{LIGOScientific:2018mvr, Abbott:2020niy} as evaluated from $f_\mathrm{low}=20$~Hz and rounding up to the next power-of-two number of seconds, are contained in the selected data segment.
The \PSD for each event is inferred directly from the same data that is to be used in the likelihood, through the parametrized model implemented in \BAYESWAVE~\cite{Littenberg:2014oda, Chatziioannou:2019zvs}.
From the inferred posterior distribution of \acp{PSD}, the median value at each frequency is then used in the final analysis~\cite{Chatziioannou:2019zvs, Biscoveanu:2020kat}.

A \ac{GW} signal emitted from a binary containing two \acp{BH} can be fully characterized by $\vec{\vartheta}$ containing a set of fifteen parameters, as introduced in Sec.~\ref{subsec:searches}, if the binary orbit is assumed to have negligible eccentricity.\footnote{See Table E1 in \cite{Romero-Shaw:2020owr} for precise definitions of all parameters used.}
The mass and spin of the post-merger remnant \ac{BH}, together with the peak \ac{GW} luminosity, are calculated from the initial binary parameters using fits to \ac{NR}~\cite{Abbott:2017vtc, Hofmann:2016yih, Jimenez-Forteza:2016oae, Healy:2016lce, JohnsonMcDaniel:2016, Keitel:2016krm}.

For binaries expected to contain at least one \ac{NS}, the time-evolution of the binary orbit is modified by the presence of matter and quantified in terms of the dimensionless quadrupole tidal deformability $\Lambda_{1,2}$, adding one more parameter for each \ac{NS}.
In addition to the quadrupole tidal effects, other matter effects are parameterized in terms of $\Lambda_{1,2}$ using \ac{EOS}-insensitive relations~\cite{Yagi:2016bkt}.
When a \ac{GW} event is assumed to contain one or more neutron star, we do not report final masses or spins for the remnant object.

\subsection{Waveform models}
\label{ss:waveforms}

The binary properties of the observed \ac{GW} events are characterized through matching against a set of waveform models.
For the events identified as \acp{BBH}, with both components inferred to have masses above $3 M_\odot$, we use the independently developed IMRPhenomXPHM~\cite{Pratten:2020ceb, Pratten:2020fqn, Garcia-Quiros:2020qpx, Garcia-Quiros:2020qlt} and SEOBNRv4PHM~\cite{Ossokine:2020kjp, Babak:2016tgq, Pan:2013rra} models.
Both waveform models capture effects from spin-induced precession of the binary orbit, as well as contributions from both the dominant and sub-dominant multipole moments of the emitted gravitational radiation.

IMRPhenomXPHM~\cite{Pratten:2020ceb} describes the \ac{GW} signal from precessing non-eccentric \acp{BBH} and is part of the fourth generation of phenomenological frequency domain models. 
Precession is implemented via a twisting-up procedure, as for its predecessors IMRPhenomPv2~\cite{Hannam:2013oca, PhenomPv2note} and IMRPhenomPv3HM~\cite{Khan:2018fmp, Khan:2019kot}. For this, an aligned-spin model defined in the co-precessing frame is mapped through a suitable frame rotation to approximate the multipolar emission of a precessing system in the inertial frame. 
The stationary phase approximation is used to obtain closed form expressions in the frequency domain~\cite{Ramos-Buades:2020noq}. 
The description for the precession dynamics is derived using a multiple scale analysis of the \PN equations of motion~\cite{Chatziioannou:2017tdw}. 
The underlying aligned spin model for IMRPhenomXPHM is IMRPhenomXHM~\cite{Pratten:2020fqn, Garcia-Quiros:2020qpx, Garcia-Quiros:2020qlt}, which calibrates the $(\ell,\, | m |) = (2,2), (2,1), (3,2), (3,3)$ and (4,4) spherical harmonic modes to hybrid waveforms constructed from \ac{NR} waveforms and information from the \ac{PN} and \ac{EOB} descriptions for the inspiral. 
IMRPhenomXHM represents the amplitudes and phases of spherical or spheroidal harmonic modes in terms of piecewise closed form expressions, with coefficients that vary across the compact binary parameter space, which results in extreme compression of the waveform information and computational efficiency.

SEOBNRv4PHM comes from another waveform family that is primarily based on the \ac{EOB} formalism where the relativistic two-body problem is mapped to motion of a single body in an effective metric. 
In this framework, analytical information from several sources, such as \PN theory and the test-particle limit, is combined in a resummed form. 
This is complemented with insights from \ac{NR} simulations that accurately model the strong-field regime and incorporated into the \ac{EOB} waveforms via a calibration procedure. 
We use the SEOBNRv4PHM~\cite{Ossokine:2020kjp, Babak:2016tgq, Pan:2013rra} model, which includes precession and modes beyond the dominant quadrupole. 
This model is based on the aligned-spin model SEOBNRv4HM~\cite{Cotesta:2018fcv} and is calibrated to \ac{NR} in that regime. 
It features full two-spin treatment of the precession equations and relies on a twisting-up procedure to map aligned spin waveforms in the co-precessing frame to the precessing waveforms in the inertial frame~\cite{Babak:2016tgq, Pan:2013rra}.

For \NAME{GW190917B_o3afin}, the less massive component is indicated to lie below $3 M_\odot$ and hence to have a strong likelihood of being a \ac{NS} instead of a \ac{BH}.
Following the discussion for \NAME{GW190814H_o3afin}{}~\cite{GW190814A}, the nature of the less massive compact object in \NAME{GW190917B_o3afin}{} cannot be discerned from the \ac{GW} data at present.
This is primarily dependent on the unequal masses~\cite{Foucart:2013psa, Kumar:2016zlj, Huang:2020pba} which will lead the merger of the binary to occur before an eventual \ac{NS} component could have been tidally disrupted for any realistic \ac{NS} \ac{EOS}~\cite{Foucart:2013psa}.
The lack of an observable \ac{NS} disruption thus removes the potential for the observed signal to contain any additional information above a point-particle baseline.
For this reason, we present results for \NAME{GW190917B_o3afin}{} and \NAME{GW190814H_o3afin}{} based on the \ac{BBH} waveform models discussed above.

For \NAME{GW190425B_o3afin}{}, the only O3a event in this catalog classified as a \BNS, we follow the analysis presented in~\cite{Abbott:2020uma, Abbott:2020niy} and report findings using the IMRPhenomP\_NRTidal waveform model~\cite{Dietrich:2017aum, Dietrich:2018uni} which is based upon the \ac{BBH} model IMRPhenomPv2~\cite{Hannam:2013oca, Husa:2015iqa, Khan:2015jqa} with the addition of \ac{EOS} dependent self-spin effects and contributions from tidal interactions tuned against \ac{NR} and tidal \ac{EOB} models.
In order to reduce computational cost for the analysis of \NAME{GW190425B_o3afin}{}, a reduced-order-quadrature method was applied to the IMRPhenomP\_NRTidal model used~\cite{Smith:2016qas, amanda_baylor_2019_3478659}.

\subsection{Sampling methods}
\label{ss:sampling}

To represent the continuous posterior probability density functions in $\vec{\vartheta}$, we draw discrete samples from those distributions using three different methods.
For analyses using IMRPhenomXPHM and IMRPhenomP\_NRTidal we use the \BILBY inference package~\cite{Ashton:2018jfp, Romero-Shaw:2020owr}, together with the nested sampling~\cite{skilling2006} method implemented in the Dynesty sampler~\cite{Speagle:2020}, or the Markov-chain Monte Carlo sampler implemented in the \LALINFERENCE package~\cite{Rover:2006ni, vanderSluys:2008qx, Veitch:2014wba, lalsuite-software}. 
For analyses using SEOBNRv4PHM, we use the \RIFT package~\cite{Pankow:2015cra, Lange:2017wki, Wysocki:2019grj} which, due to a hybrid exploration of the parameter space split into intrinsic (masses and spins) and extrinsic parameters, is better suited for use with this more computationally expensive waveform model.
The results from both analyses are collected and presented in a common format using the PESummary package~\cite{Hoy:2020vys, pesummary_zenodo}.

\subsection{Priors} 
\label{ss:priors}

The prior probability on $\vec{\vartheta}$ is defined similar to GWTC-2~\cite{Abbott:2020niy} as uniform in spin magnitudes and redshifted component masses (specified in the geocenter rest frame), and isotropic in spin orientations, sky location and orientation of the binary orbit.
We also assume uncorrelated and uniform prior probabilities for the tidal deformability parameters of the \acp{NS} in \NAME{GW190425B_o3afin}{}.
The prior on the luminosity distance follows a distribution uniform in comoving volume, using a flat $\mathrm{\Lambda{}CDM}$ cosmology with Hubble constant $H_0=67.90~\mathrm{km\,s^{-1}\,Mpc^{-1}}$ and matter density $\Omega_{\rm m}=0.3065$~\cite{Ade:2015xua}.
Masses reported in Sec.~\ref{ss:parameter-estimation-results} are defined in the rest frame of the original binary, and computed by dividing the redshifted masses by $(1+\redshift)$, with \redshift calculated from the same cosmological model.
For \NAME{GW190425B_o3afin}{} we perform two separate analyses, differing in the spin magnitudes they allow with a low-spin ($|\vec{\chi}_1| < 0.05$) and a high-spin ($|\vec{\chi}_1| < 0.89$), consistent with the choices made in GWTC-2~\cite{Abbott:2020niy} for this binary.

All analyses account for uncertainties in the reported strain calibration~\cite{Cahillane2017, Sun:2020wke}.
The calibration uncertainties are described as frequency-dependent splines, defined separately for the strain amplitude and phase~\cite{SplineCalTechNote}.
The coefficients at the spline nodes are allowed to vary alongside the binary signal parameters according to a Gaussian prior distribution set by the measured uncertainty at each node~\cite{TheLIGOScientific:2016wfe}.
For analyses performed with the \LALINFERENCE or \BILBY inference packages, calibration uncertainties are marginalized over through direct sampling of the spline coefficients whereas \RIFT analyses implement a likelihood reweighting method through importance sampling over an initial analysis where perfect calibration is assumed~\cite{Payne:2020myg}.

\subsection{Source properties}
\label{ss:parameter-estimation-results}

In this subsection we report the inferred source properties of the \NUMADDCANDIDATES{} new events reported in Table~\ref{tab:events}.
The source properties for the \ac{BBH} events from the first and second observation runs, reported in GWTC-1~\cite{LIGOScientific:2018mvr}, together with the remaining \GWTCTWOEVENTS{} events from Table~\ref{tab:events} are reported in Appendix~\ref{app:parameter-estimation-appendix}.
For the vast majority of the events reported both in this section and in Appendix~\ref{app:parameter-estimation-appendix}, the quoted source properties are taken from a set of posterior samples constructed from the two IMRPhenomXPHM and SEOBNRv4PHM analyses with each given equal weight.
For a subset of events 
(\OLDEVENTSNAME{GW151226_o3afin}, 
\NAME{GW190413A_o3afin},
\NAME{GW190413E_o3afin},
\NAME{GW190421I_o3afin},
\NAME{GW190426N_o3afin},
\NAME{GW190521B_o3afin},
\NAME{GW190602E_o3afin},
\NAME{GW190719H_o3afin},
\NAME{GW190725F_o3afin},
\NAME{GW190803B_o3afin},
\NAME{GW190814H_o3afin}, 
\NAME{GW190828A_o3afin},
\NAME{GW190828B_o3afin},
\NAME{GW190917B_o3afin}, 
\NAME{GW190926C_o3afin}{}
and \NAME{GW190929B_o3afin}) 
the respective SEOBNRv4PHM analyses were incomplete at the time of journal submission, hence we report results from the IMRPhenomXPHM only for these events.

\begin{figure*}[t]
\includegraphics[width=\textwidth]{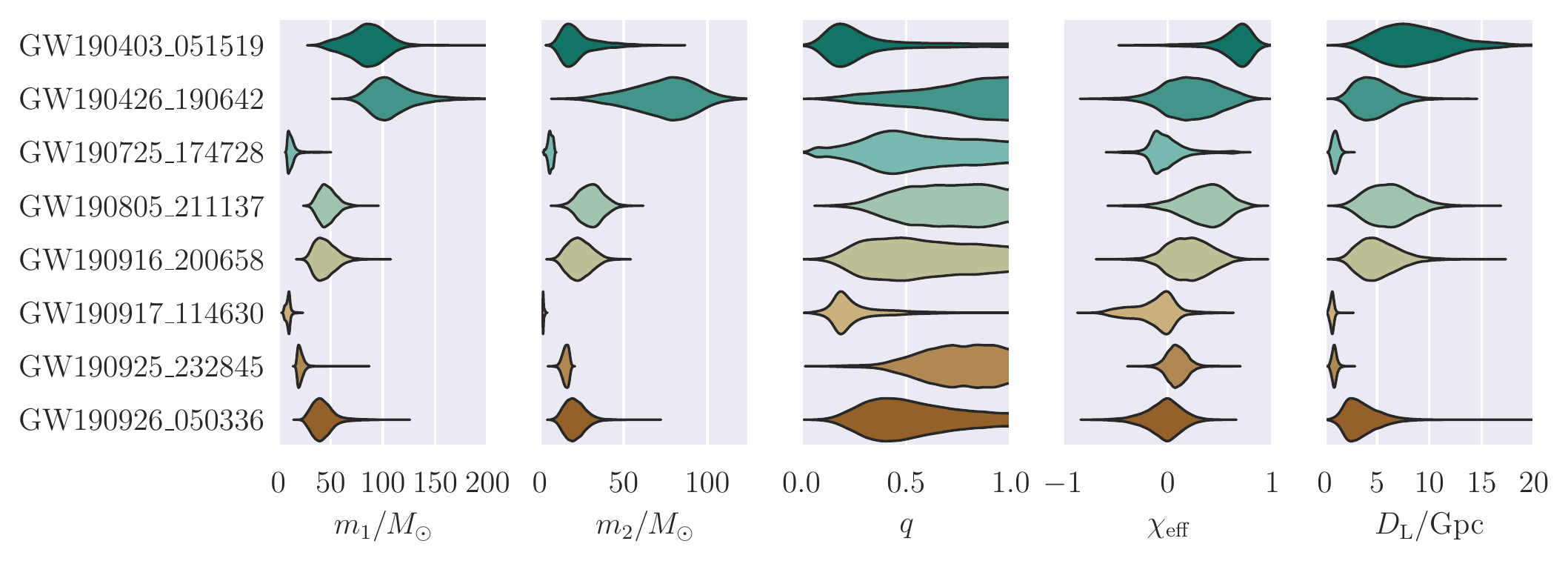}
\caption{Marginal posterior distributions on the primary mass \massone, secondary mass \masstwo, mass ratio \massratio, effective inspiral spin \chieff and luminosity distance \DL for the \NUMADDCANDIDATES{} events that are new to this catalog with $\pastro > 0.5$, highlighted in bold in Table~\ref{tab:events}.
The vertical span for each region is constructed to be proportional to the one-dimensional marginal posterior at a given parameter value for the corresponding event.
The posterior distributions are also represented numerically in terms of their one-dimensional median and $90\%$ credible intervals in Table~\ref{tab:PE_new}.}
\label{fig:PE_1d_posteriors_newEvents}
\end{figure*}

A selection of the one-dimensional marginal posterior distributions are shown in Fig.~\ref{fig:PE_1d_posteriors_newEvents}, with two-dimensional projections on the \Mtot--\massratio and \Mc--\chieff planes in Fig.~\ref{fig:PE_2d_posteriors_mtot_q} and Fig.~\ref{fig:PE_2d_posteriors_mc_chieff} respectively.
A more detailed set of results are presented in Table~\ref{tab:PE_new} in the form of median and $90\%$ credible intervals for the one-dimensional marginal posterior distributions for all \NUMADDCANDIDATES{} events.
The complete multi-dimensional posterior distributions are available as part of the public data release accompanying this paper~\cite{gwtc2p1_pe_data}, as detailed further in Sec.~\ref{sec:conclusion}.

\begin{PE_table}
\begin{table*}
    \centering
    {\rowcolors{1}{}{lightgray}
\begin{tabularx}{\textwidth}{@{\extracolsep{\fill}}p{2.85cm} U U U U V U U U U W U}
Event & $\underset{\displaystyle (M_\odot)}{M}$ & $\underset{\displaystyle (M_\odot)}{\mathcal{M}}$ & $\underset{\displaystyle (M_\odot)}{m_1}$ & $\underset{\displaystyle (M_\odot)}{m_2}$ & $\chi_{{\rm eff}}$ & $\underset{\displaystyle ({\rm Gpc})}{D_\mathrm{L}}$ & $z$ & $\underset{\displaystyle (M_\odot)}{M_\mathrm{f}}$ & $\chi_\mathrm{f}$ & $\underset{\displaystyle (\mathrm{deg}^2)}{\Delta\Omega}$ & $\mathrm{SNR}$\\ \hline
\NAME{GW190403B_o3afin} & $\totalmasssourcemed{GW190403B_o3afin}_{ -\totalmasssourceminus{GW190403B_o3afin} }^{ +\totalmasssourceplus{GW190403B_o3afin} }$ & $\chirpmasssourcemed{GW190403B_o3afin}_{ -\chirpmasssourceminus{GW190403B_o3afin} }^{ +\chirpmasssourceplus{GW190403B_o3afin} }$ & $\massonesourcemed{GW190403B_o3afin}_{ -\massonesourceminus{GW190403B_o3afin} }^{ +\massonesourceplus{GW190403B_o3afin} }$ & $\masstwosourcemed{GW190403B_o3afin}_{ -\masstwosourceminus{GW190403B_o3afin} }^{ +\masstwosourceplus{GW190403B_o3afin} }$ & $\chieffinfinityonlyprecavgmed{GW190403B_o3afin}_{ -\chieffinfinityonlyprecavgminus{GW190403B_o3afin} }^{ +\chieffinfinityonlyprecavgplus{GW190403B_o3afin} }$ & $\luminositydistancemed{GW190403B_o3afin}_{ -\luminositydistanceminus{GW190403B_o3afin} }^{ +\luminositydistanceplus{GW190403B_o3afin} }$ & $\redshiftmed{GW190403B_o3afin}_{ -\redshiftminus{GW190403B_o3afin} }^{ +\redshiftplus{GW190403B_o3afin} }$ & $\finalmasssourcemed{GW190403B_o3afin}_{ -\finalmasssourceminus{GW190403B_o3afin} }^{ +\finalmasssourceplus{GW190403B_o3afin} }$ & $\finalspinmed{GW190403B_o3afin}_{ -\finalspinminus{GW190403B_o3afin} }^{ +\finalspinplus{GW190403B_o3afin} }$ & $\skyarea{GW190403B_o3afin}$ & $\networkmatchedfiltersnrIMRPmed{GW190403B_o3afin}_{ -\networkmatchedfiltersnrIMRPminus{GW190403B_o3afin} }^{ +\networkmatchedfiltersnrIMRPplus{GW190403B_o3afin} }$\\
\NAME{GW190426N_o3afin} & $\totalmasssourcemed{GW190426N_o3afin}_{ -\totalmasssourceminus{GW190426N_o3afin} }^{ +\totalmasssourceplus{GW190426N_o3afin} }$ & $\chirpmasssourcemed{GW190426N_o3afin}_{ -\chirpmasssourceminus{GW190426N_o3afin} }^{ +\chirpmasssourceplus{GW190426N_o3afin} }$ & $\massonesourcemed{GW190426N_o3afin}_{ -\massonesourceminus{GW190426N_o3afin} }^{ +\massonesourceplus{GW190426N_o3afin} }$ & $\masstwosourcemed{GW190426N_o3afin}_{ -\masstwosourceminus{GW190426N_o3afin} }^{ +\masstwosourceplus{GW190426N_o3afin} }$ & $\chieffinfinityonlyprecavgmed{GW190426N_o3afin}_{ -\chieffinfinityonlyprecavgminus{GW190426N_o3afin} }^{ +\chieffinfinityonlyprecavgplus{GW190426N_o3afin} }$ & $\luminositydistancemed{GW190426N_o3afin}_{ -\luminositydistanceminus{GW190426N_o3afin} }^{ +\luminositydistanceplus{GW190426N_o3afin} }$ & $\redshiftmed{GW190426N_o3afin}_{ -\redshiftminus{GW190426N_o3afin} }^{ +\redshiftplus{GW190426N_o3afin} }$ & $\finalmasssourcemed{GW190426N_o3afin}_{ -\finalmasssourceminus{GW190426N_o3afin} }^{ +\finalmasssourceplus{GW190426N_o3afin} }$ & $\finalspinmed{GW190426N_o3afin}_{ -\finalspinminus{GW190426N_o3afin} }^{ +\finalspinplus{GW190426N_o3afin} }$ & $\skyarea{GW190426N_o3afin}$ & $\networkmatchedfiltersnrIMRPmed{GW190426N_o3afin}_{ -\networkmatchedfiltersnrIMRPminus{GW190426N_o3afin} }^{ +\networkmatchedfiltersnrIMRPplus{GW190426N_o3afin} }$\\
\NAME{GW190725F_o3afin} & $\totalmasssourcemed{GW190725F_o3afin}_{ -\totalmasssourceminus{GW190725F_o3afin} }^{ +\totalmasssourceplus{GW190725F_o3afin} }$ & $\chirpmasssourcemed{GW190725F_o3afin}_{ -\chirpmasssourceminus{GW190725F_o3afin} }^{ +\chirpmasssourceplus{GW190725F_o3afin} }$ & $\massonesourcemed{GW190725F_o3afin}_{ -\massonesourceminus{GW190725F_o3afin} }^{ +\massonesourceplus{GW190725F_o3afin} }$ & $\masstwosourcemed{GW190725F_o3afin}_{ -\masstwosourceminus{GW190725F_o3afin} }^{ +\masstwosourceplus{GW190725F_o3afin} }$ & $\chieffinfinityonlyprecavgmed{GW190725F_o3afin}_{ -\chieffinfinityonlyprecavgminus{GW190725F_o3afin} }^{ +\chieffinfinityonlyprecavgplus{GW190725F_o3afin} }$ & $\luminositydistancemed{GW190725F_o3afin}_{ -\luminositydistanceminus{GW190725F_o3afin} }^{ +\luminositydistanceplus{GW190725F_o3afin} }$ & $\redshiftmed{GW190725F_o3afin}_{ -\redshiftminus{GW190725F_o3afin} }^{ +\redshiftplus{GW190725F_o3afin} }$ & $\finalmasssourcemed{GW190725F_o3afin}_{ -\finalmasssourceminus{GW190725F_o3afin} }^{ +\finalmasssourceplus{GW190725F_o3afin} }$ & $\finalspinmed{GW190725F_o3afin}_{ -\finalspinminus{GW190725F_o3afin} }^{ +\finalspinplus{GW190725F_o3afin} }$ & $\skyarea{GW190725F_o3afin}$ & $\networkmatchedfiltersnrIMRPmed{GW190725F_o3afin}_{ -\networkmatchedfiltersnrIMRPminus{GW190725F_o3afin} }^{ +\networkmatchedfiltersnrIMRPplus{GW190725F_o3afin} }$\\
\NAME{GW190805J_o3afin} & $\totalmasssourcemed{GW190805J_o3afin}_{ -\totalmasssourceminus{GW190805J_o3afin} }^{ +\totalmasssourceplus{GW190805J_o3afin} }$ & $\chirpmasssourcemed{GW190805J_o3afin}_{ -\chirpmasssourceminus{GW190805J_o3afin} }^{ +\chirpmasssourceplus{GW190805J_o3afin} }$ & $\massonesourcemed{GW190805J_o3afin}_{ -\massonesourceminus{GW190805J_o3afin} }^{ +\massonesourceplus{GW190805J_o3afin} }$ & $\masstwosourcemed{GW190805J_o3afin}_{ -\masstwosourceminus{GW190805J_o3afin} }^{ +\masstwosourceplus{GW190805J_o3afin} }$ & $\chieffinfinityonlyprecavgmed{GW190805J_o3afin}_{ -\chieffinfinityonlyprecavgminus{GW190805J_o3afin} }^{ +\chieffinfinityonlyprecavgplus{GW190805J_o3afin} }$ & $\luminositydistancemed{GW190805J_o3afin}_{ -\luminositydistanceminus{GW190805J_o3afin} }^{ +\luminositydistanceplus{GW190805J_o3afin} }$ & $\redshiftmed{GW190805J_o3afin}_{ -\redshiftminus{GW190805J_o3afin} }^{ +\redshiftplus{GW190805J_o3afin} }$ & $\finalmasssourcemed{GW190805J_o3afin}_{ -\finalmasssourceminus{GW190805J_o3afin} }^{ +\finalmasssourceplus{GW190805J_o3afin} }$ & $\finalspinmed{GW190805J_o3afin}_{ -\finalspinminus{GW190805J_o3afin} }^{ +\finalspinplus{GW190805J_o3afin} }$ & $\skyarea{GW190805J_o3afin}$ & $\networkmatchedfiltersnrIMRPmed{GW190805J_o3afin}_{ -\networkmatchedfiltersnrIMRPminus{GW190805J_o3afin} }^{ +\networkmatchedfiltersnrIMRPplus{GW190805J_o3afin} }$\\
\NAME{GW190916K_o3afin} & $\totalmasssourcemed{GW190916K_o3afin}_{ -\totalmasssourceminus{GW190916K_o3afin} }^{ +\totalmasssourceplus{GW190916K_o3afin} }$ & $\chirpmasssourcemed{GW190916K_o3afin}_{ -\chirpmasssourceminus{GW190916K_o3afin} }^{ +\chirpmasssourceplus{GW190916K_o3afin} }$ & $\massonesourcemed{GW190916K_o3afin}_{ -\massonesourceminus{GW190916K_o3afin} }^{ +\massonesourceplus{GW190916K_o3afin} }$ & $\masstwosourcemed{GW190916K_o3afin}_{ -\masstwosourceminus{GW190916K_o3afin} }^{ +\masstwosourceplus{GW190916K_o3afin} }$ & $\chieffinfinityonlyprecavgmed{GW190916K_o3afin}_{ -\chieffinfinityonlyprecavgminus{GW190916K_o3afin} }^{ +\chieffinfinityonlyprecavgplus{GW190916K_o3afin} }$ & $\luminositydistancemed{GW190916K_o3afin}_{ -\luminositydistanceminus{GW190916K_o3afin} }^{ +\luminositydistanceplus{GW190916K_o3afin} }$ & $\redshiftmed{GW190916K_o3afin}_{ -\redshiftminus{GW190916K_o3afin} }^{ +\redshiftplus{GW190916K_o3afin} }$ & $\finalmasssourcemed{GW190916K_o3afin}_{ -\finalmasssourceminus{GW190916K_o3afin} }^{ +\finalmasssourceplus{GW190916K_o3afin} }$ & $\finalspinmed{GW190916K_o3afin}_{ -\finalspinminus{GW190916K_o3afin} }^{ +\finalspinplus{GW190916K_o3afin} }$ & $\skyarea{GW190916K_o3afin}$ & $\networkmatchedfiltersnrIMRPmed{GW190916K_o3afin}_{ -\networkmatchedfiltersnrIMRPminus{GW190916K_o3afin} }^{ +\networkmatchedfiltersnrIMRPplus{GW190916K_o3afin} }$\\
\NAME{GW190917B_o3afin} & $\totalmasssourcemed{GW190917B_o3afin}_{ -\totalmasssourceminus{GW190917B_o3afin} }^{ +\totalmasssourceplus{GW190917B_o3afin} }$ & $\chirpmasssourcemed{GW190917B_o3afin}_{ -\chirpmasssourceminus{GW190917B_o3afin} }^{ +\chirpmasssourceplus{GW190917B_o3afin} }$ & $\massonesourcemed{GW190917B_o3afin}_{ -\massonesourceminus{GW190917B_o3afin} }^{ +\massonesourceplus{GW190917B_o3afin} }$ & $\masstwosourcemed{GW190917B_o3afin}_{ -\masstwosourceminus{GW190917B_o3afin} }^{ +\masstwosourceplus{GW190917B_o3afin} }$ & $\chieffinfinityonlyprecavgmed{GW190917B_o3afin}_{ -\chieffinfinityonlyprecavgminus{GW190917B_o3afin} }^{ +\chieffinfinityonlyprecavgplus{GW190917B_o3afin} }$ & $\luminositydistancemed{GW190917B_o3afin}_{ -\luminositydistanceminus{GW190917B_o3afin} }^{ +\luminositydistanceplus{GW190917B_o3afin} }$ & $\redshiftmed{GW190917B_o3afin}_{ -\redshiftminus{GW190917B_o3afin} }^{ +\redshiftplus{GW190917B_o3afin} }$ & $\finalmasssourcemed{GW190917B_o3afin}_{ -\finalmasssourceminus{GW190917B_o3afin} }^{ +\finalmasssourceplus{GW190917B_o3afin} }$ & $\finalspinmed{GW190917B_o3afin}_{ -\finalspinminus{GW190917B_o3afin} }^{ +\finalspinplus{GW190917B_o3afin} }$ & $\skyarea{GW190917B_o3afin}$ & $\networkmatchedfiltersnrIMRPmed{GW190917B_o3afin}_{ -\networkmatchedfiltersnrIMRPminus{GW190917B_o3afin} }^{ +\networkmatchedfiltersnrIMRPplus{GW190917B_o3afin} }$\\
\NAME{GW190925J_o3afin} & $\totalmasssourcemed{GW190925J_o3afin}_{ -\totalmasssourceminus{GW190925J_o3afin} }^{ +\totalmasssourceplus{GW190925J_o3afin} }$ & $\chirpmasssourcemed{GW190925J_o3afin}_{ -\chirpmasssourceminus{GW190925J_o3afin} }^{ +\chirpmasssourceplus{GW190925J_o3afin} }$ & $\massonesourcemed{GW190925J_o3afin}_{ -\massonesourceminus{GW190925J_o3afin} }^{ +\massonesourceplus{GW190925J_o3afin} }$ & $\masstwosourcemed{GW190925J_o3afin}_{ -\masstwosourceminus{GW190925J_o3afin} }^{ +\masstwosourceplus{GW190925J_o3afin} }$ & $\chieffinfinityonlyprecavgmed{GW190925J_o3afin}_{ -\chieffinfinityonlyprecavgminus{GW190925J_o3afin} }^{ +\chieffinfinityonlyprecavgplus{GW190925J_o3afin} }$ & $\luminositydistancemed{GW190925J_o3afin}_{ -\luminositydistanceminus{GW190925J_o3afin} }^{ +\luminositydistanceplus{GW190925J_o3afin} }$ & $\redshiftmed{GW190925J_o3afin}_{ -\redshiftminus{GW190925J_o3afin} }^{ +\redshiftplus{GW190925J_o3afin} }$ & $\finalmasssourcemed{GW190925J_o3afin}_{ -\finalmasssourceminus{GW190925J_o3afin} }^{ +\finalmasssourceplus{GW190925J_o3afin} }$ & $\finalspinmed{GW190925J_o3afin}_{ -\finalspinminus{GW190925J_o3afin} }^{ +\finalspinplus{GW190925J_o3afin} }$ & $\skyarea{GW190925J_o3afin}$ & $\networkmatchedfiltersnrIMRPmed{GW190925J_o3afin}_{ -\networkmatchedfiltersnrIMRPminus{GW190925J_o3afin} }^{ +\networkmatchedfiltersnrIMRPplus{GW190925J_o3afin} }$\\
\NAME{GW190926C_o3afin} & $\totalmasssourcemed{GW190926C_o3afin}_{ -\totalmasssourceminus{GW190926C_o3afin} }^{ +\totalmasssourceplus{GW190926C_o3afin} }$ & $\chirpmasssourcemed{GW190926C_o3afin}_{ -\chirpmasssourceminus{GW190926C_o3afin} }^{ +\chirpmasssourceplus{GW190926C_o3afin} }$ & $\massonesourcemed{GW190926C_o3afin}_{ -\massonesourceminus{GW190926C_o3afin} }^{ +\massonesourceplus{GW190926C_o3afin} }$ & $\masstwosourcemed{GW190926C_o3afin}_{ -\masstwosourceminus{GW190926C_o3afin} }^{ +\masstwosourceplus{GW190926C_o3afin} }$ & $\chieffinfinityonlyprecavgmed{GW190926C_o3afin}_{ -\chieffinfinityonlyprecavgminus{GW190926C_o3afin} }^{ +\chieffinfinityonlyprecavgplus{GW190926C_o3afin} }$ & $\luminositydistancemed{GW190926C_o3afin}_{ -\luminositydistanceminus{GW190926C_o3afin} }^{ +\luminositydistanceplus{GW190926C_o3afin} }$ & $\redshiftmed{GW190926C_o3afin}_{ -\redshiftminus{GW190926C_o3afin} }^{ +\redshiftplus{GW190926C_o3afin} }$ & $\finalmasssourcemed{GW190926C_o3afin}_{ -\finalmasssourceminus{GW190926C_o3afin} }^{ +\finalmasssourceplus{GW190926C_o3afin} }$ & $\finalspinmed{GW190926C_o3afin}_{ -\finalspinminus{GW190926C_o3afin} }^{ +\finalspinplus{GW190926C_o3afin} }$ & $\skyarea{GW190926C_o3afin}$ & $\networkmatchedfiltersnrIMRPmed{GW190926C_o3afin}_{ -\networkmatchedfiltersnrIMRPminus{GW190926C_o3afin} }^{ +\networkmatchedfiltersnrIMRPplus{GW190926C_o3afin} }$\\
\hline
\end{tabularx}

    }
    \caption{Median and 90\% symmetric credible intervals for the one-dimensional marginal posterior distributions on selected source parameters for the \NUMADDCANDIDATES{} events that are new to this catalog with $\pastro > 0.5$, highlighted in bold in Table~\ref{tab:events}. 
        The columns show source total mass \Mtot, chirp mass \Mc and component masses \masscomponent, dimensionless effective inspiral spin \chieff, luminosity distance \DL, redshift \redshift, final mass \Mf, final spin \chif, sky localization \skyareasymbol and the network matched-filter \ac{SNR}.
      The sky localization is the area of the 90\% credible region. 
      All quoted results are calculated from a set of posterior samples drawn with equal weight from the IMRPhenomXPHM and SEOBNRv4PHM analyses, with the exception of the \acp{SNR} that are taken from the IMRPhenomXPHM analysis alone (as \RIFT, which was used for the SEOBNRv4PHM analysis, does not output that quantity).
      Additionally, following Sec.~\ref{ss:parameter-estimation-results}, the results presented for \NAME{GW190426N_o3afin}, \NAME{GW190725F_o3afin}, \NAME{GW190917B_o3afin}{} and \NAME{GW190926C_o3afin}{} are taken from an analysis using the IMRPhenomXPHM model only.
      A subset of the one-dimensional posterior distributions are visualized in Fig.~\ref{fig:PE_1d_posteriors_newEvents}. 
      Two-dimensional projections of the $90\%$ credible regions in the \Mtot--\massratio and \Mc--\chieff planes are shown in Fig.~\ref{fig:PE_2d_posteriors_mtot_q} and Fig.~\ref{fig:PE_2d_posteriors_mc_chieff}.
      }
    \label{tab:PE_new}
\end{table*}
\end{PE_table}

\subsubsection{Masses}
\label{sss:PE_masses}

The masses inferred for the \NUMADDCANDIDATES{} events presented in this section are generally comparable to, or higher, than the binaries reported in GWTC-2~\cite{LIGOScientific:2018mvr, Abbott:2020niy}, as shown in Fig.~\ref{fig:PE_2d_posteriors_mtot_q}.
We find that the most massive \ac{BBH} in GWTC-2.1 is \NAME{GW190426N_o3afin}{} with a total mass of $\totalmasssourcemed{GW190426N_o3afin}_{ -\totalmasssourceminus{GW190426N_o3afin} }^{ +\totalmasssourceplus{GW190426N_o3afin} }\Msun$ and a remnant mass of $\finalmasssourcemed{GW190426N_o3afin}_{ -\finalmasssourceminus{GW190426N_o3afin} }^{ +\finalmasssourceplus{GW190426N_o3afin} }\Msun$; it probably supersedes the previous most massive \ac{BBH} \NAME{GW190521B_o3afin}{}\footnote{In GWTC-2, \NAME{GW190521B_o3afin}{} was inferred to have a total mass of $163.9^{ +39.2}_{ -23.5}\Msun$ and remnant mass of $156.3^{ + 36.8}_{ -22.4}\Msun$~\cite{Abbott:2020niy}} with total mass of $\totalmasssourcemed{GW190521B_o3afin}_{ -\totalmasssourceminus{GW190521B_o3afin} }^{ +\totalmasssourceplus{GW190521B_o3afin} }\Msun$ and a remnant mass of $\finalmasssourcemed{GW190521B_o3afin}_{ -\finalmasssourceminus{GW190521B_o3afin} }^{ +\finalmasssourceplus{GW190521B_o3afin} }\Msun$ as reported in Appendix~\ref{app:pe_other_events_o3a}.
Both \NAME{GW190426N_o3afin}{} and \NAME{GW190403B_o3afin}{} join \NAME{GW190519J_o3afin}, \NAME{GW190521B_o3afin}, \NAME{GW190602E_o3afin}{} and \NAME{GW190706F_o3afin}{} in a population of \acp{BBH} with over $50\%$ posterior support for total mass $\Mtot > 100 \Msun$~\cite{Abbott:2020niy}.

While the majority of the new events show a preference for mass ratios near unity, following the trend already observed in GWTC-2~\cite{LIGOScientific:2018mvr,Abbott:2020niy}, both \NAME{GW190403B_o3afin}{} and \NAME{GW190917B_o3afin}{} recover posteriors with median $\massratio \sim 1/5$ with \evoneq and \evoneqC respectively.
As shown in Fig.~\ref{fig:PE_2d_posteriors_mtot_q}, this constraint for unequal masses is robust at the $90\%$ credible level for both \NAME{GW190403B_o3afin}{} and \NAME{GW190917B_o3afin}.
Although the contour indicating the  $90\%$ credible region for \NAME{GW190403B_o3afin}{} includes support at $\massratio \sim 0$ in Fig.~\ref{fig:PE_2d_posteriors_mtot_q}, this is an artefact of the bounded kernel density estimation used to construct the contours, and for this event there are no samples at the prior boundary of $q = 0.05$.

\subsubsection{Spins}
\label{sss:PE_spins}

The best measured spin parameter for \acp{CBC} with observable inspiral signals tends to be the effective inspiral spin \chieff~\cite{Purrer:2015nkh, Vitale:2016avz, Ng:2018neg}, introduced in Eq.~\eqref{eq:chiEff}, which is approximately conserved under spin-induced precession of the binary orbit~\cite{Blanchet:2013haa, Racine:2008qv, Kesden:2014sla, Gerosa:2015tea}.
Consequently, the angles between the spin-vectors and the orbital angular momentum at a formally infinite separation are well defined~\cite{Gerosa:2015tea}.
We therefore report \chieff, as well as the spin tilt angles themselves, at this fiducial reference point of infinite binary separation, or equivalently at an infinite time before the binary merger.
The spins are evolved to infinite separation~\cite{Johnson-McDaniel:2021rvv} using a precession-averaged evolution scheme~\cite{Gerosa:2015tea,Chatziioannou:2017tdw} where the orbital angular momentum is computed using higher-order \ac{PN} expressions. 

The posterior distributions for \chieff for all \NUMADDCANDIDATES{} events are shown in Fig.~\ref{fig:PE_1d_posteriors_newEvents} and Fig.~\ref{fig:PE_2d_posteriors_mc_chieff}.
Again, the majority of the binaries are consistent with containing two non-spinning \acp{BH} with only \NAME{GW190403B_o3afin}{} and \NAME{GW190805J_o3afin}{} recovering a non-zero \chieff at $90\%$ credibility.
Both binaries report predominantly positive \chieff, further strengthening the pattern of a surplus of events with $\chieff > 0$ relative to those with $\chieff < 0$ reported in GWTC-2~\cite{Abbott:2020niy} and investigated further in a companion paper~\cite{o3apop}.

Similar to the compact objects reported in GWTC-2~\cite{LIGOScientific:2018mvr, Abbott:2020niy}, the majority of the compact-object spins reported in GWTC-2.1 have magnitudes consistent with zero.
Two of the new events show evidence for large \ac{BH} spins.
In the case of \NAME{GW190403B_o3afin}{}, \ProbSpinEitherHighA of the posterior probability lies in a region where at least one of the component spin magnitudes is above $0.8$ whereas for \NAME{GW190805J_o3afin}{} this holds for \ProbSpinEitherHighB of the posterior probability.

For binaries with very unequal masses, measurements of \chieff can translate into strong measurement constraints of \spinone, the spin magnitude of the more massive object, whose spin angular momentum dominates over the secondary.
This is the case for \NAME{GW190403B_o3afin}{}, whose primary dimensionless spin is measured to be \evonechione. 
This represents the most nearly-extremal spin observed using GWs. 
Similarly, \NAME{GW190805J_o3afin}{} is recovered with \evonechioneB and \NAME{GW190917B_o3afin}{} with \evonechioneC.
Both \NAME{GW190403B_o3afin}{} and \NAME{GW190805J_o3afin}{} are recovered as strongly preferring large \spinone, with the inferred posterior distributions railing against the extremal \ac{BH}-spin bound at $\spinone=1$.
Hence, we also report the one-sided $90\%$ lower bounds of \evonechioneOnesided for \NAME{GW190403B_o3afin}{} and \evonechioneBOnesided for \NAME{GW190805J_o3afin}.
The posterior distributions for the spin magnitudes and tilt angles for these three events are shown in Fig.~\ref{fig:spin_disks}.

\subsubsection{Three-Dimensional Localization}
\label{sss:PE_localization}

As the \NUMADDCANDIDATES{} new events are all detected at relatively modest \acp{SNR}, together with several identifications as high-mass \acp{BBH}, the inferred luminosity distances \DL are generally larger than the binaries from GWTC-2~\cite{LIGOScientific:2018mvr, Abbott:2020niy}.
\NAME{GW190403B_o3afin}{} is identified as probably the most distant event, with a recovered $\DL = \luminositydistancemed{GW190403B_o3afin}_{ -\luminositydistanceminus{GW190403B_o3afin} }^{ +\luminositydistanceplus{GW190403B_o3afin} }~\Gpc$ corresponding to a redshift $\redshift = \redshiftmed{GW190403B_o3afin}_{ -\redshiftminus{GW190403B_o3afin} }^{ +\redshiftplus{GW190403B_o3afin} }$ approximately twice as distant as the most distant events that were reported in GWTC-2~\cite{LIGOScientific:2018mvr,Abbott:2020niy} as also shown in Appendix~\ref{app:pe_other_events_o3a}.
In addition \NAME{GW190426N_o3afin}, \NAME{GW190805J_o3afin}, \NAME{GW190916K_o3afin}{} and \NAME{GW190926C_o3afin}{} all have inferred distances comparable to, or larger than, \NAME{GW190413E_o3afin}{}, further highlighting the access to the distant Universe provided in GWTC-2.1.

Another effect of the modest \ac{SNR} of the new events is their comparatively poor localization on the sky.
The best localized event is \NAME{GW190805J_o3afin}{} with a $90\%$ credible region of $\skyareasymbol = \skyarea{GW190805J_o3afin}~\mathrm{deg}^2$.
The credible intervals for the inferred distances and sky areas are shown in Table~\ref{tab:PE_new}.
The inferred localizations for all events are available as part of the accompanying data release to this paper, detailed further in Sec.~\ref{sec:conclusion}.

\subsubsection{Waveform comparisons -- Model systematics}
\label{sss:PE_waveformSystematics}

The use of both the IMRPhenomXPHM~\cite{Pratten:2020ceb, Pratten:2020fqn, Garcia-Quiros:2020qpx, Garcia-Quiros:2020qlt} and SEOBNRv4PHM~\cite{Ossokine:2020kjp, Babak:2016tgq, Pan:2013rra} models in the analyses of these events are motivated by the need to capture, and account for, potential differences in the inferred source parameters caused by the different methods used in the constructions of the models themselves.
The vast majority of the posterior distributions reported in this section are constructed by combining an equal number of samples drawn from each of the IMRPhenomXPHM and SEOBNRv4PHM analyses~\cite{TheLIGOScientific:2016wfe}.
For the majority of the \NUMADDCANDIDATES{} new events, the differences between the two single-model analyses, as well as to the combined-model results, are found to be comparable to the impact of model systematics effects identified in GWTC-2~\cite{LIGOScientific:2018mvr, Abbott:2020niy} being generally subdominant to the statistical uncertainty caused by the noisy data. 
For \NAME{GW190403B_o3afin}{}
there are, however, slight differences identified between the IMRPhenomXPHM and SEOBNRv4PHM analyses, most noticeably in the shape and structure of the marginal posterior distribution of some of the recovered mass and spin parameters.
In these cases, the differences between analyses using either the IMRPhenomXPHM or SEOBNRv4PHM models are dominating over the other systematic uncertainties of the analysis, such as the estimation of the noise \ac{PSD}.
A deeper investigation into the broader impact of these model systematic effects, and their impact on the inferred source parameters for the population of \ac{GW} events presented here, is left for a future study.

\subsubsection{Comparison to 3-OGC}
\label{sss:PE_3OGC}

Out of the \NUMADDCANDIDATES{} new events presented in this section, \NAME{GW190725F_o3afin}{}, \NAME{GW190916K_o3afin}{}, \NAME{GW190925J_o3afin}{} and \NAME{GW190926C_o3afin}{} were also independently identified and analyzed as part of 3-OGC~\cite{Nitz:2021uxj} using the PyCBC Inference package~\cite{Biwer:2018osg} and the IMRPhenomXPHM waveform model.
We compare the inferred source properties for these events as presented in 3-OGC~\cite{3OGC_PEsamples} and, to minimize potential model systematic effects, the IMRPhenomXPHM analysis performed for GWTC-2.1 presented here.
Overall, we find a broad agreement between the two analyses. 
While there are differences found in the two sets of posterior distributions, they appear consistent within expectations from the differing choices of the analysis configurations and the assumed prior distributions between the two analyses for low \ac{SNR} signals~\cite{Huang:2020ysn}.

\begin{figure*}[h]
\includegraphics[width=\textwidth]{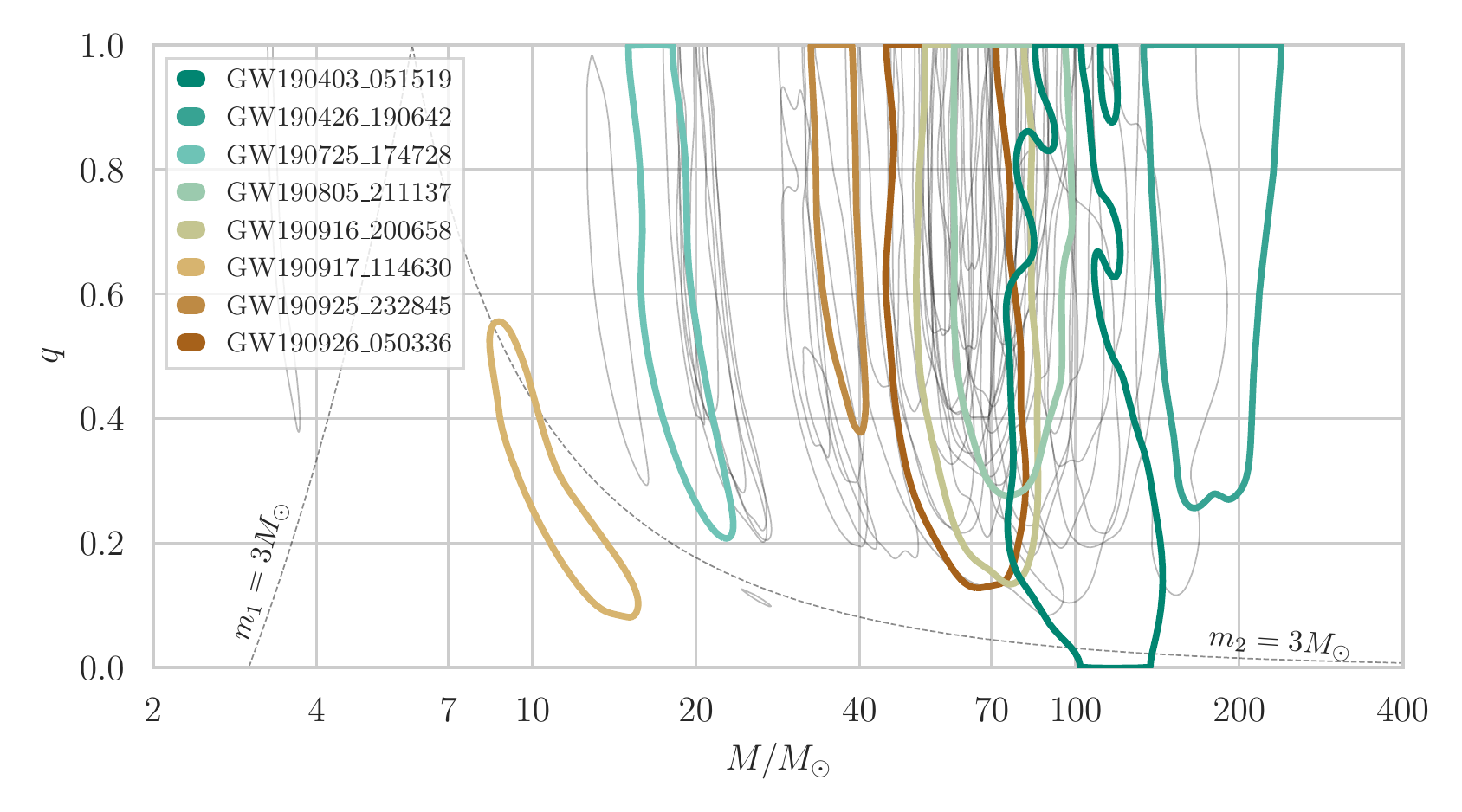}
\caption{Contours representing the $90\%$ credible regions in the total mass \Mtot and mass ratio \massratio plane for all events reported in this catalog.
The events that are new to this catalog with $\pastro > 0.5$, highlighted in bold in Table~\ref{tab:events}, are highlighted in this figure following the same color scheme used in Fig.~\ref{fig:PE_1d_posteriors_newEvents}.
The dashed lines act to separate regions where the primary and secondary binary component can have a mass below $3 \Msun$.
}
\label{fig:PE_2d_posteriors_mtot_q}
\end{figure*}

\begin{figure*}[h]
\includegraphics[width=\textwidth]{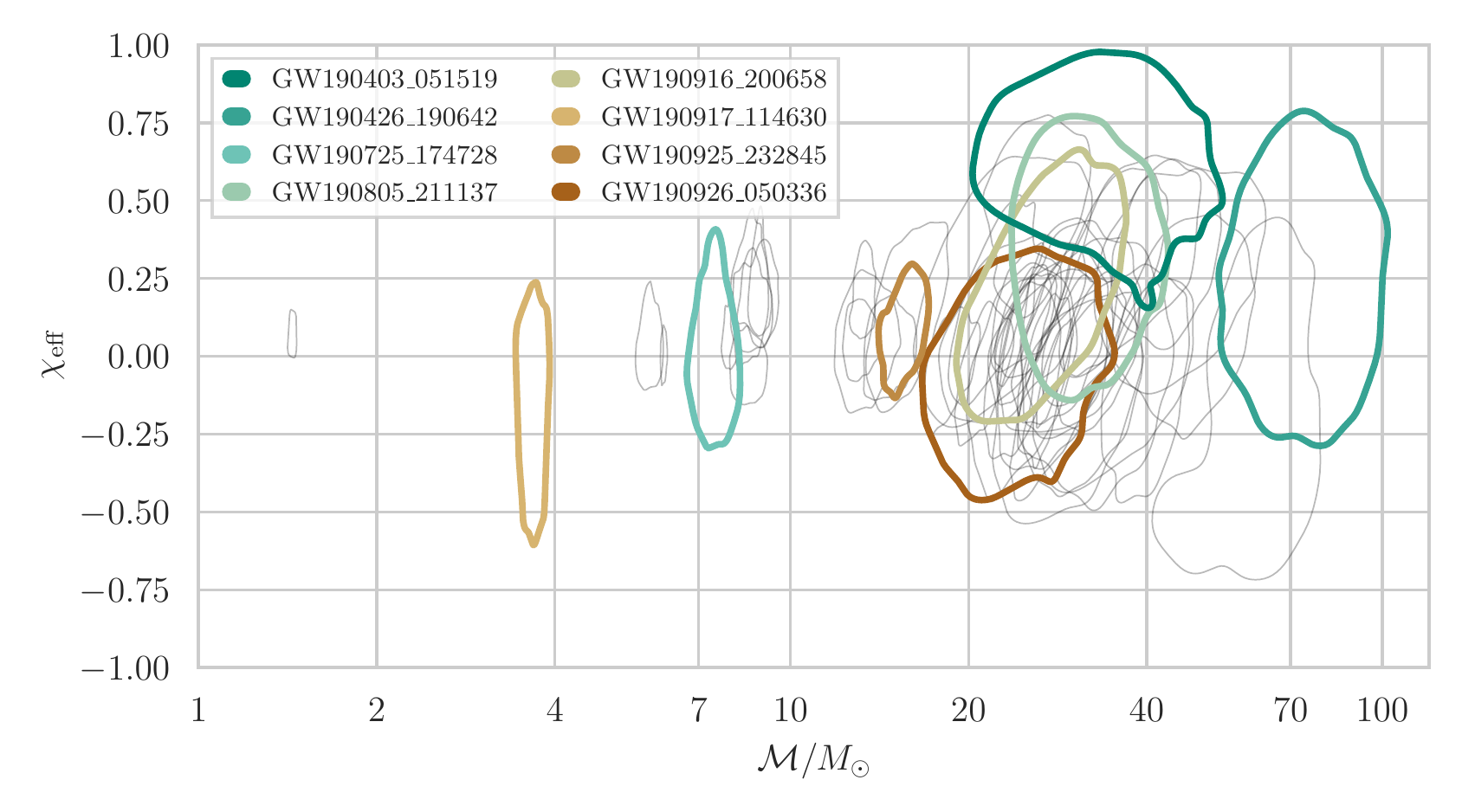}
\caption{Contours representing the $90\%$ credible regions in the plane of chirp mass \Mc and effective inspiral spin \chieff for all events reported in this catalog.
The events that are new to this catalog with $\pastro > 0.5$, highlighted in bold in Table~\ref{tab:events}, are highlighted in this figure following the same color scheme used in Fig.~\ref{fig:PE_1d_posteriors_newEvents}.
}
\label{fig:PE_2d_posteriors_mc_chieff}
\end{figure*}

\begin{figure*}[t]
\centering
\begin{subfloat}
  \centering
  \includegraphics[width=0.3\textwidth]{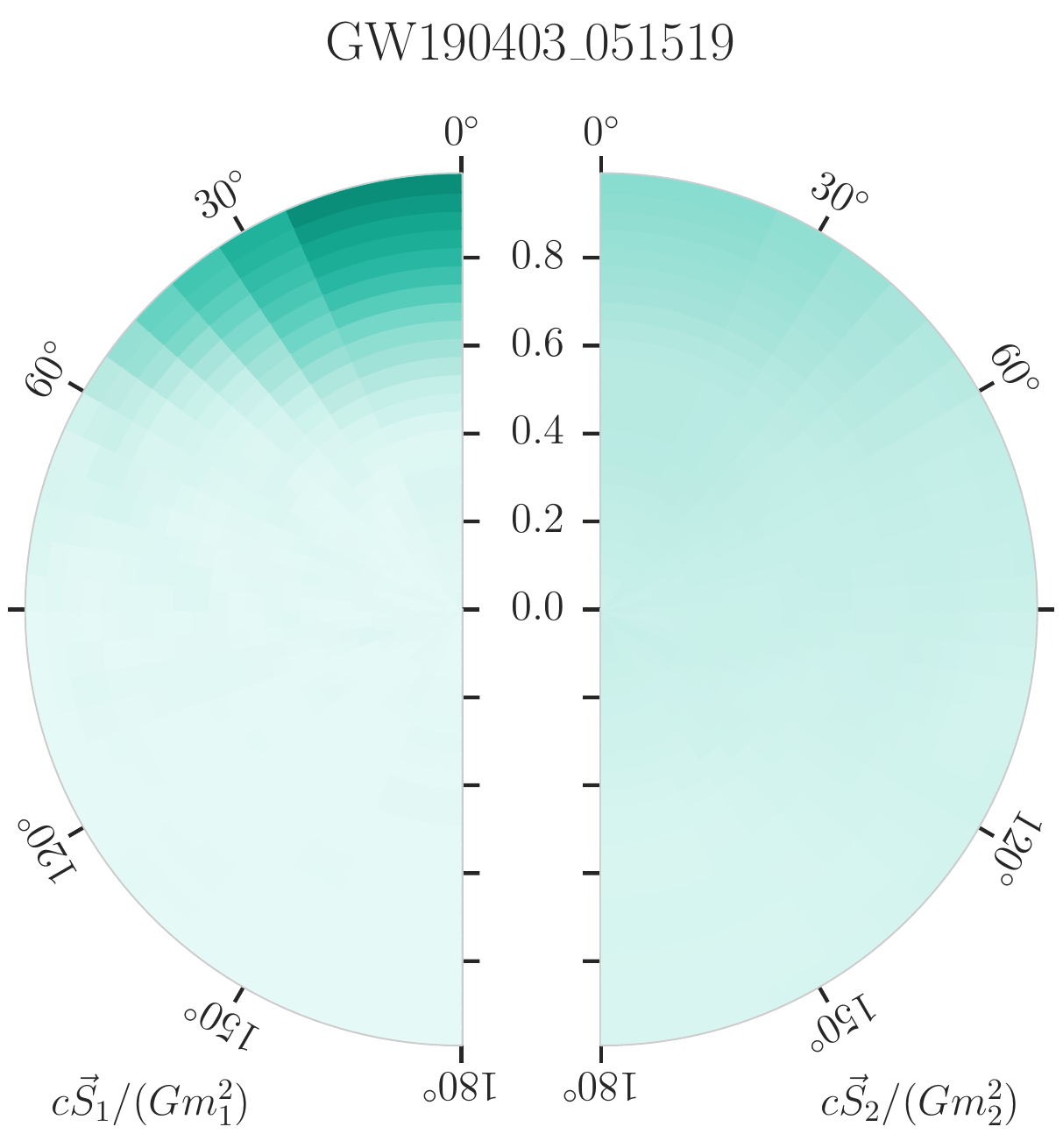}
\end{subfloat}
\hfill
\begin{subfloat}
  \centering
  \includegraphics[width=0.3\textwidth]{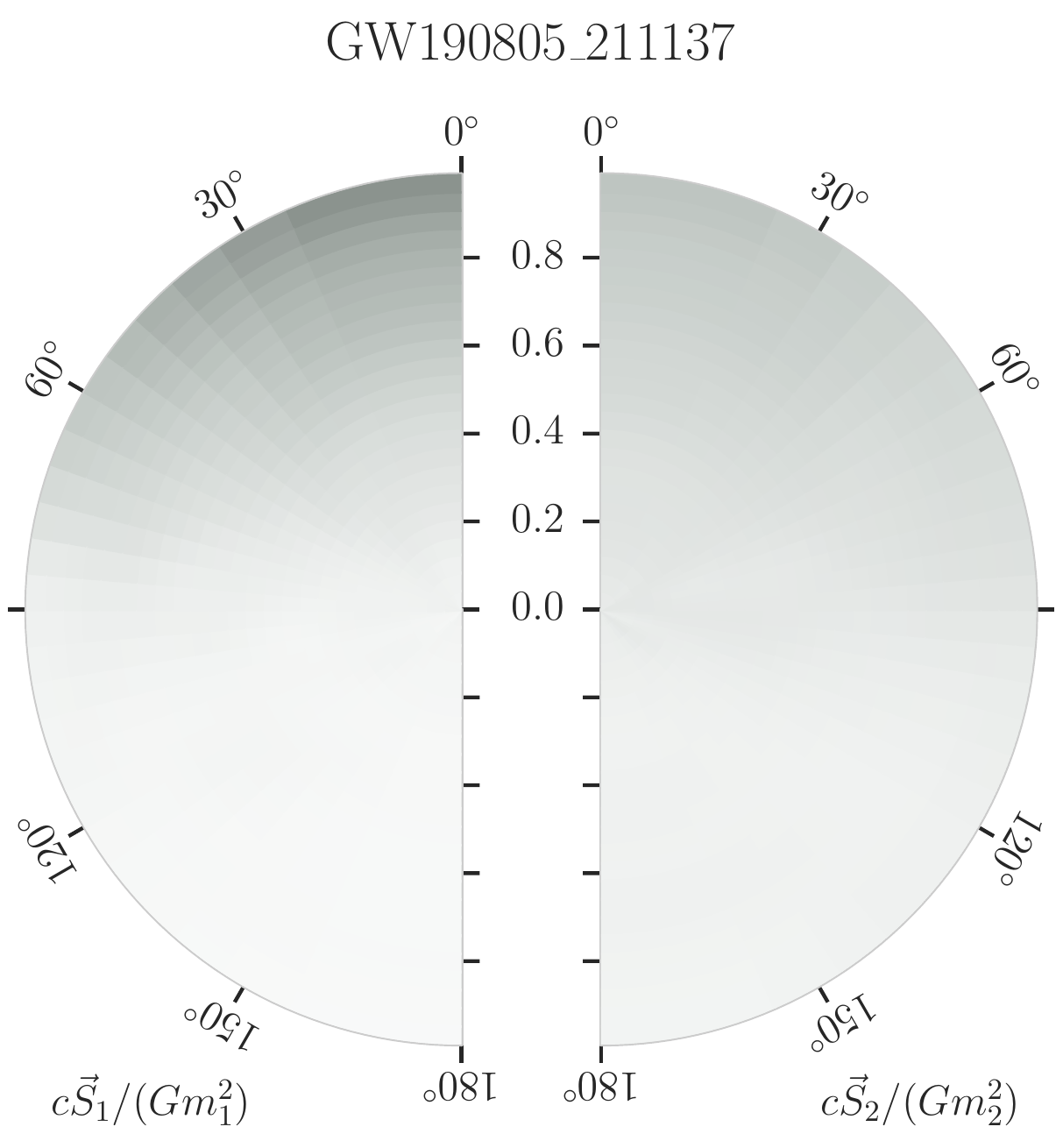}
\end{subfloat}
\hfill
\begin{subfloat}
  \centering
  \includegraphics[width=0.3\textwidth]{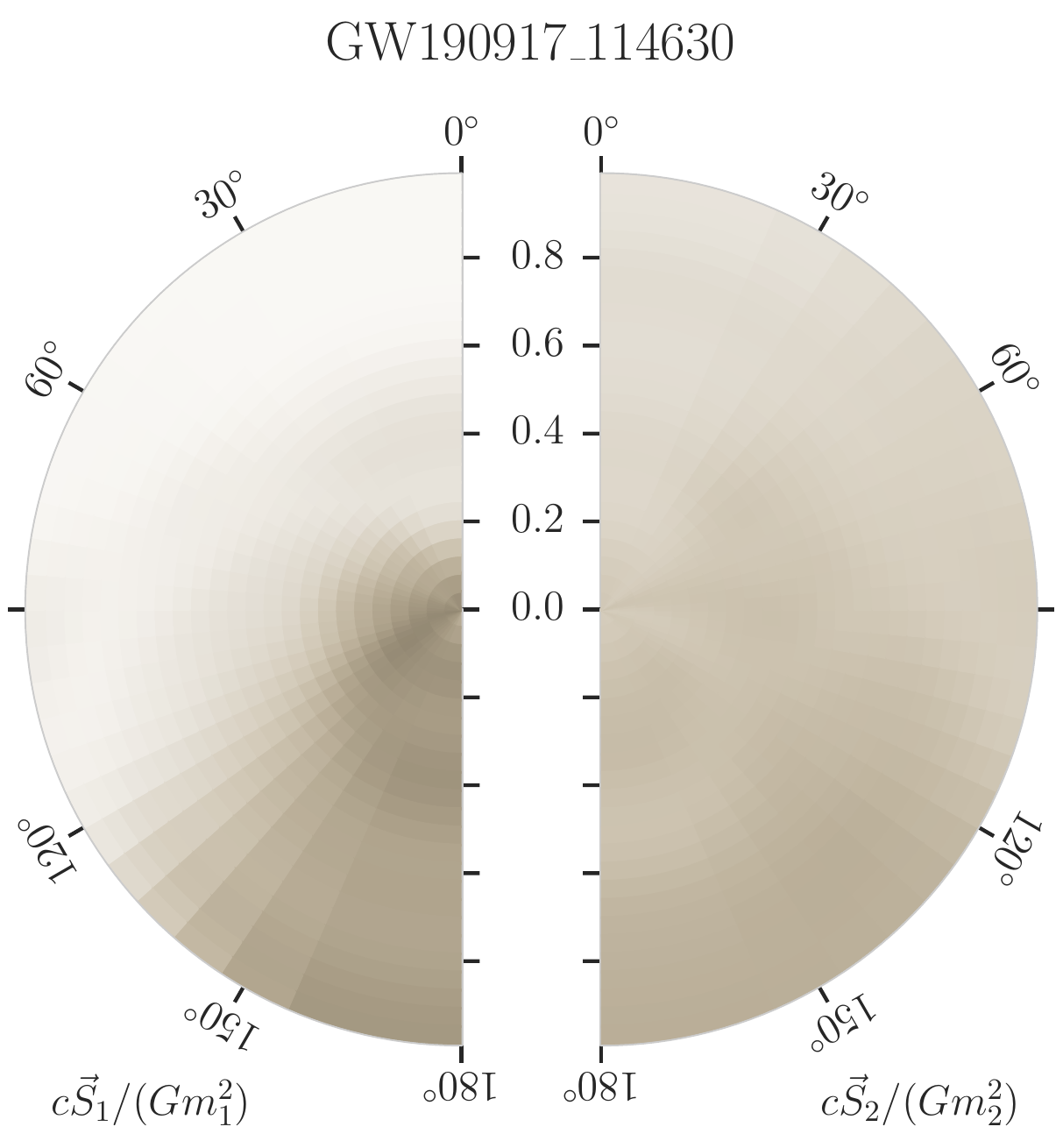}
\end{subfloat}
\caption{The dimensionless spin parameters $\vec{\chi}_{i}=c\vec{S}_i/(Gm_{i}^2)$ estimated for individual binary components of selected sources.
The radial distance of a given pixel on the left (right) of each disk, away from the center of the circle, corresponds to $|\vec{\chi}|$ for the more (less) massive compact object. 
Each pixel's angle from the vertical axis represents $\theta_{\rm LS}$, the angle between the spin vector $\vec{S}$ and the Newtonian orbital angular momentum.
All pixels have equal prior probability with the shading denoting the relative posterior probability of the pixels, after marginalization over azimuthal angles.
The events follow the same color scheme used in Fig.~\ref{fig:PE_1d_posteriors_newEvents}.}
\label{fig:spin_disks}
\end{figure*}

\section{Astrophysical implications}
\label{sec:astro-guess}

Our analysis reports \fixme{\NUMADDCANDIDATES{}} new candidates with $\pastro >0.5$ in at
least one pipeline. 
None of these candidates have \pastro{} equal to $1$ (Table~\ref{tab:events}).
\SINGLEPIPELINEEVENTS{} of them were found only by a single
analysis, and
none
were detected by all the pipelines (Table~\ref{tab:events}).  
As discussed above in Sec.~\ref{subsec:searches}, \pastro{} values are subject to 
statistical uncertainties, and are also subject to uncertainties arising from the true rate 
and distribution of signals.  Such uncertainties are larger for events which, if 
astrophysical, fall within populations with few or zero significant detections. 
Here, we highlight such uncertainties for specific candidates, and 
discuss possible astrophysical implications under the hypothesis that the candidates
do originate from compact object mergers.

Parameter estimation indicates that two of the new candidates, \NAME{GW190403B_o3afin}{} and \NAME{GW190426N_o3afin}{}, 
if astrophysical, have sources with a large total mass ($\gtrsim 100\,\Msun$, Table~\ref{tab:PE_new}). Both were found only by the 
\PYCBCBBH{} analysis with a low \ac{SNR} and relatively low \pastro{}.  They were also not recovered 
as significant events in the focused search of O3 data for intermediate-mass BH binaries \citep{LIGOScientific:2021tfm}.
Since there is only one significant detection to date of a comparable \ac{BBH} system, GW190521 \citep{GW190521Adiscovery,GW190521Aastro}, the
calculation of \pastro{} for these candidates is subject to significant potential systematic error.  These events are confidently above the break mass in the broken power law mass distribution model, at \POWERLAWPOP{}, or the Gaussian in the {\sc Power Law + Peak} model at \GAUSSIANPOP{} \citep{2017ApJ...851L..25F,2018ApJ...856..173T,o3apop}. 
The estimated primary component masses, assuming astrophysical origin, are both above the lower edge of the pair-instability mass gap $m_{\rm low}$ \citep{2021arXiv210402566W,2021arXiv210402969L,2021arXiv210407783E,2021arXiv210402685B}, even considering the large uncertainties about its value ($\approx{}40$--$70 \Msun{}$, \cite{2017ApJ...836..244W,2019ApJ...878...49W,2019ApJ...882..121S,2019ApJ...887...53F,2020ApJ...902L..36F,2020ApJ...888...76M,2020A&A...640L..18M,2021MNRAS.501.4514C,2021MNRAS.502L..40F}). Adopting a rather conservative estimate of $m_{\rm low}=65 \Msun$, the primary component of \NAME{GW190403B_o3afin}{} ($m_1=\massonesourcemed{GW190403B_o3afin}_{ -\massonesourceminus{GW190403B_o3afin} }^{ +\massonesourceplus{GW190403B_o3afin} } \Msun{}$) has a probability \PROBABELOWGAP{} of being below $m_{\rm low}$ with our standard mass prior, while the primary and secondary components of \NAME{GW190426N_o3afin}{} ($m_1=\massonesourcemed{GW190426N_o3afin}_{ -\massonesourceminus{GW190426N_o3afin} }^{ +\massonesourceplus{GW190426N_o3afin} } \Msun{}$ and $m_2=\masstwosourcemed{GW190426N_o3afin}_{ -\masstwosourceminus{GW190426N_o3afin} }^{ +\masstwosourceplus{GW190426N_o3afin} } \Msun$) are below $m_{\rm low}$ with probabilities of  \PROBBBELOWGAP{} and \PROBBSECONDARYBELOWGAP{}, respectively. 
The upper edge of the mass gap is even more uncertain, with theoretical predictions suggesting $m_{\rm up}\approx 120\,\Msun$ \citep{2017MNRAS.470.4739S,2020A&A...640A..56R}. The primary mass component of \NAME{GW190403B_o3afin}{} (\NAME{GW190426N_o3afin}{}) has a probability \PROBAABOVEGAP{} (\PROBBABOVEGAP{}) of being above this value of $m_{\rm up}$. 
Thus, if astrophysical, \NAME{GW190403B_o3afin}{} and \NAME{GW190426N_o3afin}{} lie in the same group with \NAME{GW190521B_o3afin}: their primary components might be either inside or above the mass gap.
Moreover, the estimated final mass of the merger remnant of \NAME{GW190426N_o3afin}{} ($\Mf=\finalmasssourcemed{GW190426N_o3afin}_{ -\finalmasssourceminus{GW190426N_o3afin} }^{ +\finalmasssourceplus{GW190426N_o3afin} }{}\Msun$) is in the intermediate-mass black hole regime ($10^2$--$10^5 \Msun{}$).

These features are suggestive of a dynamical formation channel, such as the hierarchical merger of smaller \acp{BH} \citep{2002MNRAS.330..232C,2016ApJ...831..187A,2017PhRvD..95l4046G,2017ApJ...840L..24F,2019MNRAS.486.5008A,2019PhRvD.100d3027R,2020ApJ...900..177K,2020ApJ...902L..26F,2020ApJ...893...35D,2020arXiv200910688P,2020PhRvD.102d3002B,2021arXiv210305016M} or repeated stellar collisions in dense star clusters \citep{2019MNRAS.487.2947D,2020MNRAS.497.1043D,2020ApJ...903...45K,2020ApJ...904L..13R}. In active galactic nuclei, the dense gaseous disk surrounding the central BH also triggers the hierarchical assembly of \acp{BH} \citep{2012MNRAS.425..460M,2017ApJ...835..165B,2017MNRAS.464..946S,2018ApJ...866...66M,2019PhRvL.123r1101Y,2020ApJ...899...26T,2021ApJ...908..194T}. Alternatively, extreme gas accretion from a dense gaseous disk \citep{2019A&A...632L...8R,2020ApJ...903L..21S,2021ApJ...908...59R} or from a stellar companion \citep{2020ApJ...897..100V} might assist the growth of BH mass above the pair-instability threshold.   Finally, primordial \acp{BH} might also have masses in the pair-instability gap \citep{2020arXiv200706481C,2021PhRvL.126e1101D}. However, even the formation of \acp{BH} in this mass range from stellar collapse cannot be excluded, given the large uncertainties in stellar-evolution models \citep{2020ApJ...902L..36F,2020A&A...636A.104B,2020PhRvD.102k5024C,2021MNRAS.501.4514C,2021MNRAS.502L..40F,2021arXiv210307933W}. For example, very massive ($\gtrsim{}230\,\Msun$) extremely metal-poor ($Z<10^{-4}$) stars might turn into \acp{BH} with mass above the pair-instability gap \cite{2001ApJ...551L..27M,2002RvMP...74.1015W,2020arXiv201007616T,2021MNRAS.501L..49K}.

Parameter-estimation analysis indicates a large positive value of the effective inspiral spin $\chi_{\rm eff}=\chieffmed{GW190403B_o3afin}_{ -\chieffminus{GW190403B_o3afin} }^{ +\chieffplus{GW190403B_o3afin} }$ and of the primary's spin magnitude \evonechione{} for \NAME{GW190403B_o3afin}{}.
From a theoretical perspective, BH spin magnitudes are highly uncertain \cite{2020A&A...636A.104B,2020A&A...635A..97B}, with some models \citep{2019MNRAS.485.3661F,2019ApJ...881L...1F} predicting very low spins ($\sim{0.01}$) for single \acp{BH} because of efficient angular momentum transport in the stellar interior \citep{2002A&A...381..923S}. Observations of high-mass X-ray binaries in the local Universe indicate that BH spins can be nearly maximal \citep{2020arXiv201108948R,2021Sci...371.1046M}, while the majority of mergers in GWTC-2 are associated with low values of \chieff, with a slight preference for positive values \citep{o3apop}.
 Even if single stars form \acp{BH} with low spins~\cite{2019ApJ...881L...1F}, \acp{BH} in binaries may still develop high spins because of mass transfer \citep{2019ApJ...870L..18Q}, tidal interactions \citep{2017ApJ...842..111H,2018A&A...616A..28Q,2020A&A...635A..97B}, or chemically homogeneous evolution \citep{2016A&A...588A..50M,2016MNRAS.458.2634M}. 
 Alternatively, \acp{BH} born from the merger of two smaller \acp{BH} are expected to have high natal spins ($\sim 0.7$--$0.9$, \cite{JohnsonMcDaniel:2016,Hofmann:2016yih,Jimenez-Forteza:2016oae}). This might suggest that the primary component of \NAME{GW190403B_o3afin}{} is a second-generation BH, which is also consistent with its large mass \citep{2017PhRvD..95l4046G,2017ApJ...840L..24F,2020RNAAS...4....2K,2021arXiv210305016M,2021arXiv210503439G}. However, the positive effective inspiral spin \chieff of \NAME{GW190403B_o3afin}{} indicates a significant alignment of the spin vectors of (any of) the two components with the orbital angular momentum vector of the BBH. Nearly aligned spins are preferentially associated with isolated binary evolution \citep{2000ApJ...541..319K,2018PhRvD..98h4036G}, while dynamically formed binaries tend to have an isotropically distributed spin orientations \citep{2016ApJ...832L...2R,2017Natur.548..426F}.

 Finally,  \NAME{GW190403B_o3afin}{} is associated with a comparatively
 small mass ratio \massratio{} (Fig.~\ref{fig:PE_1d_posteriors_newEvents}). Such low values of the mass ratio are  unusual in isolated binary evolution, especially for the chemically homogeneous evolution \citep{2016A&A...588A..50M,2016MNRAS.460.3545D} but also for the common-envelope scenario \citep{2012ApJ...759...52D,2016MNRAS.462.3302E,2018MNRAS.480.2011G,2019MNRAS.490.3740N,2020A&A...636A.104B}. In contrast, low mass ratios are expected if the primary and secondary components are a second- and a  first-generation BH, respectively \citep{2019PhRvD.100d3027R,2020ApJ...900..177K,2020ApJ...893...35D}, or if the primary BH is the result of a stellar merger in a young star cluster \citep{2020MNRAS.497.1043D}.

\NEWBULK{} of the other new candidates (\NAME{GW190805J_o3afin}{}, \NAME{GW190916K_o3afin}{}, \NAME{GW190925J_o3afin}{},  \NAME{GW190926C_o3afin}{}) fall in the mass range of the bulk of GWTC-2 \acp{BBH}, while the secondary component of \NAME{GW190725F_o3afin}{} has a \PROBLOWGAP{} probability of lying in the lower mass gap ($\sim 2$--$5\,{}\Msun$). The existence of a lower mass gap was inferred from observations of Galactic X-ray binaries \cite{1998ApJ...499..367B,2010ApJ...725.1918O,2011ApJ...741..103F}, but there are a few observations of \acp{BH} with mass $\approx 3$--$4{}\,{}\Msun{}$ in non-interacting binary systems \cite{2020Sci...368.4356T,2021MNRAS.504.2577J} and microlensing surveys find no evidence for a mass gap between \acp{NS} and \acp{BH} \citep{2016MNRAS.458.3012W,2020A&A...636A..20W}. GWTC-2 BBH observations also suggest a dearth of systems between $2.6\,{}\Msun$ and $6\,{}\Msun$ \citep{o3apop,2020ApJ...899L...8F}. The only confirmed GW event in GWTC-2 with a component in the lower mass gap is GW190814 \cite{GW190814A}.  Numerical and theoretical models do not exclude the formation of compact objects in this mass range from a core-collapse supernova \cite{2012ApJ...749...91F,2018ApJS..237...13L,2020ApJ...899L...1Z,2020MNRAS.499.3214M}. Other scenarios to explain the formation of binary compact objects in this mass range include mergers in multiple systems \cite{2020PhRvD.101j3036G,2021MNRAS.502.2049L,2021MNRAS.500.1817L,2021ApJ...908L..38A}, primordial \acp{BH} \cite{2020arXiv200706481C,2020PhRvD.102f1301V} and mass accretion onto a neutron star \cite{2021ApJ...907L..24S}.

Finally, \NAME{GW190917B_o3afin}{} has component masses consistent with an
\ac{NSBH} ($m_1=\massonesourcemed{GW190917B_o3afin}_{
-\massonesourceminus{GW190917B_o3afin} }^{
+\massonesourceplus{GW190917B_o3afin} } {}\Msun$,
$m_2=\masstwosourcemed{GW190917B_o3afin}_{
-\masstwosourceminus{GW190917B_o3afin} }^{
+\masstwosourceplus{GW190917B_o3afin} }{}\Msun$),
 but
was identified only as a BBH candidate, with $p_{\rm NSBH}=0$ and
\fixme{$p_{\rm BBH}=\GSTLALALLSKYPASTRO{GW190917B_o3afin}$}, by the pipeline
that detected it (\GSTLAL{}).  Since \NAMEMAR{GW190426H_o3asub}{} is a
marginal candidate in this catalog, due to its low $p_{\rm astro}$ (Table~\ref{tab:marginal-events}),
\NAME{GW190917B_o3afin}{} is the only high-probability candidate with mass
components in the NSBH range. However, as discussed in Sec.~\ref{ssec:hs_candidates}, had it been classified as an \ac{NSBH}
to begin with, its \pastro{} measured by \GSTLAL{} would have been smaller due
to the lower foreground rate of \acp{NSBH} as compared to \acp{BBH} in the
detection pipelines, and not passed the threshold of 0.5 considered by the follow-up
pipelines. As with the unusually high-mass \ac{BBH} candidates, the assignment of \pastro{} for \acp{NSBH} is subject to potential
systematic error since no \ac{NSBH} events have been confidently detected in the data set up to \ac{O3a} used here, although see ~\cite{Abbott_nsbh_2021, LIGOScientific:2021djp} for \ac{NSBH} discoveries in \ac{O3b}. 
The masses and effective inspiral spin of this candidate are consistent with prior expectations for \ac{NSBH} 
systems \cite{2002ApJ...572..407B,2012ApJ...759...52D,2019MNRAS.482..870E,2019MNRAS.487....2M,2020MNRAS.497.1563R,2020arXiv201113503C,2020arXiv200906655D,2021arXiv210302608B}. 
Inferring the impact on the overall population of binary compact objects of the new candidates, including those with non-negligible probability of noise origin, requires a more involved analysis which is beyond this scope of this work~\cite{Gaebel:2018poe, Galaudage:2019jdx}.

\section{Conclusion}\label{sec:conclusion}
We have presented GWTC-2.1, which includes results from a refined search
for \acp{CBC} in the first part of the third observing run of the Advanced LIGO
and Advanced Virgo detectors. This is an extension to the previous \ac{GW}
catalog, GWTC-2~\cite{Abbott:2020niy}, over the same data, and provides a deeper list of \ac{GW}
candidates. The search we presented here was carried out using three
matched-filter pipelines, \MBTA{}, \GSTLAL{}, and \PYCBC{}, and includes a list
of candidates that have a \ac{FAR} less than 2 per day in any of the pipelines.
We provide detailed source properties of the \NUMADDCANDIDATES{} events
that have \pastro{} greater than 0.5 and were not present in GWTC-2. In
addition, the source properties of previously reported events with \pastro{}
greater than 0.5 are presented in 
Appendix~\ref{app:parameter-estimation-appendix}.

Out of the \NUMADDCANDIDATES{} new candidates presented here, all events have masses
consistent with \ac{BBH} sources with the
exception of \NAME{GW190917B_o3afin}{}, whose source masses are consistent with being an
\ac{NSBH} (Sec.~\ref{ss:parameter-estimation-results}). 
If astrophysical, these events expand the scope of observed  \acp{BBH}, with
several binaries inferred at larger distances than previous
detections and with both a new broader range of recovered \ac{BH} masses and
the addition of two binaries with significantly unequal masses.  The
primary components of two of the new candidates (\NAME{GW190403B_o3afin}{} and
\NAME{GW190426N_o3afin}{}) lie inside or, less likely, above the
pair-instability mass gap.  \NAME{GW190403B_o3afin}{} also shows support for
high spin, unequal masses, and remnant mass in the intermediate-mass BH regime.  These
features are suggestive of dynamical formation, by hierarchical BH merger or
by stellar collisions in dense stellar clusters or active galactic nuclei.
However, we cannot exclude that \NAME{GW190403B_o3afin}{} and
\NAME{GW190426N_o3afin}{}  originated from isolated binary systems, because of
the large uncertainties in the mass range of the pair-instability mass gap.
Among the new candidates, \NAME{GW190725F_o3afin}{}  shows some support for a
secondary component mass in the lower mass gap ($2$--$5 \Msun$).
\NAME{GW190917B_o3afin}{}, the only candidate with component masses consistent
with an \ac{NSBH} was initially classified as a \ac{BBH} by the search pipeline, and
therefore the \pastro{} assigned to it is subject to systematics due to
uncertainty in classification.

The data products associated with GWTC-2.1 include candidate information
from relevant search pipeline(s) and localizations for all events that pass
a threshold of \fixme{2 per day} in any search pipeline. The information from each 
search pipeline includes the template mass and spin parameters, the SNR time
series, chi-squared values, the time and phase of coalescence in each detector, \ac{FAR}, and \pastro{} (Sec.~\ref{subsec:searches}). These data can be found at~\cite{gwtc2p1_search_data}.
The source localizations are computed using the rapid localization tool 
\BAYESTAR{}~\cite{PhysRevD.93.024013, Singer_2016}, which was also used to produce the
localizations in near real time during  the observing runs while sending out
\ac{GW} alerts. We also release the results of the search pipelines running over simulated signal sets classified as \ac{BNS}, \ac{NSBH}, and \ac{BBH}~\cite{gwtc2p1_injection_data} that were used to calculate the sensitivities shown in Table~\ref{tab:vt}. For candidates that have a $\pastro{} > 0.5$, we perform follow-up parameter estimation and also release the posterior
samples associated with these events. These are available via~\cite{gwtc2p1_pe_data}. Finally, the strain data for O3a used for the
analyses in this paper are also available~\cite{o3a_public_data}.

The \ac{LVK} have already announced the first observations from \acp{NSBH}~\cite{Abbott_nsbh_2021} in the data from \ac{O3b}, and the catalog that extends events up to \ac{O3b}, GWTC-3~\cite{LIGOScientific:2021djp}, has been released. 
GWTC-3 adds \GWTCTHREEEVENTS{} \ac{GW} candidates with \pastro{} greater than 0.5 from \ac{O3b}. O3 marks the most
sensitive \ac{GW} data published upon so far. The LIGO, Virgo, and KAGRA~\cite{Akutsu:2018axf} 
detectors are currently offline and undergoing commissioning to enhance their
sensitivities, and plan to all collect data simultaneously during the fourth observing run (O4)~\cite{Aasi:2013wya}. With further improvement in sensitivities and planning for pre-merger
\ac{BNS} detections~\cite{Magee2021, Nitz_2020, Sachdev_2020}, O4 offers
improved prospects for \ac{GW} and multimessenger astronomy, and promises to
build upon our current knowledge of binary populations.

\acknowledgments
This material is based upon work supported by NSF’s LIGO Laboratory which is a major facility
fully funded by the National Science Foundation.
The authors also gratefully acknowledge the support of
the Science and Technology Facilities Council (STFC) of the
United Kingdom, the Max-Planck-Society (MPS), and the State of
Niedersachsen/Germany for support of the construction of Advanced LIGO 
and construction and operation of the GEO600 detector. 
Additional support for Advanced LIGO was provided by the Australian Research Council.
The authors gratefully acknowledge the Italian Istituto Nazionale di Fisica Nucleare (INFN),  
the French Centre National de la Recherche Scientifique (CNRS) and
the Netherlands Organization for Scientific Research, 
for the construction and operation of the Virgo detector
and the creation and support  of the EGO consortium. 
The authors also gratefully acknowledge research support from these agencies as well as by 
the Council of Scientific and Industrial Research of India, 
the Department of Science and Technology, India,
the Science \& Engineering Research Board (SERB), India,
the Ministry of Human Resource Development, India,
the Spanish Agencia Estatal de Investigaci\'on,
the Vicepresid\`encia i Conselleria d'Innovaci\'o, Recerca i Turisme and the Conselleria d'Educaci\'o i Universitat del Govern de les Illes Balears,
the Conselleria d'Innovaci\'o, Universitats, Ci\`encia i Societat Digital de la Generalitat Valenciana and
the CERCA Programme Generalitat de Catalunya, Spain,
the National Science Centre of Poland and the Foundation for Polish Science (FNP),
the Swiss National Science Foundation (SNSF),
the Russian Foundation for Basic Research, 
the Russian Science Foundation,
the European Commission,
the European Regional Development Funds (ERDF),
the Royal Society, 
the Scottish Funding Council, 
the Scottish Universities Physics Alliance, 
the Hungarian Scientific Research Fund (OTKA),
the French Lyon Institute of Origins (LIO),
the Belgian Fonds de la Recherche Scientifique (FRS-FNRS), 
Actions de Recherche Concertées (ARC) and
Fonds Wetenschappelijk Onderzoek – Vlaanderen (FWO), Belgium,
the Paris \^{I}le-de-France Region, 
the National Research, Development and Innovation Office Hungary (NKFIH), 
the National Research Foundation of Korea,
the Natural Science and Engineering Research Council Canada,
Canadian Foundation for Innovation (CFI),
the Brazilian Ministry of Science, Technology, and Innovations,
the International Center for Theoretical Physics South American Institute for Fundamental Research (ICTP-SAIFR), 
the Research Grants Council of Hong Kong,
the National Natural Science Foundation of China (NSFC),
the Leverhulme Trust, 
the Research Corporation, 
the Ministry of Science and Technology (MOST), Taiwan,
the United States Department of Energy,
and
the Kavli Foundation.
The authors gratefully acknowledge the support of the NSF, STFC, INFN and CNRS for provision of computational resources.

{\it We would like to thank all of the essential workers who put their health at risk during the COVID-19 pandemic, without whom we would not have been able to complete this work.}

Analyses in this catalog relied upon the \LALSUITE{} software
library~\cite{lalsuite-software}.  The detection of the signals and subsequent
significance evaluations were performed with the
\GSTLAL{}-based inspiral software
pipeline~\cite{Messick:2016aqy,Sachdev:2019vvd,Hanna:2019ezx,Cannon:2020qnf},
with the \MBTA{} pipeline~\cite{Adams:2015ulm,Aubin:2020goo}, and with the
\PYCBC{}~\cite{Usman:2015kfa,Nitz:2017svb,Davies:2020tsx, pycbc-software} package. Estimates of
the noise spectra and glitch models were obtained using
\BAYESWAVE{}~\cite{Cornish:2014kda,Littenberg:2015kpb,PhysRevD.103.044006}.  Source
parameter estimation was performed with the \BILBY{}
library~\cite{Ashton:2018jfp, Romero-Shaw:2020owr} using the
\DYNESTY{} nested sampling package~\cite{Speagle:2019ivv}, the \RIFT{}
library~\cite{Pankow:2015cra,Lange:2017wki,Wysocki:2019grj} and the
\LALINFERENCE{} library~\cite{Veitch:2014wba}.
\PESUMMARY{} was used to post-process and collate parameter-estimation
results~\cite{Hoy:2020vys}.  The various stages of the parameter-estimation
analysis were managed with the \ASIMOV{} library~\cite{asimov-software}.  Plots
were prepared with \PLT{}~\cite{Hunter:2007ouj}, \SEABORN{}~\cite{Waskom2021} and
\GWPY{}~\cite{gwpy-software}.  \NUMPY{}~\cite{Harris:2020xlr} and
\SCIPY{}~\cite{Virtanen:2019joe} were used in the preparation of the
manuscript.

\appendix
\section{Estimation of source parameters}
\label{app:parameter-estimation-appendix}

\subsection{Binary black holes from the first and second observing runs}
\label{app:pe_o1o2}

In order to provide a self-consistent set of source properties, inferred using the state-of-the-art \ac{BBH} waveform models described in Sec.~\ref{ss:waveforms}, we have reanalyzed the 10 \ac{BBH} events observed during O1 and O2, and reported in GWTC-1~\cite{LIGOScientific:2018mvr}.
We present results combining samples from analyses using both the IMRPhenomXPHM and SEOBNRv4PHM, with the exception of  \OLDEVENTSNAME{GW151226_o3afin}{} which, as mentioned earlier in Sec.~\ref{ss:parameter-estimation-results}, was analyzed using IMRPhenomXPHM only.
As the \ac{BNS} models available at the time of GWTC-1 still can be considered state-of-the-art in the \ac{NS}-physics they describe, we have elected to not reanalyze the \ac{BNS} event GW170817 as part of this study.
For the source properties of GW170817, we instead refer to GWTC-1~\cite{LIGOScientific:2018mvr} and its accompanying data release~\cite{GWTC1_PEdatarelease}.

\begin{figure*}[t]
\includegraphics[width=\textwidth]{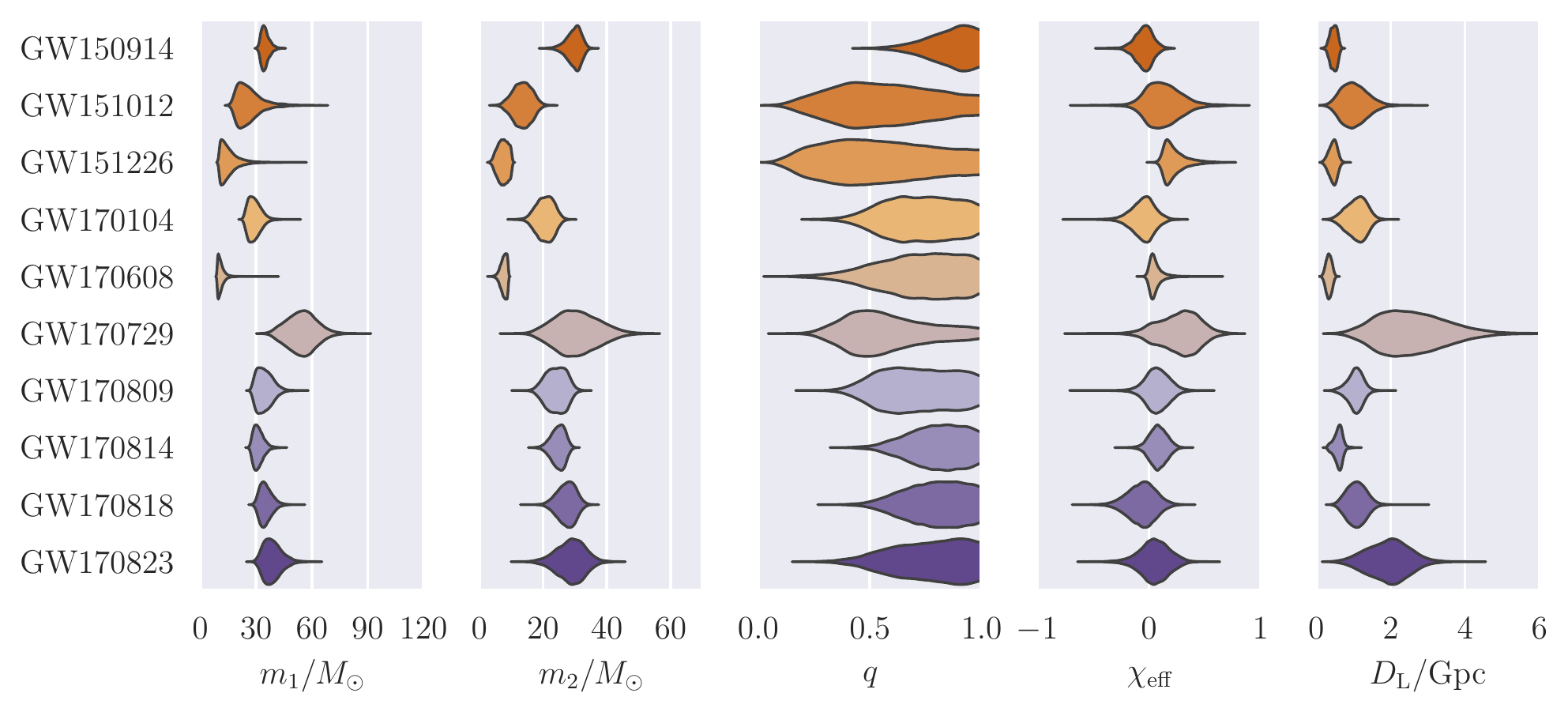}
\caption{Marginal posterior distributions on the primary mass \massone, secondary mass \masstwo, mass ratio \massratio, effective inspiral spin \chieff and luminosity distance \DL for the 10 \ac{BBH} events observed during O1 and O2.
The vertical span for each region is constructed to be proportional to the one-dimensional marginal posterior at a given parameter value for the corresponding event.
The posterior distributions are also represented numerically in terms of their one-dimensional median and $90\%$ credible intervals in Table~\ref{tab:PE_o1o2}.}
\label{fig:PE_1d_posteriors_o1o2}
\end{figure*}

The source properties for the 10 \ac{BBH} events from the O1 and O2 are reported in Table~\ref{tab:PE_o1o2}, with a selection of the one-dimensional marginal posterior distributions shown in Fig.~\ref{fig:PE_1d_posteriors_o1o2}.
The two-dimensional projections on the \Mtot--\massratio and \Mc--\chieff planes are shown as light-grey contours in Fig.~\ref{fig:PE_2d_posteriors_mtot_q} and Fig.~\ref{fig:PE_2d_posteriors_mc_chieff} respectively.
The full 15-dimensional posterior distributions are available as part of the public data release accompanying this paper~\cite{gwtc2p1_pe_data}, as detailed further in Sec.~\ref{sec:conclusion}.

Generally, the inferred source properties for these 10 \acp{BBH} are consistent with those presented in GWTC-1~\cite{LIGOScientific:2018mvr}, but there are some new features worth highlighting.
Where most binaries have a nominal support for $\chieff=0$, \OLDEVENTSNAME{GW151226_o3afin}{} was in GWTC-1 identified to exclude this value at $>90\%$ probability~\cite{Abbott:2016nmj, LIGOScientific:2018mvr}, a conclusion which is strengthened further as of the analysis presented here in GWTC-2.1.
The other \ac{BBH} in GWTC-1 with only marginal support for $\chieff=0$, \OLDEVENTSNAME{GW170729_o3afin}{}, is now found to include support for negative \chieff in its $90\%$ credible interval while also simultaneously preferring \ac{BH} components with more unequal masses relative to what was inferred in GWTC-1.

Independent analyses of these 10 events with the IMRPhenomXPHM model were previously presented in~\cite{Mateu-Lucena:2021siq}, showing broad consistency with the results presented in this section.

\begin{PE_table}
\begin{table*}
    \centering
    {\rowcolors{1}{}{lightgray}
\begin{tabularx}{\textwidth}{@{\extracolsep{\fill}}p{2.85cm} U U U U V U U U U W U}
Event & $\underset{\displaystyle (M_\odot)}{M}$ & $\underset{\displaystyle (M_\odot)}{\mathcal{M}}$ & $\underset{\displaystyle (M_\odot)}{m_1}$ & $\underset{\displaystyle (M_\odot)}{m_2}$ & $\chi_{{\rm eff}}$ & $\underset{\displaystyle ({\rm Gpc})}{D_\mathrm{L}}$ & $z$ & $\underset{\displaystyle (M_\odot)}{M_\mathrm{f}}$ & $\chi_\mathrm{f}$ & $\underset{\displaystyle (\mathrm{deg}^2)}{\Delta\Omega}$ & $\mathrm{SNR}$\\ \hline
\OLDEVENTSNAME{GW150914_o3afin} & $\totalmasssourcemed{GW150914_o3afin}_{ -\totalmasssourceminus{GW150914_o3afin} }^{ +\totalmasssourceplus{GW150914_o3afin} }$ & $\chirpmasssourcemed{GW150914_o3afin}_{ -\chirpmasssourceminus{GW150914_o3afin} }^{ +\chirpmasssourceplus{GW150914_o3afin} }$ & $\massonesourcemed{GW150914_o3afin}_{ -\massonesourceminus{GW150914_o3afin} }^{ +\massonesourceplus{GW150914_o3afin} }$ & $\masstwosourcemed{GW150914_o3afin}_{ -\masstwosourceminus{GW150914_o3afin} }^{ +\masstwosourceplus{GW150914_o3afin} }$ & $\chieffinfinityonlyprecavgmed{GW150914_o3afin}_{ -\chieffinfinityonlyprecavgminus{GW150914_o3afin} }^{ +\chieffinfinityonlyprecavgplus{GW150914_o3afin} }$ & $\luminositydistancemed{GW150914_o3afin}_{ -\luminositydistanceminus{GW150914_o3afin} }^{ +\luminositydistanceplus{GW150914_o3afin} }$ & $\redshiftmed{GW150914_o3afin}_{ -\redshiftminus{GW150914_o3afin} }^{ +\redshiftplus{GW150914_o3afin} }$ & $\finalmasssourcemed{GW150914_o3afin}_{ -\finalmasssourceminus{GW150914_o3afin} }^{ +\finalmasssourceplus{GW150914_o3afin} }$ & $\finalspinmed{GW150914_o3afin}_{ -\finalspinminus{GW150914_o3afin} }^{ +\finalspinplus{GW150914_o3afin} }$ & $\skyarea{GW150914_o3afin}$ & $\networkmatchedfiltersnrIMRPmed{GW150914_o3afin}_{ -\networkmatchedfiltersnrIMRPminus{GW150914_o3afin} }^{ +\networkmatchedfiltersnrIMRPplus{GW150914_o3afin} }$\\
\OLDEVENTSNAME{GW151012_o3afin} & $\totalmasssourcemed{GW151012_o3afin}_{ -\totalmasssourceminus{GW151012_o3afin} }^{ +\totalmasssourceplus{GW151012_o3afin} }$ & $\chirpmasssourcemed{GW151012_o3afin}_{ -\chirpmasssourceminus{GW151012_o3afin} }^{ +\chirpmasssourceplus{GW151012_o3afin} }$ & $\massonesourcemed{GW151012_o3afin}_{ -\massonesourceminus{GW151012_o3afin} }^{ +\massonesourceplus{GW151012_o3afin} }$ & $\masstwosourcemed{GW151012_o3afin}_{ -\masstwosourceminus{GW151012_o3afin} }^{ +\masstwosourceplus{GW151012_o3afin} }$ & $\chieffinfinityonlyprecavgmed{GW151012_o3afin}_{ -\chieffinfinityonlyprecavgminus{GW151012_o3afin} }^{ +\chieffinfinityonlyprecavgplus{GW151012_o3afin} }$ & $\luminositydistancemed{GW151012_o3afin}_{ -\luminositydistanceminus{GW151012_o3afin} }^{ +\luminositydistanceplus{GW151012_o3afin} }$ & $\redshiftmed{GW151012_o3afin}_{ -\redshiftminus{GW151012_o3afin} }^{ +\redshiftplus{GW151012_o3afin} }$ & $\finalmasssourcemed{GW151012_o3afin}_{ -\finalmasssourceminus{GW151012_o3afin} }^{ +\finalmasssourceplus{GW151012_o3afin} }$ & $\finalspinmed{GW151012_o3afin}_{ -\finalspinminus{GW151012_o3afin} }^{ +\finalspinplus{GW151012_o3afin} }$ & $\skyarea{GW151012_o3afin}$ & $\networkmatchedfiltersnrIMRPmed{GW151012_o3afin}_{ -\networkmatchedfiltersnrIMRPminus{GW151012_o3afin} }^{ +\networkmatchedfiltersnrIMRPplus{GW151012_o3afin} }$\\
\OLDEVENTSNAME{GW151226_o3afin} & $\totalmasssourcemed{GW151226_o3afin}_{ -\totalmasssourceminus{GW151226_o3afin} }^{ +\totalmasssourceplus{GW151226_o3afin} }$ & $\chirpmasssourcemed{GW151226_o3afin}_{ -\chirpmasssourceminus{GW151226_o3afin} }^{ +\chirpmasssourceplus{GW151226_o3afin} }$ & $\massonesourcemed{GW151226_o3afin}_{ -\massonesourceminus{GW151226_o3afin} }^{ +\massonesourceplus{GW151226_o3afin} }$ & $\masstwosourcemed{GW151226_o3afin}_{ -\masstwosourceminus{GW151226_o3afin} }^{ +\masstwosourceplus{GW151226_o3afin} }$ & $\chieffinfinityonlyprecavgmed{GW151226_o3afin}_{ -\chieffinfinityonlyprecavgminus{GW151226_o3afin} }^{ +\chieffinfinityonlyprecavgplus{GW151226_o3afin} }$ & $\luminositydistancemed{GW151226_o3afin}_{ -\luminositydistanceminus{GW151226_o3afin} }^{ +\luminositydistanceplus{GW151226_o3afin} }$ & $\redshiftmed{GW151226_o3afin}_{ -\redshiftminus{GW151226_o3afin} }^{ +\redshiftplus{GW151226_o3afin} }$ & $\finalmasssourcemed{GW151226_o3afin}_{ -\finalmasssourceminus{GW151226_o3afin} }^{ +\finalmasssourceplus{GW151226_o3afin} }$ & $\finalspinmed{GW151226_o3afin}_{ -\finalspinminus{GW151226_o3afin} }^{ +\finalspinplus{GW151226_o3afin} }$ & $\skyarea{GW151226_o3afin}$ & $\networkmatchedfiltersnrIMRPmed{GW151226_o3afin}_{ -\networkmatchedfiltersnrIMRPminus{GW151226_o3afin} }^{ +\networkmatchedfiltersnrIMRPplus{GW151226_o3afin} }$\\
\OLDEVENTSNAME{GW170104_o3afin} & $\totalmasssourcemed{GW170104_o3afin}_{ -\totalmasssourceminus{GW170104_o3afin} }^{ +\totalmasssourceplus{GW170104_o3afin} }$ & $\chirpmasssourcemed{GW170104_o3afin}_{ -\chirpmasssourceminus{GW170104_o3afin} }^{ +\chirpmasssourceplus{GW170104_o3afin} }$ & $\massonesourcemed{GW170104_o3afin}_{ -\massonesourceminus{GW170104_o3afin} }^{ +\massonesourceplus{GW170104_o3afin} }$ & $\masstwosourcemed{GW170104_o3afin}_{ -\masstwosourceminus{GW170104_o3afin} }^{ +\masstwosourceplus{GW170104_o3afin} }$ & $\chieffinfinityonlyprecavgmed{GW170104_o3afin}_{ -\chieffinfinityonlyprecavgminus{GW170104_o3afin} }^{ +\chieffinfinityonlyprecavgplus{GW170104_o3afin} }$ & $\luminositydistancemed{GW170104_o3afin}_{ -\luminositydistanceminus{GW170104_o3afin} }^{ +\luminositydistanceplus{GW170104_o3afin} }$ & $\redshiftmed{GW170104_o3afin}_{ -\redshiftminus{GW170104_o3afin} }^{ +\redshiftplus{GW170104_o3afin} }$ & $\finalmasssourcemed{GW170104_o3afin}_{ -\finalmasssourceminus{GW170104_o3afin} }^{ +\finalmasssourceplus{GW170104_o3afin} }$ & $\finalspinmed{GW170104_o3afin}_{ -\finalspinminus{GW170104_o3afin} }^{ +\finalspinplus{GW170104_o3afin} }$ & $\skyarea{GW170104_o3afin}$ & $\networkmatchedfiltersnrIMRPmed{GW170104_o3afin}_{ -\networkmatchedfiltersnrIMRPminus{GW170104_o3afin} }^{ +\networkmatchedfiltersnrIMRPplus{GW170104_o3afin} }$\\
\OLDEVENTSNAME{GW170608_o3afin} & $\totalmasssourcemed{GW170608_o3afin}_{ -\totalmasssourceminus{GW170608_o3afin} }^{ +\totalmasssourceplus{GW170608_o3afin} }$ & $\chirpmasssourcemed{GW170608_o3afin}_{ -\chirpmasssourceminus{GW170608_o3afin} }^{ +\chirpmasssourceplus{GW170608_o3afin} }$ & $\massonesourcemed{GW170608_o3afin}_{ -\massonesourceminus{GW170608_o3afin} }^{ +\massonesourceplus{GW170608_o3afin} }$ & $\masstwosourcemed{GW170608_o3afin}_{ -\masstwosourceminus{GW170608_o3afin} }^{ +\masstwosourceplus{GW170608_o3afin} }$ & $\chieffinfinityonlyprecavgmed{GW170608_o3afin}_{ -\chieffinfinityonlyprecavgminus{GW170608_o3afin} }^{ +\chieffinfinityonlyprecavgplus{GW170608_o3afin} }$ & $\luminositydistancemed{GW170608_o3afin}_{ -\luminositydistanceminus{GW170608_o3afin} }^{ +\luminositydistanceplus{GW170608_o3afin} }$ & $\redshiftmed{GW170608_o3afin}_{ -\redshiftminus{GW170608_o3afin} }^{ +\redshiftplus{GW170608_o3afin} }$ & $\finalmasssourcemed{GW170608_o3afin}_{ -\finalmasssourceminus{GW170608_o3afin} }^{ +\finalmasssourceplus{GW170608_o3afin} }$ & $\finalspinmed{GW170608_o3afin}_{ -\finalspinminus{GW170608_o3afin} }^{ +\finalspinplus{GW170608_o3afin} }$ & $\skyarea{GW170608_o3afin}$ & $\networkmatchedfiltersnrIMRPmed{GW170608_o3afin}_{ -\networkmatchedfiltersnrIMRPminus{GW170608_o3afin} }^{ +\networkmatchedfiltersnrIMRPplus{GW170608_o3afin} }$\\
\OLDEVENTSNAME{GW170729_o3afin} & $\totalmasssourcemed{GW170729_o3afin}_{ -\totalmasssourceminus{GW170729_o3afin} }^{ +\totalmasssourceplus{GW170729_o3afin} }$ & $\chirpmasssourcemed{GW170729_o3afin}_{ -\chirpmasssourceminus{GW170729_o3afin} }^{ +\chirpmasssourceplus{GW170729_o3afin} }$ & $\massonesourcemed{GW170729_o3afin}_{ -\massonesourceminus{GW170729_o3afin} }^{ +\massonesourceplus{GW170729_o3afin} }$ & $\masstwosourcemed{GW170729_o3afin}_{ -\masstwosourceminus{GW170729_o3afin} }^{ +\masstwosourceplus{GW170729_o3afin} }$ & $\chieffinfinityonlyprecavgmed{GW170729_o3afin}_{ -\chieffinfinityonlyprecavgminus{GW170729_o3afin} }^{ +\chieffinfinityonlyprecavgplus{GW170729_o3afin} }$ & $\luminositydistancemed{GW170729_o3afin}_{ -\luminositydistanceminus{GW170729_o3afin} }^{ +\luminositydistanceplus{GW170729_o3afin} }$ & $\redshiftmed{GW170729_o3afin}_{ -\redshiftminus{GW170729_o3afin} }^{ +\redshiftplus{GW170729_o3afin} }$ & $\finalmasssourcemed{GW170729_o3afin}_{ -\finalmasssourceminus{GW170729_o3afin} }^{ +\finalmasssourceplus{GW170729_o3afin} }$ & $\finalspinmed{GW170729_o3afin}_{ -\finalspinminus{GW170729_o3afin} }^{ +\finalspinplus{GW170729_o3afin} }$ & $\skyarea{GW170729_o3afin}$ & $\networkmatchedfiltersnrIMRPmed{GW170729_o3afin}_{ -\networkmatchedfiltersnrIMRPminus{GW170729_o3afin} }^{ +\networkmatchedfiltersnrIMRPplus{GW170729_o3afin} }$\\
\OLDEVENTSNAME{GW170809_o3afin} & $\totalmasssourcemed{GW170809_o3afin}_{ -\totalmasssourceminus{GW170809_o3afin} }^{ +\totalmasssourceplus{GW170809_o3afin} }$ & $\chirpmasssourcemed{GW170809_o3afin}_{ -\chirpmasssourceminus{GW170809_o3afin} }^{ +\chirpmasssourceplus{GW170809_o3afin} }$ & $\massonesourcemed{GW170809_o3afin}_{ -\massonesourceminus{GW170809_o3afin} }^{ +\massonesourceplus{GW170809_o3afin} }$ & $\masstwosourcemed{GW170809_o3afin}_{ -\masstwosourceminus{GW170809_o3afin} }^{ +\masstwosourceplus{GW170809_o3afin} }$ & $\chieffinfinityonlyprecavgmed{GW170809_o3afin}_{ -\chieffinfinityonlyprecavgminus{GW170809_o3afin} }^{ +\chieffinfinityonlyprecavgplus{GW170809_o3afin} }$ & $\luminositydistancemed{GW170809_o3afin}_{ -\luminositydistanceminus{GW170809_o3afin} }^{ +\luminositydistanceplus{GW170809_o3afin} }$ & $\redshiftmed{GW170809_o3afin}_{ -\redshiftminus{GW170809_o3afin} }^{ +\redshiftplus{GW170809_o3afin} }$ & $\finalmasssourcemed{GW170809_o3afin}_{ -\finalmasssourceminus{GW170809_o3afin} }^{ +\finalmasssourceplus{GW170809_o3afin} }$ & $\finalspinmed{GW170809_o3afin}_{ -\finalspinminus{GW170809_o3afin} }^{ +\finalspinplus{GW170809_o3afin} }$ & $\skyarea{GW170809_o3afin}$ & $\networkmatchedfiltersnrIMRPmed{GW170809_o3afin}_{ -\networkmatchedfiltersnrIMRPminus{GW170809_o3afin} }^{ +\networkmatchedfiltersnrIMRPplus{GW170809_o3afin} }$\\
\OLDEVENTSNAME{GW170814_o3afin} & $\totalmasssourcemed{GW170814_o3afin}_{ -\totalmasssourceminus{GW170814_o3afin} }^{ +\totalmasssourceplus{GW170814_o3afin} }$ & $\chirpmasssourcemed{GW170814_o3afin}_{ -\chirpmasssourceminus{GW170814_o3afin} }^{ +\chirpmasssourceplus{GW170814_o3afin} }$ & $\massonesourcemed{GW170814_o3afin}_{ -\massonesourceminus{GW170814_o3afin} }^{ +\massonesourceplus{GW170814_o3afin} }$ & $\masstwosourcemed{GW170814_o3afin}_{ -\masstwosourceminus{GW170814_o3afin} }^{ +\masstwosourceplus{GW170814_o3afin} }$ & $\chieffinfinityonlyprecavgmed{GW170814_o3afin}_{ -\chieffinfinityonlyprecavgminus{GW170814_o3afin} }^{ +\chieffinfinityonlyprecavgplus{GW170814_o3afin} }$ & $\luminositydistancemed{GW170814_o3afin}_{ -\luminositydistanceminus{GW170814_o3afin} }^{ +\luminositydistanceplus{GW170814_o3afin} }$ & $\redshiftmed{GW170814_o3afin}_{ -\redshiftminus{GW170814_o3afin} }^{ +\redshiftplus{GW170814_o3afin} }$ & $\finalmasssourcemed{GW170814_o3afin}_{ -\finalmasssourceminus{GW170814_o3afin} }^{ +\finalmasssourceplus{GW170814_o3afin} }$ & $\finalspinmed{GW170814_o3afin}_{ -\finalspinminus{GW170814_o3afin} }^{ +\finalspinplus{GW170814_o3afin} }$ & $\skyarea{GW170814_o3afin}$ & $\networkmatchedfiltersnrIMRPmed{GW170814_o3afin}_{ -\networkmatchedfiltersnrIMRPminus{GW170814_o3afin} }^{ +\networkmatchedfiltersnrIMRPplus{GW170814_o3afin} }$\\
\OLDEVENTSNAME{GW170818_o3afin} & $\totalmasssourcemed{GW170818_o3afin}_{ -\totalmasssourceminus{GW170818_o3afin} }^{ +\totalmasssourceplus{GW170818_o3afin} }$ & $\chirpmasssourcemed{GW170818_o3afin}_{ -\chirpmasssourceminus{GW170818_o3afin} }^{ +\chirpmasssourceplus{GW170818_o3afin} }$ & $\massonesourcemed{GW170818_o3afin}_{ -\massonesourceminus{GW170818_o3afin} }^{ +\massonesourceplus{GW170818_o3afin} }$ & $\masstwosourcemed{GW170818_o3afin}_{ -\masstwosourceminus{GW170818_o3afin} }^{ +\masstwosourceplus{GW170818_o3afin} }$ & $\chieffinfinityonlyprecavgmed{GW170818_o3afin}_{ -\chieffinfinityonlyprecavgminus{GW170818_o3afin} }^{ +\chieffinfinityonlyprecavgplus{GW170818_o3afin} }$ & $\luminositydistancemed{GW170818_o3afin}_{ -\luminositydistanceminus{GW170818_o3afin} }^{ +\luminositydistanceplus{GW170818_o3afin} }$ & $\redshiftmed{GW170818_o3afin}_{ -\redshiftminus{GW170818_o3afin} }^{ +\redshiftplus{GW170818_o3afin} }$ & $\finalmasssourcemed{GW170818_o3afin}_{ -\finalmasssourceminus{GW170818_o3afin} }^{ +\finalmasssourceplus{GW170818_o3afin} }$ & $\finalspinmed{GW170818_o3afin}_{ -\finalspinminus{GW170818_o3afin} }^{ +\finalspinplus{GW170818_o3afin} }$ & $\skyarea{GW170818_o3afin}$ & $\networkmatchedfiltersnrIMRPmed{GW170818_o3afin}_{ -\networkmatchedfiltersnrIMRPminus{GW170818_o3afin} }^{ +\networkmatchedfiltersnrIMRPplus{GW170818_o3afin} }$\\
\OLDEVENTSNAME{GW170823_o3afin} & $\totalmasssourcemed{GW170823_o3afin}_{ -\totalmasssourceminus{GW170823_o3afin} }^{ +\totalmasssourceplus{GW170823_o3afin} }$ & $\chirpmasssourcemed{GW170823_o3afin}_{ -\chirpmasssourceminus{GW170823_o3afin} }^{ +\chirpmasssourceplus{GW170823_o3afin} }$ & $\massonesourcemed{GW170823_o3afin}_{ -\massonesourceminus{GW170823_o3afin} }^{ +\massonesourceplus{GW170823_o3afin} }$ & $\masstwosourcemed{GW170823_o3afin}_{ -\masstwosourceminus{GW170823_o3afin} }^{ +\masstwosourceplus{GW170823_o3afin} }$ & $\chieffinfinityonlyprecavgmed{GW170823_o3afin}_{ -\chieffinfinityonlyprecavgminus{GW170823_o3afin} }^{ +\chieffinfinityonlyprecavgplus{GW170823_o3afin} }$ & $\luminositydistancemed{GW170823_o3afin}_{ -\luminositydistanceminus{GW170823_o3afin} }^{ +\luminositydistanceplus{GW170823_o3afin} }$ & $\redshiftmed{GW170823_o3afin}_{ -\redshiftminus{GW170823_o3afin} }^{ +\redshiftplus{GW170823_o3afin} }$ & $\finalmasssourcemed{GW170823_o3afin}_{ -\finalmasssourceminus{GW170823_o3afin} }^{ +\finalmasssourceplus{GW170823_o3afin} }$ & $\finalspinmed{GW170823_o3afin}_{ -\finalspinminus{GW170823_o3afin} }^{ +\finalspinplus{GW170823_o3afin} }$ & $\skyarea{GW170823_o3afin}$ & $\networkmatchedfiltersnrIMRPmed{GW170823_o3afin}_{ -\networkmatchedfiltersnrIMRPminus{GW170823_o3afin} }^{ +\networkmatchedfiltersnrIMRPplus{GW170823_o3afin} }$\\
\hline
\end{tabularx}

    }
    \caption{Median and 90\% symmetric credible intervals for the one-dimensional marginal posterior distributions on selected source parameters for the 10 \ac{BBH} events observed during the O1 and O2.
        These binaries were reported in GWTC-1~\cite{LIGOScientific:2018mvr}.
        The columns show source total mass \Mtot, chirp mass \Mc and component masses \masscomponent, dimensionless effective inspiral spin \chieff, luminosity distance \DL, redshift \redshift, final mass \Mf, final spin \chif, sky localization \skyareasymbol and the network matched-filter \ac{SNR}.
      The sky localization is the area of the 90\% credible region. 
      All quoted results are calculated from a set of posterior samples drawn with equal weight from the IMRPhenomXPHM and SEOBNRv4PHM analyses, with the exception of the \acp{SNR} that are taken from the IMRPhenomXPHM analysis alone (as \RIFT, which was used for the SEOBNRv4PHM analysis, does not output that quantity).
      Additionally, following Sec.~\ref{ss:parameter-estimation-results}, the results presented for \OLDEVENTSNAME{GW151226_o3afin}{} are taken from an analysis using the IMRPhenomXPHM model only.
      A subset of the one-dimensional posterior distributions are visualized in Fig.~\ref{fig:PE_1d_posteriors_o1o2}. 
      Two-dimensional projections of the $90\%$ credible regions in the \Mtot--\massratio and \Mc--\chieff planes are shown in grey in Fig.~\ref{fig:PE_2d_posteriors_mtot_q} and Fig.~\ref{fig:PE_2d_posteriors_mc_chieff}.
      }
    \label{tab:PE_o1o2}
\end{table*}
\end{PE_table}

\subsection{Previously reported binaries from the first half of the third observing run}
\label{app:pe_other_events_o3a}

The high-significance events from O3a are reported in Table~\ref{tab:events}.
Out of these events, \GWTCTWOEVENTS{} were included in GWTC-2~\cite{Abbott:2020niy} with its accompanying data release~\cite{GWTC2_PEdatarelease}.
Again, to ensure a self-consistent set of inferred source properties available for all \ac{CBC} events observed by Advanced LIGO and Advanced Virgo, we provide a reanalysis of these \GWTCTWOEVENTS{} events using the \ac{BBH} waveform models described in Sec.~\ref{ss:waveforms}. 
We present results combining samples from analyses using both the IMRPhenomXPHM and SEOBNRv4PHM, with the exception of \NAME{GW190413A_o3afin},
\NAME{GW190413E_o3afin},
\NAME{GW190421I_o3afin},
\NAME{GW190521B_o3afin},
\NAME{GW190602E_o3afin},
\NAME{GW190719H_o3afin},
\NAME{GW190803B_o3afin},
\NAME{GW190814H_o3afin}, 
\NAME{GW190828A_o3afin},
\NAME{GW190828B_o3afin}{} 
and \NAME{GW190929B_o3afin}{} which, as mentioned earlier in Sec.~\ref{ss:parameter-estimation-results}, were analysed using IMRPhenomXPHM only.
As also described in Sec.~\ref{ss:waveforms}, for the \ac{BNS} event \NAME{GW190425B_o3afin}, the IMRPhenomP\_NRTidal waveform model~\cite{Dietrich:2017aum, Dietrich:2018uni} was used.
The analyses of these events also used the \ac{GW} strain data described in Sec.~\ref{sec:calibration}, an additional improvement over the analyses presented in GWTC-2~\cite{Abbott:2020niy}.
For the events listed in Table~\ref{tab:detchar_events} all analyses made use of data which included glitch subtraction or a reduction in the bandwidth available for astrophysical inference.

\begin{figure*}[t]
\includegraphics[width=\textwidth]{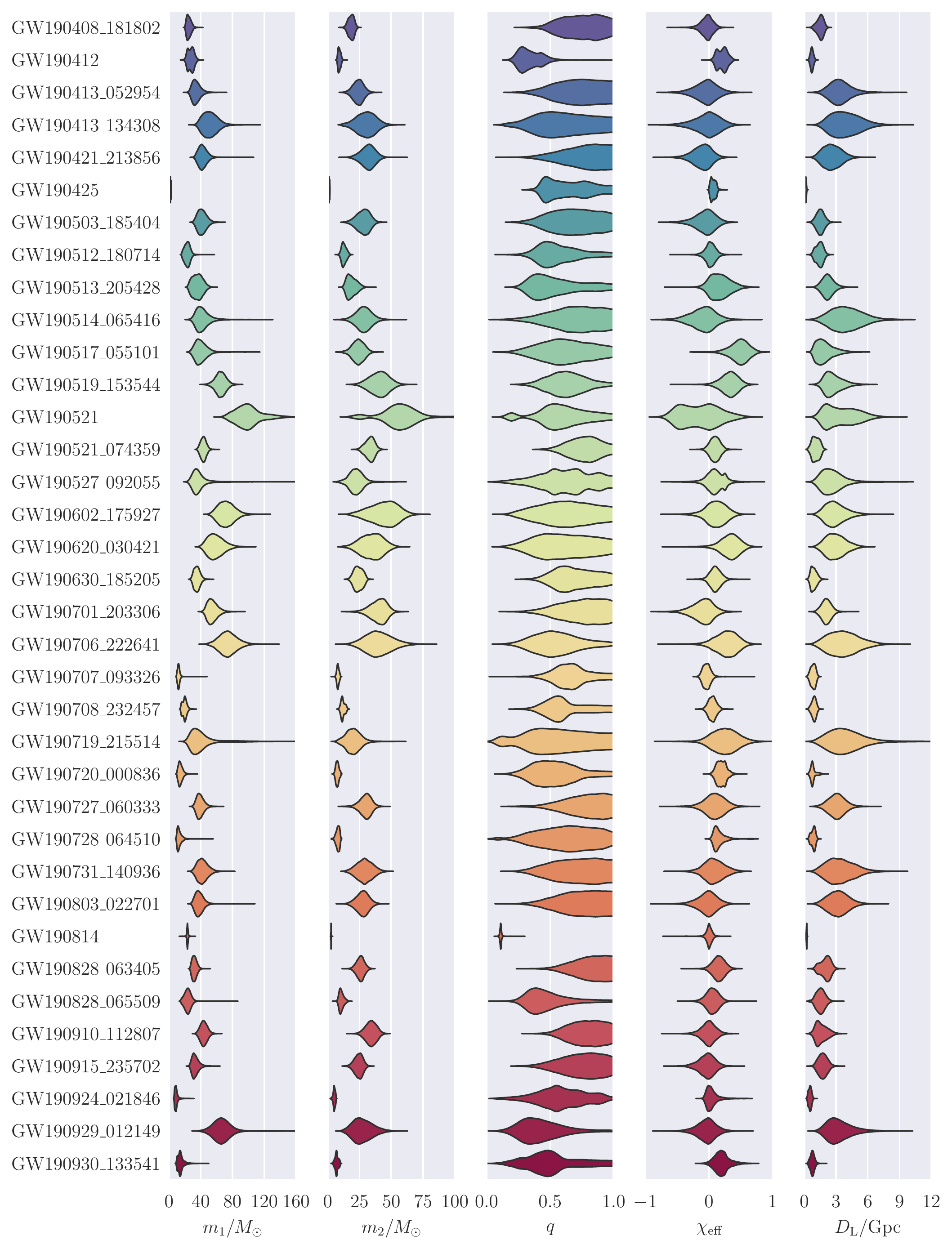}
\caption{Marginal posterior distributions on the primary mass \massone, secondary mass \masstwo, mass ratio \massratio, effective inspiral spin \chieff and luminosity distance \DL for the  \GWTCTWOEVENTS{} events from Table~\ref{tab:events} that were not shown in Fig.~\ref{fig:PE_1d_posteriors_newEvents}.
The vertical span for each region is constructed to be proportional to the one-dimensional marginal posterior at a given parameter value for the corresponding event.
The posterior distributions are also represented numerically in terms of their one-dimensional median and $90\%$ credible intervals in Table~\ref{tab:PE_OtherO3aEvents}.}
\label{fig:PE_1d_posteriors_OtherO3aEvents}
\end{figure*}

The source properties for the \GWTCTWOEVENTS{} events from O3a are reported in Table~\ref{tab:PE_OtherO3aEvents}, with a selection of the one-dimensional marginal posterior distributions shown in Fig.~\ref{fig:PE_1d_posteriors_OtherO3aEvents}.
The two-dimensional projections on the \Mtot--\massratio and \Mc--\chieff planes are shown as light-grey contours in Fig.~\ref{fig:PE_2d_posteriors_mtot_q} and Fig.~\ref{fig:PE_2d_posteriors_mc_chieff} respectively.
The full multi-dimensional posterior distributions are available as part of the public data release accompanying this paper~\cite{gwtc2p1_pe_data}, as detailed further in Sec.~\ref{sec:conclusion}.

Similar to the results presented in Sec.~\ref{app:pe_o1o2}, the vast majority of the inferred source properties for these \GWTCTWOEVENTS{} binaries are consistent with those presented in GWTC-2~\cite{Abbott:2020niy}.
For a subset of binaries, their inferred masses have changed nominally with \NAME{GW190706F_o3afin}{} as of the GWTC-2.1 analysis preferring a higher total mass whereas both \NAME{GW190521B_o3afin}{} and \NAME{GW190929B_o3afin}{} now are recovered as less massive than in GWTC-2.
Additionally, \NAME{GW190929B_o3afin}{} as recovered in GWTC-2 had a comparatively broad and multimodal posterior distribution for its primary mass.
The higher-mass mode is no longer present in the GWTC-2.1 analysis, which together with the secondary mass of \NAME{GW190929B_o3afin}{} remaining largely unchanged between the GWTC-2 and GWTC-2.1 analyses also leads to larger support for a more equal-mass binary.
\NAME{GW190708M_o3afin}{} on the other hand is now identified more predominantly with an unequal \massratio distribution as compared to the broad support, with stronger preference for equal masses, reported in GWTC-2.

Independent results with the IMRPhenomXPHM model for many of these events were previously presented in 3-OGC~\cite{Nitz:2021uxj};
other groups have also presented results with either the IMRPhenomXPHM, SEOBNRv4PHM or other precessing higher-mode models for, most prominently, the events \NAME{GW190412B_o3afin}{}~\cite{Islam:2020reh, Colleoni:2020tgc, Zevin:2020gxf, Gamba:2021ydi}
and GW190521~\cite{Nitz:2020mga, Capano:2021etf, Estelles:2021jnz, Olsen:2021qin}.
While there is general agreement for the overall inferred source properties from many of these studies, there are significant differences present between them.
These differences can however, as also explicitly stated in the studies themselves, be predominantly attributed to different prior assumptions or analysis configurations across the spread of the individual studies, in addition to the variance induced by waveform differences.
This further highlights the need for the clear and public dissemination of both the exact analysis configurations used and the generated datasets containing the source properties inferred in order to encourage reproducibility and further model comparisons, especially as more events are added to the population of observed \acp{CBC}.

\begin{PE_table}
\begin{table*}
    \centering
    {\rowcolors{1}{}{lightgray}


    }
    \caption{Median and 90\% symmetric credible intervals for the one-dimensional marginal posterior distributions on selected source parameters for the \GWTCTWOEVENTS{} events from Table~\ref{tab:events} that were not reported in Table~\ref{tab:PE_new}.
        The columns show source total mass \Mtot, chirp mass \Mc and component masses \masscomponent, dimensionless effective inspiral spin \chieff, luminosity distance \DL, redshift \redshift, final mass \Mf, final spin \chif, sky localization \skyareasymbol and the network matched-filter \ac{SNR}.
      The sky localization is the area of the 90\% credible region. 
      The results for the \acp{BBH} are calculated from a set of posterior samples drawn with equal weight from the IMRPhenomXPHM and SEOBNRv4PHM analyses, with the exception of the \acp{SNR} that are taken from the IMRPhenomXPHM analysis alone (as \RIFT, which was used for the SEOBNRv4PHM analysis, does not output that quantity).
      Additionally, following Sec.~\ref{ss:parameter-estimation-results}, the results for \NAME{GW190413A_o3afin}, \NAME{GW190413E_o3afin}, \NAME{GW190421I_o3afin}, \NAME{GW190521B_o3afin}, \NAME{GW190602E_o3afin}, \NAME{GW190719H_o3afin}, \NAME{GW190803B_o3afin}, \NAME{GW190814H_o3afin}, \NAME{GW190828A_o3afin}, \NAME{GW190828B_o3afin}{} and \NAME{GW190929B_o3afin}{} are from analyses using the IMRPhenomXPHM model only.
      For \NAME{GW190425B_o3afin}{}, we report results from the high-spin ($|\vec{\chi}_1| < 0.89$) analysis, and since the calculation of the \ac{BH} remnant properties is only valid for \ac{BBH} model input those properties are excluded for this \ac{BNS} signal.
      A subset of the one-dimensional posterior distributions are visualized in Fig.~\ref{fig:PE_1d_posteriors_OtherO3aEvents}. 
      Two-dimensional projections of the $90\%$ credible regions in the \Mtot--\massratio and \Mc--\chieff planes are shown in grey in Fig.~\ref{fig:PE_2d_posteriors_mtot_q} and Fig.~\ref{fig:PE_2d_posteriors_mc_chieff}.
      }
    \label{tab:PE_OtherO3aEvents}
\end{table*}
\end{PE_table}

\clearpage

\bibliography{o3catalog}

\end{document}